\documentclass[usenatbib,fleqn]{mnras}
\usepackage{graphicx,hyperref,amsmath,amssymb,xspace,longtable}
\usepackage[dvipsnames]{xcolor}
\usepackage{etoolbox}
\makeatletter
\patchcmd\@combinedblfloats{\box\@outputbox}{\unvbox\@outputbox}{}{\errmessage{\noexpand patch failed}}
\makeatother

\title[Globular Clusters]
{Proper motions and dynamics of the Milky Way globular cluster system from \textit{Gaia} DR2}
\author[E. Vasiliev]{Eugene Vasiliev$^{1,2}$\thanks{E-mail: eugvas@lpi.ru}\\
$^1$Institute of Astronomy, Madingley road, Cambridge, CB3 0HA, UK\\
$^2$Lebedev Physical Institute, Leninsky prospekt 53, Moscow, 119991, Russia}

\newcommand{\Gaia}{\textit{Gaia}\xspace}
\newcommand{\HST}{\textit{HST}\xspace}
\newcommand{\kms}{km\:s$^{-1}$\xspace}
\newcommand{\masyr}{mas\:yr$^{-1}$\xspace}
\defcitealias{Helmi2018}{H18}

\begin{document}
\date{Accepted 2019 January 11. Received 2018 December 4; in original form 2018 August 6}
\maketitle

\begin{abstract}
We use \Gaia Data Release 2 to determine the mean proper motions for 150 Milky Way globular clusters (almost the entire known population), with a typical uncertainty of 0.05~\masyr limited mainly by systematic errors.
Combining them with distance and line-of-sight velocity measurements from the literature, we analyze the distribution of globular clusters in the 6d phase space, using both position/velocity and action/angle coordinates. The population of clusters in the central 10 kpc has a mean rotational velocity reaching $50-80$~\kms, and a nearly isotropic velocity dispersion $100-120$~\kms, while in the outer galaxy the cluster orbits are strongly radially anisotropic. We confirm a concentration of clusters at high radial action in the outer region of the Galaxy.
Finally, we explore a range of equilibrium distribution function-based models for the entire globular cluster system, and the information they provide about the potential of the Milky Way. The dynamics of clusters is best described by models with the circular velocity between 10 and 50~kpc staying in the range $210-240$~\kms.
\end{abstract}


\section{Introduction}

With the second data release (DR2) of the \Gaia mission in April 2018 \citep{Brown2018}, a new era in dynamical astronomy has begun. Astrometric data -- parallaxes and proper motions (PM) -- are now available for $>10^9$ stars, with a typical parallax uncertainty ranging from 0.1~mas at the bright end to 1~mas for fainter stars down to magnitude $G=21$, and a similar PM uncertainty in \masyr. A few million stars also have measured line-of-sight velocities, mostly within heliocentric distance $1-3$~kpc, thus providing a full six-dimensional phase-space view on the distribution and kinematics of stars in the Solar neighbourhood. 

For more distant stars, the uncertainties in individual parallax and PM measurements are not negligible and translate into larger errors in the distance and the transverse velocity. However, mean PM of compact bound stellar systems such as globular clusters or satellite galaxies may still be computed with high precision by averaging over many stars belonging to each object. These data, together with the distances and line-of-sight velocities obtained by other methods, can be used to study the orbits of clusters and to put constraints on the total potential of the Milky Way.

Ground-based PM measurements are available for a large fraction of the Milky Way globular clusters ($60-120$ in various catalogues), but as will be shown later, their accuracy is inferior to the space-based ones, due to intrinsic difficulties in measuring stellar positions accurately in the presence of atmospheric aberrations. \textit{Hubble} space telescope (\HST) has been used to determine the PM of more that two dozed clusters (\citealt{Sohn2018} and other studies). In both cases, the internal (relative) PM are more easily obtained than the absolute PM; they are suitable for membership filtering, but not for the determination of cluster orbits. To measure the absolute PM of a cluster, one needs to anchor the relative PM to some objects with known absolute PM -- for instance, distant quasars or stars from the \textit{Hipparcos} catalogue \citep{Perryman1997}; however, these objects are rare, and only a few may be present in each field of view.
The first data release of \Gaia contained a relatively small number of absolute astrometric measurements for stars previously observed by \textit{Hipparcos}, and has been used to determine the mean PM of several closest globular clusters \citep{Watkins2017}; however, with only a few stars per cluster, the uncertainties are large.
On the other hand, \Gaia DR2 provides the PM with much higher precision for a vastly larger number of stars over the entire sky in the absolute reference frame, although not without systematic errors, which will be discussed further below.

\Gaia collaboration (\citealt{Helmi2018}, hereafter H18) determined the mean PM for 75 galactic globular clusters -- roughly a half of the known population listed in the \citet{Harris1996,Harris2010} catalogue, mostly lying within a heliocentric distance of 15~kpc and having a sufficient number of bright stars with accurate astrometric measurements. However, this does not imply that the remaining ones are unsuitable for analysis.
More recently (after the submission of this paper), \citet{Baumgardt2019} published independent measurements of mean PM for almost all Milky Way globular clusters based on \Gaia DR2 data. 

In this paper, we use the \Gaia data to measure the mean PM for 150 globular clusters -- twice larger number than provided by \citetalias{Helmi2018}. One novel feature compared to the two other \Gaia-based studies is the treatment of spatially correlated systematic errors, using the approach presented in a companion paper \citep{Vasiliev2018b}.
Section~\ref{sec:pm} presents this new catalogue and compares it with other existing measurements, both ground- and space-based. 
We then combine the PM with the distances and line-of-sight velocity measurements available in the literature, to obtain the full 6d phase-space distribution of globular clusters. In section~\ref{sec:kin}, we analyze various trends in this distribution (e.g., velocity dispersions, anisotropy and rotation) in the position/velocity phase space. We also examine this distribution in the action/angle space, which makes easier the identification of various sub-populations with different kinematics, and also hints at possible selection biases.
Then in Section~\ref{sec:dyn} we model the population of globular clusters by an equilibrium distribution function (DF) in the action space. We determine the range of parameters of both the DF and the gravitational potential which are consistent with the observed dynamics.
Section~\ref{sec:summary} summarizes our results. \hyperref[sec:gmm]{Technical details} about our PM measurement procedure, \hyperref[sec:notes]{remarks} about individual clusters, and the \hyperref[tab:pm]{table} containing PM and other parameters, are deferred to the Appendix, while the supplementary material contains the plots of distribution of cluster members and field stars in the sky plane, PM space, and colour--magnitude diagrams for all 150 clusters.

\section{Proper motions}  \label{sec:pm}

\subsection{Data}  \label{sec:pm_data}

We use the following selection criteria on the input sample:

\begin{itemize}
\item Select all stars with full astrometric data (position, parallax $\varpi$ and PM $\mu_\alpha, \mu_\delta$) within a certain angular distance $R_\mathrm{max}$ from the cluster centre. The distance is typically chosen to be several times larger than the half-light radius (a few arcminutes for most clusters), to ensure that a sufficient number of likely non-member stars are found in the region. These are needed for our mixture modelling approach: in the absense of a reliable field population, we would be forced to attribute each source to the cluster, even if its PM is grossly inconsistent with the mean value.
\item Retain only sources with measured parallax $\varpi < 1/D + 3\,\epsilon_\varpi$, where $D$ is the distance to the cluster, and $\epsilon_\varpi$ is the quoted parallax uncertainty. This removes most nearby stars which lie on the line of sight.
\item Remove sources with \texttt{astrometric\_excess\_noise} parameter greater than 1, or with renormalized unit weight error exceeding 1.2 (this parameter is defined in \citealt{Lindegren2018TN}). Larger values of these parameters indicate that the astrometric solution might be unreliable (in particular, the formal uncertainties might be underestimated), which happens mostly for unresolved binaries or faints sources in crowded fields \citep{Lindegren2018}.
\item For all but a few clusters, we also eliminated stars with significant colour excess: \texttt{phot\_bp\_rp\_excess\_factor}${}>1.3 + 0.06\;\mbox{\texttt{bp\_rp}}^2$, as suggested by the above paper. This mostly affects fainter stars with strong contamination by nearby brighter sources in denser central regions of clusters, and reduces the number of sources in each cluster by a factor between 2 and 10. \Gaia DR2 does not have any special treatment for blended or contaminated sources (although such treatment should be implemented in future data releases, see Sections 2 and 3 in \citealt{Pancino2017} for a thorough discussion), and their quoted PM uncertainties are also likely underestimated. Even though this cut dramatically reduces the number of available stars, it has only a minor effect on the precision of mean PM measurements (but should improve their reliability), because the omitted sources typically have large uncertainties anyway. However, we decided not to apply this cut in several clusters that would otherwise retain only a few stars: AM~1, Crater, Pal~3, Pal~4, Terzan~6.
\end{itemize}

We chose not to use the \Gaia BP and RP colour measurements in the fitting procedure.
The current stellar population models, although sophisticated, still do not describe well the population of blue horizontal branch stars. Likewise, the ages, metallicities and extinction coefficients are not well known for some of the clusters, and sometimes the values quoted in the catalogue clearly do not match the reddening-corrected colour--magnitude diagram (CMD) seen in the \Gaia data. These uncertainties mean that filtering the CMD based on the proximity of observed magnitudes to the theoretical isochrone curve would leave out some, or sometimes even most sources.
Instead, we use the \Gaia colours as an independent consistency check, inspecting the CMD of stars classified as likely cluster members based on their PM and position. In most cases these stars align well with the theoretical isochrone curve, or at least clearly stand out from the field population. Exceptions mostly occur in heavily extincted clusters, where the colour information would not help anyway.

\subsection{Method}  \label{sec:pm_method}

The distribution of sources in the PM space%
\footnote{Throughout the paper, we denote $\mu_\alpha\equiv \mathrm d\alpha/\mathrm d t\;\cos\delta$.}
$\{\mu_\alpha, \mu_\delta\}$ typically has a well-defined clump corresponding to the cluster members, and a broader distribution of field stars. Quite often, though, there is no clear separation between the two, and we employ a probabilistic Gaussian mixture model to select member stars and to infer the intrinsic (error-deconvolved) parameters of the distributions of both member and non-member stars. In doing so, we use a spatially-dependent prior for the membership probability.
A detailed description of the procedure is presented in \hyperref[sec:gmm]{Appendix A}.

Some of the most massive or nearby clusters contain tens of thousands stars in the \Gaia catalogue, allowing one to measure not only the average PM for the entire cluster, but also, to some degree, its internal kinematics. The radial profiles of sky-plane rotation have been measured for some clusters by \citet{Bianchini2018}, and the PM dispersion profiles -- by \citet{Baumgardt2019}.
In the present study, we use a simplified description of the internal kinematics, ignoring the rotation and assuming that the PM dispersion is isotropic and follows a particular radial profile with an amplitude freely adjusted during the fit.
The companion paper \citep{Vasiliev2018b} uses the same membership determination procedure, but a more flexible description of the internal kinematics, to measure the PM dispersion and rotation profiles, taking into account the systematic errors.

\begin{figure}
\includegraphics{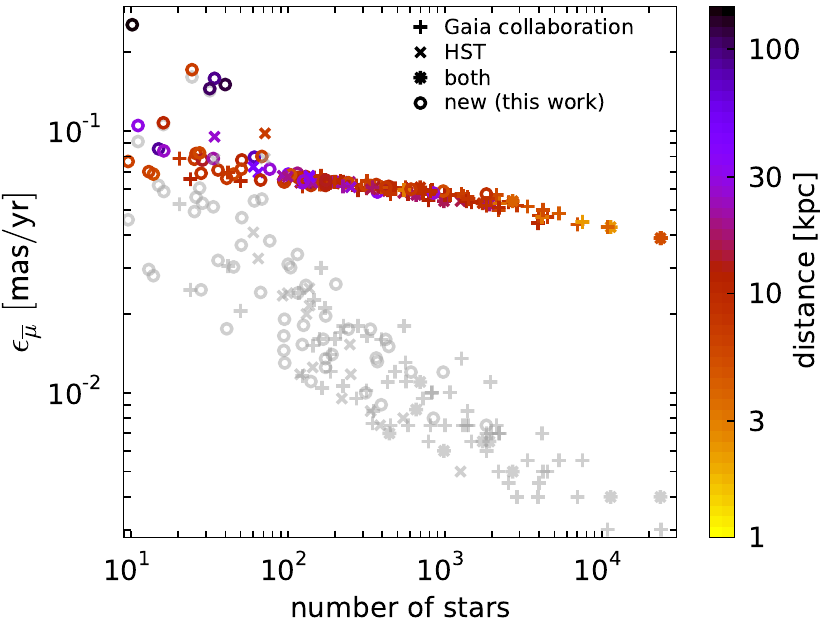}
\caption{
Uncertainty of mean PM (average of $\epsilon_{\overline{\mu_\alpha}}, \epsilon_{\overline{\mu_\delta}}$) as a function of the number of cluster members $N_\star$ remaining after applying all quality filters.
Gray symbols show the statistical uncertainty alone, which broadly follows the $N^{-1/2}$ trend expected for Poisson noise. Coloured symbols show the combined statistical and systematic uncertainty, which is dominated by the latter for nearly all clusters with more than a few dozen members. The systematic uncertainty is $\sim0.07$~\masyr for very compact clusters and decreases with the angular size of an object, hence is lower for the closest and richest clusters. Different symbols denote the new clusters in this work (circles) or clusters with previously determined PM (from \citetalias{Helmi2018}, \HST, or both). Points are coloured according to the distance to the clusters as given in the Harris catalogue.
\vspace*{-1mm}
} \label{fig:pm_uncertainty}
\end{figure}

The fitting procedure produces the estimate of the mean PM $\{\overline{\mu_\alpha}, \overline{\mu_\delta}\}$ and its associated uncertainty covariance matrix, which may equivalently be represented by two standard deviations $\{\epsilon_{\overline{\mu_\alpha}}, \epsilon_{\overline{\mu_\delta}}\}$ and the correlation coefficient $r_{\overline{\mu_\alpha}\,\overline{\mu_\delta}}$.
In doing so, we may consider only the statistical uncertainties, which decrease from few$\times10^{-1}$ to few$\times10^{-3}$~\masyr as the number of member stars increases (Figure~\ref{fig:pm_uncertainty}, gray symbols). However, for the majority of clusters, the dominant source of uncertainty are the spatially correlated systematic errors. Our final uncertainty estimates take these systematic errors into account (coloured symbols), as sketched in \citet{Lindegren2018TN} and explained in more detail in the companion paper and in the \hyperref[sec:gmm]{Appendix}.

\subsection{Results}  \label{sec:pm_results}

\begin{figure*}
\includegraphics{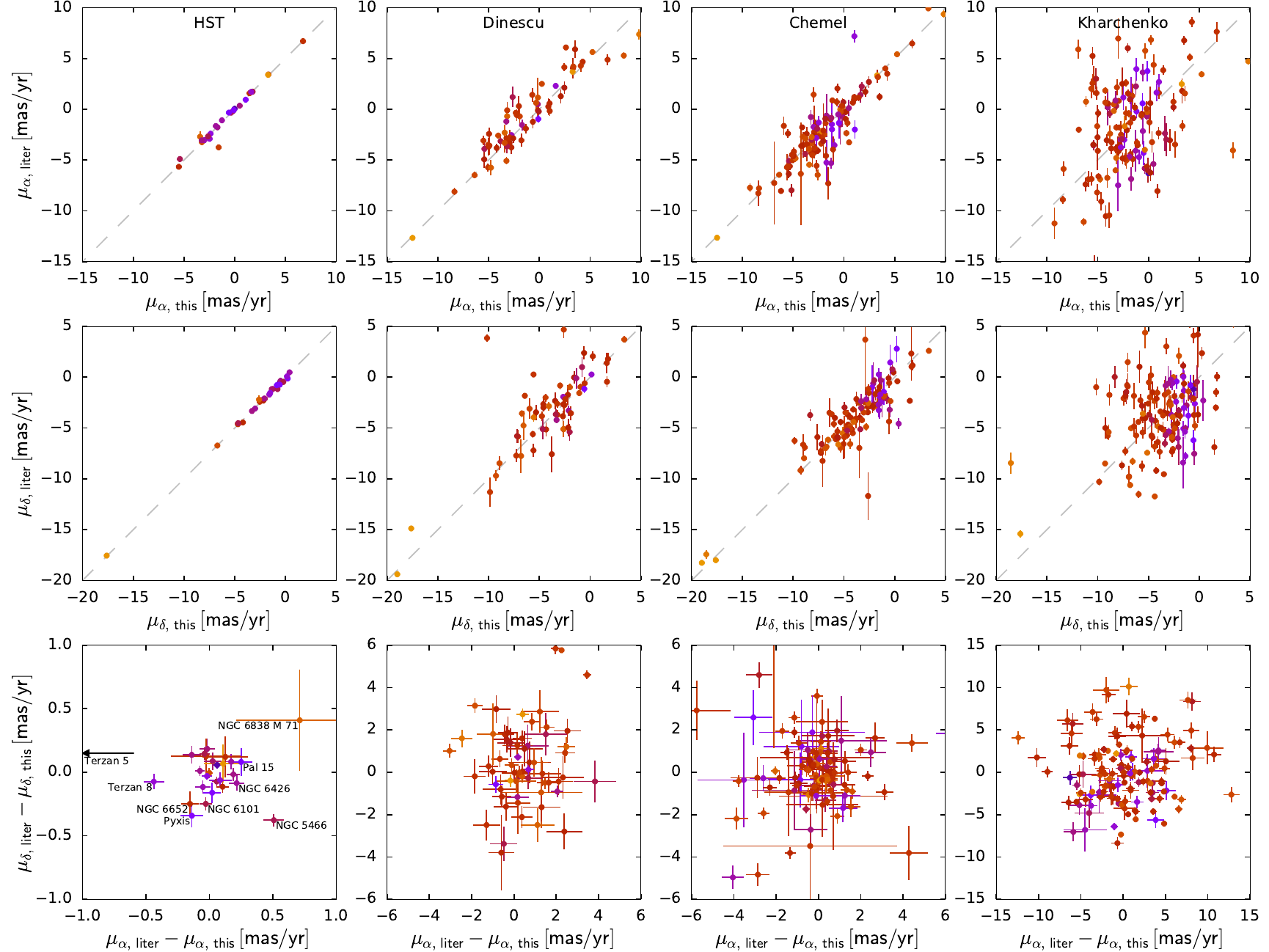}
\caption{
Comparison of PM measurements between different catalogues and this work. First column: \HST (several studies, primarily \citealt{Sohn2018}). Second column: \citet{Dinescu1999} and several later papers. Third column: \citet{Chemel2018}. Last column: \citet{Kharchenko2013}.
Top row: comparison between $\overline{\mu_\alpha}$ from the literature and from this work; middle row: same for $\overline{\mu_\delta}$; bottom row: difference between the literature measurements and this work (two PM components plotted against each other, note the different scale for each column). Colours denote the distance to the clusters (same colour scheme as in the previous plot).
\vspace*{-1mm}
} \label{fig:pm_comparison}
\end{figure*}

Table~\ref{tab:pm} lists the derived PM, together with several other parameters, for all 150 clusters in this study. We were not able to measure it only for a few clusters from the \citet{Harris2010} catalogue -- either very distant ones with less than a few stars on the red giant branch detectable by \Gaia (Ko~1, Ko~2 and AM~4), or heavily extincted clusters in the disc plane that were discovered in infrared surveys (2MASS-GC01, 2MASS-GC02, GLIMPSE1, GLIMPSE2, Liller 1 and UKS1). We added two more clusters not originally present in the Harris catalogue: Crater (Laevens 1), discovered by \citet{Belokurov2014} and \citet{Laevens2014}, and FSR~1716, recently confirmed as an old globular cluster by \citet{Minniti2017}.

We compared our PM with several existing catalogues (Figure~\ref{fig:pm_comparison}). 
Half of our sample has already been analyzed by the \Gaia collaboration, using somewhat different approach for membership determination \citepalias{Helmi2018}. For the vast majority of these clusters, our measurements agree with the ones from that paper to better than 0.05~\masyr. The two clusters with the largest difference ($\sim 0.15-0.25$~\masyr) are NGC~6626 (M~28) and NGC~6266 (M~62), which are both located in rather dense regions on the sky, and the coordinates of cluster centres used by \citepalias{Helmi2018} are offset by some 1.5 arcmin from the true values. A visual inspection of PM diagrams confirms the validity of mean values obtained by our Gaussian mixture procedure.

\citet{Baumgardt2019} independently measured PM for all clusters in our catalogue, also using \Gaia data but a different approach for membership determination. Our measurements agree to better than 0.05~\masyr, except for a few clusters: FSR~1735 (0.75~\masyr), Ton~2 (0.25), Terzan~1 (0.2), AM~1 (0.4), Pal~2 (0.35), Eridanus (0.3). The first three are located in heavily extincted regions close to the disc plane, and the last three are fairly distant clusters with only faint stars. Overall, the agreement between the three catalogues based on \Gaia data is excellent.

Absolute PM have been obtained with \HST for 27 clusters in our sample: 20 distant clusters in \citet{Sohn2018}, NGC~362 in \citet{Libralato2018b}, NGC~5139 ($\omega$~Cen) in \citet{Libralato2018a}, NGC~6397 in \citet{Milone2006}, NGC~6652 in \citet{Sohn2015}, NGC~6681 (M~70) in \citet{Massari2013}, NGC~6838 (M~71) in \citet{Cadelano2017}, and Terzan~5 in \citet{Massari2015}. Of these, only Terzan~5 differs from our value by about 2~\masyr in $\mu_\alpha$, and the remaining measurements agree to better than 1~\masyr. Terzan~5 lies in a dense field within the Galactic bulge, and in order to establish the absolute PM, \citet{Massari2015} used bulge stars as a reference, not distant quasars as typically done for other \HST clusters, which may explain the offset. The good agreement between two space-based missions is encouraging, and lends further credibility to the \Gaia results.

On the other hand, a comparison with ground-based PM measurements demonstrates that there are often significant differences, up to a few \masyr. 
The catalogues of \citet{CasettiDinescu2007,CasettiDinescu2010,CasettiDinescu2013} and \citet{Chemel2018} still demonstrate a rather good correlation with the \Gaia measurements (second and third columns in Figure~\ref{fig:pm_comparison}), and an independent Pal~5 measurement by \citet{Fritz2015} is only $\sim0.6$~\masyr off. However, the catalogue of \citet{Kharchenko2013} bears almost no resemblance to the space-based measurements, with differences exceeding 5~\masyr (last column). Unfortunately, it is the default source of PM values listed in SIMBAD.

We performed several internal validation procedures, comparing the PM determined from different subsets of stars, or under different assumptions about the internal structure and kinematics (such as more flexible models from the companion paper). For instance, we removed stars with magnitudes $G>20$ as in \citetalias{Helmi2018}, or skipped the cut in \texttt{phot\_bp\_rp\_excess\_factor}, or changed the maximum angular distance from the cluster centre. The results were robust to these changes, with differences typically far smaller than the quoted uncertainty. We also verified that the mean PM values obtained with and without account for systematic errors are compatible (of course, the uncertainties are far smaller in the latter case, but they do not tell the whole story). We therefore believe that the mean cluster PM provided in this study are not only precise (corresponding to transverse velocity errors less than 10~\kms for most clusters except a few distant ones), but also accurate insofar as the \Gaia data themselves.

\section{Kinematics of the globular cluster system}  \label{sec:kin}

\subsection{Velocity distribution}  \label{sec:kin_velocity}

\begin{figure}
\includegraphics{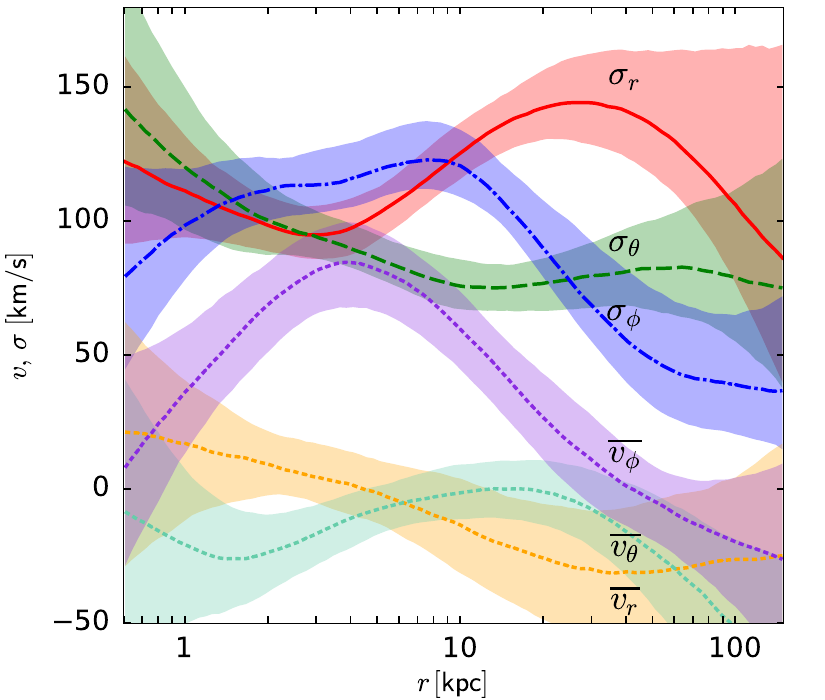}
\caption{
Kinematics of the Milky Way globular cluster system.
Plotted are the velocity dispersions $\sigma$ in spherical coordinates and the mean velocities $\overline v$ as functions of spherical radius, with 68\% uncertainty bands shaded in respective colour. The significantly non-zero azimuthal velocity $\overline{v_\phi}$ indicates the overall prograde rotation of the ensemble of clusters, while $\overline{v_r}$ and $\overline{v_\theta}$ are consistent with zero, as expected in a dynamical equilibrium. At large radii, the distribution of clusters in velocity space becomes radially anisotropic ($\sigma_r$ is significantly higher than the other two components of velocity dispersion). Clusters associated with the Sgr galaxy are excluded (they would have inflated $\sigma_\theta$ and shifted $\overline{v_\theta}$ around $r=30$~kpc).
} \label{fig:veldisp}
\end{figure}

Having precise PM for most of Milky Way globular clusters, we are ready to analyze their distribution in the full 6d phase space, combining the \Gaia PM with the distances and line-of-sight velocities from the literature. 
The velocities are taken from the catalogue of \citet{Baumgardt2019}, and have typical uncertainties of only $1-2$~\kms.
We use the distances from the \citet{Harris2010} catalogue, and assume an error of 0.1 in distance modulus, corresponding to a relative error of 0.046 in the distance. This is probably a rather optimistic choice: for some clusters, the variation in several independent distance estimates from the literature could exceed 10\%. Even so, the distance appears to be the largest source of uncertainty for the majority of clusters, except a few outermost ones with large PM errors. Unfortunately, \Gaia parallaxes currently cannot be used to improve the distance uncertainty, due to a substantial and spatially varying systematic offset $\epsilon_\varpi\sim -0.03\pm0.05$~mas (\citealt{Lindegren2018}, \citetalias{Helmi2018}). As most clusters are located at distances $\gtrsim 5$~kpc, corresponding to the true parallax $\lesssim 0.2$~mas, this unknown systematic offset has a very substantial impact on the measurement. Only in some special cases, when a sufficient number of suitable background reference objects are available, this offset could be measured and subtracted \citep[e.g.,][]{Chen2018}. By the end of the mission, the parallax and PM uncertainties are expected to reach 1\% level for clusters within 15~kpc \citep{Pancino2017}.

The uncertainties in the data are propagated to the subsequent models as follows: for each cluster, we draw $N_\mathrm{samples}=1000$ Monte Carlo samples from the distribution of errors in PM (taking into account their covariance), distance and line-of-sight velocity. We then convert these samples to Galactic coordinates, assuming the solar position within the Galaxy at $R_0=8.2\pm 0.1$~kpc, $z_0=25\pm 5$~pc, and the solar velocity vector $\{v_R,v_\phi,v_z\}=\{-10\pm 1, 248\pm 3, 7\pm 1\}$~\kms \citep{BlandHawthorn2016}.
The distribution of Monte Carlo samples for each cluster is often significantly elongated in position/velocity space, and sometimes does not resemble a simple Gaussian at all. By using the full ensemble of samples for each cluster, rather than just their covariance matrix, we ensure a correct convolution with observational errors throughout our modelling procedure. 

We excluded some clusters from the subsequent analysis for various reasons.
The few clusters for which we were unable to determine the mean PM could be included into the analysis by assigning them a large formal uncertainty (as in \citealt{Eadie2017} or \citealt{Binney2017}), however, they add very little information, and omitting them makes no difference to the modelling results.
A few clusters listed below are associated with the Sagittarius stream, and hence do not constitute independent samples; we omitted them from the computation of velocity dispersion profiles and the dynamical modelling of Section~\ref{sec:dyn}, but kept them in the phase-space plots.

To measure the density and 1d velocity distributions, and the radial profiles of mean velocity and its dispersion, we use a non-parametric penalized spline fitting approach, described in \hyperref[sec:nonparafit]{Appendix B}, which produces smooth estimates of these quantities and associated confidence intervals.

\begin{figure*}
\includegraphics{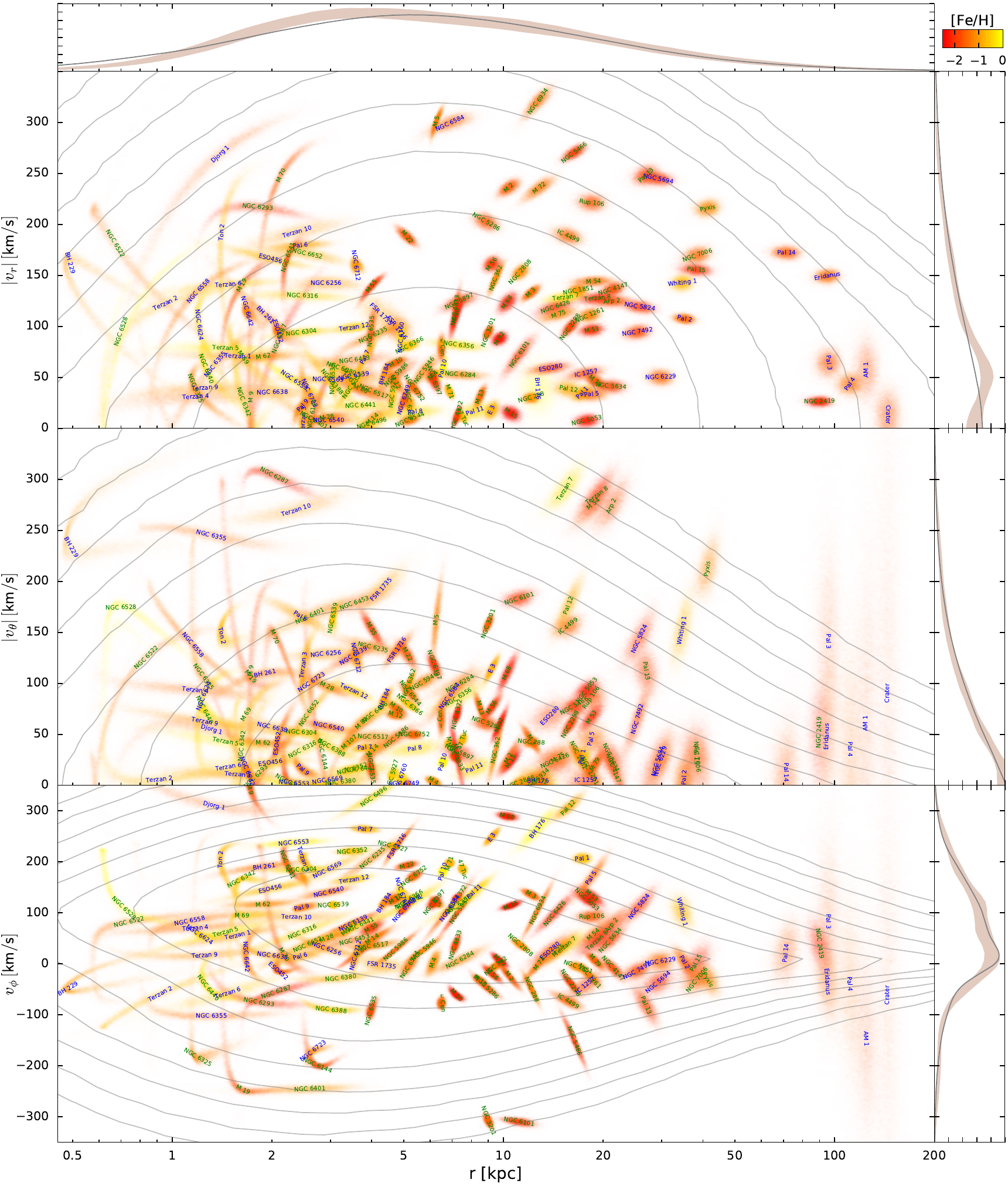}
\caption{The onion diagram showing the distribution of globular clusters in galactocentric radius and galactocentric velocity in spherical coordinates (one panel per velocity component).
Each cluster is shown as a cloud of Monte Carlo samples representing the uncertainties in its position and velocity, coloured according to the metallicity as quoted in the \citet{Harris2010} catalogue. Clusters with previously existing space-based PM measurements (from the \Gaia collaboration or from various \HST studies) have their names shown in green; the new measurements in this paper -- in blue.
Gray contour lines show the projection of the model DF (Section~\ref{sec:dyn_df}) into the coordinate/velocity space, averaged over many acceptable models in the MCMC chain, all computed in the \citet{McMillan2017} potential. Contours are spaced logarithmically, each successive line corresponding to a twice lower probability than the previous one. 
Side panels show the 1d projections of the distributions: shaded bands -- actual data with uncertainties, solid gray -- model.
} \label{fig:veldistr}
\end{figure*}

Figure~\ref{fig:veldisp} shows the radial profiles of velocity dispersions and mean velocities of the entire population of clusters excluding the ones associated with the Sagittarius stream (see below). Several trends are apparent: 
\begin{itemize} 
\item Clusters within $r\lesssim 10$~kpc move mostly on prograde orbits, with mean velocity reaching $70-80$~\kms at $4-7$~kpc.
\item The velocity dispersion tensor is close to isotropic with $\sigma\simeq 100$~\kms in the inner Galaxy ($r\lesssim 10$~kpc), and becomes radially anisotropic further out. This follows similar patterns observed in the population of halo stars \citep[e.g.,][]{Deason2012}, and will be discussed in more detail in the next section.
\item Mean radial and polar velocities are consistent with zero within uncertainties, as expected in a steady state. However, if we include the Sagittarius stream clusters, the mean polar velocity becomes somewhat negative around $\sim20-30$~kpc, and its dispersion is increased.
\end{itemize}

A more detailed illustration is presented in Figure~\ref{fig:veldistr}, which shows the distribution of clusters in 2d spaces of radius and velocity (separately for each velocity component), coloured by metallicity. It highlights various features in the cluster kinematics, which have been known previously:
\begin{itemize}
\item Metal-rich clusters (so-called disc population, \citealt{Zinn1985}) are predominantly found inside 10~kpc and on preferentially corotating orbits, with typical value of azimuthal velocity $v_\phi \gtrsim 100$~\kms. These are responsible for the peak in $\overline{v_\phi}$ in the inner Galaxy, seen in the previous plot.
\item There are several clusters on significantly retrograde orbits (e.g., NGC~3201 and NGC~6101), which may have been accreted from an infalling satellite galaxy. The now available accurate kinematic information complements the age and metallicity measurements and will assist in distinguishing accreted clusters from those formed in situ \citep[e.g.,][]{Forbes2010, Kruijssen2018}.
\item Some kinematic structures are readily identified, such as the four clusters in the core of the Sagittarius stream, which all have similar radii ($\sim 20$~kpc) and high $|v_\theta|\gtrsim 250$~\kms: NGC~6715 (M~54), Terzan~7, Terzan~8 and Arp~2. However, this is not at all apparent for two other clusters that are on very similar orbits (Pal~12 and Whiting~1, with the PM of the latter first measured in this study); to see their relation to the stream, one would need to compute the orbits or the integrals of motion. This requires the knowledge of the potential.
\end{itemize}

\begin{figure*}
\includegraphics{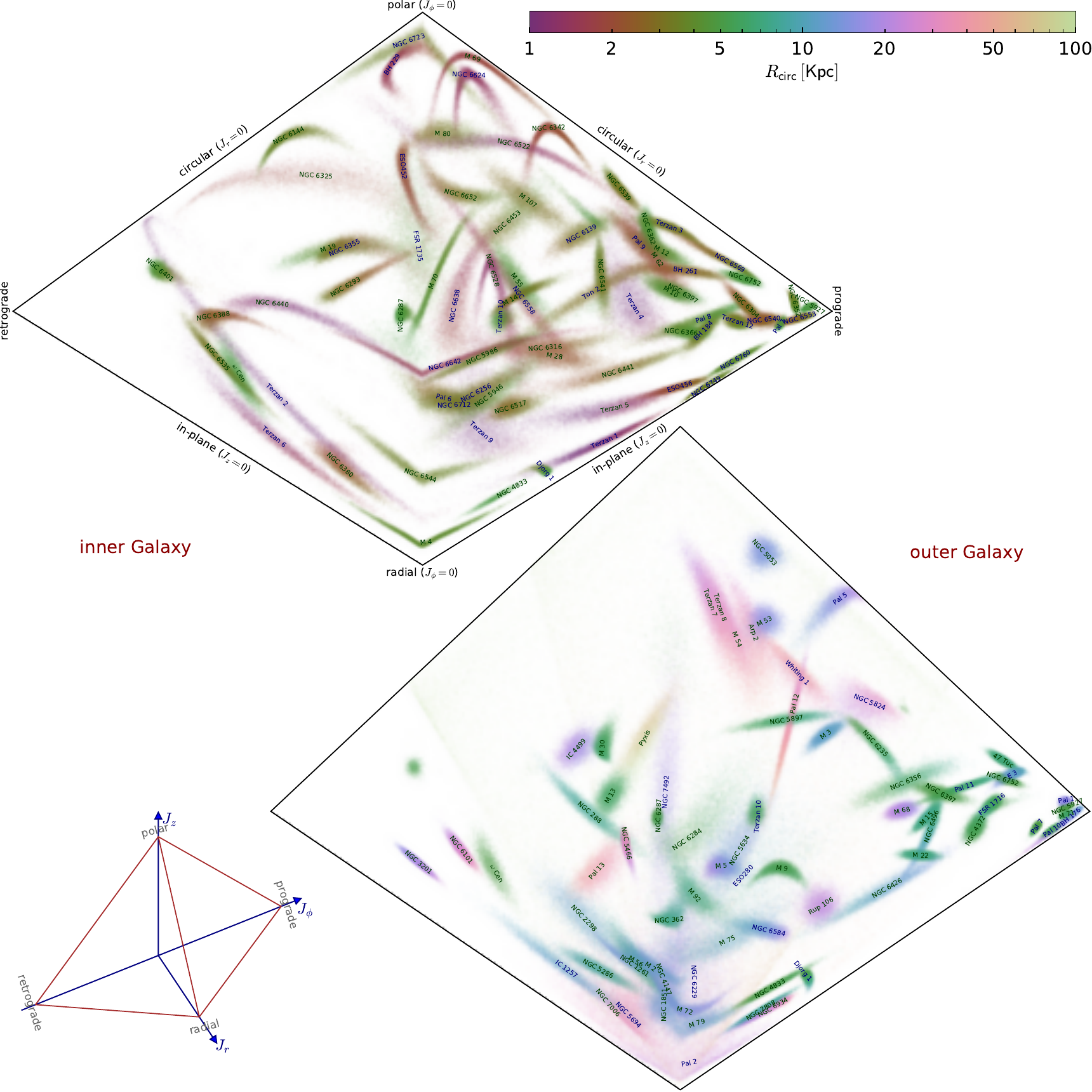}
\caption{The action-space map of Milky Way globular clusters in the \citet{McMillan2017} potential. \protect\\
Each cluster is shown as a cloud of Monte Carlo samples representing the uncertainties in its actions, coloured by the radius of a circular orbit $R_\mathrm{circ}(E)$ with the given total energy. Clusters in the inner Galaxy, with $R_\mathrm{circ}<5$~kpc, are plotted on the top left panel, and the remaining ones -- on the bottom right panel; this choice splits the sample into roughly equal halves.\protect\\
The horizontal coordinate is $\eta \equiv J_\phi / L_\mathrm{circ}(E)$ -- the normalized $z$-component of angular momentum, ranging from $-1$ in the left corner to $+1$ in the right corner. 
The vertical coordinate $\zeta \equiv (J_z-J_r) / L_\mathrm{circ}(E)$ quantifies the relative weight of radial vs.\ vertical motion; it range depends on the energy and is narrower than $-1$ to $+1$ because the maximum possible values of $J_r,J_z$ at a fixed energy are smaller than $L_\mathrm{circ}$.
The upper boundary at a given $\eta$ is occupied by circular or thin-shell orbits ($J_r=0$), and the lower -- by radial orbits.
The horizontal axis roughly corresponds to the orbit inclination, and the vertical to the eccentricity (0 along the top boundary, 1 in the bottom corner), which are not true integrals of motion.
Bottom-left inset shows the approximate constant-energy surface in the 3d action space (cf.\ Figure~1 in \citealt{Binney2014}); the four corners of this rhombus correspond to the four extrema in the $\eta,\zeta$ plane.\protect\\
As in the previous figure, clusters with previously measured PM are shown by green labels, and the new ones -- by blue; a few clusters with too large uncertainties are not labelled. 
Objects that are both close in these 2d coordinates and in colour may belong to a common dynamical structure, such as the group of several clusters associated with the Sagittarius stream in the top of the right panel, coloured in pink.
} \label{fig:actions}
\end{figure*}

\begin{figure}
\includegraphics{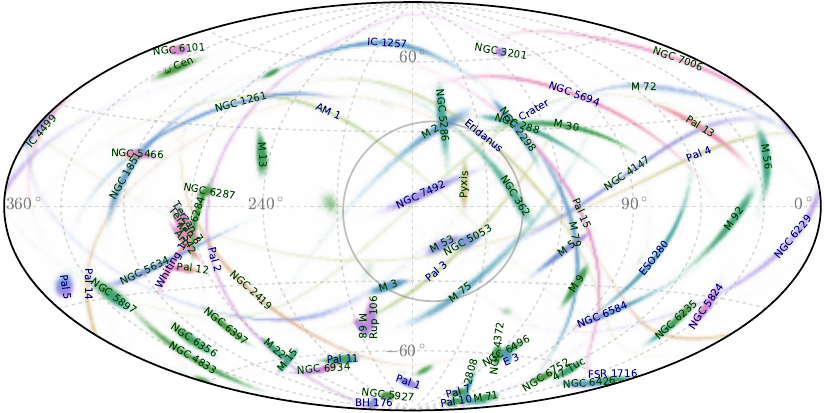}
\caption{
Distribution of orbital poles (directions of angular momentum) in the galactocentric coordinates. Shown are clusters from the outer galaxy, with $R_\mathrm{circ}\ge 5$~kpc, coloured by energy as in the the bottom panel of the previous figure. The gray circle near the centre of the image denotes the location of the ``Vast Plane of Satellites'' (VPOS) from \citet{Pawlowski2012}, same as in Figure~3 of \citet{Fritz2018}. There is no obvious clumping of orbital poles in this direction, although the orbital plane of the Sagittarius stream and associated clusters is clearly seen some $90^\circ$ to the left of it, coloured in pink.
} \label{fig:poles}
\end{figure}

\begin{figure}
\includegraphics{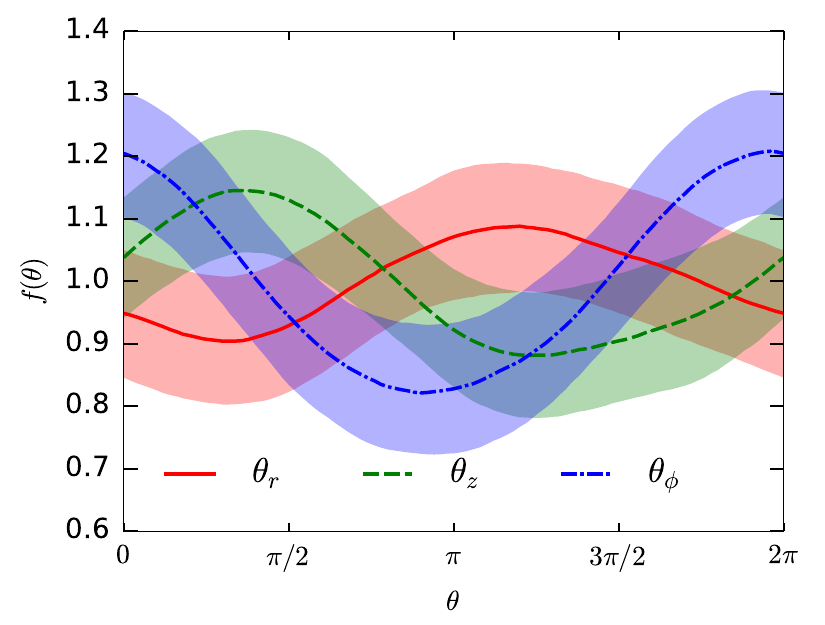}
\caption{
Distribution of observed globular clusters in phase angles (canonically conjugate variables to the actions), evaluated in the \citet{McMillan2017} potential.
Only the distribution in $\theta_\phi$ is significantly non-uniform, possibly indicating incompleteness of the catalogue at the more distant side of the Galaxy (behind the bulge/disc).
} \label{fig:angles}
\end{figure}

\subsection{Action-angle space distribution}  \label{sec:kin_actions}

To get further insight into the observed distribution of globular clusters, we convert their positions and velocities into actions and angles. For this, one needs to specify the potential, and we use the best-fit model of \citet{McMillan2017} as our default choice. The actions and angles are estimated using the St\"ackel fudge approach (\citealt{Binney2012}, see also \citealt{Sanders2016} for an overview of action computation methods), as implemented in the \textsc{Agama} galaxy modelling framework \citep{Vasiliev2019}.

Figure~\ref{fig:actions} illustrates the distribution of clusters in the action space: two coordinates on the plot specify the relative weight of each of the three actions, and the third dimension -- the total energy $E$, expressed equivalently as the radius of a circular orbit $R_\mathrm{circ}(E)$ -- is shown by colour. To avoid clutter, we split the entire population into two roughly equal halves, with $R_\mathrm{circ}=5$~kpc as the boundary (a few clusters appear in both panels because of uncertainties in their energy). This plot shows various sub-populations more clearly than the velocity-space diagram. For instance, all six Sagittarius stream clusters are grouped together in the upper half of the right panel (pink colours), indicating that they are on nearly polar orbits. A few more pairs of clusters are on similar orbits: NGC~7078 (M~15) and NGC~6656 (M~22), NGC~6838 (M~71) and Pal~10, and with a stretch of imagination, NGC~5024 (M~53) and NGC~5053. The latter pair happens to be very close in physical space at present, suggesting a possible link between them \citep{Chun2010}; their metallicities, PM and orbital planes are also similar, but the line-of-sight velocities differ by 100~\kms.

Another interesting feature is the clumping of several clusters in the bottom part of the right panel, indicating that they have rather eccentric orbits with large radial actions. This population was identified by \citet{Myeong2018} and associated with a similar radially-biased population in the stellar halo. The eight clusters listed in that paper are NGC~1851, NGC~1904 (M~79), NGC~2298, NGC~2808, NGC~5286, NGC~6779 (M~56), NGC~6864 (M~75) and NGC~7089 (M~2). Possible other candidates in the same region with newly measured PM include NGC~6584 and IC~1257, although no two of the entire sample of high-$J_r$ clusters appear to be on similar orbits in physical space. 

Figure~\ref{fig:poles} shows the distribution of orbital poles (instantaneous directions of angular momenta) for the outer galaxy clusters; the coordinates are the same as in Figure~3 of \citet{Fritz2018}. The clumping of Sagittarius clusters is obvious in the left side of this plot, but the orbital planes of clusters from \citet{Myeong2018} are scattered across the sky. Some of them and a few other clusters (e.g., the outermost ones: Pyxis, Eridanus, AM~1, Crater, Pal~3, Pal~4) lie close to the direction of the ``Vast Plane of Satellites'' \citep{Pawlowski2012}, which was recently confirmed to contain a large fraction of Milky Way satellite galaxies \citep{Fritz2018}; however, most globular clusters do not appear to be associated with this structure. On the other hand, the angular momentum vector is not an integral of motion, as the orbital planes precess and nutate in a non-spherical potential, thus a single progenitor cannot be excluded even for clusters whose orbital planes do not align at present.
A more detailed classification of the cluster population, its relation to the stellar halo and to the accretion history of the Milky Way would need to take into account chemical composition and star formation history \citep[e.g.,][]{Kruijssen2018}, in addition to the now available kinematic information.

The use of actions, as opposed to any other integrals of motion, also allows one to examine the distribution of clusters in phase angles, which are canonically conjugate variables to actions. We expect a uniform distribution in angles for a well phase-mixed population. Figure~\ref{fig:angles} shows the distribution of observed clusters (taking into account the errors) in three phase angles, computed using an amount of smoothing calibrated by cross-validation (suppressing likely insignificant fluctuations), as explained in \hyperref[sec:nonparafit]{Appendix B}.
The phase angle of azimuthal motion $\theta_\phi$, which describes the location with respect to the solar position, is distributed significantly non-uniformly: the higher probability of finding clusters around 0 or $2\pi$ suggests that the catalogue is deficient in clusters located at the other side of the Galaxy, which may have escaped detection because of dust obscuration. A similar asymmetry is observed in the distribution of azimuthal angle $\phi$ itself -- see Figure~7 in \citet{Binney2017}, who propose that it could arise either from missing clusters hidden by dust, or from systematic underestimation of cluster distances (which would place them preferentially on the near side of the Galaxy).
The other two phase angles are consistent with being uniformly distributed.

\section{Dynamical modelling}  \label{sec:dyn}

\subsection{Method}  \label{sec:dyn_method}

We now explore the dynamical properties of the globular cluster system, and the information it provides about the Milky Way potential, following the standard assumption that this is a steady-state equilibrium configuration. The population of clusters is described by a certain distribution function (DF) $f(\boldsymbol{w})$, which specifies the probability of observing a particular combination of position and velocity $\boldsymbol{w} \equiv \{\boldsymbol{x},\boldsymbol{v}\}$ for any cluster. According to Jeans' theorem, in a steady state the DF must be a function of integrals of motion $\boldsymbol{I}(\boldsymbol{w};\,\Phi)$, which themselves depend on the potential $\Phi$. The best-fit parameters of the DF, and possibly the potential, are found by maximizing the likelihood of drawing the observed positions and velocities of clusters from this DF -- the same approach was used in \citet{Binney2017} and \citet{Posti2019}.

The log-likelihood of the model is given by the sum of log-likelihoods for each cluster, convolved with their respective error distributions:
\begin{equation}  \label{eq:likelihood}
\ln \mathcal{L} = \sum_{i=1}^{N_\mathrm{clusters}} \ln \frac{ S(\boldsymbol{w}_i)\; 
\int \mathrm{d} \boldsymbol{w}'\; E(\boldsymbol{w}_i\,|\,\boldsymbol{w}')\; 
f(\boldsymbol{I}[\boldsymbol{w}';\,\Phi] ) }
{\int \mathrm{d} \boldsymbol{w}'\; S(\boldsymbol{w}')\; 
f(\boldsymbol{I}[\boldsymbol{w}';\,\Phi] ) } .
\end{equation}

Here $E(\boldsymbol{w}_i\,|\,\boldsymbol{w}')$ is the probability of measuring the phase-space coordinates of $i$-th cluster as $\boldsymbol{w}_i$ given the true values $\boldsymbol{w}'$. This is not the same as the multivariate normal distribution for the observable quantities $\boldsymbol{u}$ (sky coordinates $\alpha,\delta$, heliocentric distance $D$, PM $\mu_\alpha,\mu_\delta$, and line-of-sight velocity $v_\mathrm{los}$), but includes a Jacobian factor for transformation from $\boldsymbol{u}$ to $\boldsymbol{w}$, equal to $D^4\,\cos\delta$. However, for relatively small adopted distance uncertainty, its role is negligible (apart from shifting the total likelihood by an unimportant constant).

$S(\boldsymbol{w})$ is the selection function, which specifies the probability of observing a cluster at a given position (and possibly velocity) in the survey. For clusters, this function is close to unity, except in some regions in the central Galaxy or beyond, where the strong extinction along the line of sight might have prevented their discovery or observation, or in the very distant outer regions, where the clusters are too faint. Empirically, we could not locate 6\% (9 out of 157) clusters from the \citet{Harris2010} catalogue in the Gaia data, which suggests that the incompleteness of our sample with respect to known cluster population is rather minor. However, the non-uniformity of cluster distribution in phase angles, identified in the previous section, hints at a possible existence of yet undiscovered clusters. Despite this, in the remaining analysis we assume $S=1$. \citet{Binney2017} found that inclusion of an extinction-dependent selection function had little effect on their results.

The integral in the numerator of Equation~\ref{eq:likelihood} is the convolution of the DF with the error distribution for each cluster. We replace this integral by a sum over the Monte Carlo samples $\boldsymbol{w}'_{i,k}$, $k=1..N_\mathrm{samples}$, defined in the previous section. Note that this sum is inside the argument of the logarithm, so that even if some of the samples lie in a physically inaccessible region (e.g., have positive energy) and hence have zero probability, this only moderately affects the likelihood of the given cluster, provided that there are enough samples for which $f(\boldsymbol{w}'_{i,k})$ has a non-negligible value. Of course, the set of Monte Carlo samples must remain the same for all models, to compensate the Poisson fluctuations in the likelihood \citep{McMillanBinney2013}.

The integral in the denominator is the overall normalization factor, identical for all clusters. In the absense of a non-trivial selection function, it can be computed directly in the $\boldsymbol{I}$ space, and gives the total number of clusters.

In addition, the posterior likelihood of a model with the given parameters of the DF and the potential may have a contribution from a prior probability of these parameters. In practice, we specify non-trivial priors for a couple of parameters, and simply restrict other parameters to lie in some finite intervals, to ensure that the prior probability is normalizable. For dimensional parameters such as scale radius, we adopt uninformative priors in their logarithms.

\subsection{Ingredients}  \label{sec:dyn_ingredients}

Similarly to \citet{Binney2017} and \citet{Posti2019}, we express the DF as a function of actions: the radial action $J_r$, the vertical action $J_z$, and the azimuthal action $J_\phi$, equivalent to the $z$-component of angular momentum if the potential is axisymmetric. The St\"ackel fudge in the current implementation only works for oblate axisymmetric potentials, hence we have to neglect the dynamical effect of the bar and possibly other non-axisymmetric features in the inner Galaxy, and restrict our halo shape to be spherical or oblate. For this reason, we do not expect the model to offer an accurate description of the dynamics of clusters in the innermost few kpc.

We assume that the entire population of globular clusters is described by a single DF, i.e., do not make any distinction between metal-poor halo and metal-rich disky populations. It is possible to generalize this approach by considering a mixture of two DFs and attribute each cluster to either of them probabilistically, just as we did to distinguish cluster members from field stars in the PM space. This would add a few more free parameters to the model without changing the overall procedure. However, our choice for the functional form of the single DF is already quite flexible, and it can simultaneously describe both populations reasonably well. It is similar to the double-power-law DF families used by \citet{Posti2015}, \citet{Binney2017}, or \citet{Williams2015}:
\begin{equation}  \label{eq:df}
\begin{aligned}
f(\boldsymbol{J}) &= \frac{M}{(2\pi\, J_0)^3}
\left[1 + \left(\frac{J_0}{h(\boldsymbol{J})}\right)^\eta \right]^{\Gamma/\eta}  
\left[1 + \left(\frac{g(\boldsymbol{J})}{J_0}\right)^\eta \right]^{-\mathrm{B}/\eta} 
\makebox[-1cm]{}\\
&\times \bigg(1 + \tanh\frac{\varkappa J_\phi}{J_r + J_z + |J_\phi|}\bigg).
\end{aligned}
\end{equation}
Here
\begin{equation*}
\begin{aligned}
g(\boldsymbol{J}) &\equiv g_r J_r + g_z J_z\, + (3-g_r-g_z)\, |J_\phi|, \\
h(\boldsymbol{J}) &\equiv h_r J_r + h_z J_z   + (3-h_r-h_z)   |J_\phi|
\end{aligned}
\end{equation*}
are linear combinations of actions, with dimensionless coefficients controlling the spatial flattening and velocity anisotropy of the model in the outer region (above the break action $J_0$) and the inner region (below $J_0$), respectively. 
The power-law indices $\mathrm{B}$ and $\Gamma$ control the outer and inner slopes of the density profile (although the relation between these the slopes of the DF and the density also depends on the potential). The parameter $\eta$ determines the steepness of the transition between the two regimes. The rotation is introduced by the parameter $\varkappa$ in a way that is roughly constant across all energies, since we normalize $J_\phi$ by the sum of all three actions. This is different from the convention used, e.g., in \citet{Binney2014}, \citet{Binney2017} or \citet{Posti2019}, which had a non-rotating core because $J_\phi$ in the argument of tanh was normalized by a fixed constant.
Overall, there are 9 free parameters in the DF (the total mass is fixed by the normalization constraint), which are restricted to physically acceptable ranges (B${}>3$, $0<\Gamma<3$, $0.5\le\eta\le2$, and positive coefficients in the linear combinations of actions).

The DF determines both the density profile and the velocity distribution of the tracer population of clusters. Of course, if the functional form of the DF and its parameters are adequate, the resulting density profile would match the actual spatial distribution of clusters, but we do not fit for it independently.

For the potential, we use the following three components: a central bulge with a truncated power-law profile, an exponential disc, and a rather flexible functional form for the halo density profile. The bulge profile is identical to that of \citet{McMillan2017}, and for simplicity, we replace four separate disc components from that study by a single exponential disc with a scale radius 3~kpc and scale height 0.3~kpc. The rotation curve produced by such a disc is very similar to the combination of four separate ones. The halo density profile follows the \citet{Zhao1996} $\alpha\beta\gamma$ model: 
\begin{equation}  \label{eq:zhao}
\rho(r) = \rho_\mathrm{h}\,\left(\frac{r}{r_\mathrm{h}}\right)^{-\gamma}
\left[ 1 + \left(\frac{r}{r_\mathrm{h}}\right)^\alpha \right] ^ { (\gamma-\beta) / \alpha }.
\end{equation}
The case $\alpha=1,\beta=3,\gamma=1$ corresponds to the NFW profile, often used in other studies. However, we wish to avoid possible biases resulting from a restricted functional form of the halo potential, hence our model is very flexible -- the inner density slope is controlled by $\gamma$, the outer by $\beta$, and $\alpha$ determines the steepness of transition between two asymptotic regimes. We experimented with non-spherical (oblate) shapes of the halo, but the fit always preferred the axis ratio close to 1, hence we assume the halo to be spherical.
Overall, the potential has seven free parameters: the density normalization $\rho_\mathrm{h}$ and scale radius $r_\mathrm{h}$ of the halo, its three dimensionless slope parameters, and masses of the bulge and the disc. We allow the latter two to vary in a rather limited range, inspired by the best-fit potential suggested in \citet{McMillan2017}: $M_\mathrm{bulge}=9\times 10^9\,M_\odot$, $M_\mathrm{disc}=5.5\times 10^{10}\,M_\odot$, both with relative uncertainty of 10\%. The distribution of posterior values closely follows the prior, but the added flexibility allows us to propagate the uncertainty into the halo component. In addition, we put a prior on the amplitude of circular velocity at the solar radius: $v_\mathrm{circ}(R_0) = 235\pm10$~\kms. This value is consistent with both the best-fit potential of \citet{McMillan2017} and the observed total azimuthal solar velocity of $248\pm 3$~\kms after subtracting the peculiar motion of $11\pm 2$~\kms \citep{BlandHawthorn2016}. We neglect the contribution of globular clusters to the total mass of the Galaxy.

\subsection{Monte Carlo simulations}  \label{sec:dyn_mc}

We explore the parameter space with the Monte Carlo Markov Chain (MCMC) method, implemented in the \texttt{emcee} package \citep{ForemanMackey2013}. We run 50 walkers for several thousand steps, monitoring the convergence of the posterior distribution, and use the median, 68\% and 95\% percentiles to display the results and their $1\sigma$ and $2\sigma$ confidence intervals.

We first run the fitting procedure, fixing the parameters of the potential to the values of \citet{McMillan2017} best-fit model, and only varying the DF parameters. 
The resulting models successfully reproduce the cluster density profile and principal features in their velocity distribution (Figures~8 and 9, discussed in the next section).
As it turns out, the range of acceptable DF parameters is rather insensitive to whether we vary the potential or not, so we discuss the properties of the DF later, after describing the entire simulation suite.

We then tested the ability of our machinery to recover the gravitational potential, if we let it free. For this purpose, we took one of the DFs from the MCMC chain, and created two types of mock datasets by drawing 150 samples from the DF in the known potential. In the first case, we use the measured positions of the observed clusters $\boldsymbol{x}_i$ and randomly assign the velocities from the conditional velocity distribution $f(\boldsymbol{x}=\boldsymbol{x}_i,\, \boldsymbol{v})$. By using the measured positions, we test the possible biases arising from a mismatch between the actual density profile and the one generated by the DF. In the other case, we draw both $\boldsymbol{x}$ and $\boldsymbol{v}$ from the DF. We then converted the position and velocity of each cluster to the observable coordinates (sky position, distance, PM and line-of-sight velocity) and added a normally distributed random error to these values, drawn from the actual error estimates for each cluster. These mock datasets were then analyzed by the same pipeline, but this time the parameters of the potential are also allowed to vary.
By repeating this experiment several times with different choices of DF, we found that the potential is well recovered, with a typical uncertainty on the rotation curve in the range $15-25$~\kms at large radii. The median circular velocity drops somewhat faster with radius at large distances than the true one, but the $1\sigma$ confidence interval encloses the true curve.

Finally, we proceed with modelling the original dataset with both the DF and potential parameters varied in the fitting process. The posterior distributions do not show any significant correlation between the DF and potential parameters, and only a few parameters in each of these two groups are substantially interdependent (e.g., the outer slope B of the DF and its scale action $J_0$). Several parameters are not well constrained, such as the outer slope of the halo density profile $\beta$ and the steepness of the transition between inner and outer halo density slopes $\alpha$. We let them vary in the fit to avoid possible biases due to a more constrained functional form of our models, but in this case the results become more dependent on the prior. We choose flat priors in $\ln\alpha$, $\ln\beta$ within the ranges $0.3\le\alpha\le2$ and $2.5\le\beta\le5$, which favour smaller values of these parameters (closer to the NFW profile).

We explored the robustness of our results to various changes in the input data, for instance,  keeping the PM data only for the 91 clusters previously measured by \Gaia collaboration and \HST (the sample used in \citealt{Posti2019} and \citealt{Watkins2018}). The range of acceptable models was slightly wider for the more restricted input sample, but the general trends remained the same.
We also checked the influence of Sagittarius stream clusters, which are excluded from our main sample: their inclusion increases the inferred circular velocity by $10-15$~\kms outside 20~kpc, which is well within the $1\sigma$ confidence interval.

\subsection{Results for the DF}  \label{sec:dyn_df}

\begin{figure}
\includegraphics{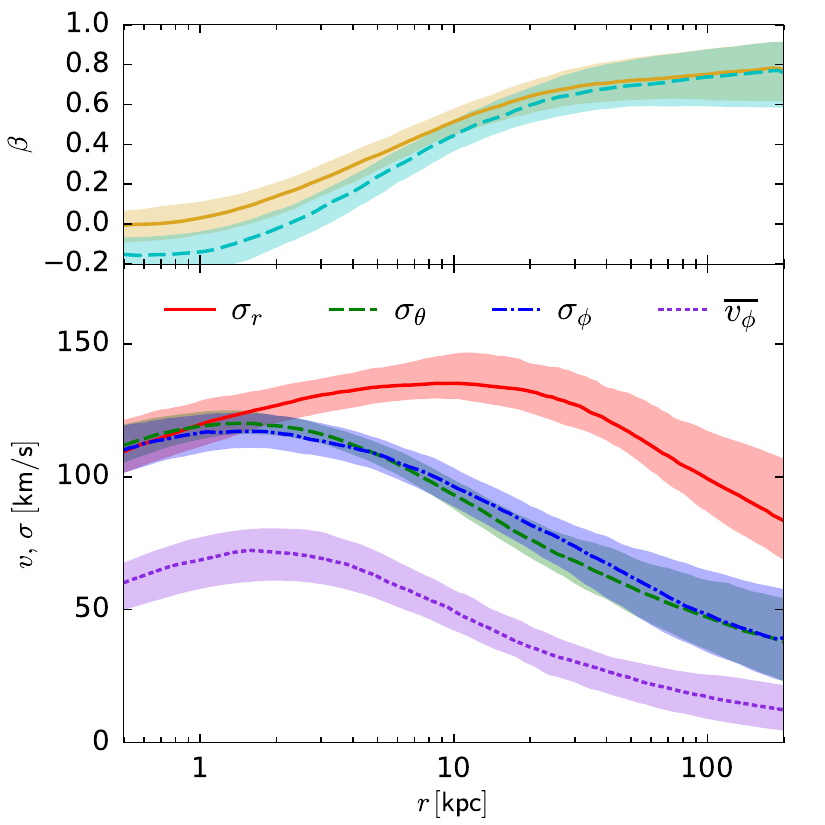}
\caption{
Bottom panel: velocity dispersion and mean azimuthal velocity profiles for an ensemble of models from the MCMC run. Colour coding is the same as in Figure~\ref{fig:veldisp}, and 68\% uncertainty bands are shaded. Top panel: velocity anisotropy coefficient $\beta_\mathrm{a} = 1 - (\sigma_\theta^2+\sigma_\phi^2)/(2\sigma_r^2)$ (solid yellow), and a modified version with $\sigma_\phi^2$ replaced by $\sigma_\phi^2+\overline{v_\phi}^2$, taking into account the kinetic energy of rotation (dashed cyan).
} \label{fig:veldisp_model}
\end{figure}

\begin{figure}
\includegraphics{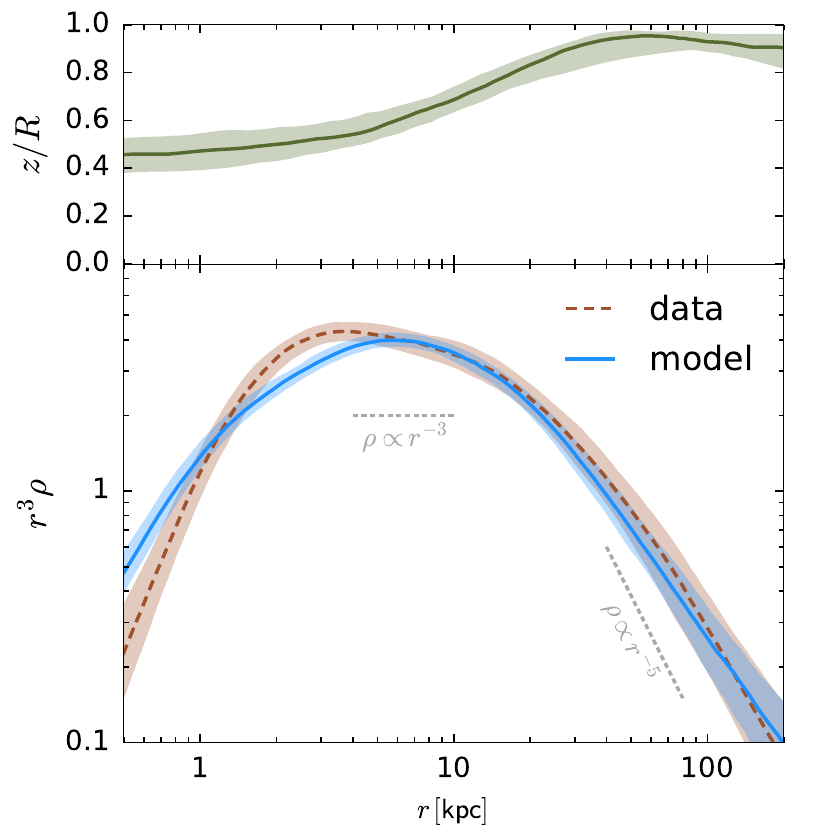}
\caption{
Bottom panel: spatial density profile of the clusters averaged over many models from the MCMC run (solid blue), and estimated from the input data (brown dashed). To compress the dynamical range, the plotted quantity is $r^3\,\rho(r)$, in arbitrary units. Top panel: vertical flattening (axis ratio) of the density profile from models. 
} \label{fig:density}
\end{figure}

The chosen functional form of the DF proves to be flexible enough to reproduce the main trends in the observed velocity distribution (Figure~\ref{fig:veldisp_model}). The velocity dispersion is close to isotropic in the central few kpc and radially biased further out, with the anisotropy coefficient $\beta_\mathrm{a}\equiv 1 - (\sigma_\theta^2+\sigma_\phi^2)/(2\sigma_r^2)$ reaching the range $0.6-0.8$ at large radii (not to be confused with the outer slope of the halo density profile $\beta$).
The mean rotation velocity in the central few kpc peaks at a similar amplitude as the data, although at a somewhat smaller galactocentric distance.
The marginalized 1d velocity distributions and the density profile matches the data fairly well, as shown on the side panels of Figure~\ref{fig:veldistr}.

The density profile is also well reproduced, as shown in Figure~\ref{fig:density}, except in the very centre, where both the input catalogue is likely incomplete, and the model is not expected to be very accurate. The spherically-averaged cluster density produced by the DF does not have an explicit expression, but is well approximated by the $\alpha\beta\gamma$ profile (\ref{eq:zhao}) with $\alpha=0.5$, $\beta=6$, $\gamma=0$, and scale radius 6~kpc. The logarithmic slope $\mathrm{d}\,\log \rho/\mathrm{d}\,\log r$ gradually changes from close to $-3$ around the solar radius to $-5$ around 100~kpc. The density is significantly flattened in the central few kpc, with axis ratio $z/x\simeq 0.4-0.6$.

\citet{Binney2017} and \citet{Posti2019} used a similar DF-based approach as in our study, but divided the entire population into a disc-like and halo-like subset, with two separate DFs. The functional form of their halo DF was also somewhat different from the one in the present study, preventing a direct comparison between parameters. Instead, we focus on more directly observable features, such as the anisotropy and rotation. \citet{Binney2017} report a similar amount of radial anisotropy in the outer halo as we find, and a weak prograde rotation of the halo component. Their disc component had a much higher mean azimuthal velocity $\overline{v_\phi}$, reaching 185~\kms at 5~kpc, but with a much lower dispersion $\sigma_\phi \lesssim 50$~\kms, whereas $\sigma_z$ was around 100~\kms. When combining both components, they find $\overline{v_\phi} \simeq 70\pm30$~\kms, consistent with our values. The azimuthal velocity dispersion $\sigma_\phi$ then must necessarily increase, and is likely in the same range as we have. The cluster density at large radii declined as $r^{-5}$ or even steeper, and was quite flattened in the central part - the axis ratio $z/R$ is $\sim 1/3$ for the disc component and $\sim 2/3$ for the halo component, although they do not quote it for the entire population. This is also similar to our results for the overall density profile.
\citet{Posti2019}, on the other hand, find the halo component to be only mildly radially anisotropic ($\beta_\mathrm{a}\simeq 0.2$), and the density profile to roughly follow a power law with slope $-3.3$, although this probably refers to a smaller range of radii than our (steeper) asymptotic slope.

\citet{Sohn2018} examined the distribution and kinematics of 20 globular clusters with galactocentric distances ranging from 10 to 40~kpc, for which the PM is measured by \HST. They find a density profile $\rho\propto r^{-3.5}$, and anisotropy $\beta_\mathrm{a}\sim 0.4-0.7$, similar to our estimates for this range of radii. \citet{Watkins2018} augmented this sample with \Gaia-derived PM for 34 clusters with apocentre distances greater than 6~kpc. They find their density to be well approximated by a broken power law with inner slope $-2$, outer slope $-3.5$, and break radius around 4~kpc. Their estimate for velocity anisotropy for the entire \Gaia{}+\HST sample is $\beta_\mathrm{a}\sim 0.5\pm0.15$. These results are compatible with ours.

Interestingly, the radial velocity anisotropy of the globular cluster population at large radii parallels that of the stellar halo found in some recent studies.
\citet{Deason2012} determined $\beta_\mathrm{a}\simeq 0.5^{+0.1}_{-0.2}$ at $16\le r \le 50$~kpc from the kinematics blue horizontal branch (BHB) stars, and
\citet{Kafle2014} inferred $\beta_\mathrm{a}=0.4\pm 0.2$ beyond 25~kpc from the kinematics of K-giants and BHB stars in the SDSS survey. 
On the other hand, \citet{Das2016} find only mild radial anisotropy ($\beta_\mathrm{a}\lesssim 0.3$) from these data. 
The above studies used only the line-of-sight velocities, but \citet{Hattori2017} demonstrated that this could lead to substantial biases in the anisotropy coefficient, and more recent studies used full 3d velocity information.
\citet{Bird2019} combined the line-of-sight velocities of K-giants from the LAMOST survey with their PM from \Gaia DR2, and found $\beta_\mathrm{a}=0.3-0.8$ at $25\le r \le 100$~kpc.
\citet{Cunningham2019} measured $\beta_\mathrm{a}\sim 0.6$ around 25~kpc from the \HST-based PM and Keck spectroscopy of main-sequence turnoff stars.
\citet{Wegg2019} deduced $\beta_\mathrm{a}\gtrsim0.7$ for RR Lyrae stars at $5\le r \le 20$~kpc from \Gaia DR2 PM (without line-of-sight velocity information).
The radially-anisotropic velocity distribution in the halo is also observed in cosmological galaxy formation simulations \citep[e.g.,][]{Fattahi2019,Mackereth2019}, and has been associated with an ancient major accretion event \citep{Belokurov2018, Helmi2018b}.
The connection between the kinematics, chemistry and age for both the halo stars and the globular clusters, and its implications for the ``galactic archeology'', will certainly remain a hot topic for future studies.

\subsection{Results for the potential}  \label{sec:dyn_pot}

\begin{figure}
\includegraphics{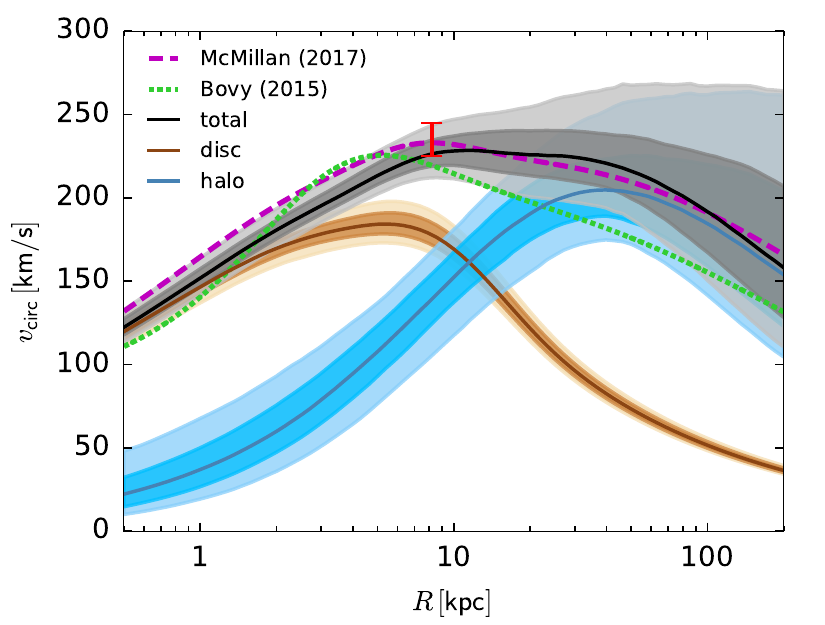}
\caption{
Rotation curves of the ensemble of models from the MCMC run.
Solid lines show the median values of circular velocity at a given radius, shaded regions -- 68\% (darker) and 95\% (lighter) confidence regions. Orange is the contribution of disc and bulge, cyan is the halo, and gray is the total. For comparison, the rotation curves of the best-fit potential of \citet{McMillan2017} is shown in dashed magenta line, and the one from \citet{Bovy2015} -- in dotted green. The red error bar shows our prior on the circular velocity at $R=8.2$~kpc ($235\pm10$~\kms). The radial distribution of tracers covers roughly the range $1-100$~kpc.
} \label{fig:rotcurve}
\end{figure}

We now discuss the constraints on the potential obtained by allowing its parameters (mainly the halo) to vary during the fit, together with the DF parameters. When we consider the properties of the potential alone, this implies a marginalization over the DF parameters (cf.\ \citealt{Magorrian2014} for a motivation, although, of course, our DF comes from a restricted family of models and does not explore all possibilities).

Rather than showing the posterior distribution for each parameter, we plot the median value and $1\sigma/2\sigma$ confidence intervals on the rotation curve in Figure~\ref{fig:rotcurve}.
For comparison, we also show the circular velocity of our default potential, taken from \citet{McMillan2017}, and the one from \citet{Bovy2015} (\texttt{MWPotential2014}). The former  potential is within one--two $\sigma$ from the most-likely models in our chain, but the latter one is clearly inconsistent with the rotation curve beyond 10~kpc. However, the potential at large radii is not well constrained by our models.

The density profile of the halo preferred by the fitting routine is shallower than $\rho\propto r^{-1}$ in the inner part of the Galaxy. The outer slope $\beta$ is not well constrained, and almost the entire range $2.5\le \beta\le 5$ allowed by our prior is equally likely. Therefore, the formally defined virial mass and radius%
\footnote{We define $r_\mathrm{virial}$ such that the mean density within this radius is 100 times higher than the cosmic density of matter: $r_\mathrm{virial} = (M_\mathrm{virial} / 10^{12}\,M_\odot)^{1/3}\times 260$~kpc (cf. Equation~3 in \citealt{BlandHawthorn2016}).}
have large uncertainties: $M_\mathrm{virial} = 1.2^{+1.5}_{-0.5}\times 10^{12}\,M_\odot$,  $r_\mathrm{virial} = 280^{+80}_{-50}$~kpc. However, it makes little sense to compare these extrapolated values, and more meaningful is the total enclosed mass (including $6.5\times10^{10}\,M_\odot$ in baryons) within a given radius probed by the tracer population ($\lesssim 100$~kpc). For our models, the mass within 50~kpc is $0.54^{+0.11}_{-0.08}\times 10^{12}\,M_\odot$, and within 100~kpc is $0.85^{+0.33}_{-0.20}\times 10^{12}\,M_\odot$. As mentioned above, the inclusion of Sagittarius clusters increases these values by $10-15$\%, within the confidence intervals. The enclosed mass profile agrees rather comfortably with most estimates in the literature (see Table~8 in \citealt{BlandHawthorn2016} or Figure~6 in \citealt{Eadie2019} for a compilation of results).

We restrict the scope of comparison of potential estimates to papers using the kinematics of globular clusters only. Our DF-based approach is most similar to that of \citet{Posti2019}, but their assumptions about the form of gravitational potential make a direct comparison rather difficult. We put a strong prior on the local circular velocity, but a very broad prior on the outer halo density profile, which is not well constrained in our models. They assumed an NFW profile for the halo with a fixed concentration, varying its mass and radius simultaneously, and also allowed it to be non-spherical. However, it should be noted that the prolate shape of their best-fit potential is not compatible with the St\"ackel fudge method for action computation, possibly leading to a biased inference on the shape. As mentioned above, we restrict the halo to be either oblate or spherical, and find that a spherical shape is always preferred by the fit. Nevertheless, our median values for the virial mass and radius are close to theirs, but have much larger uncertainties.

From the analysis of \HST-derived PM for 20 outer globular clusters, \citet{Sohn2018} find the circular velocity at 40~kpc to be $259^{+35}_{-26}$~\kms, corresponding to the enclosed mass $0.61^{+0.18}_{-0.12}\times10^{12}\,M_\odot$. \citet{Watkins2018}, with an expanded \Gaia{}+\HST sample, determined the mass enclosed within 40~kpc to be $0.44^{+0.07}_{-0.06}\times10^{12}\,M_\odot$, or $v_\mathrm{circ}=220^{+17}_{-16}$~\kms.
Our estimate of circular velocity is roughly constant within 50~kpc ($225\pm15$~\kms), closer to the latter study but consistent with both. As discussed in Section~\ref{sec:pm_results}, our PM agree well with the \HST data (and, of course, with the previously derived \Gaia PM), hence the agreement with the results obtained by a rather different modelling method is reassuring.

\citet{Eadie2017} also used kinematics of the cluster population, although relying on the older, ground-based PM measurements, and obtained a substantially lower enclosed mass -- $(0.34\pm0.04)\times10^{12}\,M_\odot$ within 50~kpc, see Figure~1 in the erratum of that paper.
More recently, \citet{Eadie2019} used our PM catalogue, and obtained the enclosed mass profile that is still considerably lower than ours:
$(0.38\pm0.05)\times10^{12}\,M_\odot$ within 50~kpc, and $(0.54\pm0.08)\times10^{12}\,M_\odot$ within 100~kpc. Hence the difference is unlikely to be attributed to the data, but rather to a different method: they assumed a power-law profile for both the tracer DF and the total potential, and in the latter study, only used clusters with galactocentric radii $>15$~kpc. They tested their method on simulated galaxies \citep{Eadie2018} and found that it may underestimate the enclosed mass (see Figures~4 and 12 in that paper).

Of course, a more elaborate approach would take into account other dynamical constraints on the potential, for instance, from the motion of satellite galaxies, stars in the halo, or tidal streams. The goal of this section was to demonstrate that globular clusters also provide a useful dataset to work with, and their constraints on the potential seem reasonable (although not very tight).

\section{Summary}  \label{sec:summary}

\begin{itemize}
\item We provide the catalogue of PM for nearly all Milky Way globular clusters, with estimated uncertainties less than 0.1~\masyr for most objects (dominated by systematic errors in \Gaia PM, which are fully taken into account by our approach). This catalogue complements the measurements provided by the \Gaia collaboration \citepalias{Helmi2018}, the independent \Gaia-based catalogue of \citet{Baumgardt2019}, and various \HST-based determinations \citep[e.g.,][]{Sohn2018}, and is in excellent agreement with these studies. By contrast, we find that existing ground-based measurements are not only much less precise than the space-based ones, but often deviate from them by a far larger amount than implied by their quoted uncertainties.
\item By analyzing the 6d spatial and velocity distribution of galactic globular clusters, we confirm that their population in the inner Galaxy has a significant rotation, with mean azimuthal velocity in the range $50-80$~\kms within 10~kpc. The velocity dispersion is largely isotropic in this region, with $\sigma\simeq 100-120$~\kms. At large galactocentric distances, the distribution becomes radially anisotropic, with the value of anisotropy parameter $\beta_\mathrm{a}\gtrsim 0.6$  similar to that of the stellar halo.
\item We also explore the 6d distribution of clusters in the action/angle space. It illustrates more clearly several kinematically distinct populations, for instance, the 6 clusters associated with the Sagittarius stream (confirming some of the candidate members suggested in \citealt{Law2010}), or the group of outer halo clusters with little net rotation and high radial action, identified in \citet{Myeong2018}.
\item We find that the distribution of clusters in phase angles deviates from a uniform one, with a deficit of clusters at the opposite side of the Galaxy. Since a dynamically relaxed population is expected to be randomly spread in orbital phases, this hints at a possible observational selection bias, naturally arising from the dust obscuration within the galactic plane. 
\item We model the observed 6d phase-space distribution of clusters by a combination of an action-space DF $f(\boldsymbol{J})$ and the total gravitational potential $\Phi$, using a likelihood-based approach with full account of observational errors. By exploring the parameters of the DF and the potential in an MCMC simulation, we find a range of acceptable models, producing a rotation curve that stays at a level $210-240$~\kms in the range $8-50$~kpc. This is consistent with our default choice of potential from \citet{McMillan2017}; however, potentials with a lower circular velocity in this range of radii, such as \texttt{MWPotential2014} from \citet{Bovy2015} are disfavoured by the data.
\end{itemize}

I thank J.~Binney, V.~Belokurov, D.~Erkal, T.~Fritz, S.~Koposov, G.~Myeong, J.~Sanders, and the anonymous referee for valuable comments.
This work uses the data from the European Space Agency mission \Gaia (\url{https://www.cosmos.esa.int/gaia}), processed by the \Gaia Data Processing and Analysis Consortium (\url{https://www.cosmos.esa.int/web/gaia/dpac/consortium}).
This work was supported by the European Research council under the 7th Framework programme (grant No.\ 308024). 


\appendix

\section{Gaussian mixture modelling}  \label{sec:gmm}

In this section we describe our method for deriving membership probabilities and parameters of the distribution from a mixture of cluster and field stars in the PM space, which is used in Section~\ref{sec:pm_method}. 
We specialize to the case of two components (members and contaminants, with $c=1$ being the cluster and $c=2$ the field population) and two dimensions ($\mu_\alpha$, $\mu_\delta$), but present the approach more universally.

We assume that the intrinsic (noise-free) distribution of each component $c$ in the mixture is a Gaussian with some mean value $\overline{\boldsymbol\mu_\mathrm{c}}$ and a symmetric covariance matrix $\mathsf{\Sigma}_\mathrm{c}$ (in this section, vectors are denoted by boldface and matrices -- by sans-serif font). The $D$-dimensional multivariate Gaussian probability distribution is
\begin{equation}
\mathcal{N}(\boldsymbol{\mu} \:|\: \overline{\boldsymbol{\mu}_c}, \mathsf{\Sigma}_c) 
\equiv \frac{ \exp\big[ -{\textstyle\frac{1}{2}}
(\boldsymbol{\mu}-\overline{\boldsymbol{\mu}_c})^T\,
\mathsf{\Sigma}_c^{-1}\, 
(\boldsymbol{\mu}-\overline{\boldsymbol{\mu}_c}) \big] }
{(2\pi)^{D/2}\, \sqrt{\det \mathsf{\Sigma}_c}} .
\end{equation}
The overall distribution function of the mixture is a weighted sum of $K$ Gaussian components:
\begin{equation}
f(\boldsymbol{\mu}) = \sum_{c=1}^K q_c\,\mathcal{N}(\boldsymbol{\mu} \:|\: \overline{\boldsymbol{\mu}_c}, \mathsf{\Sigma}_c) , \quad
q_c \ge 0, \; \sum_{c=1}^K q_c = 1 .
\end{equation}

The actual values of $\mu_i$ for each $i$-th star are not measured exactly, but with some observational error, which we assume to be normally distributed with a zero mean and an uncertainty covariance matrix $\mathsf{E}_i$ (different for each star). Thus the probability of drawing a value $\mu_i$ from the mixture distribution is a convolution of $f$ with the error distribution for this datapoint. For the Gaussian distribution, the convolution is trivial and produces another Gaussian with the same mean and a covariance matrix $\mathsf{\Sigma}_c + \mathsf{E}_i$ for each component $c$.

The total log-likelihood of this Gaussian mixture model, given the set of $N$ measured values $\boldsymbol{\mu}_i$ and their error estimates $\mathsf{E}_i$, is given by
\begin{align}  \label{eq:loglike}
\ln\mathcal{L} = \sum_{i=1}^N \ln\Big[ \sum_{c=1}^K
q_c\:\mathcal{N}(\boldsymbol{\mu}_i \:|\: 
\overline{\boldsymbol{\mu}_c}, \mathsf{\Sigma}_c+\mathsf{E}_i) \Big] .
\end{align}

The elements of matrices $\mathsf{\Sigma}_c$ and vectors $\boldsymbol{\mu}_c$, and $K-1$ weights $q_c$, are all varied during the fit to maximize the total log-likelihood of the entire ensemble of stars. These quantities may have the same value for all stars, or more generally, be some functions of other attributes of each star besides $\boldsymbol{\mu}_i$ and $\mathsf{E}_i$, with parameters being optimized during the fit.
For instance, the intrinsic PM dispersion of cluster stars (the matrix $\mathsf\Sigma_c$ for $c=1$), or the prior probability of membership $q_{c;\,i}$ of each $i$-th star, may depend on the distance $R_i$ of the given star from the cluster centre. 

After the best-fit parameters that maximize the total log-likelihood (\ref{eq:loglike}) are determined, one may compute the posterior probability of $i$-th star to belong to the $k$-th component:
\begin{equation}  \label{eq:posterior}
p_{k;\,i} = \frac{\quad\;  q_{k;\,i}\; \mathcal{N}(\boldsymbol{\mu}_i \:|\: 
\overline{\boldsymbol{\mu}_{k;\,i}},\; \mathsf{\Sigma}_{k;\,i}+\mathsf{E}_i) }
{\sum\limits_{c=1}^{K} q_{c;\,i}\; \mathcal{N}(\boldsymbol{\mu}_i \:|\: 
\overline{\boldsymbol{\mu}_{c;\,i}},\; \mathsf{\Sigma}_{c;\,i}+\mathsf{E}_i) }.
\end{equation}
Here we made explicit that the parameters of the Gaussian mixture model (means, dispersions and weights) may be different for each star.

The standard approach for finding the maximum of (\ref{eq:loglike}) is the iterative expectation--maximization (EM) algorithm \citep[e.g.,][Chapter 16.1]{NumRec}. It proceeds by repeating the following two steps: 
\begin{itemize}
\item The probability $p_{c;\,i}$ of each datapoint to belong to each Gaussian component (Equation~\ref{eq:posterior}) is computed from the current parameters of the model, treating them as fixed quantities (expectation step).
\item The parameters of the mixture model are updated to maximize the log-likelihood (Equation~\ref{eq:loglike}), treating the current membership probabilities as fixed quantities (maximization step). When these parameters are the same for all stars, this amounts to computing the mean values and covariance matrices of each component from the data points, weighting each data point in accordance to its expected membership probability.
\end{itemize} 
This process is guaranteed to converge to a stationary point, although it may well be a local, not the global maximum of likelihood. This EM algorithm was generalized by \citet{Bovy2011} to the case of noisy data, under the name Extreme Deconvolution.
However, we find that the convergence of the EM algorithm is often slow, especially when it is close to a strongly degenerate maximum. Moreover, the maximization step is far more complicated when the parameters of the model are not identical for all stars, but are nonlinear functions of other free parameters and additional attributes of each star. Therefore, we use the standard Nelder--Mead (\textsc{amoeba}) algorithm \citep[Chapter 10.5]{NumRec} for locating the maximum of Equation~\ref{eq:loglike} directly.

We adopt a fixed functional form for the distance-dependent prior membership probability, assuming that the surface density of the cluster members follows a spherical Plummer profile, but allow its scale radius $R_\mathrm{scale}$ to be adjusted during the fit.
Recently \citet{Pace2018} used a similar mixture model to determine the mean PM and membership of ultrafaint dwarf satellites from \Gaia DR2, but they fixed the parameters of their (elliptical) Plummer profiles to the literature values. Of course, structural parameters are also known for most globular clusters, but we do not rely on these data, as the density profile of stars in the \Gaia data (after applying all quality cutoffs) is quite different from the overall density profile (usually the former has a significantly larger scale radius). Hence we do not attach any physical significance to the inferred scale radius, treating it simply as a nuisance parameter that enhances the contrast between members (more numerous in the central parts) and non-members (uniformly spread across the field of view).
As another example, \citet{Walker2009} used a two-component mixture model for the line-of-sight velocity distribution, with the surface density profile of cluster stars represented non-parametrically as an arbitrary non-increasing function of the distance $R_i$, which was determined during the fit.

We also assume that the intrinsic PM dispersion of cluster members is an isotropic Gaussian, with a diagonal matrix $\mathsf{\Sigma}_{1;\,i} = \mathrm{diag}\big[\sigma^2(R_i)\big]$. Here $\sigma(R)$ is a distance-dependent one-dimensional PM dispersion profile, for which we adopt a form that corresponds to a spherical isotropic Plummer model with a scale radius $R_\sigma$: $\sigma(R) = \sigma(0) / \big[ 1 + (R/R_\sigma)^2 \big]^{1/4}$. As mentioned above, the spatial density of member stars is also approximated by a Plummer profile, but we found that their scale radii are typically different. Instead of using two independent free parameters, we fix $R_\sigma = 0.5\,R_\mathrm{scale}$, which adequately represents the intrinsic PM dispersion profiles computed from \textsl{a posteriori} cluster members. The amplitude $\sigma(0)$ remains a free parameter during the fit, but we add a penalty term to the total likelihood (\ref{eq:loglike}) that discourages $\sigma(0)$ from exceeding the central line-of-sight velocity dispersion as quoted in the catalogue of \citet{Baumgardt2019}.
In the present study, we use a single value of $\overline{\boldsymbol{\mu}_1}$ (mean PM of the cluster) and a fixed functional form of $\sigma(R)$, but a more general model for the intrinsic kinematics (flexible rotation and PM dispersion profiles) is explored in the companion paper \citep{Vasiliev2018b}.

After the maximum-likelihood solution has been found, we need to estimate the uncertainties on the derived parameters, in particular, $\overline{\boldsymbol{\mu}_1}$. Since the log-likelihood function (\ref{eq:loglike}) is quadratic near its maximum, we compute its Hessian matrix $\mathsf{H}\equiv \mathrm{d}^2 \ln\mathcal{L} / \mathrm{d}\xi_\alpha\,\mathrm{d}\xi_\beta$, where the vector $\boldsymbol{\xi}$ contains all model parameters. The inverse of the Hessian (with negative sign) is the covariance matrix of parameter uncertainties, of which we are only interested in the $2\times2$ fragment corresponding to $\overline{\boldsymbol{\mu}_1}$. The uncertainties on the mean values of each component in a Gaussian mixture are larger than in the case of a single Gaussian (when they could be computed from data error estimates $\mathsf{E_i}$), because they also encompass the uncertainty of attributing each data point to either component. We also experimented with inferring the uncertainties from a MCMC simulation, and the results were similar.

The above procedure computes the statistical uncertainties, but for the majority of clusters, these are smaller than the systematic errors in \Gaia PM. Unfortunately, it seems impossible to combine the probabilistic membership determination with a proper treatment of spatially correlated systematic errors. Hence we estimate the latter with the method described in the Appendix of the companion paper, using only stars with membership probability $p>0.8$, and list the larger of the two uncertainty estimates (statistical or systematic) in the final catalogue.

The Gaussian mixture modelling is superior to the more conventional techniques such as $\sigma$-clipping, not only because it is a statistically well-defined inference procedure and not just a prescription (cf.\ \citealt{Hogg2010}), but also because it can more robustly isolate a not-too-narrow peak in a crowded background population. However, for it to work correctly, one must include a sufficient number of non-member stars in the input sample. The probability of each data point to belong to either cluster or field component depends on several factors: the geometric distance from the cluster centre, the distance from each component's centre in the PM space (normalized by the inverse of $\mathsf{\Sigma}_c+\mathsf{E}_i$), and the total weight of each component.
Consider an example where the PM distribution of both populations are centered at origin, cluster stars have intrinsic PM dispersion 0.1, and field stars -- 5. If the fraction of field stars is $10^{-3}$, then a star with a unit measurement error would need to be at a distance $\gtrsim 4.2$ from the origin, in order for its membership probability to drop below 0.5; however, if the fraction of contaminants is 50\%, then this star will be classified as a likely field star already when it is $1.8\sigma$ off. Hence, somewhat counter-intuitively, a larger contaminant fraction leads to a cleaner sample. In addition, the implicit assumption is that the field population has a broad enough distribution to capture any non-compliant measurements; however, if there are very few truly non-member stars, the parameters of the field distribution cannot be reliably determined, again distorting the classification.
This is why we sometimes had to consider a rather large area on the sky compared to the cluster half-mass radius, especially for clusters in very sparsely populated regions in the halo.

We provide a \textsc{python} script that implements our approach for membership determination, measures the mean PM and computes various auxiliary quantities (such as actions and peri/apocentre distances) at 
\url{https://github.com/GalacticDynamics-Oxford/GaiaTools}.

\section{Non-parametric estimates of 1d distributions}  \label{sec:nonparafit}

This section explains the method for estimating 1d distributions of various quantities from the observed cluster positions and velocities, taking into account both their measurement uncertainties and finite-sample effects, which is used in Section~\ref{sec:kin}. 
We employ a non-parametric approach similar to the one from \citet{Merritt1997}, and described in detail in \citet[][Section~A.2]{Vasiliev2018a}. The function $f(x;\,\boldsymbol{c})$ to be estimated is represented as a cubic spline with a sufficient number $M$ of pre-defined control points $\{x_m\}_{m=1}^{M}$, and the same number of free parameters $\{c_m\}$ specifying its value at these points, or by some other linear combination of basis functions. The parameters are chosen to maximize the cost function 
\begin{equation}
F(\boldsymbol{c}) \equiv \ln\mathcal{L}(\mathrm{data};\,\boldsymbol{c}) +
\mathcal{P}(\boldsymbol{c}),
\end{equation}
which consists of two terms: one describes the log-likelihood of the observed data, and the other is the penalty for non-smoothness, introduced to prevent overfitting \citep[e.g.,][]{Green1994}.

For the purpose of probability density estimation (e.g., distribution of particles in velocity space or in log-radius, as in side panels of Figure~\ref{fig:veldistr}), it is convenient to let $f(x)$ be the logarithm of the probability density $p(x)$: in this way its positiveness is ensured by construction, and the maximum-likelihood solution is found by unconstrained minimization. Of course, $p(x)$ must be normalized to unity, which may be ensured by adding a normalization term to $\mathcal{L}$:
\begin{equation}
\ln\mathcal{L} \equiv \sum_{i=1}^{N_\mathrm{data}}
f(x_i;\,\boldsymbol{c})
- N_\mathrm{data}\:\int \exp\big[ f(x;\,\boldsymbol{c}) \big] \:\mathrm{d}x .
\end{equation}
Because of the normalization constraint, only $M-1$ of parameters $c_m$ are independent; the remaining one may be fixed e.g.\ to zero. Note that for a spline function $f$, the first term is linear in $\boldsymbol{c}$, but the normalization condition is not.

For estimating the distribution function $f(\theta)$ in phase angle (Figure~\ref{fig:angles}), it is more convenient to represent it a Fourier series with free parameters $\{c_m,s_m\}_{m=1}^M$: 
\begin{equation}
\ln\mathcal{L} \equiv \sum_{i=1}^{N_\mathrm{data}} 
\ln \left[ \frac{1}{2\pi} + \sum_{m=1}^M (c_m\,\cos m\theta \,+\, s_m\,\sin m\theta) \right].
\end{equation}
This automatically ensures correct normalization, but the non-negativity constraint must be imposed explicitly in the fitting procedure.

Finally, for a non-parametric recovery of mean velocity $\overline{v}$ and its dispersion $\sigma$ as functions of radius (Figure~\ref{fig:veldisp}), they are represented as cubic splines in $\ln r$, and the log-likelihood is written as
\begin{equation}
\ln\mathcal{L} \equiv \sum_{i=1}^{N_\mathrm{data}}
\left( -\ln\big[\sqrt{2\pi}\,\sigma(\ln r_i)\big] -
\frac{[v_i-\overline{v}(\ln r_i)]^2}{2\,\sigma(\ln r_i)^2} \right).
\end{equation}
In this case the probability function for data points is a Gaussian, i.e., already normalized, but we need to impose a non-negativity constraint on $\sigma(r)$.

In all these cases, the uncertainties in the data are dealt with in the same way: instead of taking the likelihood of a single data point at a time, we take the average likelihood over all Monte Carlo samples representing its measurement errors.

The penalty term prevents the solution from becoming too wiggly; a standard choice is
\begin{equation}
\mathcal{P}(\boldsymbol{c}) = -\lambda\,\int [f''(x;\,\boldsymbol{c})]^2\,\mathrm{d}x ,
\end{equation}
where $f$ are spline functions for the log-density, or the distribution function in angles, or the radial profile of velocity dispersion.
The smoothing parameter $\lambda$ is determined by the leave-one-out cross-validation procedure. For each value of $\lambda$, we perform $N_\mathrm{data}$ separate maximizations of $F$ for all data points except $i$-th one, and then compute the likelihood of this omitted data point, using the best-fit parameters $\boldsymbol{c}^{(-i)}$ inferred from the remaining points. The cross-validation score is the sum of these likelihoods for all data points, and it typically depends on the parameter $\lambda$ in a non-monotonic way: zero or very small smoothing produces too wiggly fitting functions $f(x;\,\boldsymbol{c}^{(-i)})$, which are poor predictors for the omitted data points, and conversely too much smoothing ignores the data altogether. Therefore, there is an optimal value of $\lambda$ that maximizes the cross-validation score. For this optimal choice, the best-fit function will be rather smooth and nearly independent of the number of free parameters $M$ (terms in the Fourier series or control points in splines), provided that $M$ is large enough to represent all significant variations (not driven by noise). For our 150 clusters, 10 terms are enough in all estimators.

After fixing the value of $\lambda$, we finally find the best-fit parameters $\boldsymbol{c}$ from the entire dataset. The uncertainties due to the finite number of data points are estimated by computing the Hessian matrix $\mathrm{d}^2F/\mathrm{d}c_m\,\mathrm{d}c_n$, the inverse of which (with negative sign) is the covariance matrix of fit parameters $\boldsymbol{c}$. 
As these parameters are not of much physical relevance themselves, we instead plot the median value of the fit function $f(x)$ and its 68\% uncertainty intervals. Namely, we sample the parameters $\boldsymbol{c}$ from a multivariate Gaussian distribution with the given mean (best-fit values) and covariance matrix, plot $f(x;\,\boldsymbol{c})$ for each choice of $c$, and at each value of coordinate $x$ take the interval of values of $f$ containing 16\% to 84\% percentile of plotted curves as the confidence interval.
Alternatively, we computed the uncertainties by running a MCMC simulation and taking the ensemble of parameters $\boldsymbol{c}$ from the chain, instead of sampling them from the inverse Hessian; the results are very similar.

We stress that these non-parametric estimates are used to visualize the trends in the data and the significance of various features, but in comparing the parametric models of Section~\ref{sec:dyn} to the data, we again use the original measurements and their uncertainties represented by Monte Carlo samples, not the fitted non-parametric curves.

\section{The catalogue}  \label{sec:notes}

Table~\ref{tab:pm} lists the mean PM and other parameters for all 150 globular clusters in this study, which are discussed in Section~\ref{sec:pm_results}. The rest of this section contains notes on several individual objects with some particular features.

\textbf{AM 1}, \textbf{Pal 3} and \textbf{Pal 4} are very distant halo clusters ($D\simeq 100$~kpc) in sparsely populated regions on the sky. \Gaia detects several dozen stars in each of them, mostly belonging to the horizontal branch ($G$ magnitudes aroung $20.5-21$, with correspondingly large PM errors), and a few brighter ones in the upper red giant branch. Almost all of them are also identified as possible member stars by \citet{Hilker2006} for the first two clusters, based on photometry, and by \citet{Frank2012} for the last one, based on line-of-sight velocities. The large distance and small number of member stars mean that the statistical errors in transverse velocities are quite large, of order 100~\kms. Interestingly, \citet{Zonoozi2017} predicted an eccentric orbit for Pal~4, based on evolutionary $N$-body simulations, and their prediction for PM is consistent with the actual measurement.

\textbf{BH 176} did not have line-of-sight velocity listed in either \citet{Harris2010} or \citet{Baumgardt2019} catalogues; we took the value from \citet{Dias2016}.

\textbf{Crater (Laevens 1)}, discovered independently by \citet{Belokurov2014} and \citet{Laevens2014}, is the most distant globular cluster in our sample, not listed in the  \citet{Harris2010} catalogue. Follow-up spectroscopic studies of \citet{Kirby2015} and \citet{Voggel2016} determined the line-of-sight velocity ($\sim 149 \pm 1.5$~\kms) and provided the list of likely members, of which 5 are found in \Gaia data with PM close to zero (as could be expected for such a distant object). We identified several more candidate members based on their PM, but the estimate of the statistical uncertainty for the mean PM is still very large, corresponding to an error of $200$~\kms in the transverse velocity. Our PM agree fairly well with the measurement of \citet{Fritz2018}, which is also based on \Gaia.

\textbf{Djorg 1} and \textbf{Terzan 10} are strongly extincted clusters in the direction of Galactic centre. The distances listed in the Harris catalogue are replaced with values measured by \citet{Ortolani2019}, and the line-of-sight velocity for the latter cluster is taken from Geisler et al.(in prep.). 

\textbf{FSR 1716} is another object not present in the \citet{Harris2010} catalogue, which was classified as a globular cluster by \citet{Minniti2017}. It is also located in a strongly extincted region but is firmly detected in the PM space.

\textbf{FSR 1735 (2MASS-GC03)} is a strongly extincted cluster close to the plane of the Galaxy. \citet{CarballoBello2016} measured the metallicity [Fe/H]${}=-0.9\pm 0.2$ and line-of-sight velocity $V_\mathrm{los}=-78\pm12$~\kms from near-infrared spectroscopy of 10 candidate member stars. These stars are also detected in \Gaia. However, the five candidate members suggested in that paper have very disparate PM, hence cannot belong to the cluster. Instead, another four stars (numbered 04, 05, 09 and 10 in their Table 1) are clustered both in the PM space (around $\mu_\alpha=-5, \mu_\delta=0$) and in the colour--magnitude diagram near the tip of the red giant branch; three of them have measured line-of-sight velocities consistent with $V_\mathrm{los}=5\pm 10$~\kms, which we take as the mean value for the cluster. The Gaussian mixture analysis picks up a dozen more stars which likely belong to the cluster and have PM consistent with that tentative value; they also line up nicely in the image plane.

\begin{table*}
\caption{Catalog of PM for the Milky Way globular clusters.\protect\\
Names of clusters with no previous space-based PM measurements (except the independent catalogue of \citealt{Baumgardt2019}) are highlighted in italic.
Coordinates $\alpha$, $\delta$ and the distance $D$ are taken mostly from the \citet{Harris2010} catalogue, with modifications highlighted in italic.
The line-of-sight velocity $V_\mathrm{los}$ and its error estimate $\epsilon_{V_\mathrm{los}}$ are taken from \citet{Baumgardt2019} or other recent papers.
Mean PM $\overline{\mu_\alpha} \equiv [\mathrm{d}\alpha/\mathrm{d}t]\,\cos\delta$, $\overline{\mu_\delta} \equiv \mathrm{d}\delta/\mathrm{d}t$ are derived in this work.
Their uncertainty estimates $\epsilon_{\overline{\mu_\alpha}}$, $\epsilon_{\overline{\mu_\delta}}$ already take into account systematic errors, and $r_{\overline{\mu_\alpha}\,\overline{\mu_\delta}}$ is the correlation coefficient (normalized non-diagonal element in the error covariance matrix).
Last two columns list the angular scale radius $R_\mathrm{scale}$ of the best-fit Plummer profile of member stars in the filtered \Gaia catalogue, and their number.
}
\label{tab:pm}
\begin{tabular}{p{3.2cm} rrrrrrrrrrrr}
Name & $\alpha\;\;\;\;$ & $\delta\;\;\;\;$ & $D\;$ & $V_\mathrm{los}\;\;$ & $\!\epsilon_{V_\mathrm{los}}$ & $\overline{\mu_\alpha}\;\;$ & $\overline{\mu_\delta}\;\;$ & $\epsilon_{\overline{\mu_\alpha}}\;\;$ & $\epsilon_{\overline{\mu_\delta}}\;\;$ & $\!\!r_{\overline{\mu_\alpha}\,\overline{\mu_\delta}}$ & \makebox[5mm]{$R_\mathrm{scale}$} & $N_\star$ \\
     & \multicolumn{2}{c}{\;\footnotesize deg} & \footnotesize kpc & \multicolumn{2}{c}{\footnotesize\kms} & \multicolumn{2}{c}{\footnotesize\masyr} & \multicolumn{2}{c}{\footnotesize\masyr} &  & \makebox[2mm]{\!\!\footnotesize arcmin}  & \\
\hline
NGC 104 (47 Tuc) & 6.024 & -72.081 & 4.5 & -17.21 & 0.18 & 5.237 & -2.524 & 0.039 & 0.039 & -0.002 & 18.3 & 23917 \\
NGC 288 & 13.188 & -26.583 & 8.9 & -44.83 & 0.13 & 4.267 & -5.636 & 0.054 & 0.053 & 0.018 & 6.0 & 2555 \\
NGC 362 & 15.809 & -70.849 & 8.6 & 223.26 & 0.28 & 6.730 & -2.535 & 0.053 & 0.052 & -0.004 & 7.8 & 1741 \\
\textsl{Whiting 1} & 30.738 & -3.253 & 30.1 & -130.41 & 1.79 & -0.234 & -1.782 & 0.115 & 0.094 & -0.050 & 0.5 & 11 \\
\textsl{NGC 1261} & 48.068 & -55.216 & 16.3 & 71.36 & 0.24 & 1.632 & -2.038 & 0.057 & 0.057 & 0.009 & 4.1 & 541 \\
\textsl{Pal 1} & 53.334 & 79.581 & 11.1 & -75.41 & 0.21 & -0.171 & 0.070 & 0.074 & 0.081 & 0.003 & 0.7 & 51 \\
\textsl{E 1 (AM 1)} & 58.760 & -49.615 & \!\!\!123.3 & 118.00 & \!\!\!14.14 & 0.357 & -0.424 & 0.128 & 0.169 & -0.108 & 0.4 & 39 \\
\textsl{Eridanus} & 66.185 & -21.187 & 90.1 & -23.79 & 1.07 & 0.493 & -0.402 & 0.084 & 0.087 & -0.133 & 0.8 & 15 \\
\textsl{Pal 2} & 71.525 & 31.381 & 27.2 & -135.97 & 1.55 & 1.034 & -1.557 & 0.075 & 0.068 & 0.009 & 1.5 & 76 \\
NGC 1851 & 78.528 & -40.047 & 12.1 & 320.30 & 0.25 & 2.120 & -0.589 & 0.054 & 0.054 & -0.007 & 5.5 & 786 \\
NGC 1904 (M 79) & 81.046 & -24.525 & 12.9 & 205.84 & 0.19 & 2.467 & -1.573 & 0.057 & 0.058 & -0.001 & 4.5 & 583 \\
NGC 2298 & 102.248 & -36.005 & 10.8 & 146.18 & 0.71 & 3.316 & -2.186 & 0.061 & 0.061 & 0.001 & 3.2 & 444 \\
\textsl{NGC 2419} & 114.535 & 38.882 & 82.6 & -20.67 & 0.34 & -0.011 & -0.557 & 0.064 & 0.061 & 0.010 & 3.2 & 131 \\
\textsl{Pyxis} & 136.991 & -37.221 & 39.4 & 40.46 & 0.21 & 1.078 & 0.212 & 0.068 & 0.071 & 0.025 & 2.0 & 64 \\
NGC 2808 & 138.013 & -64.864 & 9.6 & 103.90 & 0.30 & 1.005 & 0.274 & 0.051 & 0.051 & -0.006 & 8.4 & 1837 \\
\textsl{E 3 (ESO 37-1)} & 140.238 & -77.282 & 8.1 & 4.93 & 0.42 & -2.695 & 7.115 & 0.064 & 0.064 & -0.009 & 1.9 & 191 \\
\textsl{Pal 3} & 151.383 & 0.072 & 92.5 & 94.04 & 0.80 & 0.055 & -0.085 & 0.146 & 0.171 & -0.515 & 0.7 & 34 \\
NGC 3201 & 154.403 & -46.412 & 4.9 & 494.34 & 0.14 & 8.324 & -1.991 & 0.044 & 0.044 & 0.001 & 12.5 & 7021 \\
\textsl{Pal 4} & 172.320 & 28.974 & \!\!\!108.7 & 72.40 & 0.24 & -0.135 & -0.518 & 0.106 & 0.176 & -0.471 & 0.5 & 32 \\
\textsl{Crater (Laevens 1)} & 174.067 & -10.877 & \!\!\!145.0 & 148.30 & 0.93 & 0.001 & -0.130 & 0.308 & 0.190 & -0.181 & 0.4 & 10 \\
\textsl{NGC 4147} & 182.526 & 18.543 & 19.3 & 179.52 & 0.33 & -1.705 & -2.114 & 0.065 & 0.063 & -0.014 & 2.2 & 248 \\
NGC 4372 & 186.439 & -72.659 & 5.8 & 75.59 & 0.30 & -6.378 & 3.358 & 0.050 & 0.050 & 0.001 & 7.8 & 2189 \\
\textsl{Rup 106} & 189.667 & -51.150 & 21.2 & -38.42 & 0.30 & -1.263 & 0.399 & 0.064 & 0.063 & 0.027 & 2.1 & 119 \\
NGC 4590 (M 68) & 189.867 & -26.744 & 10.3 & -92.99 & 0.22 & -2.752 & 1.762 & 0.054 & 0.053 & -0.027 & 6.7 & 1479 \\
NGC 4833 & 194.891 & -70.876 & 6.6 & 201.99 & 0.40 & -8.361 & -0.949 & 0.055 & 0.054 & 0.006 & 7.1 & 1861 \\
NGC 5024 (M 53) & 198.230 & 18.168 & 17.9 & -62.85 & 0.26 & -0.148 & -1.355 & 0.053 & 0.052 & -0.025 & 7.7 & 1924 \\
NGC 5053 & 199.113 & 17.700 & 17.4 & 42.77 & 0.25 & -0.366 & -1.248 & 0.058 & 0.056 & -0.021 & 4.0 & 657 \\
NGC 5139 ($\omega$ Cen) & 201.697 & -47.480 & 5.2 & 234.28 & 0.24 & -3.234 & -6.719 & 0.039 & 0.039 & -0.002 & 25.2 & 23635 \\
NGC 5272 (M 3) & 205.548 & 28.377 & 10.2 & -147.28 & 0.34 & -0.142 & -2.647 & 0.045 & 0.043 & -0.006 & 9.5 & 3924 \\
NGC 5286 & 206.612 & -51.374 & 11.7 & 62.38 & 0.40 & 0.207 & -0.111 & 0.059 & 0.059 & -0.002 & 4.1 & 431 \\
NGC 5466 & 211.364 & 28.534 & 16.0 & 106.93 & 0.18 & -5.412 & -0.800 & 0.053 & 0.053 & 0.010 & 4.8 & 988 \\
NGC 5634 & 217.405 & -5.976 & 25.2 & -16.07 & 0.60 & -1.724 & -1.507 & 0.064 & 0.064 & -0.002 & 2.3 & 188 \\
\textsl{NGC 5694} & 219.901 & -26.539 & 35.0 & -139.55 & 0.49 & -0.486 & -1.071 & 0.069 & 0.068 & -0.060 & 2.1 & 100 \\
\textsl{IC 4499} & 225.077 & -82.214 & 18.8 & 38.41 & 0.31 & 0.491 & -0.485 & 0.059 & 0.059 & 0.006 & 3.4 & 387 \\
\textsl{NGC 5824} & 225.994 & -33.068 & 32.1 & -25.24 & 0.52 & -1.170 & -2.226 & 0.060 & 0.058 & 0.012 & 5.0 & 373 \\
\textsl{Pal 5} & 229.022 & -0.112 & 23.2 & -58.60 & 0.21 & -2.736 & -2.646 & 0.064 & 0.064 & -0.048 & 3.3 & 203 \\
NGC 5897 & 229.352 & -21.010 & 12.5 & 101.31 & 0.22 & -5.427 & -3.438 & 0.057 & 0.056 & -0.014 & 4.8 & 1002 \\
NGC 5904 (M 5) & 229.638 & 2.081 & 7.5 & 53.70 & 0.25 & 4.078 & -9.854 & 0.047 & 0.047 & -0.008 & 10.0 & 4502 \\
NGC 5927 & 232.003 & -50.673 & 7.7 & -104.07 & 0.28 & -5.049 & -3.231 & 0.055 & 0.055 & -0.003 & 11.0 & 1829 \\
NGC 5946 & 233.869 & -50.660 & 10.6 & 137.41 & 1.42 & -5.331 & -1.614 & 0.065 & 0.064 & -0.008 & 2.9 & 217 \\
\textsl{BH 176 (ESO 224-8)} & 234.781 & -50.053 & 18.9 & \textsl{-6.00} & \!\!\!\textsl{14.00} & -4.002 & -3.064 & 0.071 & 0.067 & 0.019 & 2.3 & 115 \\
NGC 5986 & 236.512 & -37.786 & 10.4 & 101.18 & 0.43 & -4.186 & -4.604 & 0.060 & 0.059 & -0.011 & 5.6 & 812 \\
\textsl{FSR 1716} & 242.625 & -53.748 & 9.5 & -33.14 & 1.01 & -4.527 & -8.639 & 0.066 & 0.064 & -0.028 & 2.3 & 67 \\
\textsl{Pal 14 (Arp 1)} & 242.752 & 14.958 & 76.5 & 72.30 & 0.14 & -0.504 & -0.461 & 0.081 & 0.078 & 0.167 & 1.8 & 61 \\
\textsl{BH 184 (Lynga 7)} & 242.765 & -55.318 & 8.0 & 17.86 & 0.83 & -3.844 & -7.039 & 0.064 & 0.063 & -0.007 & 2.2 & 154 \\
NGC 6093 (M 80) & 244.260 & -22.976 & 10.0 & 10.93 & 0.39 & -2.931 & -5.578 & 0.061 & 0.060 & 0.001 & 4.3 & 474 \\
NGC 6121 (M 4) & 245.897 & -26.526 & 2.2 & 71.05 & 0.08 & \!\!\!-12.490 & \!\!\!-19.001 & 0.044 & 0.044 & -0.001 & 8.6 & 7526 \\
\textsl{NGC 6101} & 246.451 & -72.202 & 15.4 & 366.33 & 0.32 & 1.757 & -0.223 & 0.053 & 0.054 & 0.005 & 6.1 & 1260 \\
NGC 6144 & 246.808 & -26.023 & 8.9 & 195.74 & 0.74 & -1.772 & -2.626 & 0.061 & 0.060 & 0.002 & 2.0 & 122 \\
\textsl{NGC 6139} & 246.918 & -38.849 & 10.1 & 24.41 & 0.95 & -6.184 & -2.648 & 0.062 & 0.061 & 0.006 & 3.6 & 426 \\
\textsl{Terzan 3} & 247.167 & -35.353 & 8.2 & -135.76 & 0.57 & -5.602 & -1.690 & 0.063 & 0.062 & 0.006 & 2.9 & 365 \\
NGC 6171 (M 107) & 248.133 & -13.054 & 6.4 & -34.68 & 0.19 & -1.924 & -5.968 & 0.057 & 0.056 & 0.000 & 4.8 & 1304 \\
\textsl{ESO 452-11 (1636-283)} & 249.856 & -28.399 & 8.3 & 16.27 & 0.48 & -1.540 & -6.418 & 0.070 & 0.068 & -0.016 & 0.9 & 28 \\
NGC 6205 (M 13) & 250.422 & 36.460 & 7.1 & -244.49 & 0.43 & -3.164 & -2.588 & 0.047 & 0.047 & 0.011 & 10.3 & 3982 \\
\textsl{NGC 6229} & 251.745 & 47.528 & 30.5 & -138.64 & 0.77 & -1.192 & -0.440 & 0.062 & 0.064 & 0.009 & 2.2 & 123 \\
NGC 6218 (M 12) & 251.809 & -1.949 & 4.8 & -41.35 & 0.20 & -0.141 & -6.802 & 0.052 & 0.051 & 0.019 & 6.8 & 3356 \\
\makebox[0pt][l]{\textsl{FSR 1735 (2MASS-GC03)}} & 253.044 & -47.058 & 10.8 & \textsl{5.00} & \!\!\!\textsl{10.00} & -5.015 & -0.141 & 0.112 & 0.103 & -0.134 & 0.8 & 16 \\
\hline
\end{tabular}
\end{table*}
\begin{table*}
\contcaption{Catalog of PM for the Milky Way globular clusters}
\begin{tabular}{p{3.2cm} rrrrrrrrrrrr}
Name & $\alpha\;\;\;\;$ & $\delta\;\;\;\;$ & $D\;$ & $V_\mathrm{los}\;\;$ & $\!\epsilon_{V_\mathrm{los}}$ & $\overline{\mu_\alpha}\;\;$ & $\overline{\mu_\delta}\;\;$ & $\epsilon_{\overline{\mu_\alpha}}\;\;$ & $\epsilon_{\overline{\mu_\delta}}\;\;$ & $\!\!r_{\overline{\mu_\alpha}\,\overline{\mu_\delta}}$ & \makebox[5mm]{$R_\mathrm{scale}$} & $N_\star$ \\
\hline
NGC 6235 & 253.355 & -22.177 & 11.5 & 126.68 & 0.33 & -3.973 & -7.624 & 0.064 & 0.063 & 0.009 & 2.1 & 168 \\
NGC 6254 (M 10) & 254.288 & -4.100 & 4.4 & 74.02 & 0.31 & -4.759 & -6.554 & 0.049 & 0.048 & 0.012 & 8.2 & 5285 \\
\textsl{NGC 6256} & 254.886 & -37.121 & 10.3 & -101.37 & 1.19 & -3.664 & -1.493 & 0.068 & 0.066 & 0.005 & 1.9 & 126 \\
\textsl{Pal 15} & 254.963 & -0.539 & 45.1 & 72.27 & 1.74 & -0.580 & -0.861 & 0.076 & 0.070 & 0.027 & 1.8 & 60 \\
NGC 6266 (M 62) & 255.303 & -30.114 & 6.8 & -73.49 & 0.70 & -5.041 & -2.952 & 0.057 & 0.057 & 0.004 & 6.9 & 1169 \\
NGC 6273 (M 19) & 255.657 & -26.268 & 8.8 & 145.54 & 0.59 & -3.232 & 1.669 & 0.057 & 0.056 & 0.010 & 6.2 & 1059 \\
NGC 6284 & 256.119 & -24.765 & 15.3 & 28.26 & 0.93 & -3.196 & -2.008 & 0.065 & 0.064 & 0.009 & 2.5 & 141 \\
NGC 6287 & 256.288 & -22.708 & 9.4 & -294.74 & 1.65 & -4.977 & -1.882 & 0.063 & 0.062 & 0.003 & 2.8 & 338 \\
NGC 6293 & 257.543 & -26.582 & 9.5 & -143.66 & 0.39 & 0.890 & -4.338 & 0.064 & 0.063 & 0.007 & 2.4 & 182 \\
NGC 6304 & 258.634 & -29.462 & 5.9 & -108.62 & 0.39 & -4.051 & -1.073 & 0.063 & 0.063 & 0.004 & 2.8 & 159 \\
NGC 6316 & 259.155 & -28.140 & 10.4 & 99.81 & 0.82 & -4.945 & -4.608 & 0.066 & 0.065 & 0.012 & 2.5 & 114 \\
NGC 6341 (M 92) & 259.281 & 43.136 & 8.3 & -120.48 & 0.27 & -4.925 & -0.536 & 0.052 & 0.052 & 0.010 & 7.4 & 1847 \\
NGC 6325 & 259.497 & -23.766 & 7.8 & 29.54 & 0.58 & -8.426 & -9.011 & 0.068 & 0.067 & 0.008 & 1.7 & 147 \\
NGC 6333 (M 9) & 259.797 & -18.516 & 7.9 & 310.75 & 2.12 & -2.228 & -3.214 & 0.059 & 0.058 & 0.009 & 4.7 & 526 \\
NGC 6342 & 260.292 & -19.587 & 8.5 & 116.56 & 0.74 & -2.932 & -7.101 & 0.064 & 0.064 & 0.004 & 2.2 & 195 \\
NGC 6356 & 260.896 & -17.813 & 15.1 & 38.93 & 1.88 & -3.814 & -3.381 & 0.060 & 0.059 & 0.010 & 4.7 & 547 \\
\textsl{NGC 6355} & 260.994 & -26.353 & 9.2 & -194.13 & 0.83 & -4.657 & -0.522 & 0.066 & 0.065 & 0.014 & 1.6 & 93 \\
NGC 6352 & 261.371 & -48.422 & 5.6 & -125.63 & 1.01 & -2.172 & -4.398 & 0.056 & 0.056 & 0.003 & 5.4 & 2064 \\
\textsl{IC 1257} & 261.785 & -7.093 & 25.0 & -137.97 & 2.04 & -0.928 & -1.407 & 0.087 & 0.081 & 0.037 & 0.8 & 16 \\
\textsl{Terzan 2 (HP 3)} & 261.888 & -30.802 & 7.5 & 128.96 & 1.18 & -2.237 & -6.210 & 0.081 & 0.075 & 0.086 & 0.8 & 25 \\
NGC 6366 & 261.934 & -5.080 & 3.5 & -120.65 & 0.19 & -0.363 & -5.115 & 0.053 & 0.053 & 0.015 & 6.2 & 2226 \\
\textsl{Terzan 4 (HP 4)} & 262.663 & -31.596 & 7.2 & -39.93 & 3.76 & -5.386 & -3.361 & 0.088 & 0.080 & 0.020 & 0.6 & 26 \\
\textsl{BH 229 (HP 1)} & 262.772 & -29.982 & 8.2 & 40.61 & 1.29 & 2.462 & \!\!\!-10.142 & 0.073 & 0.071 & -0.001 & 1.4 & 35 \\
NGC 6362 & 262.979 & -67.048 & 7.6 & -14.58 & 0.18 & -5.510 & -4.750 & 0.051 & 0.052 & -0.000 & 6.5 & 2877 \\
NGC 6380 (Ton 1) & 263.617 & -39.069 & 10.9 & -6.54 & 1.48 & -2.142 & -3.107 & 0.067 & 0.066 & -0.006 & 2.1 & 181 \\
\textsl{Terzan 1 (HP 2)} & 263.949 & \textsl{-30.481} & 6.7 & 57.55 & 1.61 & -2.967 & -4.811 & 0.083 & 0.080 & 0.026 & 1.1 & 65 \\
\textsl{Ton 2 (Pismis 26)} & 264.044 & -38.553 & 8.2 & -184.72 & 1.12 & -5.912 & -0.548 & 0.067 & 0.066 & 0.013 & 1.9 & 116 \\
NGC 6388 & 264.072 & -44.736 & 9.9 & 82.85 & 0.48 & -1.331 & -2.672 & 0.057 & 0.057 & 0.004 & 5.6 & 791 \\
NGC 6402 (M 14) & 264.400 & -3.246 & 9.3 & -60.71 & 0.45 & -3.640 & -5.035 & 0.058 & 0.058 & 0.008 & 4.8 & 718 \\
NGC 6401 & 264.652 & -23.910 & 10.6 & -99.26 & 3.18 & -2.849 & 1.476 & 0.070 & 0.069 & -0.001 & 2.0 & 60 \\
NGC 6397 & 265.175 & -53.674 & 2.3 & 18.39 & 0.10 & 3.285 & \!\!\!-17.621 & 0.043 & 0.043 & 0.001 & 11.1 & 11406 \\
\textsl{Pal 6} & 265.926 & -26.223 & 5.8 & 176.28 & 1.53 & -9.256 & -5.330 & 0.078 & 0.076 & 0.038 & 1.3 & 47 \\
\textsl{NGC 6426} & 266.228 & 3.170 & 20.6 & -210.51 & 0.51 & -1.862 & -2.994 & 0.064 & 0.064 & 0.019 & 2.1 & 144 \\
\textsl{Djorg 1} & 266.868 & -33.066 & \textsl{9.3} & -359.81 & 1.98 & -5.158 & -8.323 & 0.081 & 0.076 & -0.006 & 0.9 & 28 \\
\textsl{Terzan 5 (Terzan 11)} & 267.020 & -24.779 & 6.9 & -81.40 & 1.36 & -1.560 & -4.724 & 0.106 & 0.102 & 0.072 & 2.0 & 59 \\
NGC 6440 & 267.220 & -20.360 & 8.5 & -69.39 & 0.93 & -1.070 & -3.828 & 0.069 & 0.068 & 0.025 & 2.9 & 154 \\
NGC 6441 & 267.554 & -37.051 & 11.6 & 17.27 & 0.93 & -2.568 & -5.322 & 0.059 & 0.059 & 0.001 & 6.2 & 247 \\
\textsl{Terzan 6 (HP 5)} & 267.693 & -31.275 & 6.8 & 137.15 & 1.70 & -5.634 & -7.042 & 0.186 & 0.156 & 0.155 & 1.1 & 24 \\
NGC 6453 & 267.715 & -34.599 & 11.6 & -91.16 & 3.08 & 0.165 & -5.895 & 0.071 & 0.070 & 0.006 & 1.8 & 30 \\
NGC 6496 & 269.765 & -44.266 & 11.3 & -134.72 & 0.26 & -3.037 & -9.239 & 0.061 & 0.061 & 0.004 & 2.8 & 350 \\
\textsl{Terzan 9} & 270.412 & -26.840 & 7.1 & 29.31 & 2.96 & -2.197 & -7.451 & 0.073 & 0.070 & 0.041 & 0.9 & 50 \\
\textsl{Djorg 2 (ESO 456-38)} & 270.455 & -27.826 & 6.3 & -148.05 & 1.38 & 0.515 & -3.052 & 0.079 & 0.074 & -0.036 & 0.9 & 9 \\
NGC 6517 & 270.461 & -8.959 & 10.6 & -37.07 & 1.68 & -1.498 & -4.221 & 0.066 & 0.066 & 0.016 & 2.5 & 237 \\
\textsl{Terzan 10} & \textsl{270.742} & -26.067 & \textsl{10.4} & \textsl{208.00} & \!\!\!\textsl{3.60} & -6.912 & -2.409 & 0.085 & 0.080 & 0.074 & 1.6 & 27 \\
NGC 6522 & 270.892 & -30.034 & 7.7 & -13.90 & 0.71 & 2.618 & -6.431 & 0.072 & 0.071 & 0.003 & 1.9 & 38 \\
NGC 6535 & 270.960 & -0.298 & 6.8 & -214.85 & 0.46 & -4.249 & -2.900 & 0.064 & 0.064 & 0.007 & 2.2 & 318 \\
NGC 6528 & 271.207 & -30.056 & 7.9 & 210.31 & 0.75 & -2.327 & -5.527 & 0.074 & 0.071 & 0.018 & 0.9 & 17 \\
NGC 6539 & 271.207 & -7.586 & 7.8 & 35.69 & 0.55 & -6.865 & -3.477 & 0.061 & 0.061 & 0.009 & 3.2 & 563 \\
\textsl{NGC 6540 (Djorg 3)} & 271.536 & -27.765 & 5.3 & -17.98 & 0.84 & -3.760 & -2.799 & 0.077 & 0.076 & 0.029 & 0.8 & 15 \\
NGC 6544 & 271.836 & -24.997 & 3.0 & -38.12 & 0.76 & -2.349 & \!\!\!-18.557 & 0.060 & 0.060 & 0.001 & 5.2 & 551 \\
NGC 6541 & 272.010 & -43.715 & 7.5 & -163.97 & 0.46 & 0.349 & -8.843 & 0.056 & 0.055 & -0.007 & 6.3 & 1375 \\
\textsl{ESO 280-06} & 272.275 & -46.423 & 21.4 & 93.20 & 0.34 & -0.552 & -2.724 & 0.082 & 0.075 & 0.085 & 1.1 & 33 \\
\textsl{NGC 6553} & 272.323 & -25.909 & 6.0 & 0.72 & 0.40 & 0.246 & -0.409 & 0.063 & 0.063 & 0.003 & 3.5 & 325 \\
\textsl{NGC 6558} & 272.573 & -31.764 & 7.4 & -195.70 & 0.70 & -1.810 & -4.133 & 0.067 & 0.066 & 0.003 & 1.1 & 42 \\
\textsl{Pal 7 (IC 1276)} & 272.684 & -7.208 & 5.4 & 155.06 & 0.69 & -2.577 & -4.374 & 0.059 & 0.059 & 0.009 & 4.2 & 938 \\
\textsl{Terzan 12} & 273.066 & -22.742 & 4.8 & 94.77 & 0.97 & -6.151 & -2.679 & 0.069 & 0.068 & 0.049 & 1.5 & 104 \\
\textsl{NGC 6569} & 273.412 & -31.827 & 10.9 & -49.83 & 0.50 & -4.109 & -7.267 & 0.062 & 0.062 & 0.005 & 2.7 & 169 \\
\textsl{BH 261 (ESO 456-78)} & 273.527 & -28.635 & 6.5 & -29.38 & 0.60 & 3.590 & -3.573 & 0.077 & 0.076 & 0.037 & 0.5 & 13 \\
\textsl{NGC 6584} & 274.657 & -52.216 & 13.5 & 260.64 & 1.58 & -0.053 & -7.185 & 0.061 & 0.061 & -0.000 & 3.6 & 392 \\
\textsl{NGC 6624} & 275.919 & -30.361 & 7.9 & 54.26 & 0.45 & 0.099 & -6.904 & 0.065 & 0.064 & 0.002 & 2.0 & 162 \\
NGC 6626 (M 28) & 276.137 & -24.870 & 5.5 & 11.11 & 0.60 & -0.301 & -8.913 & 0.061 & 0.061 & -0.003 & 3.7 & 383 \\
\textsl{NGC 6638} & 277.734 & -25.497 & 9.4 & 8.63 & 2.00 & -2.550 & -4.075 & 0.066 & 0.065 & 0.013 & 1.9 & 103 \\
NGC 6637 (M 69) & 277.846 & -32.348 & 8.8 & 46.63 & 1.45 & -5.113 & -5.813 & 0.063 & 0.063 & 0.011 & 2.5 & 283 \\
\textsl{NGC 6642} & 277.975 & -23.475 & 8.1 & -33.23 & 1.13 & -0.189 & -3.898 & 0.065 & 0.065 & 0.004 & 1.8 & 93 \\
\hline
\end{tabular}
\end{table*}
\begin{table*}
\contcaption{Catalog of PM for the Milky Way globular clusters}
\begin{tabular}{p{3.2cm} rrrrrrrrrrrr}
Name & $\alpha\;\;\;\;$ & $\delta\;\;\;\;$ & $D\;$ & $V_\mathrm{los}\;\;$ & $\!\epsilon_{V_\mathrm{los}}$ & $\overline{\mu_\alpha}\;\;$ & $\overline{\mu_\delta}\;\;$ & $\epsilon_{\overline{\mu_\alpha}}\;\;$ & $\epsilon_{\overline{\mu_\delta}}\;\;$ & $\!\!r_{\overline{\mu_\alpha}\,\overline{\mu_\delta}}$ & \makebox[5mm]{$R_\mathrm{scale}$} & $N_\star$ \\
\hline
\textsl{NGC 6652} & 278.940 & -32.991 & 10.0 & -99.04 & 0.51 & -5.506 & -4.204 & 0.064 & 0.063 & 0.001 & 2.4 & 225 \\
NGC 6656 (M 22) & 279.100 & -23.905 & 3.2 & -147.76 & 0.30 & 9.833 & -5.557 & 0.047 & 0.047 & 0.008 & 11.8 & 4178 \\
\textsl{Pal 8} & 280.375 & -19.826 & 12.8 & -41.14 & 1.81 & -2.031 & -5.634 & 0.063 & 0.063 & 0.015 & 2.4 & 174 \\
NGC 6681 (M 70) & 280.803 & -32.292 & 9.0 & 216.62 & 0.84 & 1.458 & -4.688 & 0.060 & 0.060 & 0.009 & 4.0 & 686 \\
\textsl{NGC 6712} & 283.268 & -8.706 & 6.9 & -107.45 & 0.29 & 3.341 & -4.384 & 0.062 & 0.061 & 0.006 & 2.8 & 309 \\
NGC 6715 (M 54) & 283.764 & -30.480 & 26.5 & 143.06 & 0.56 & -2.711 & -1.355 & 0.058 & 0.058 & 0.009 & 7.9 & 854 \\
\textsl{NGC 6717 (Pal 9)} & 283.775 & -22.701 & 7.1 & 32.45 & 1.44 & -3.106 & -4.951 & 0.065 & 0.064 & 0.022 & 2.0 & 97 \\
\textsl{NGC 6723} & 284.888 & -36.632 & 8.7 & -94.18 & 0.26 & 1.033 & -2.445 & 0.057 & 0.057 & 0.004 & 5.2 & 1824 \\
\textsl{NGC 6749} & 286.314 & 1.901 & 7.9 & -58.44 & 0.96 & -2.865 & -5.987 & 0.061 & 0.061 & 0.011 & 4.3 & 395 \\
NGC 6752 & 287.717 & -59.985 & 4.0 & -26.28 & 0.16 & -3.170 & -4.043 & 0.042 & 0.042 & 0.004 & 12.3 & 10850 \\
\textsl{NGC 6760} & 287.800 & 1.030 & 7.4 & -0.42 & 1.63 & -1.129 & -3.561 & 0.060 & 0.059 & 0.009 & 4.2 & 572 \\
NGC 6779 (M 56) & 289.148 & 30.183 & 9.4 & -136.97 & 0.45 & -2.020 & 1.644 & 0.059 & 0.059 & 0.004 & 4.5 & 811 \\
\textsl{Terzan 7} & 289.433 & -34.658 & 22.8 & 159.45 & 0.14 & -2.999 & -1.586 & 0.068 & 0.067 & 0.032 & 1.5 & 93 \\
\textsl{Pal 10} & 289.509 & 18.572 & 5.9 & -31.70 & 0.23 & -4.250 & -6.924 & 0.064 & 0.064 & 0.005 & 2.2 & 246 \\
\textsl{Arp 2} & 292.184 & -30.356 & 28.6 & 123.01 & 0.33 & -2.381 & -1.510 & 0.067 & 0.066 & 0.013 & 1.9 & 137 \\
NGC 6809 (M 55) & 294.999 & -30.965 & 5.4 & 174.40 & 0.24 & -3.420 & -9.269 & 0.050 & 0.050 & 0.010 & 8.8 & 4224 \\
\textsl{Terzan 8} & 295.435 & -33.999 & 26.3 & 148.53 & 0.17 & -2.472 & -1.556 & 0.062 & 0.060 & -0.006 & 2.7 & 251 \\
\textsl{Pal 11} & 296.310 & -8.007 & 13.4 & -67.64 & 0.76 & -1.821 & -4.934 & 0.066 & 0.064 & 0.027 & 1.8 & 174 \\
NGC 6838 (M 71) & 298.444 & 18.779 & 4.0 & -22.27 & 0.19 & -3.415 & -2.614 & 0.054 & 0.054 & 0.005 & 5.1 & 2676 \\
NGC 6864 (M 75) & 301.520 & -21.921 & 20.9 & -189.08 & 1.12 & -0.559 & -2.798 & 0.062 & 0.061 & 0.009 & 2.8 & 224 \\
\textsl{NGC 6934} & 308.547 & 7.404 & 15.6 & -406.22 & 0.73 & -2.636 & -4.667 & 0.060 & 0.060 & 0.009 & 4.0 & 336 \\
NGC 6981 (M 72) & 313.365 & -12.537 & 17.0 & -331.39 & 1.47 & -1.233 & -3.290 & 0.062 & 0.061 & 0.016 & 3.0 & 350 \\
\textsl{NGC 7006} & 315.372 & 16.187 & 41.2 & -383.47 & 0.73 & -0.102 & -0.569 & 0.067 & 0.068 & -0.013 & 2.0 & 137 \\
NGC 7078 (M 15) & 322.493 & 12.167 & 10.4 & -106.76 & 0.25 & -0.643 & -3.763 & 0.051 & 0.051 & 0.001 & 9.9 & 2198 \\
NGC 7089 (M 2) & 323.363 & -0.823 & 11.5 & -3.72 & 0.34 & 3.518 & -2.145 & 0.054 & 0.054 & 0.013 & 6.2 & 1369 \\
NGC 7099 (M 30) & 325.092 & -23.180 & 8.1 & -185.19 & 0.17 & -0.694 & -7.271 & 0.055 & 0.054 & 0.031 & 5.8 & 1414 \\
\textsl{Pal 12} & 326.662 & -21.253 & 19.0 & 27.91 & 0.28 & -3.249 & -3.303 & 0.067 & 0.067 & 0.034 & 1.5 & 99 \\
\textsl{Pal 13} & 346.685 & 12.772 & 26.0 & 25.87 & 0.27 & 1.615 & 0.142 & 0.101 & 0.089 & 0.016 & 0.8 & 34 \\
\textsl{NGC 7492} & 347.111 & -15.611 & 26.3 & -176.70 & 0.27 & 0.799 & -2.273 & 0.065 & 0.064 & 0.009 & 1.5 & 125 \\
\hline
\end{tabular}
\end{table*}

\begin{figure*}
\caption{
Colour--magnitude diagrams (left column), spatial distribution (middle column) and PM distribution (right column) of stars in individual clusters.
Membership probability is marked by colour (red -- likely members, gray -- field stars); the number of members and the total number of stars in the sample is given in brackets.
Green crosses indicate likely member stars listed in other papers. 
Isochrone tracks for the assumed age, metallicity and distance are plotted in the left panel by blue dots (PARSEC, \citealt{Bressan2012}) and green dots (MIST, \citealt{Choi2016}), with extinction and reddening computed using the coefficients from Table~1 in \citet{Babusiaux2018}.
Cyan histogram in the central panel shows the distribution of membership probabilities $p$ in 10 bins (leftmost -- $p\le 10\%$, rightmost -- $p\ge 90\%$); a strongly bimodal distribution indicates a good separation of cluster and field stars in the PM space. Green dot-dashed circle shows the half-light radius from \citet{Baumgardt2019}, and blue dashed circle -- the scale radius $R_\mathrm{scale}$ of the Plummer profile of member stars in the filtered \Gaia catalogue, which is typically larger than the true half-light radius because the catalogue is incomplete in the central regions.
Blue ellipse in the right panel indicates the uncertainty on the mean PM, summed in quadrature with the internal PM dispersion $\sigma_\mu$. Green crosses indicate the mean PM measured by \HST, and pluses -- by \citetalias{Helmi2018}.\protect\\
The plots for all 150 clusters are presented as supplementary material.
}  \label{fig:individual_clusters}
\end{figure*}

\renewcommand{\floatpagefraction}{.66}  
\clearpage\begin{figure*}
\contcaption{}
\includegraphics{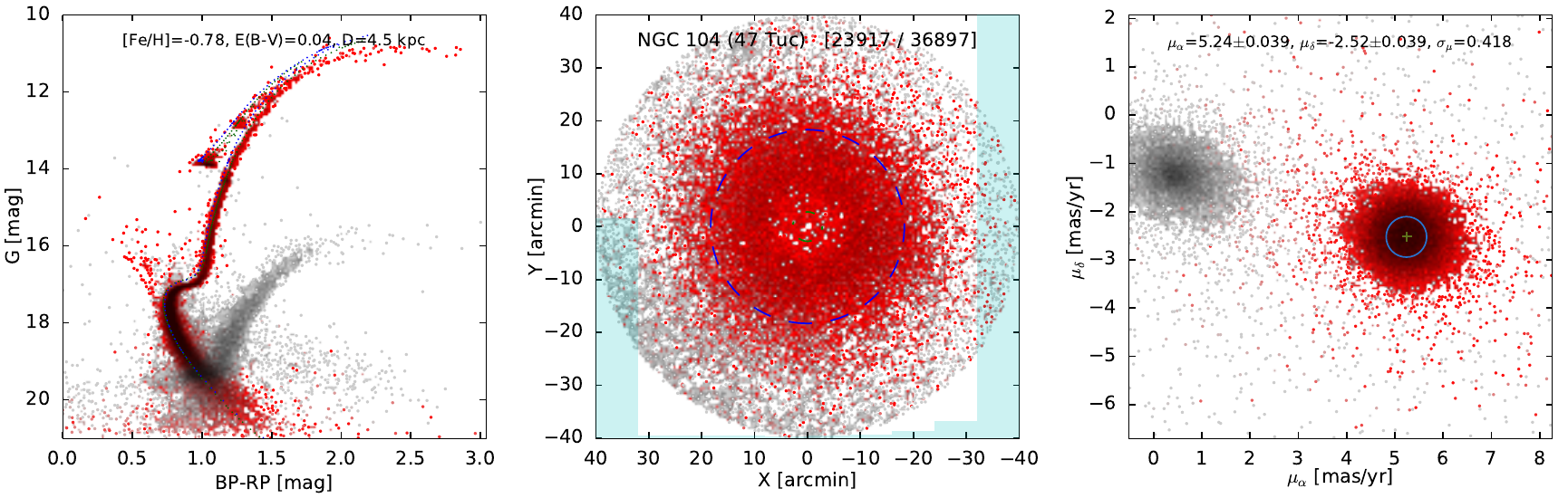}
\includegraphics{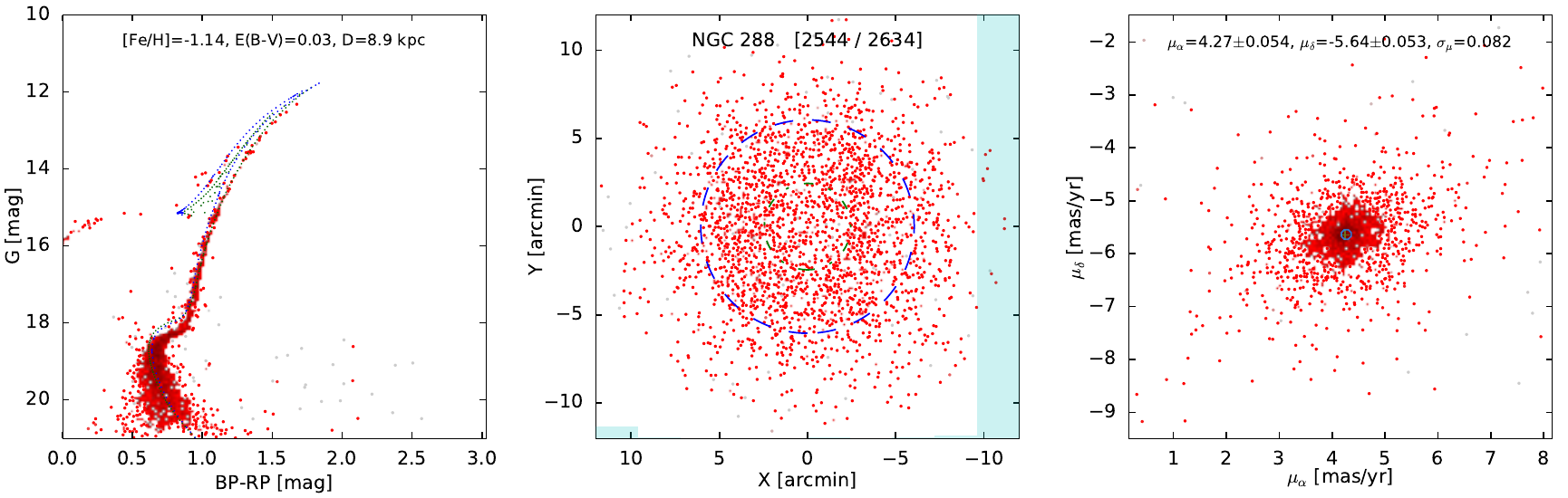}
\includegraphics{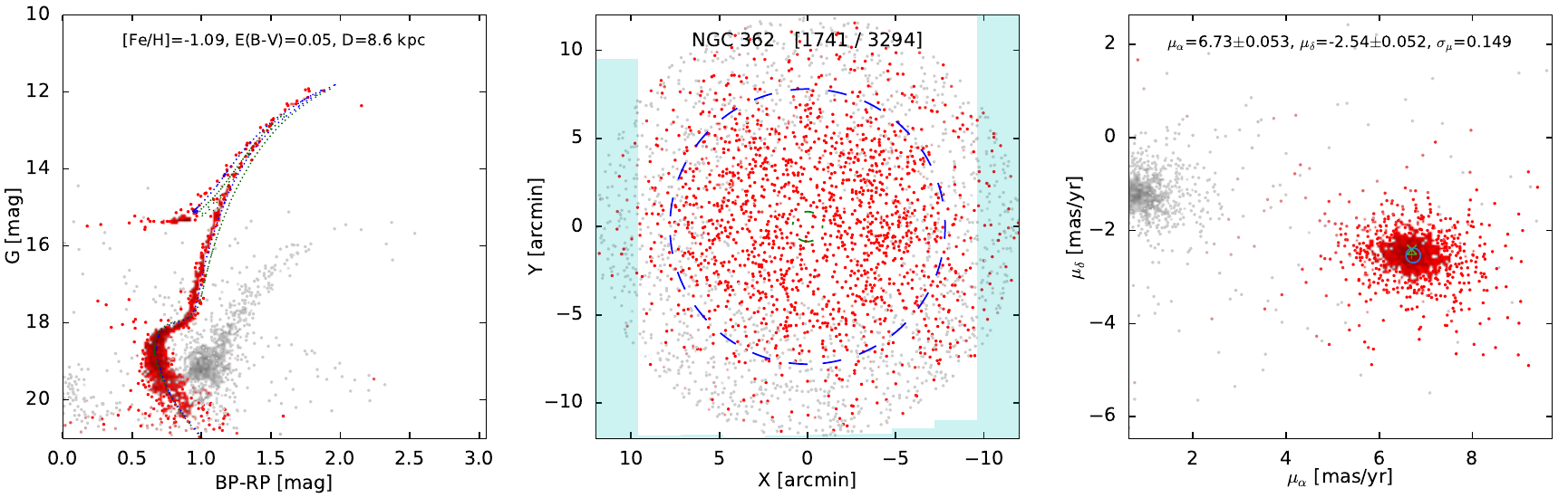}
\includegraphics{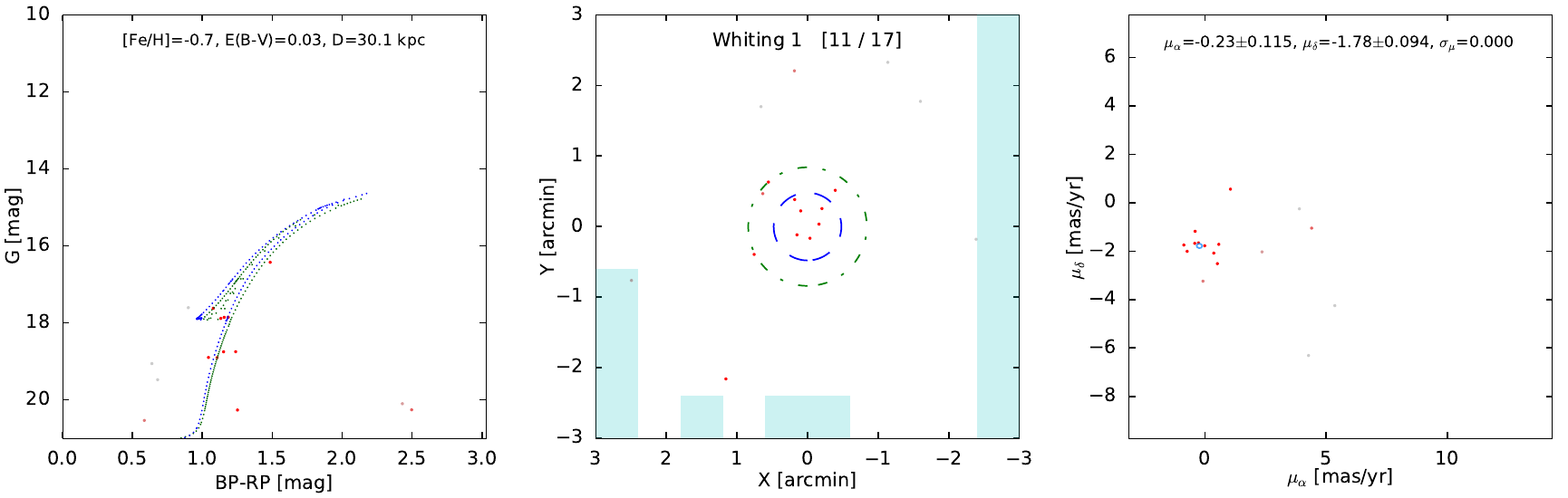}
\end{figure*}

\clearpage\begin{figure*}
\contcaption{}
\includegraphics{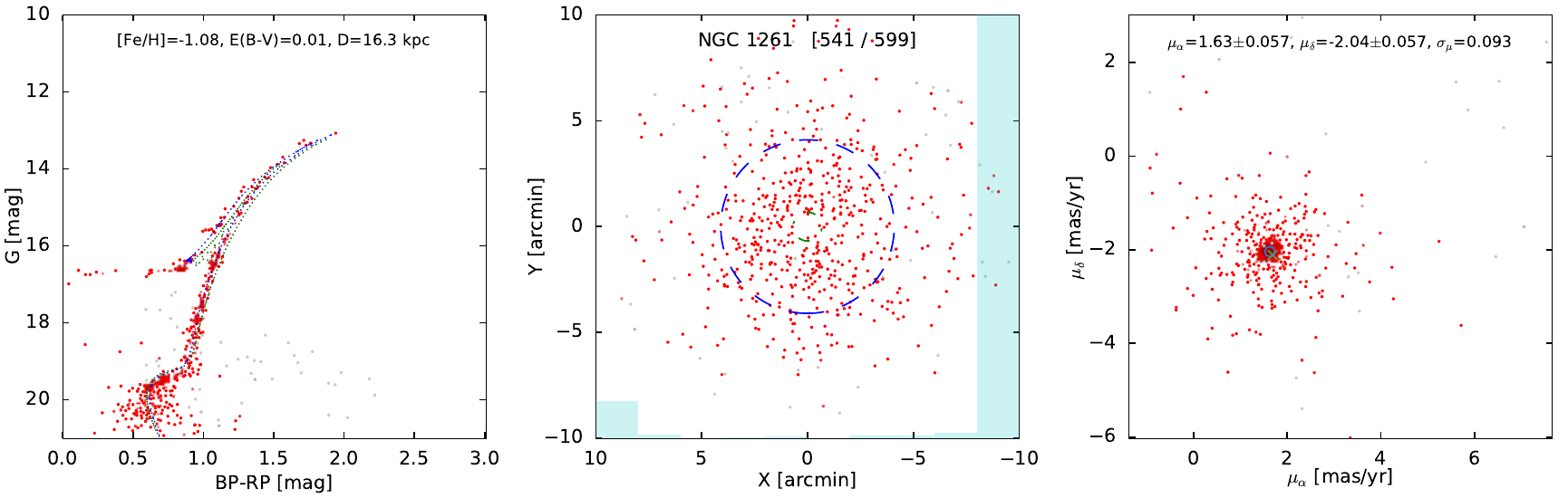}
\includegraphics{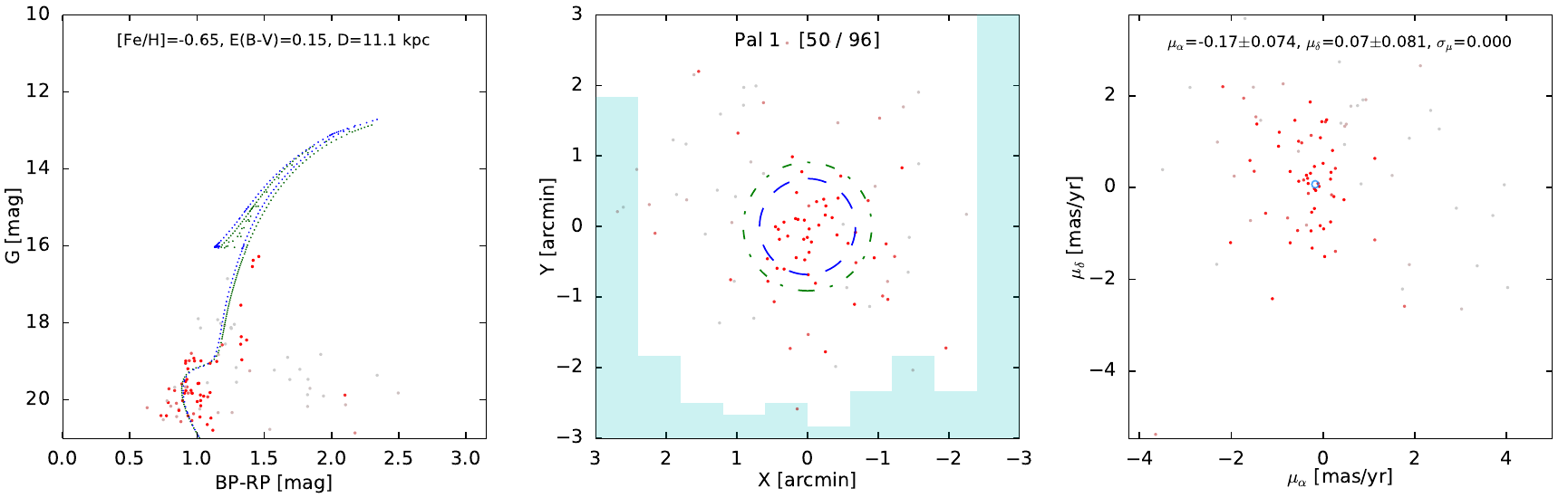}
\includegraphics{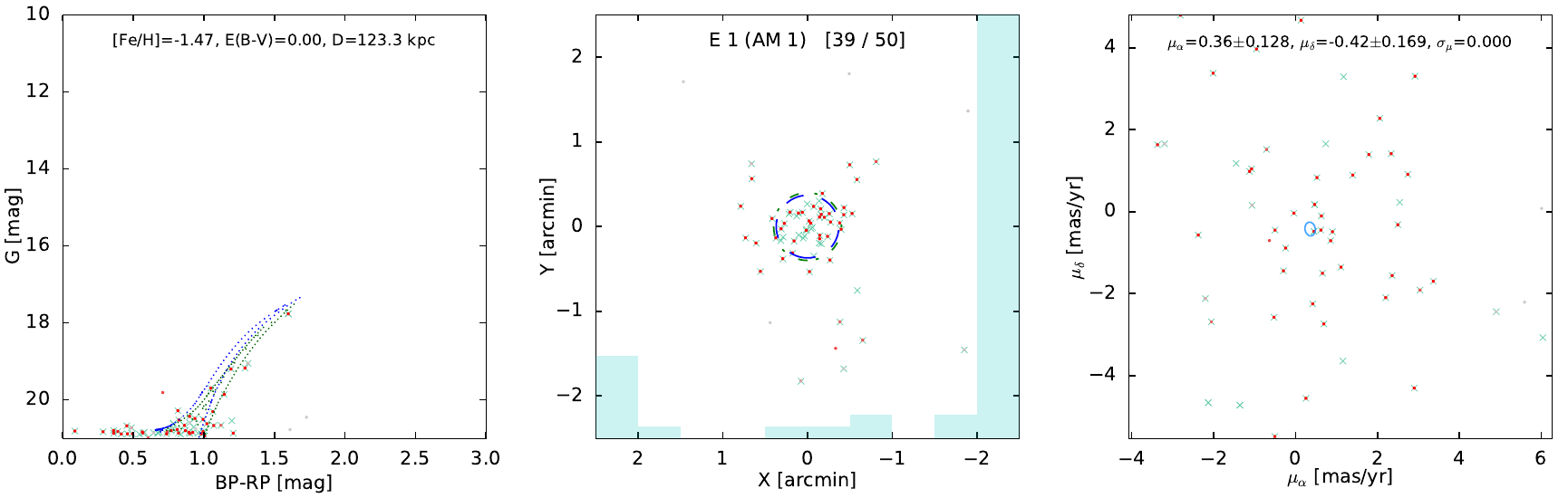}
\includegraphics{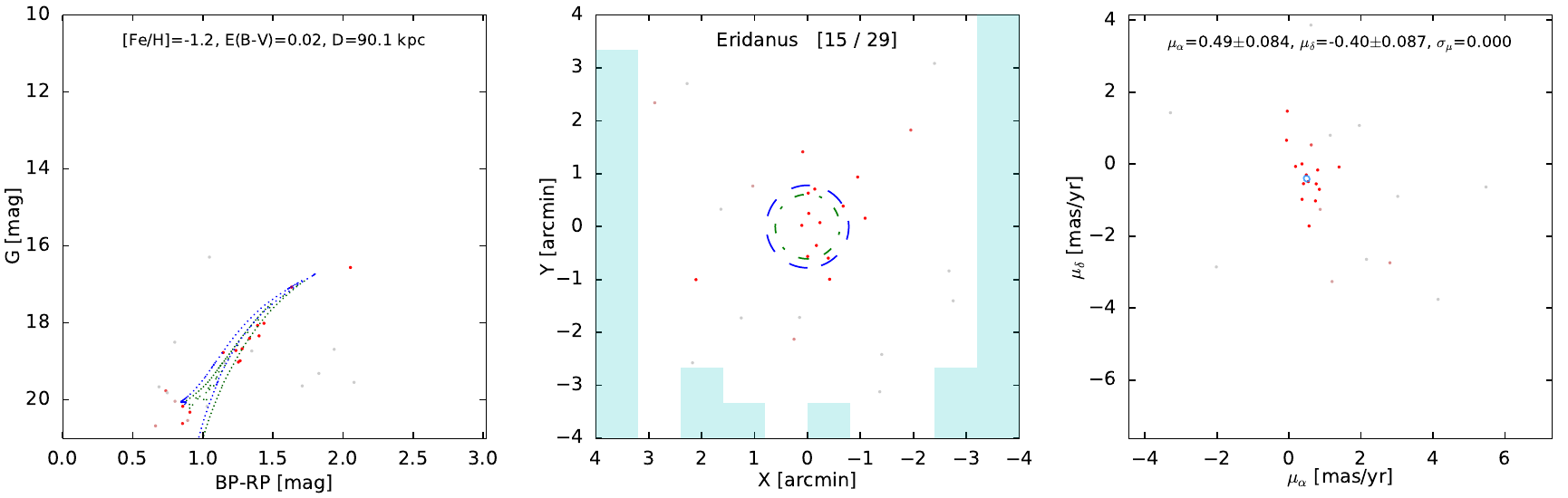}
\end{figure*}

\clearpage\begin{figure*}
\contcaption{}
\includegraphics{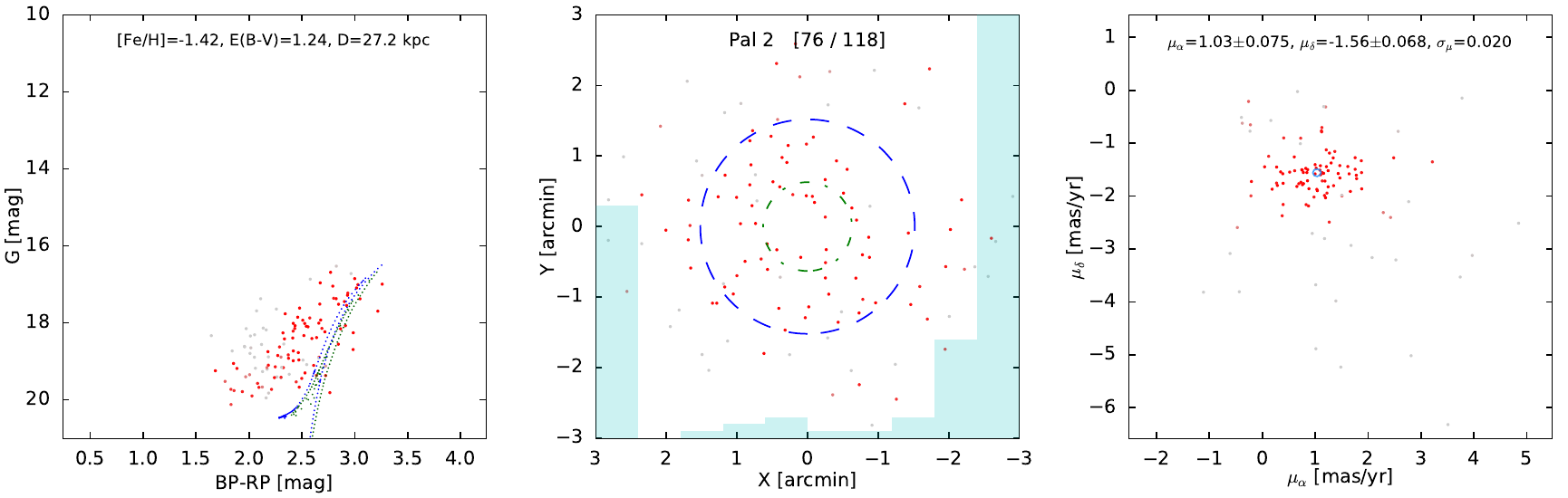}
\includegraphics{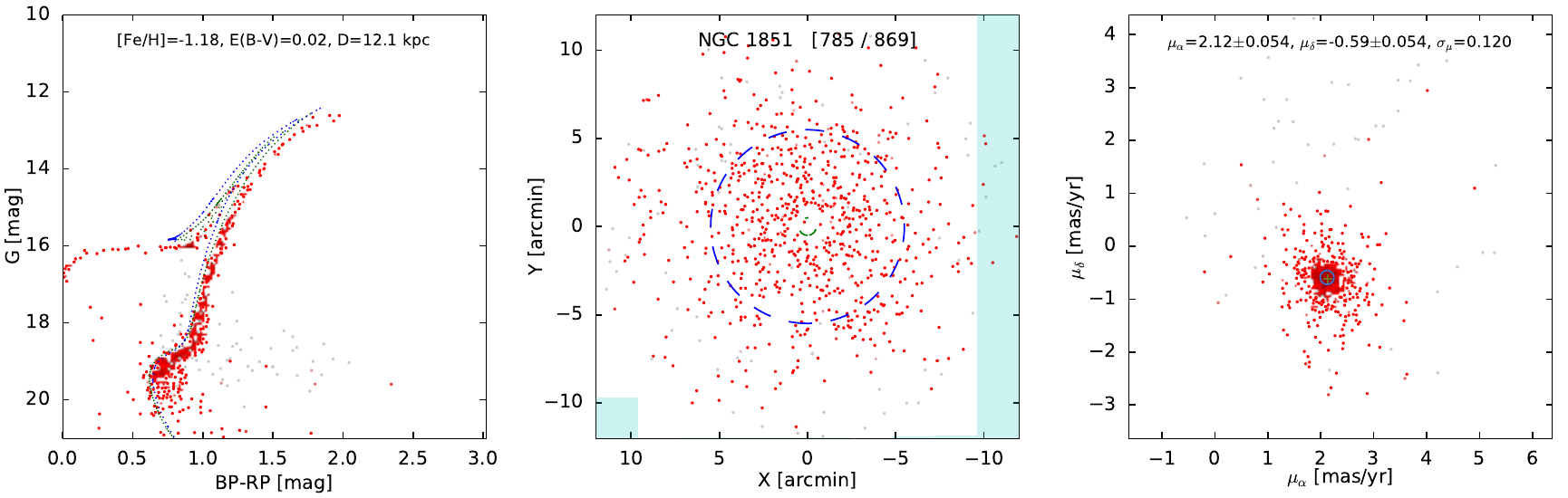}
\includegraphics{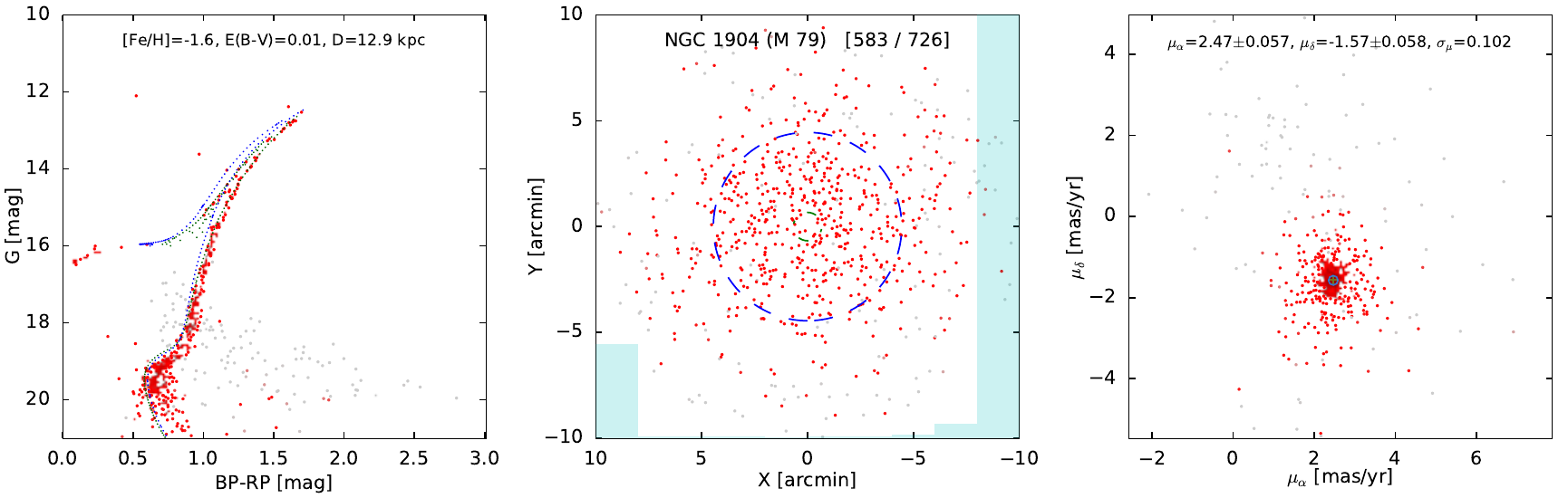}
\includegraphics{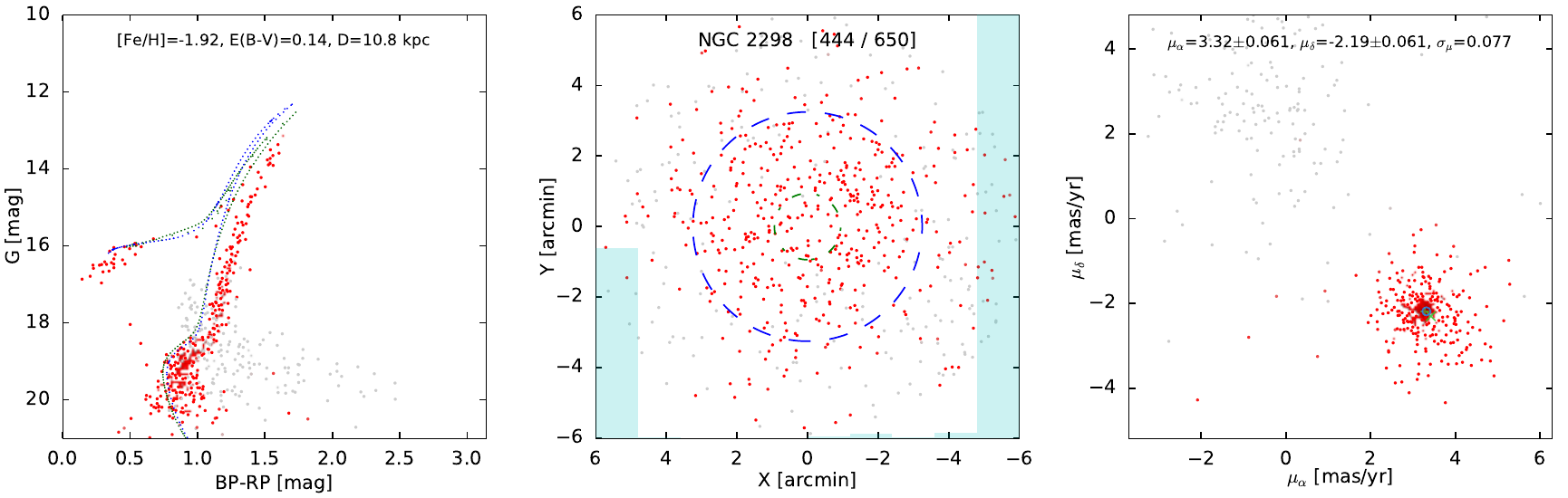}
\end{figure*}

\clearpage\begin{figure*}
\contcaption{}
\includegraphics{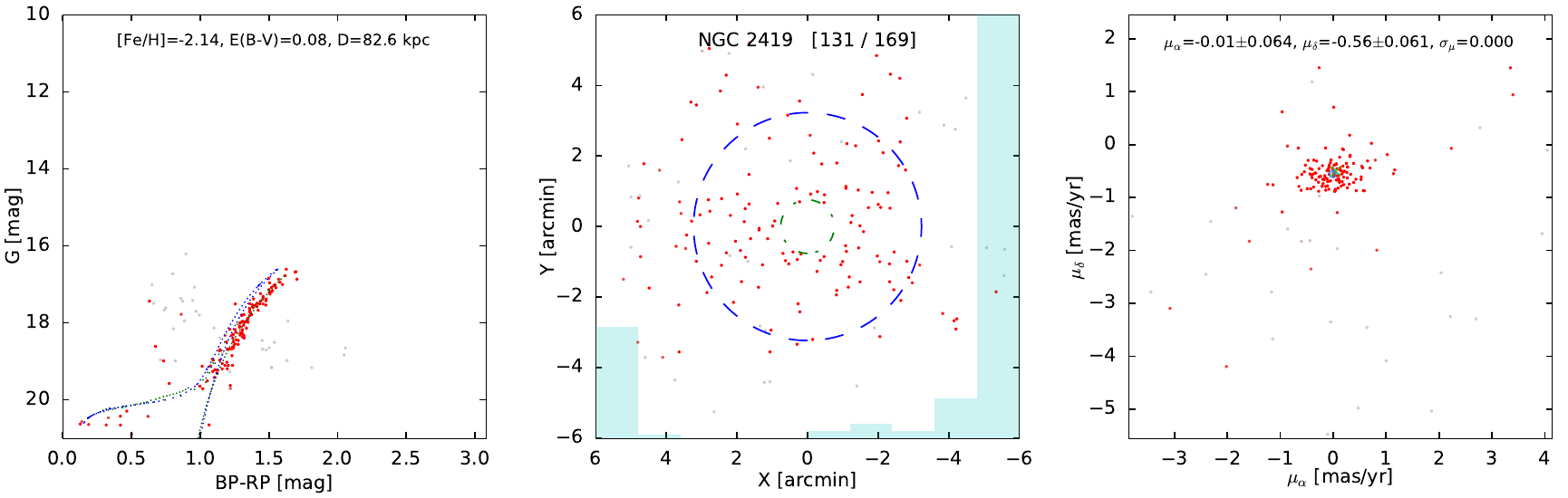}
\includegraphics{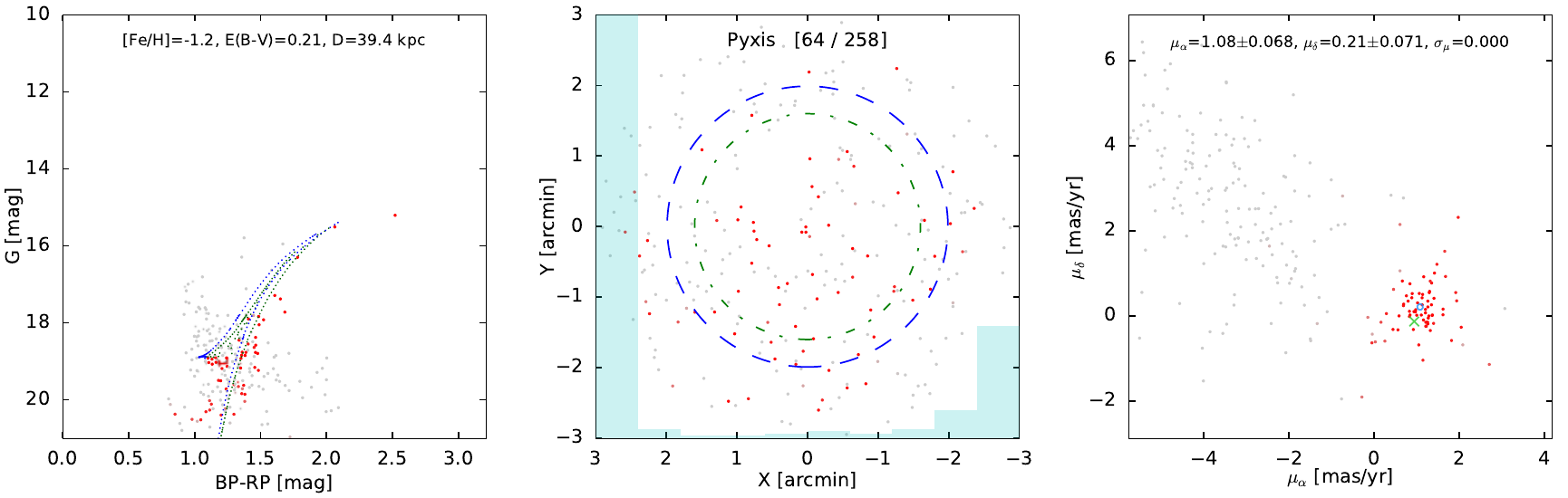}
\includegraphics{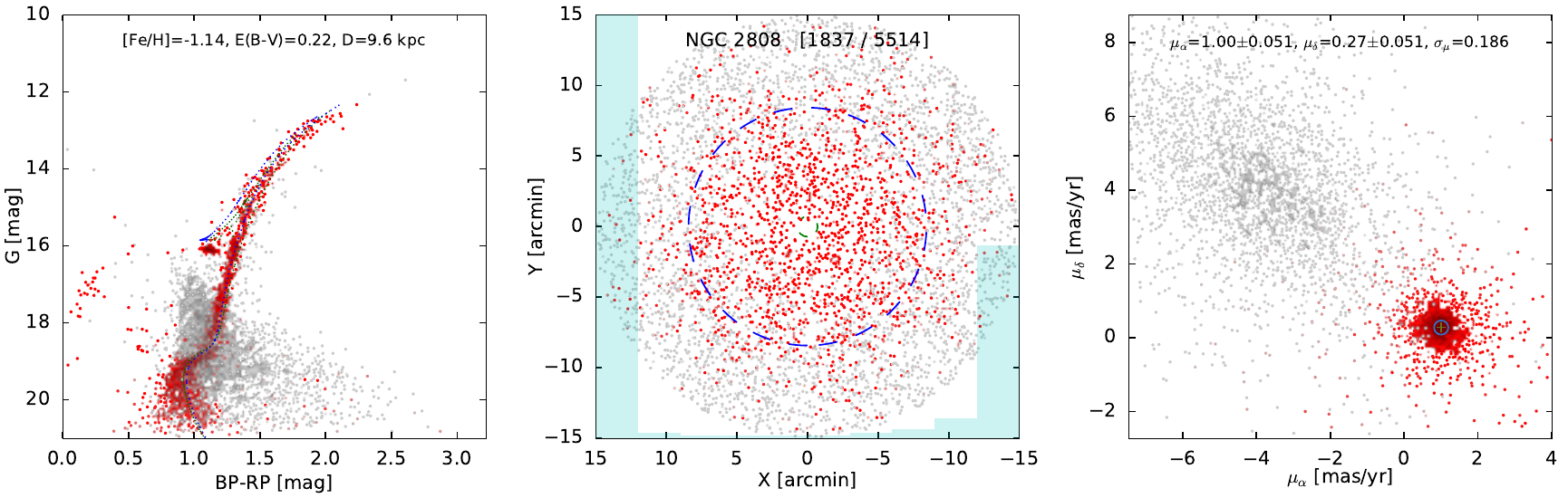}
\includegraphics{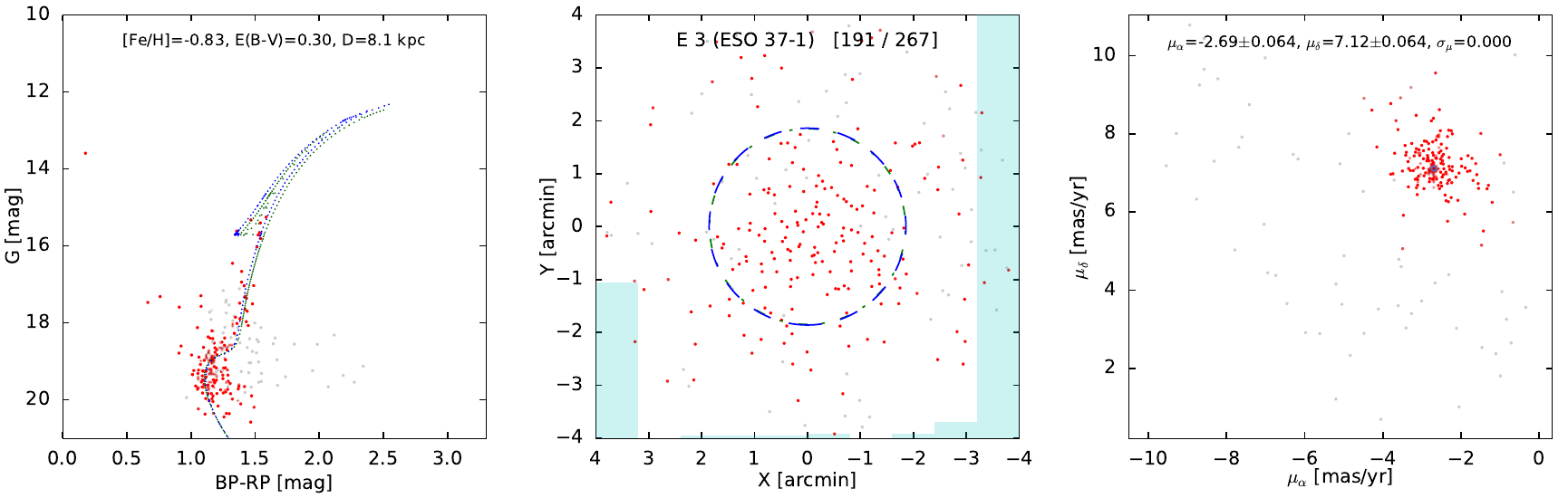}
\end{figure*}

\clearpage\begin{figure*}
\contcaption{}
\includegraphics{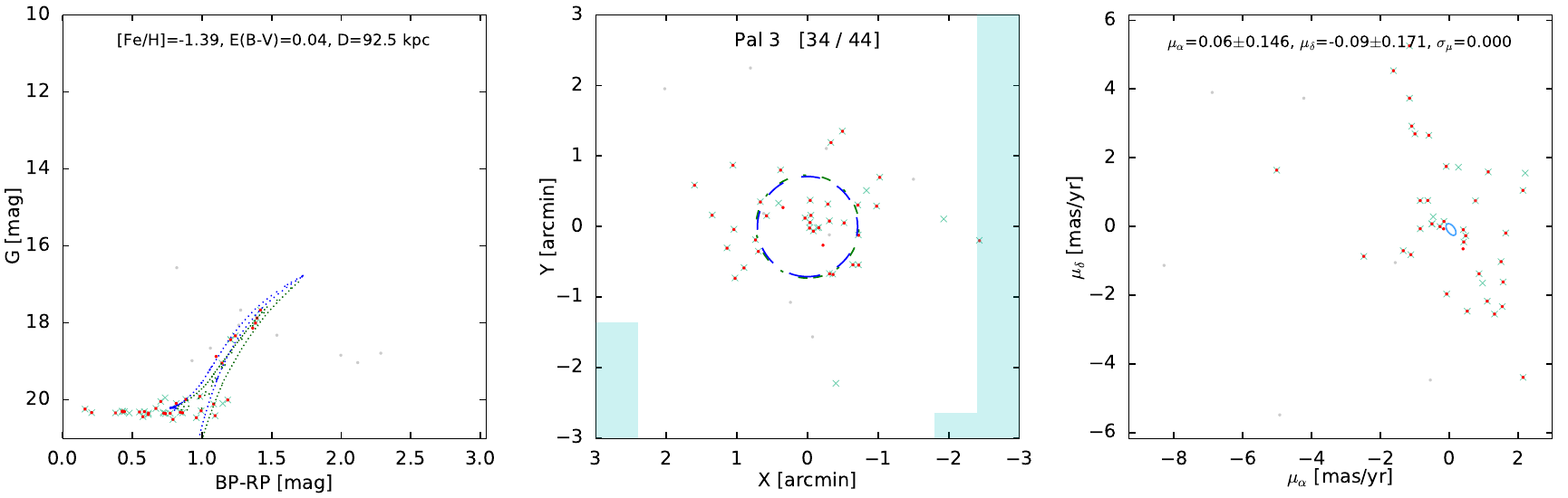}
\includegraphics{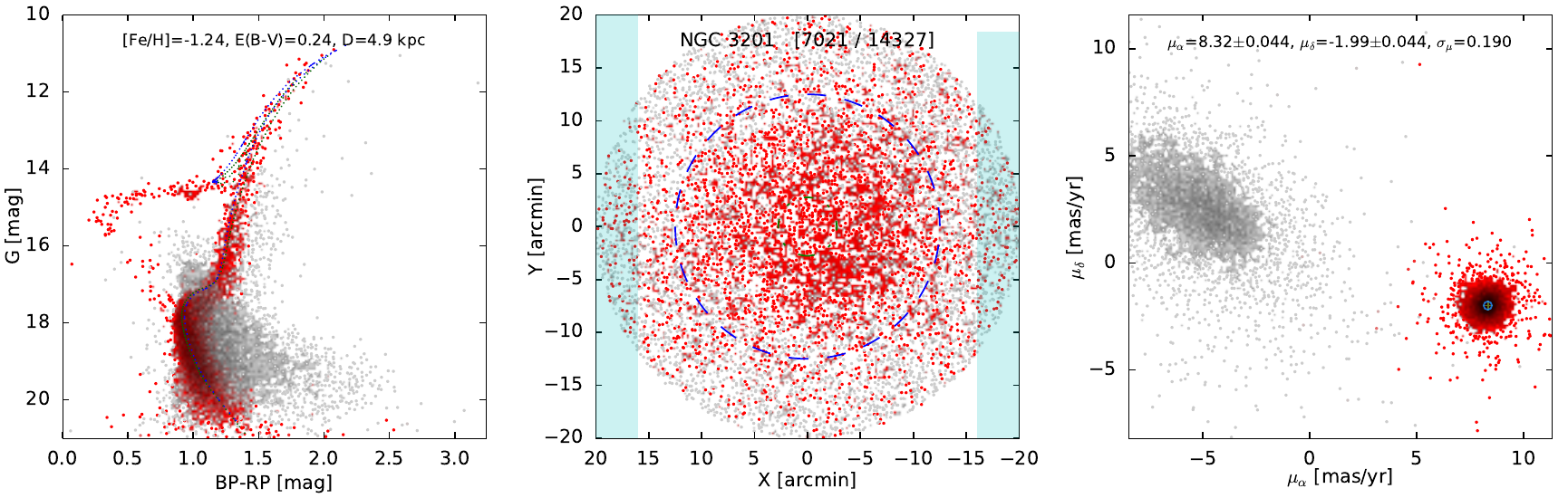}
\includegraphics{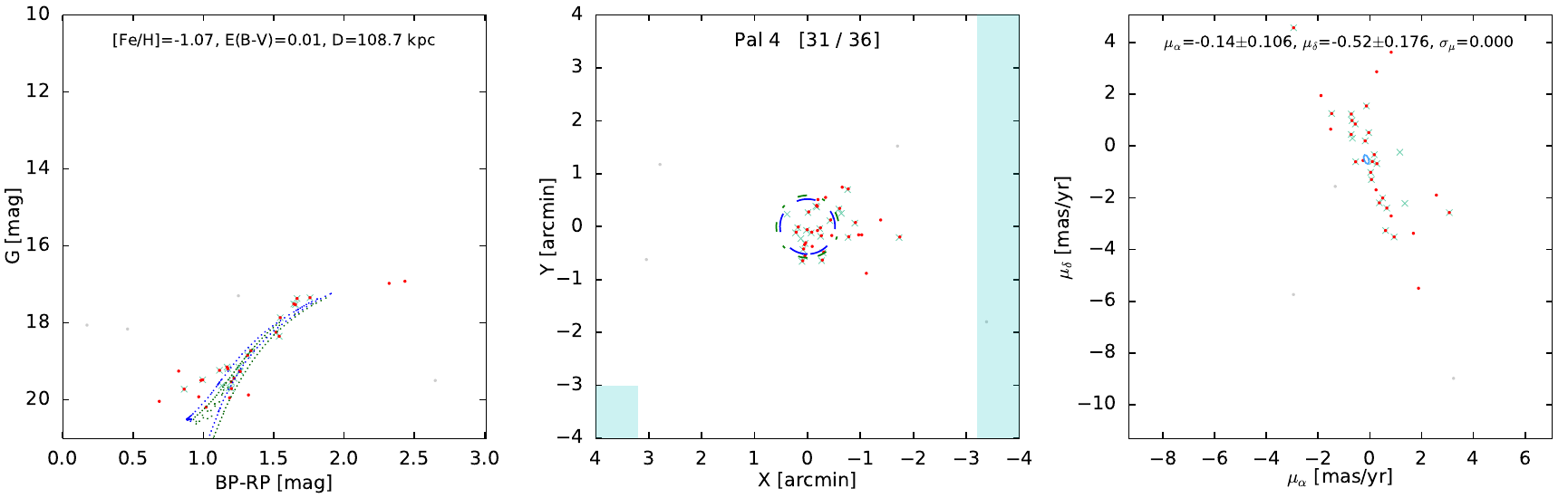}
\includegraphics{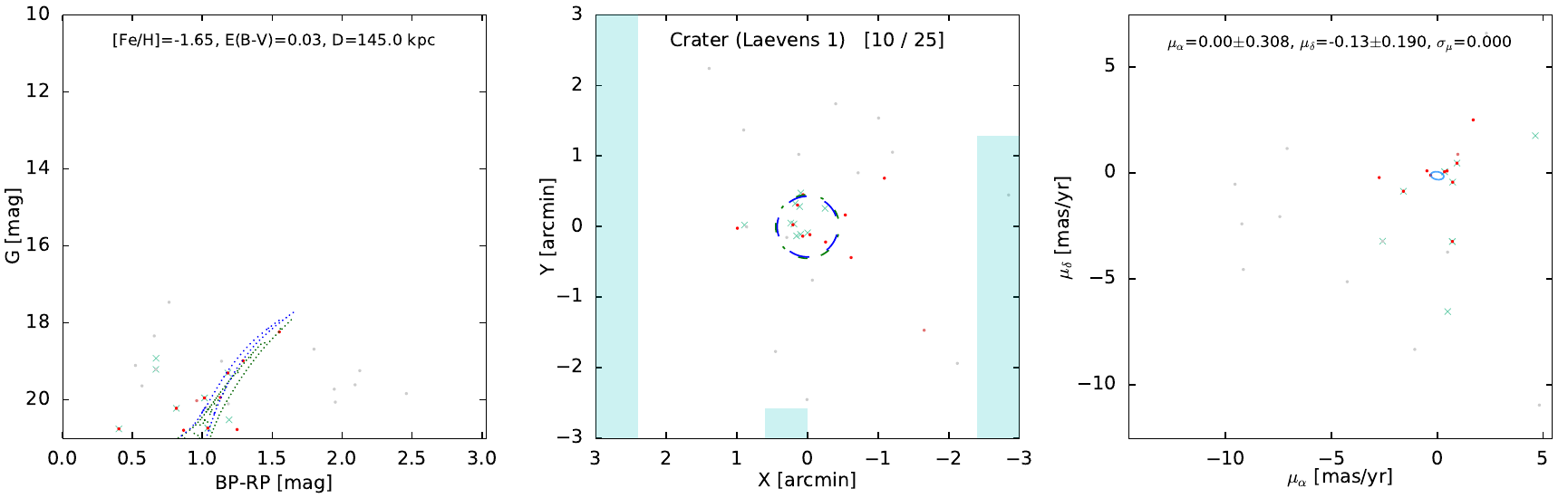}
\end{figure*}

\clearpage\begin{figure*}
\contcaption{}
\includegraphics{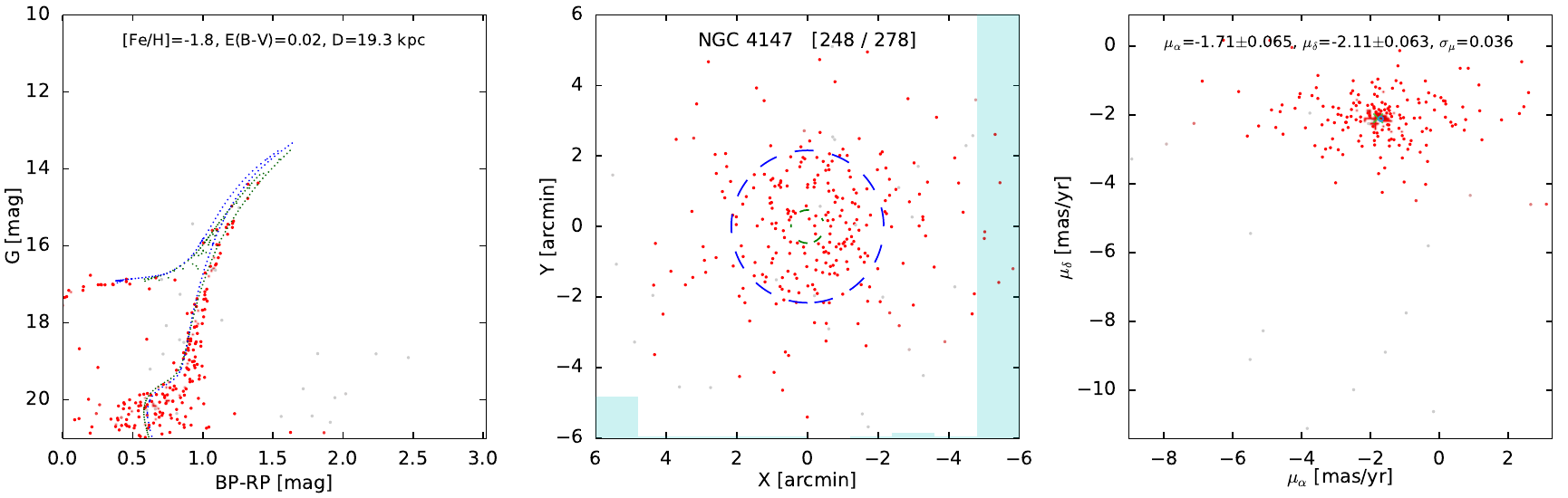}
\includegraphics{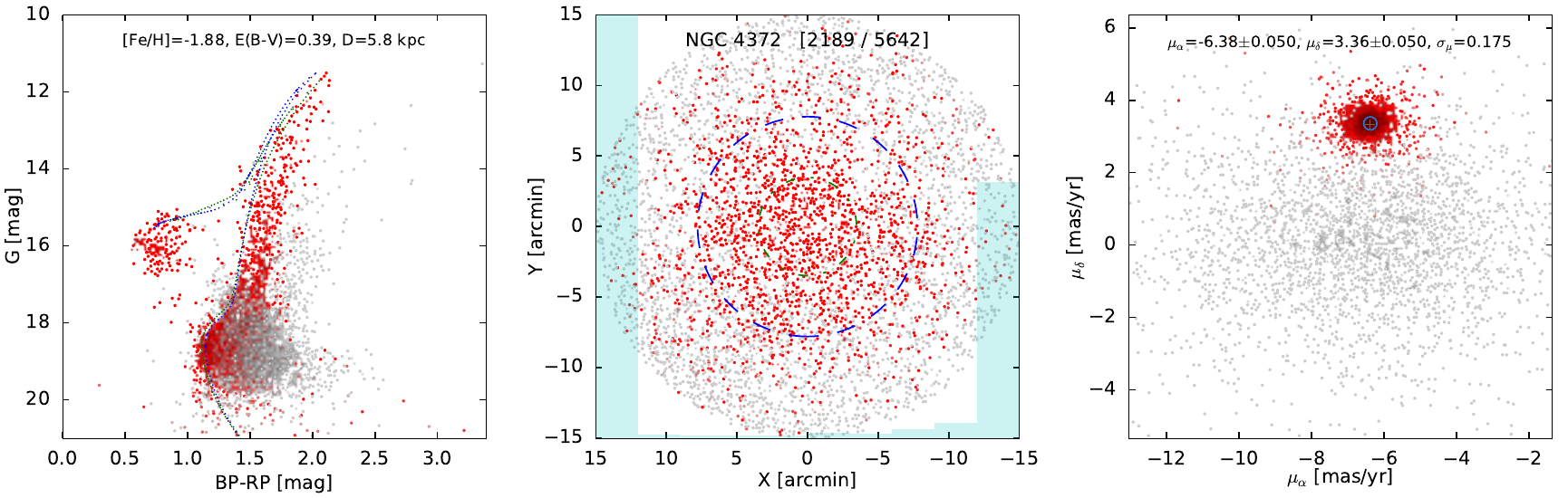}
\includegraphics{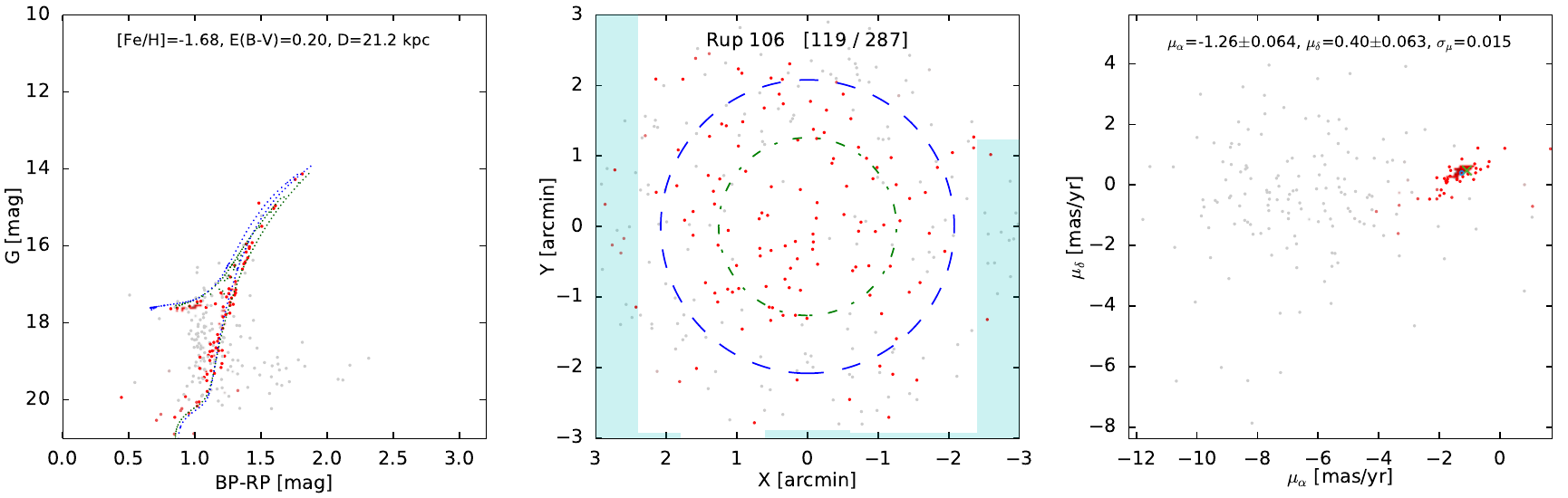}
\includegraphics{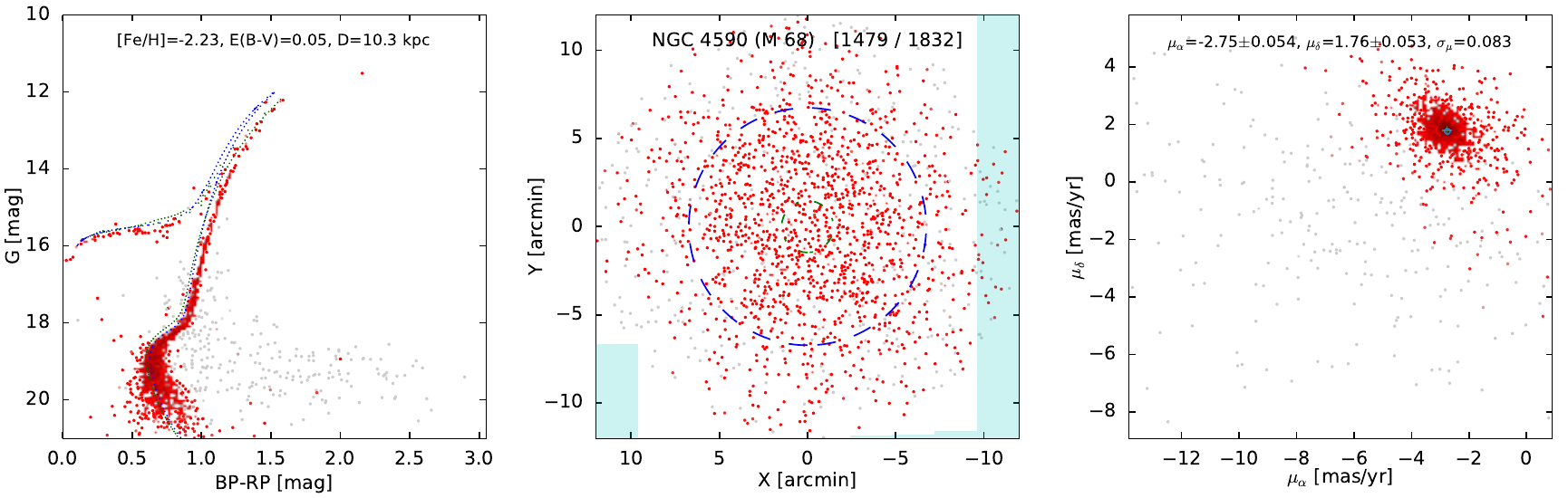}
\end{figure*}

\clearpage\begin{figure*}
\contcaption{}
\includegraphics{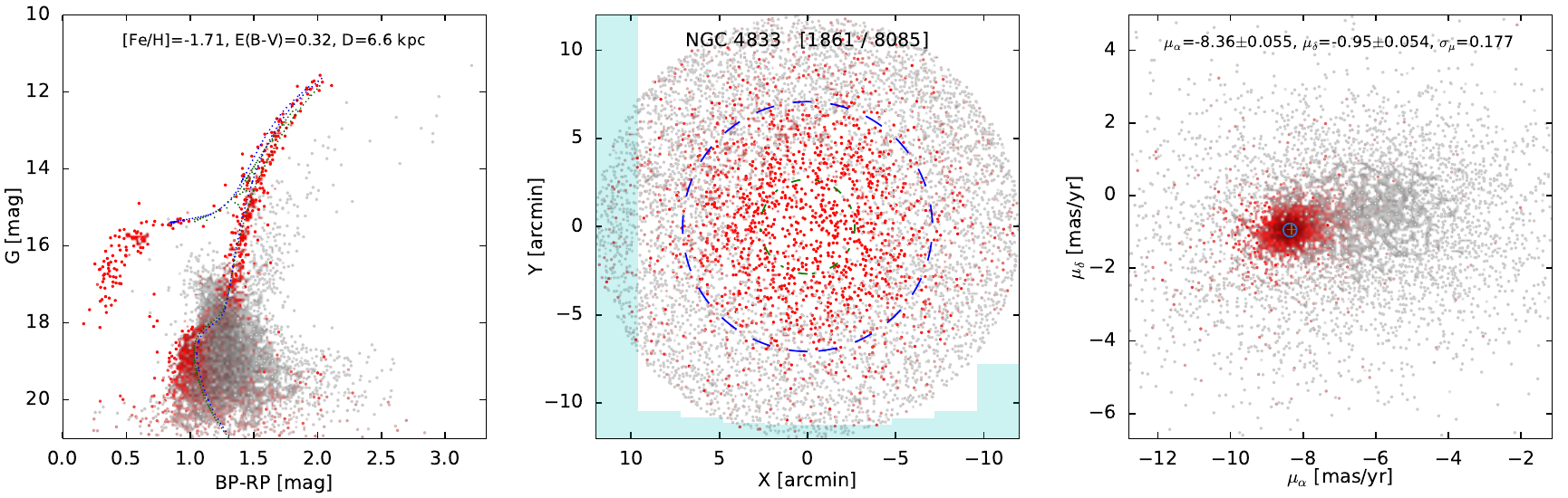}
\includegraphics{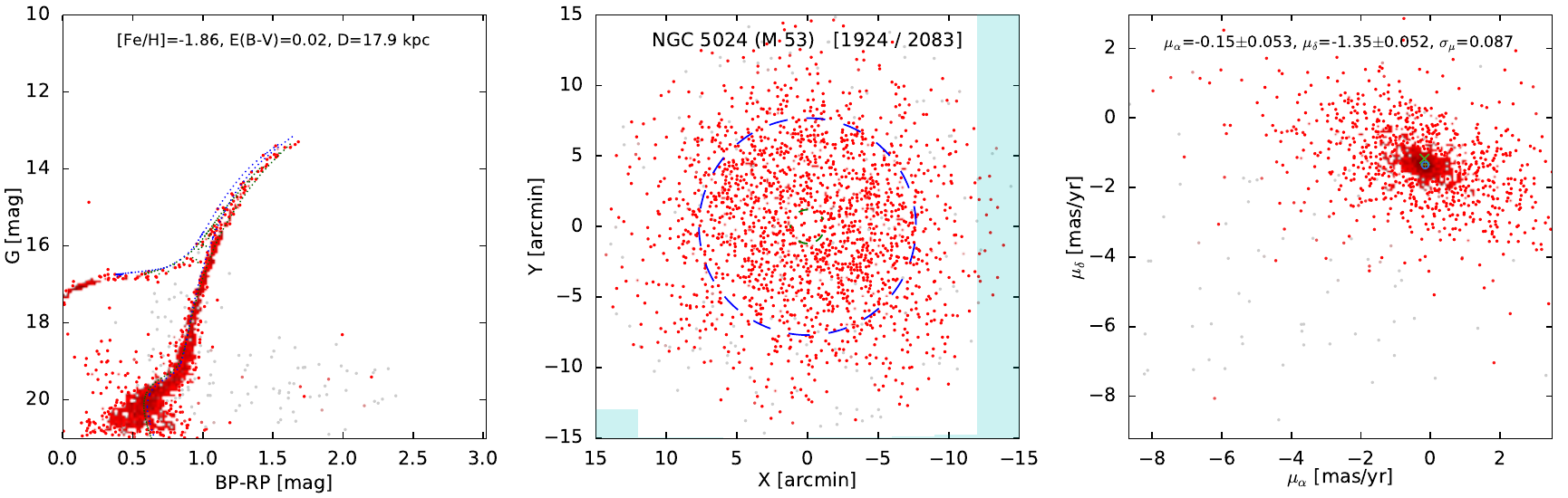}
\includegraphics{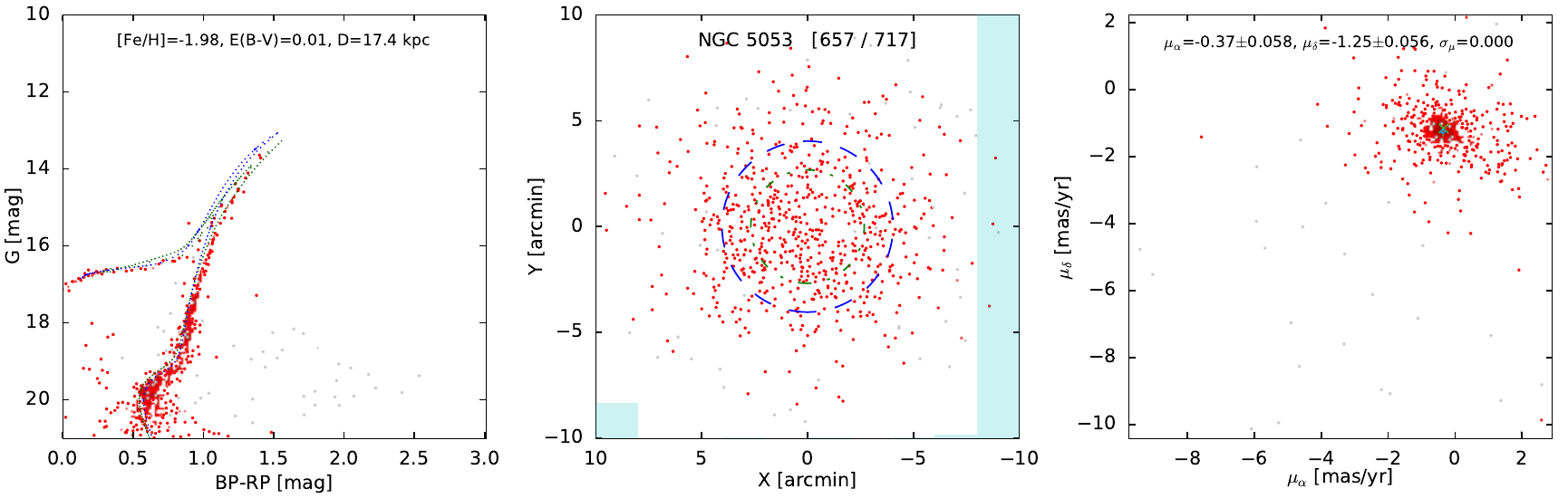}
\includegraphics{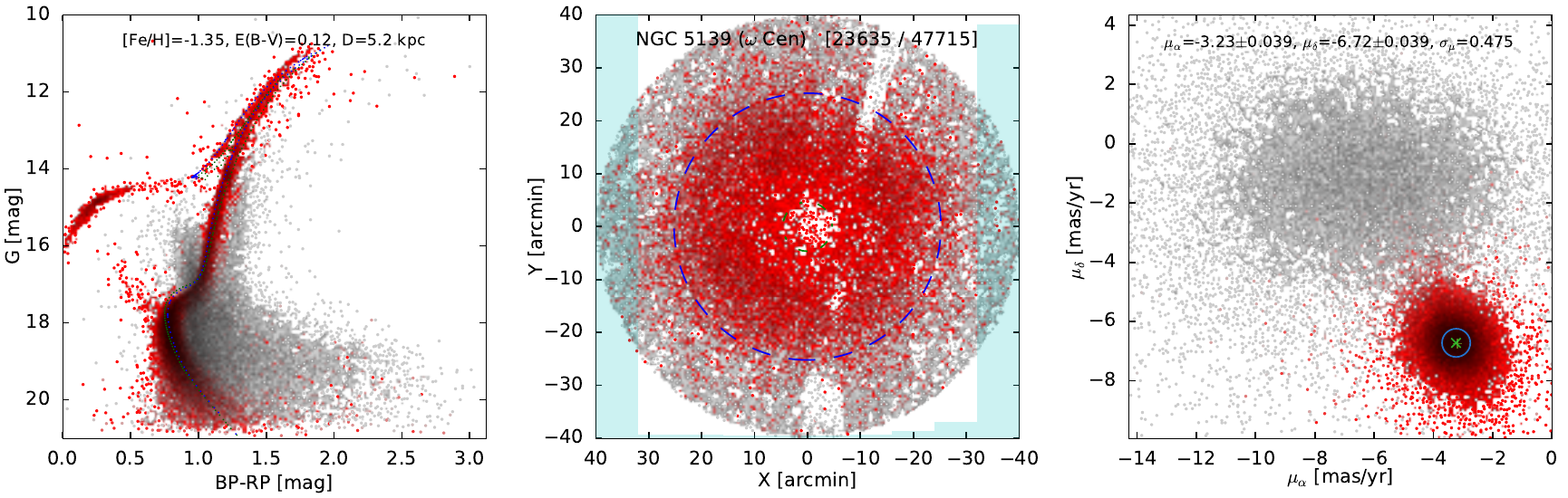}
\end{figure*}

\clearpage\begin{figure*}
\contcaption{}
\includegraphics{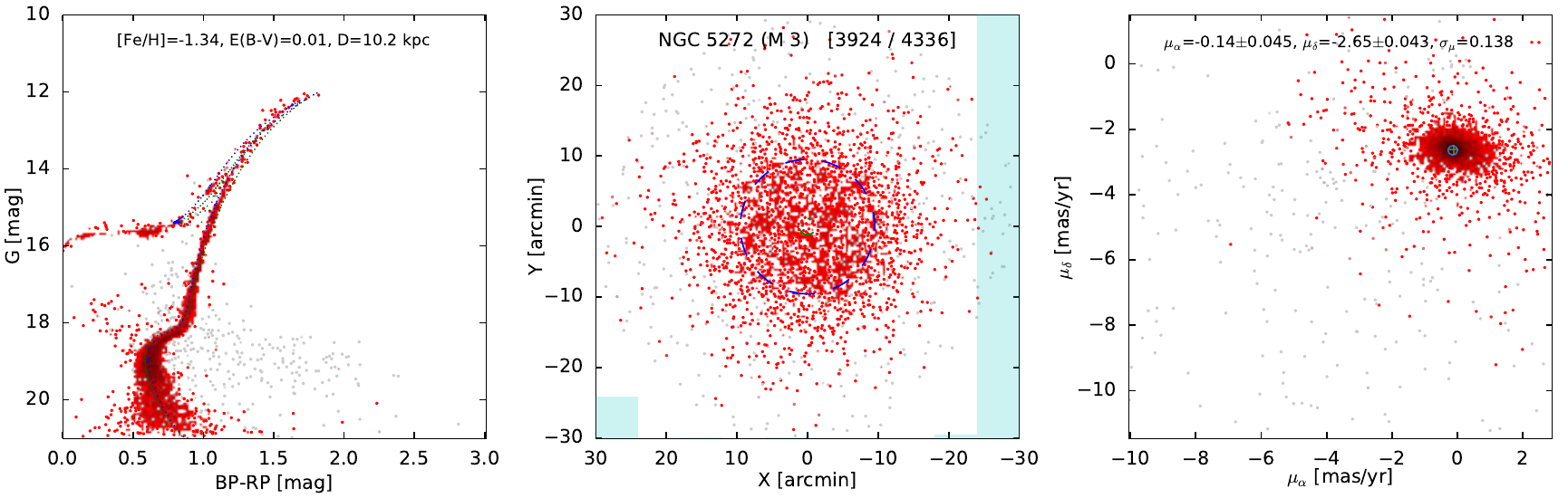}
\includegraphics{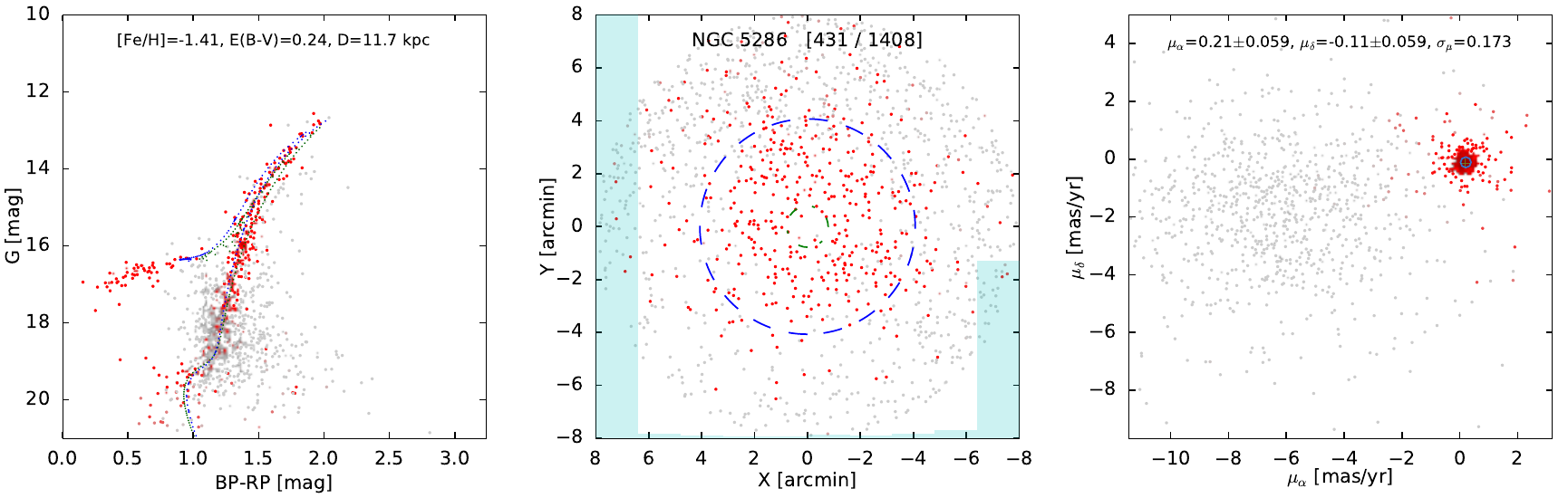}
\includegraphics{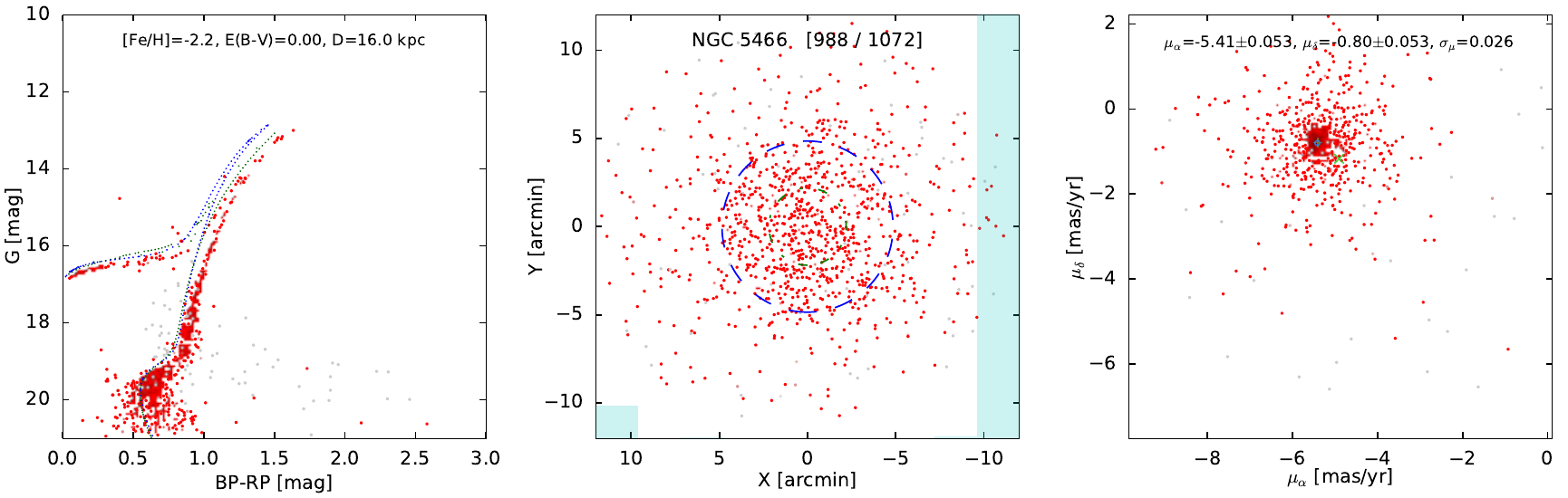}
\includegraphics{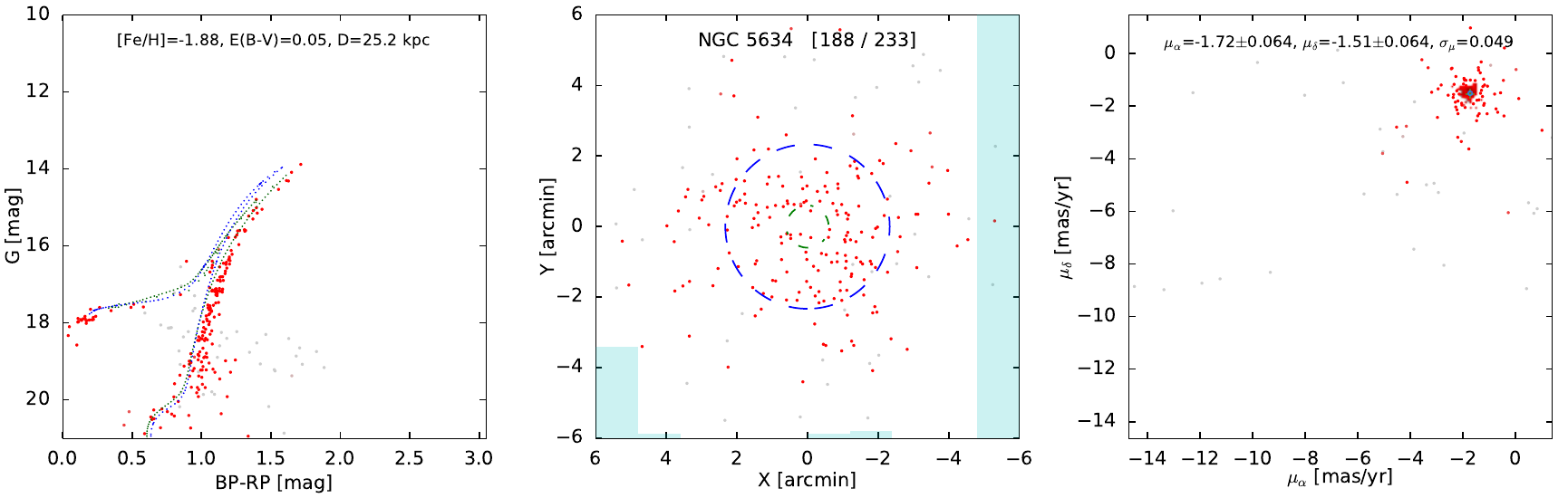}
\end{figure*}

\clearpage\begin{figure*}
\contcaption{}
\includegraphics{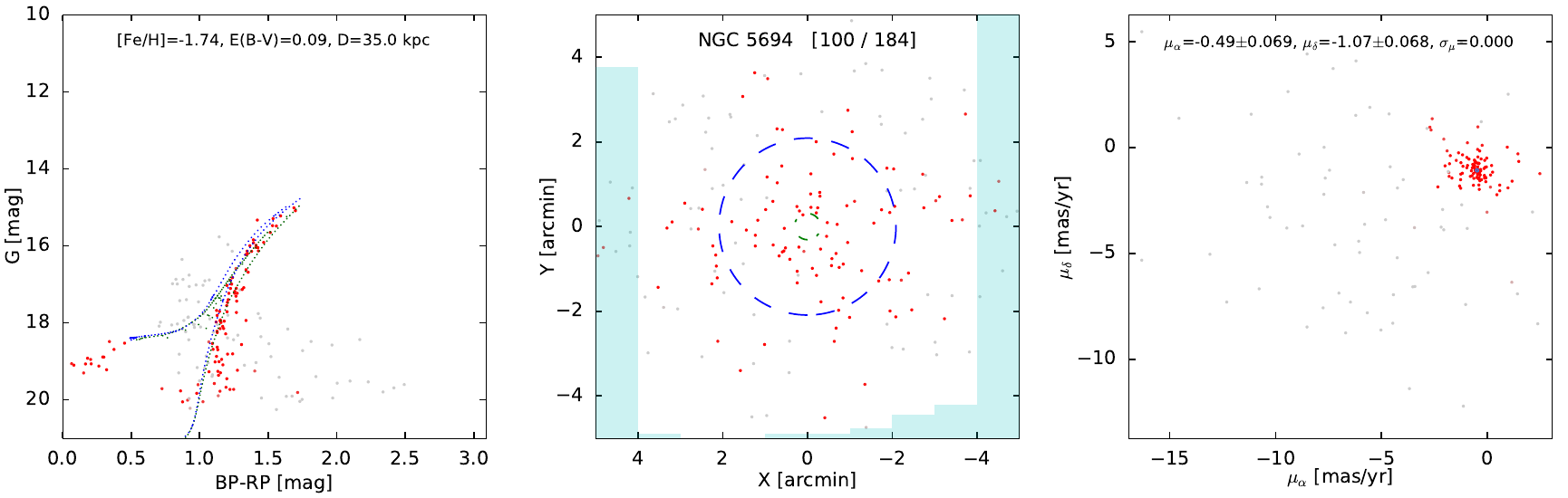}
\includegraphics{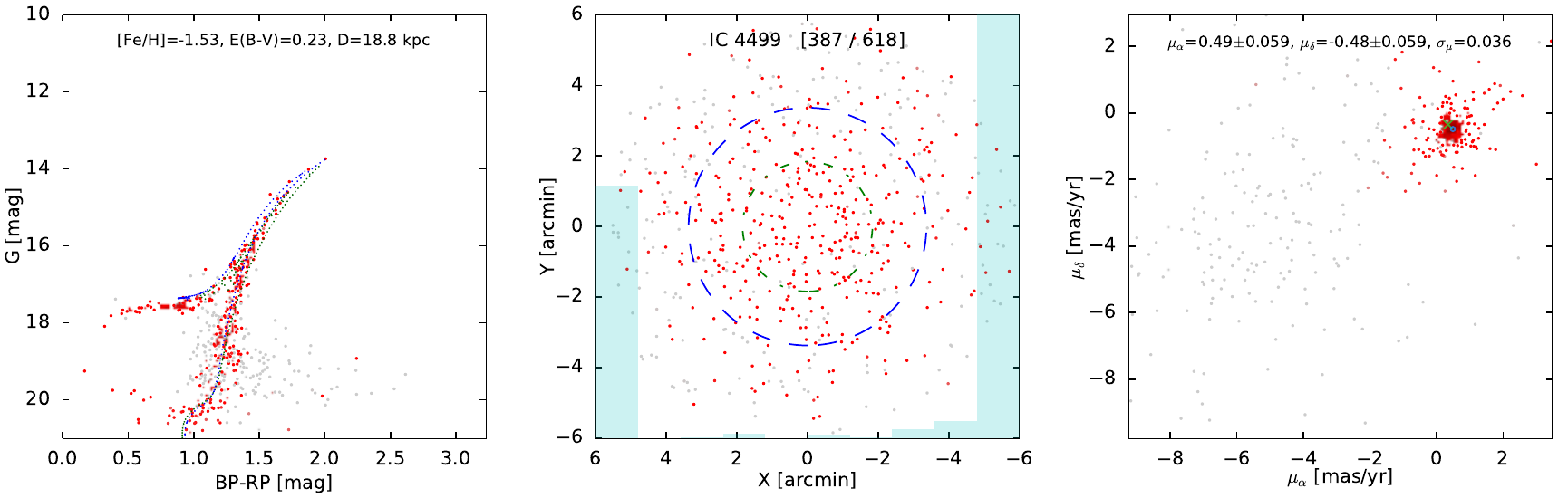}
\includegraphics{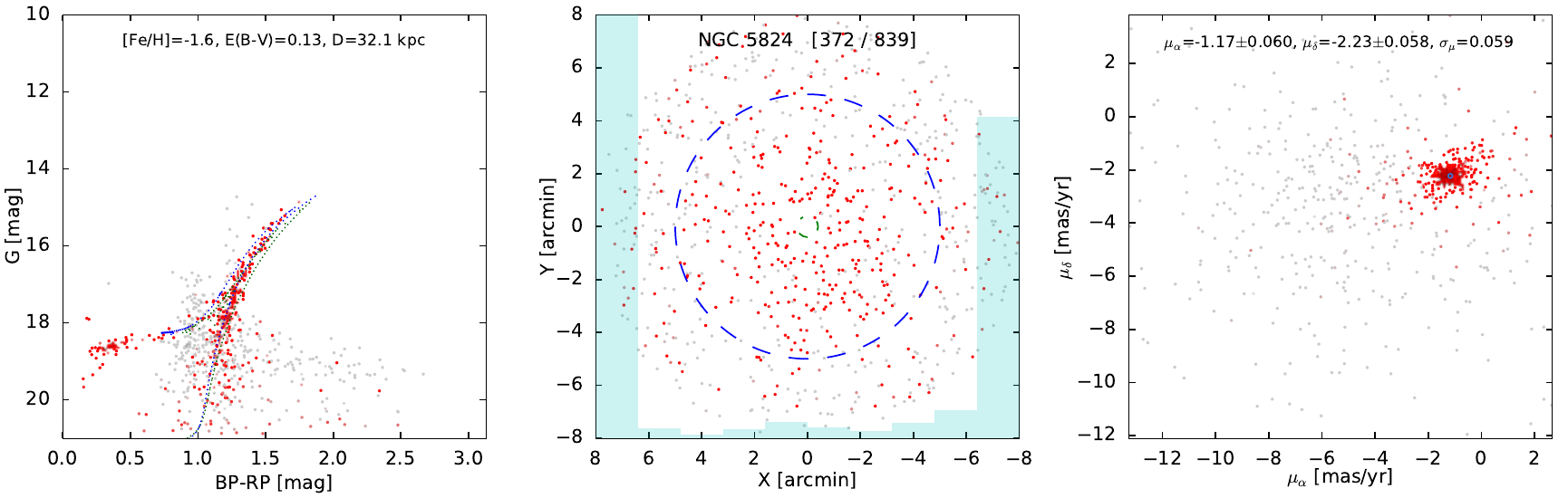}
\includegraphics{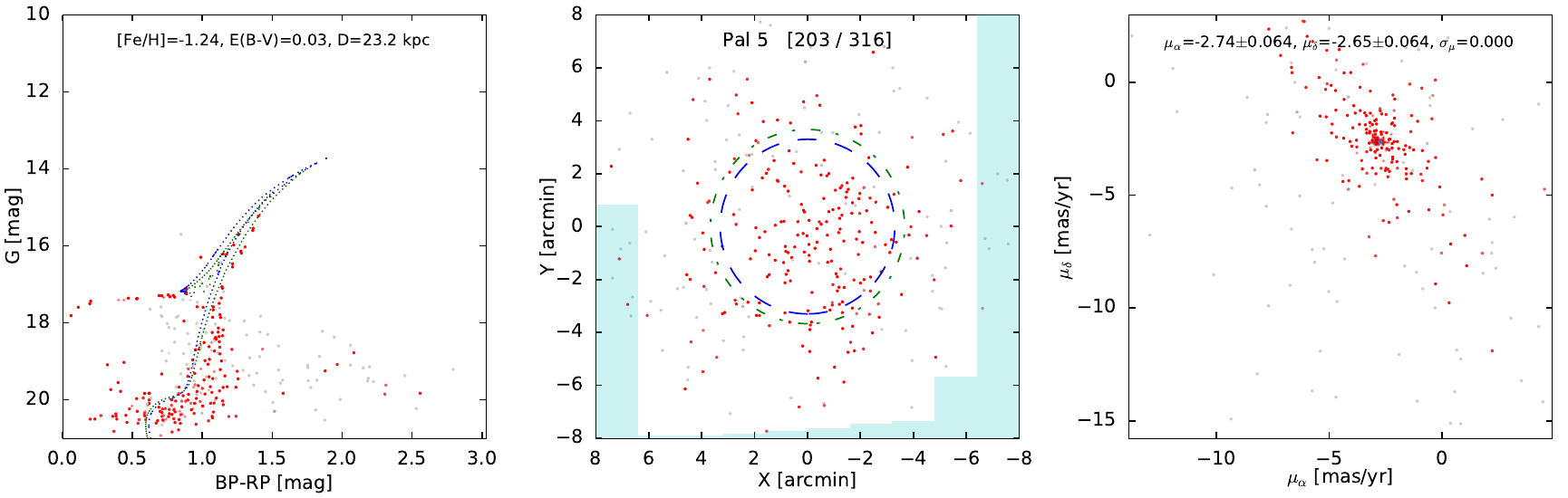}
\end{figure*}

\clearpage\begin{figure*}
\contcaption{}
\includegraphics{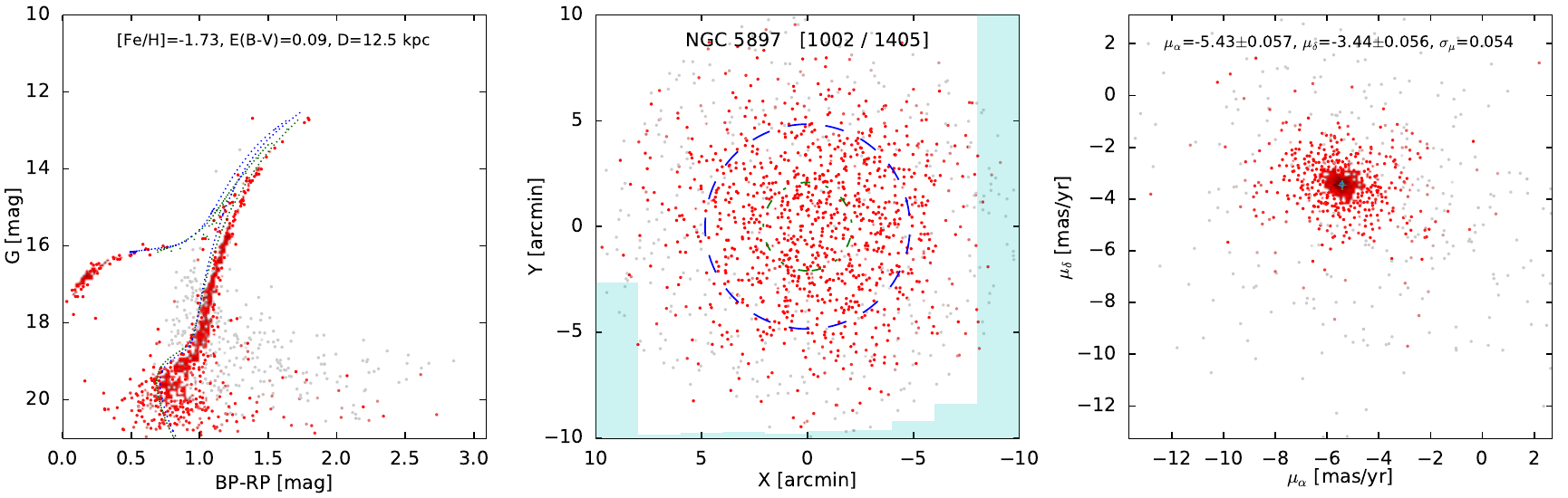}
\includegraphics{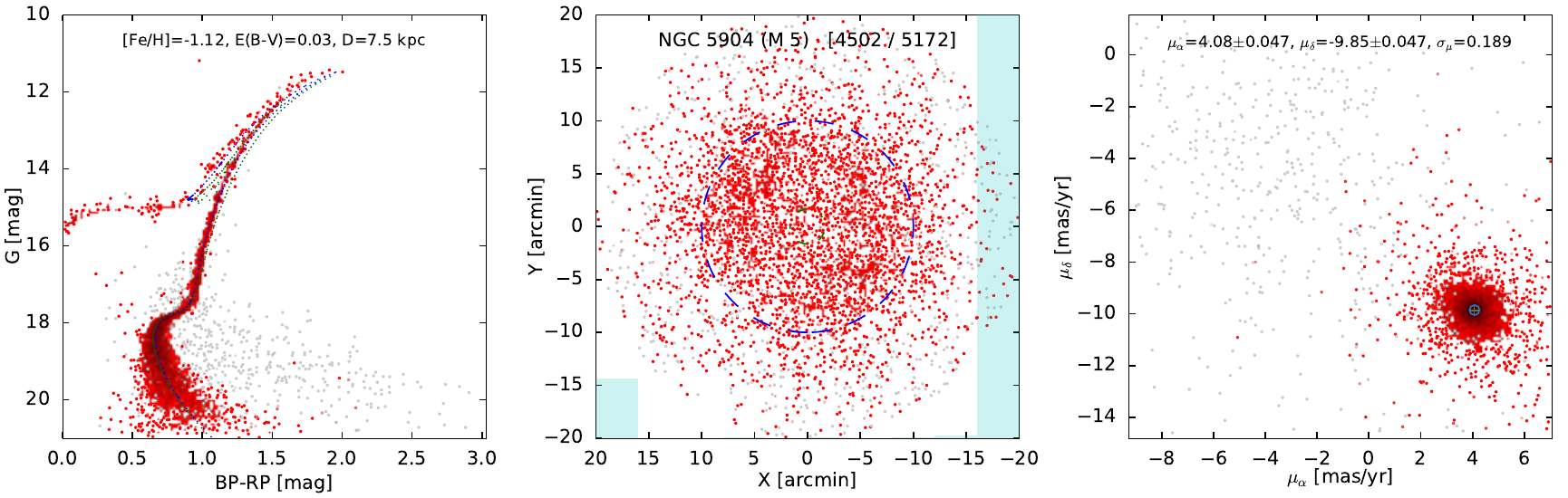}
\includegraphics{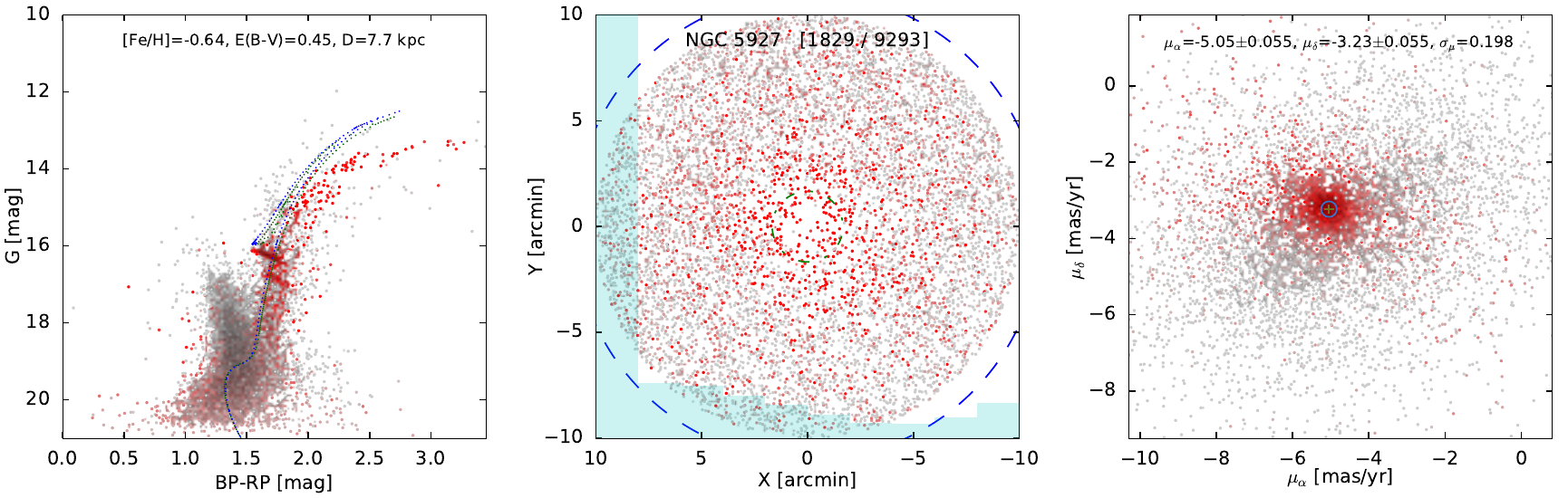}
\includegraphics{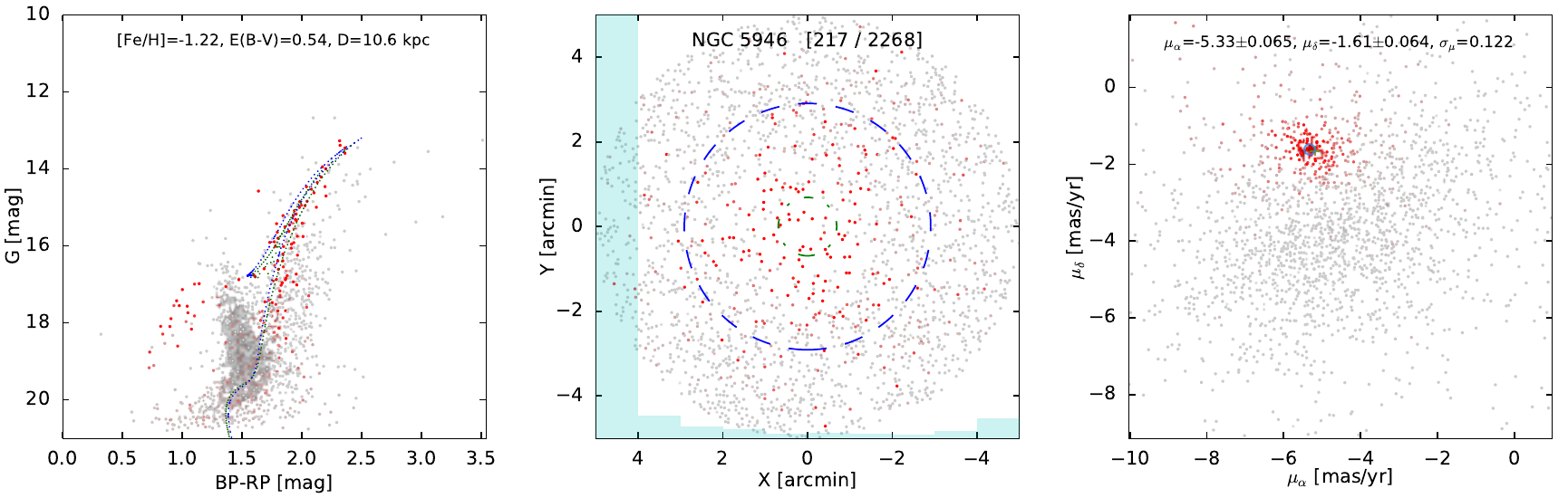}
\end{figure*}

\clearpage\begin{figure*}
\contcaption{}
\includegraphics{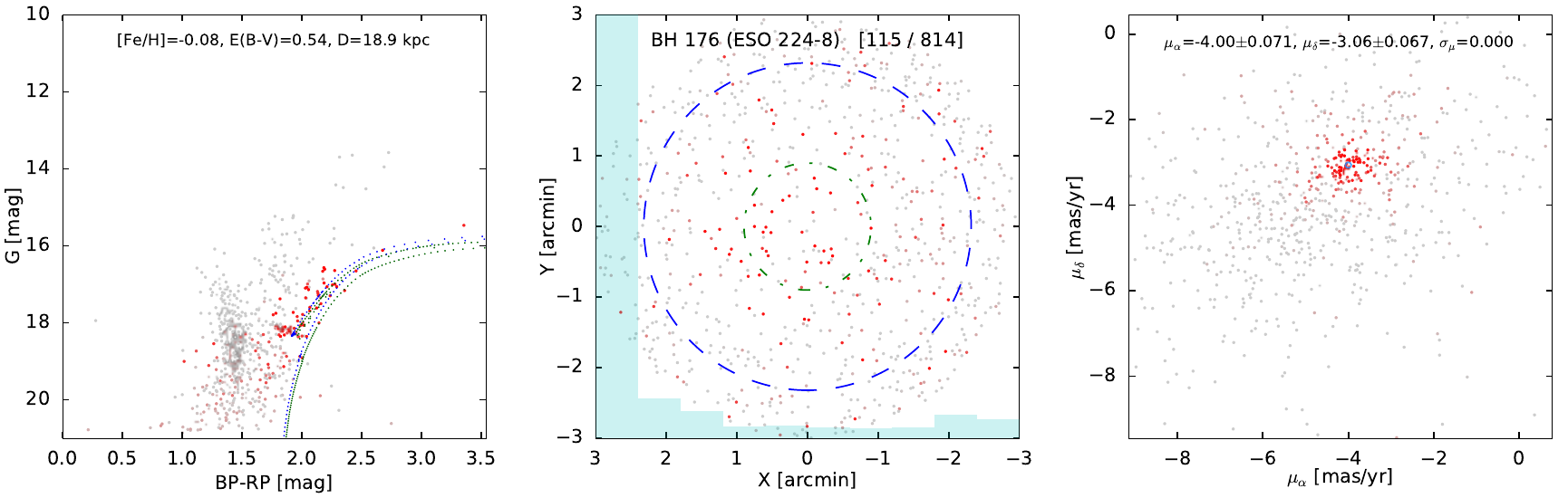}
\includegraphics{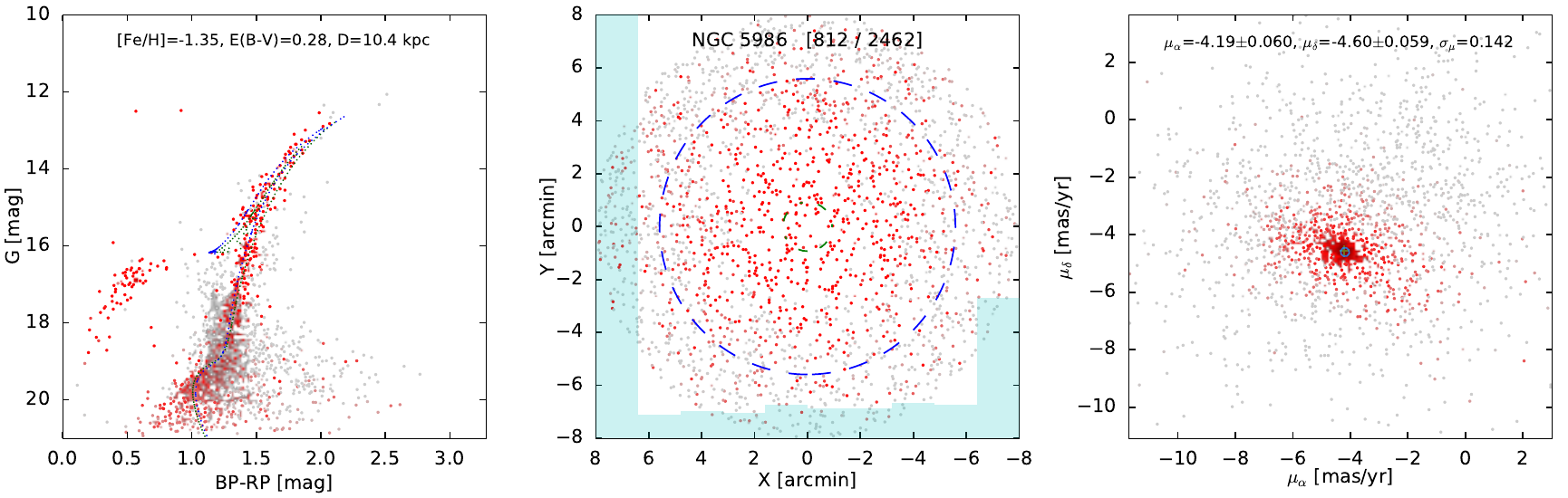}
\includegraphics{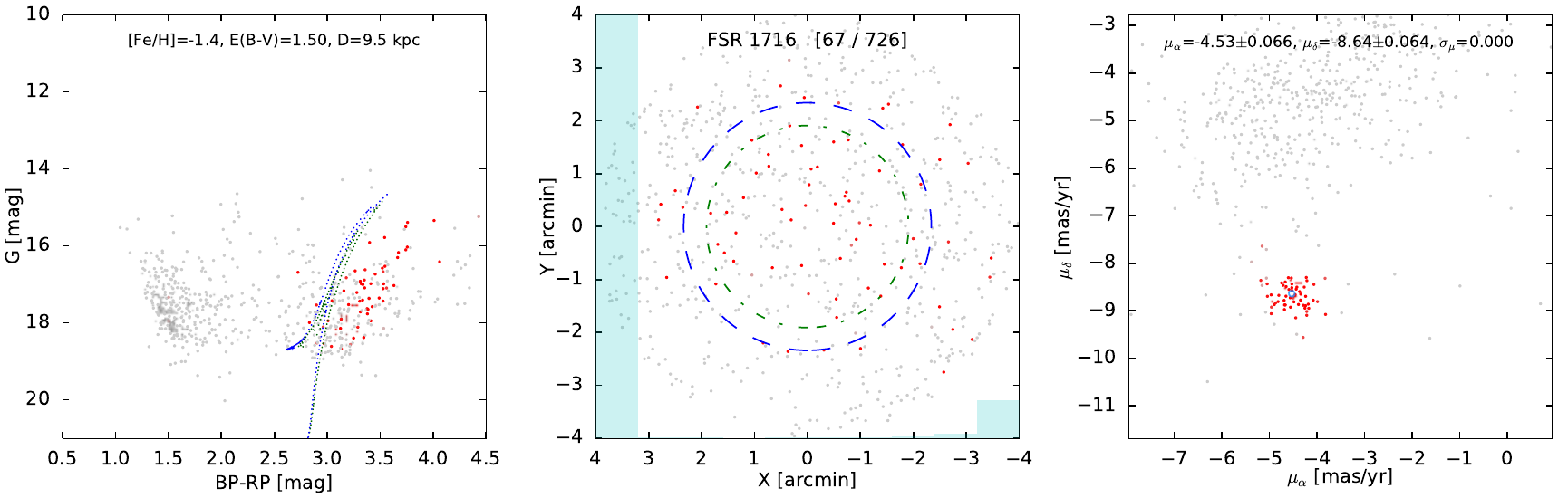}
\includegraphics{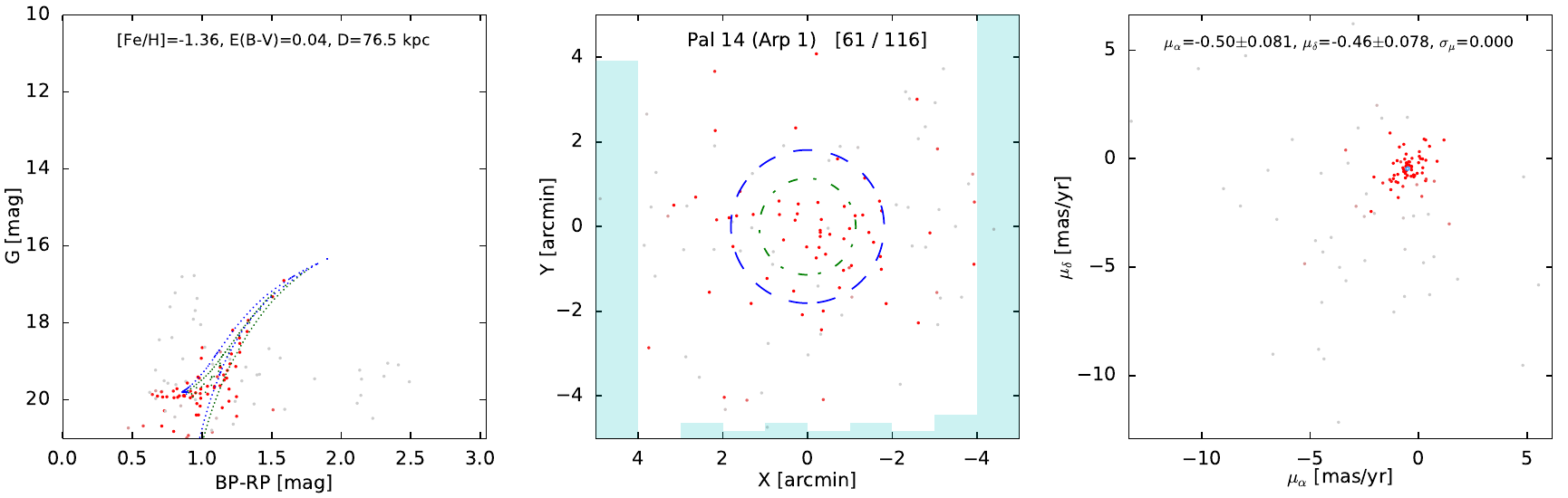}
\end{figure*}

\clearpage\begin{figure*}
\contcaption{}
\includegraphics{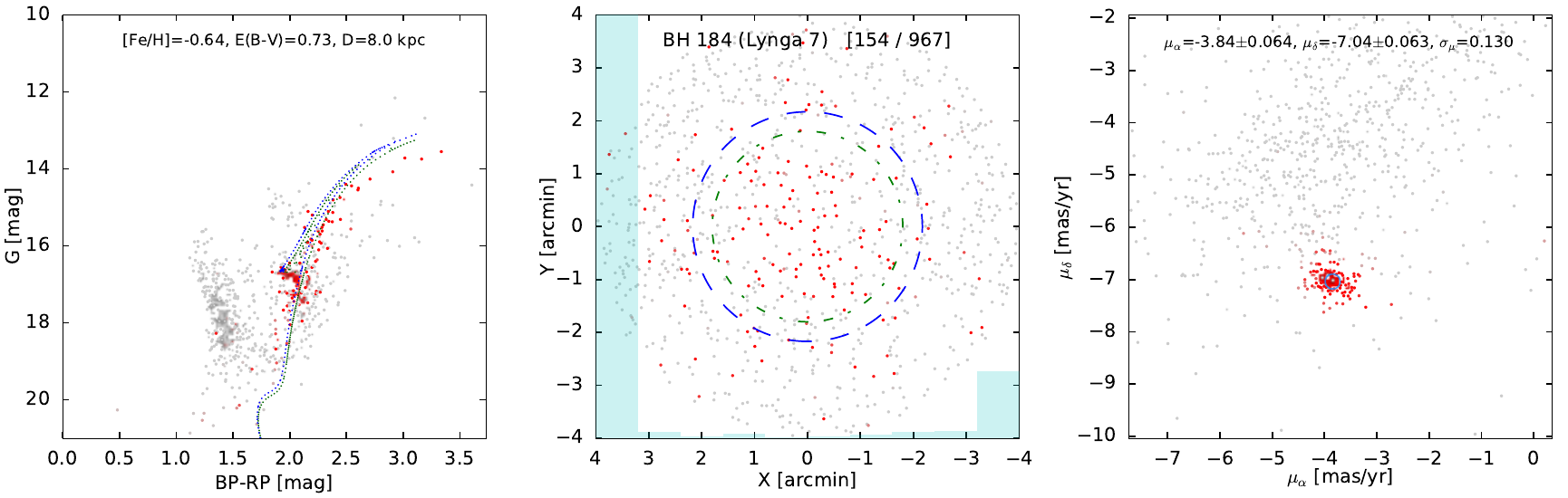}
\includegraphics{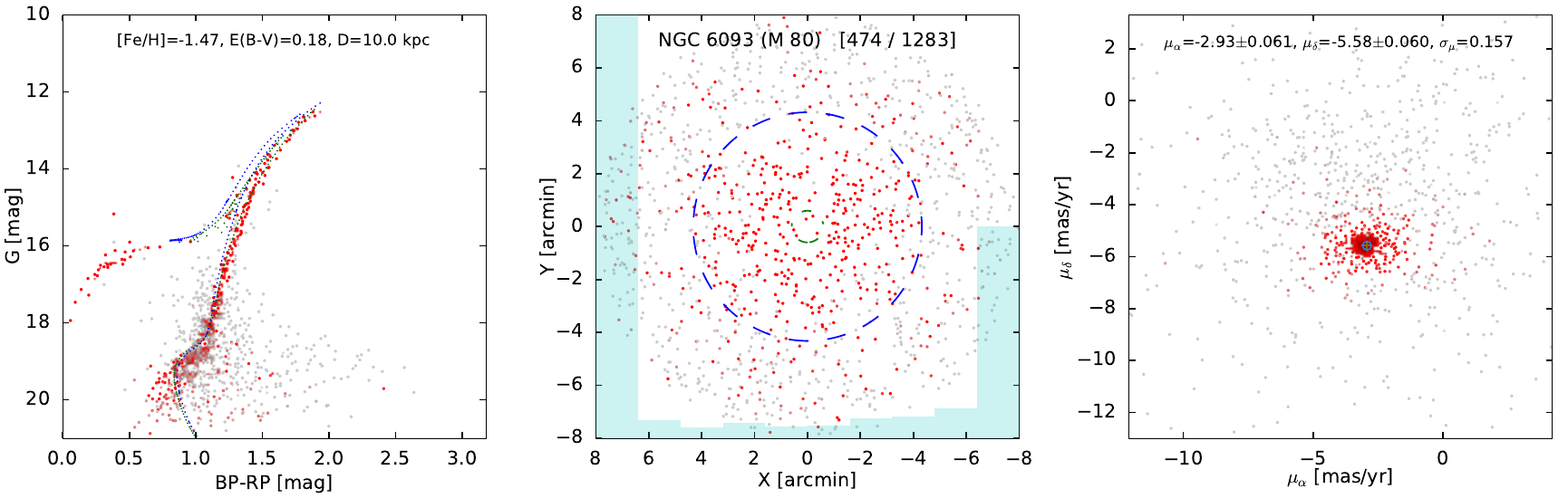}
\includegraphics{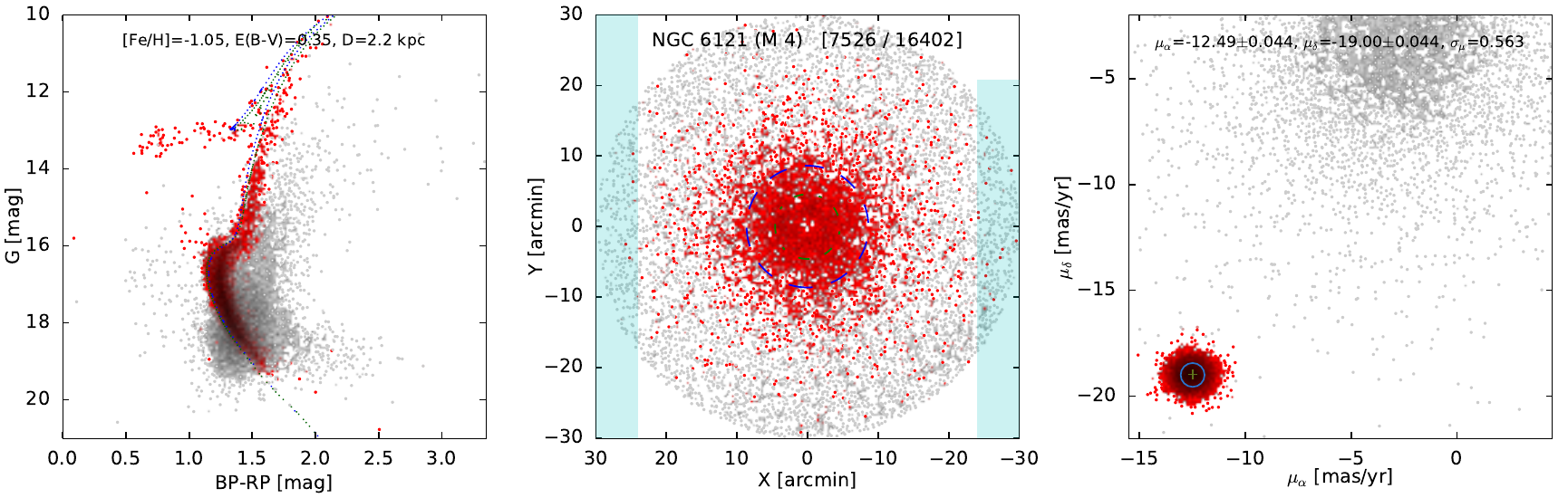}
\includegraphics{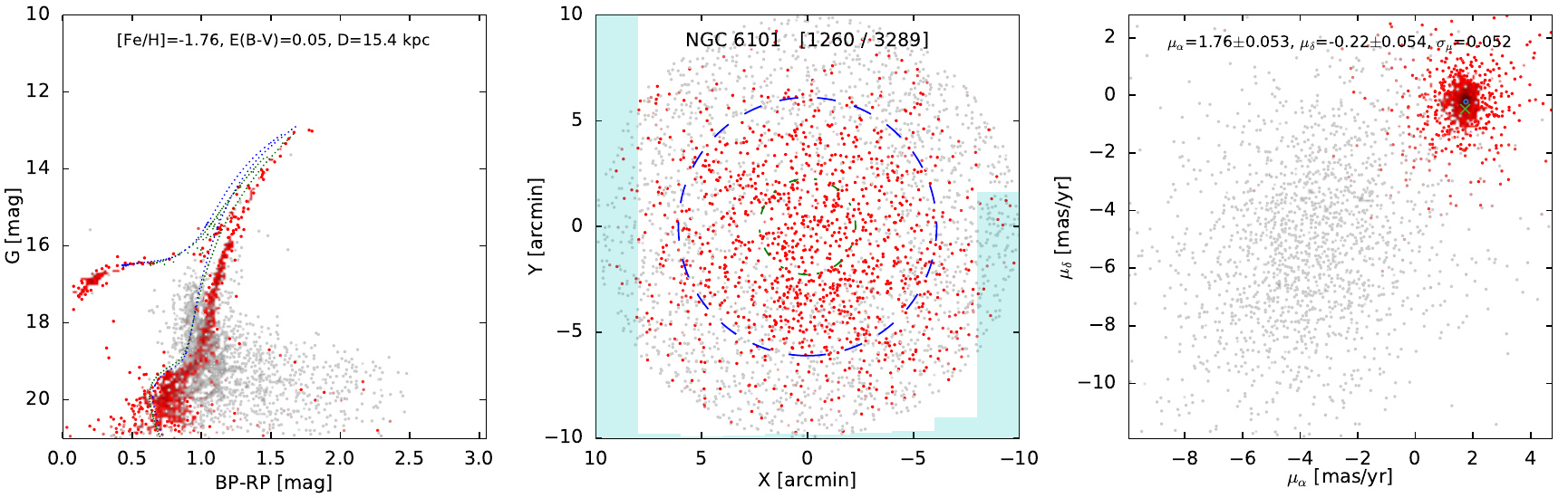}
\end{figure*}

\clearpage\begin{figure*}
\contcaption{}
\includegraphics{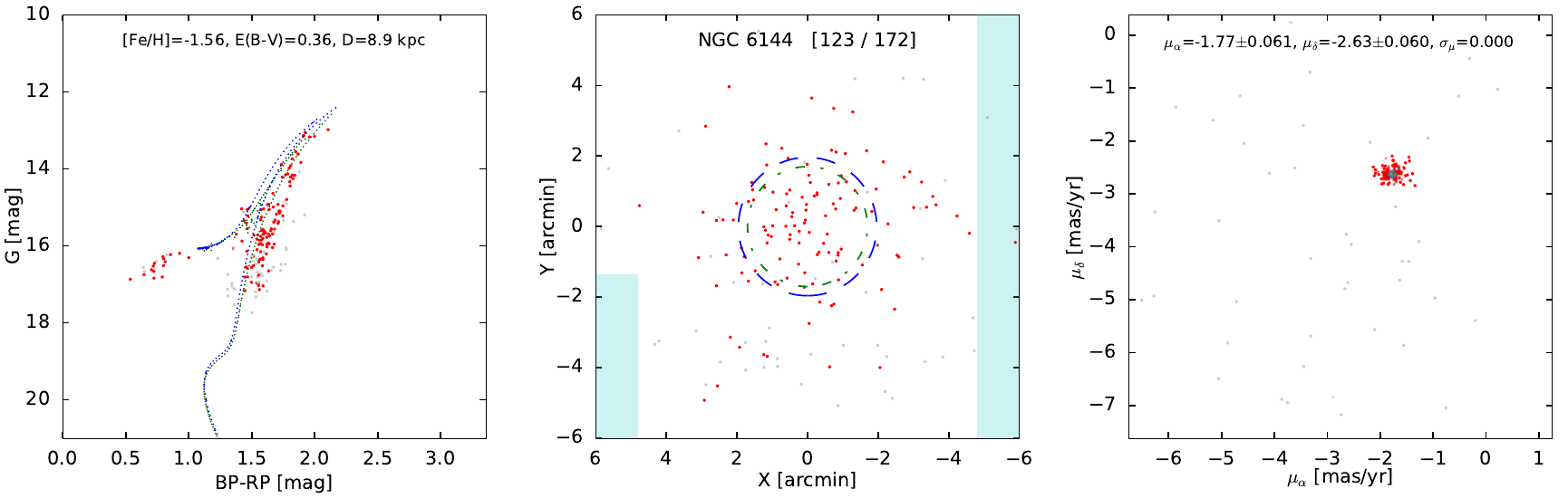}
\includegraphics{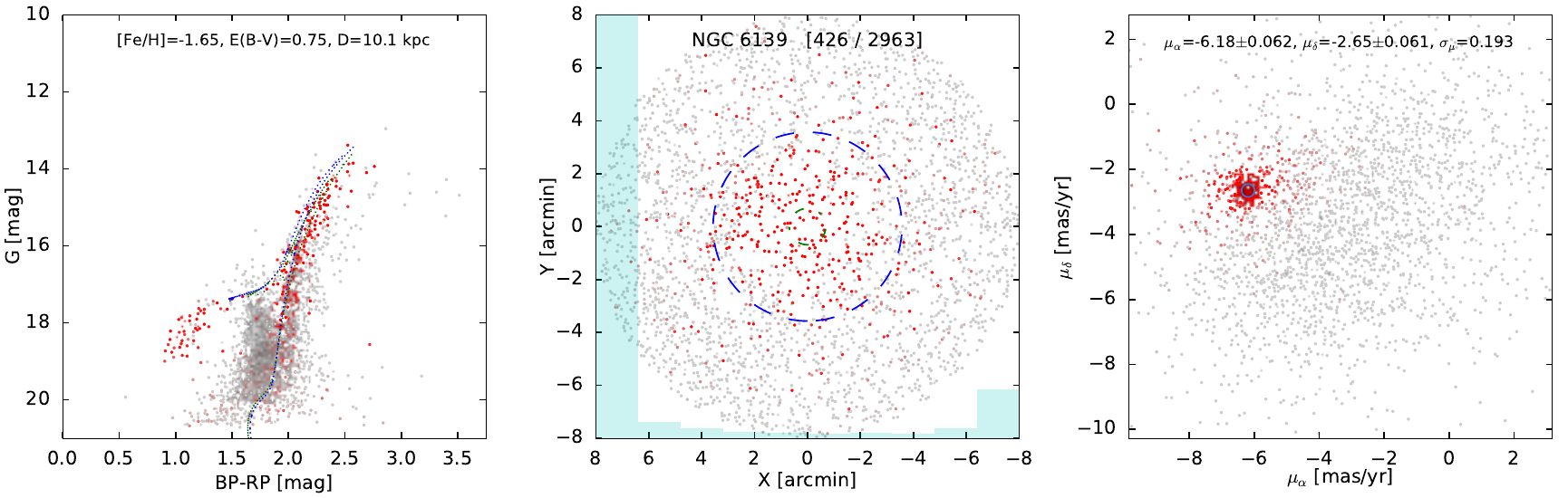}
\includegraphics{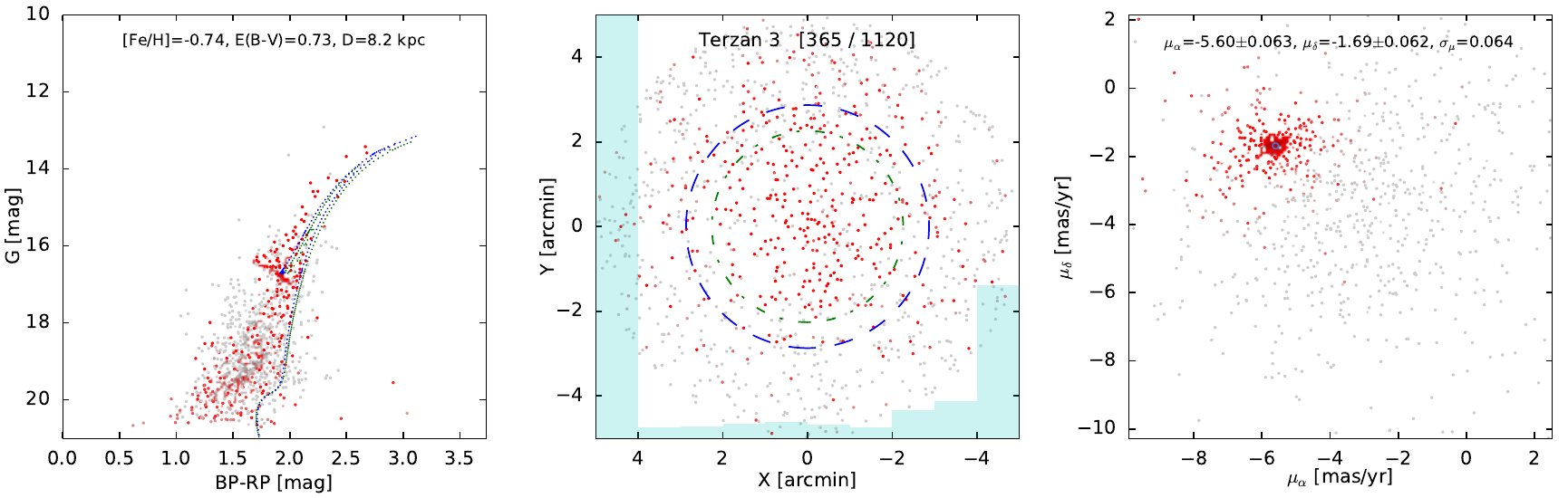}
\includegraphics{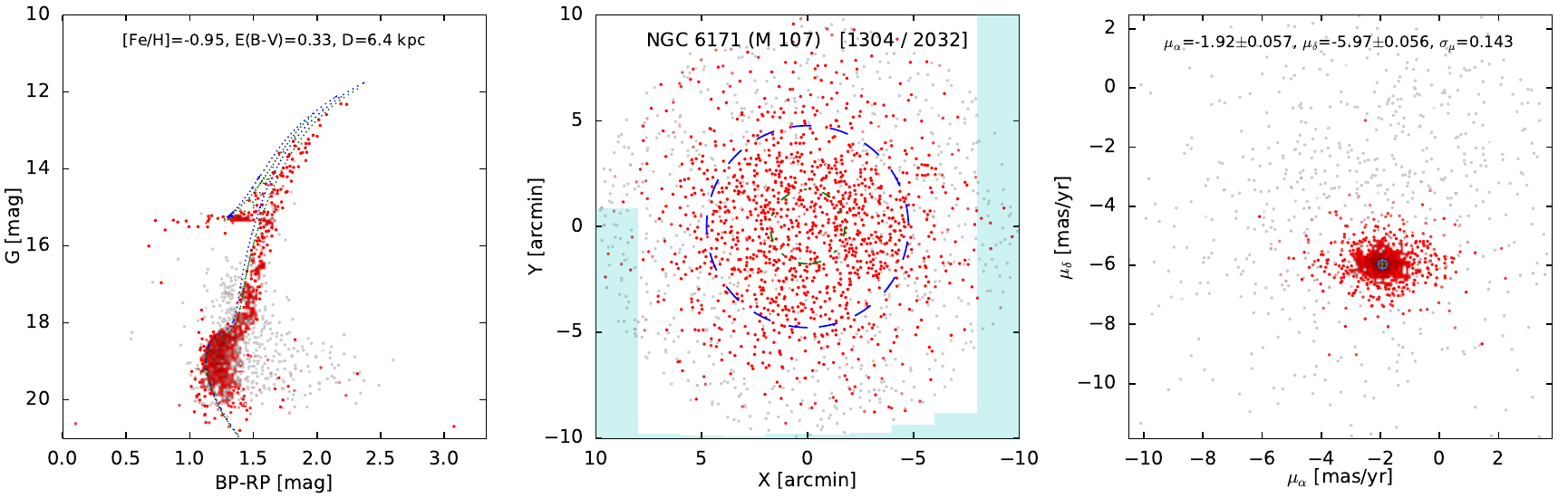}
\end{figure*}

\clearpage\begin{figure*}
\contcaption{}
\includegraphics{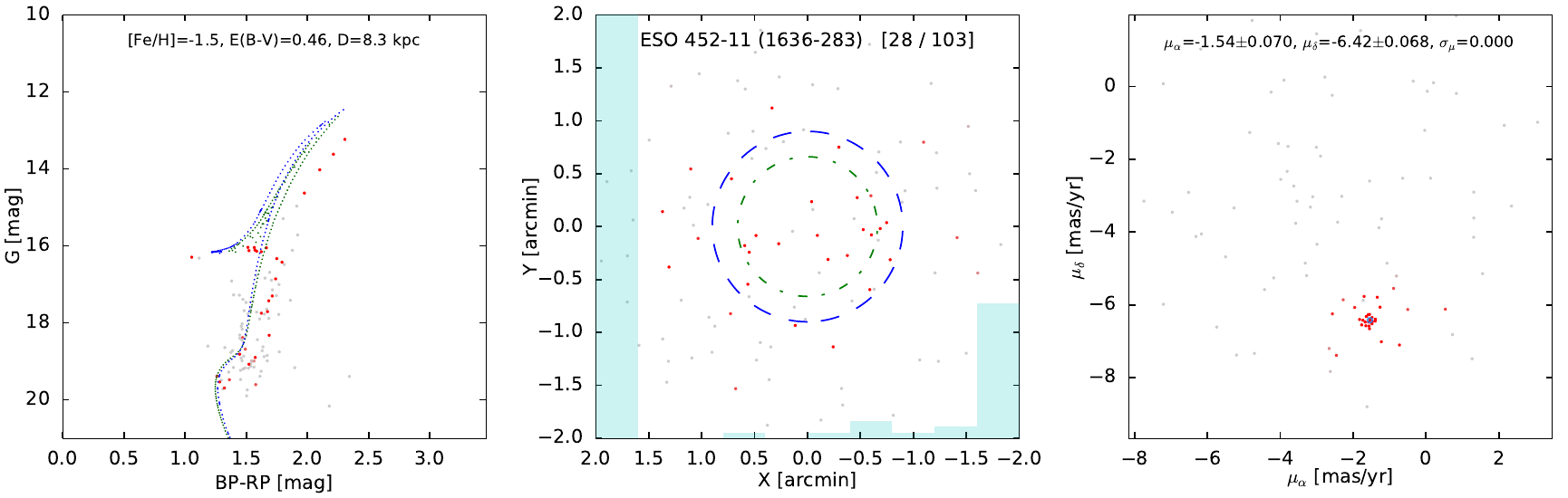}
\includegraphics{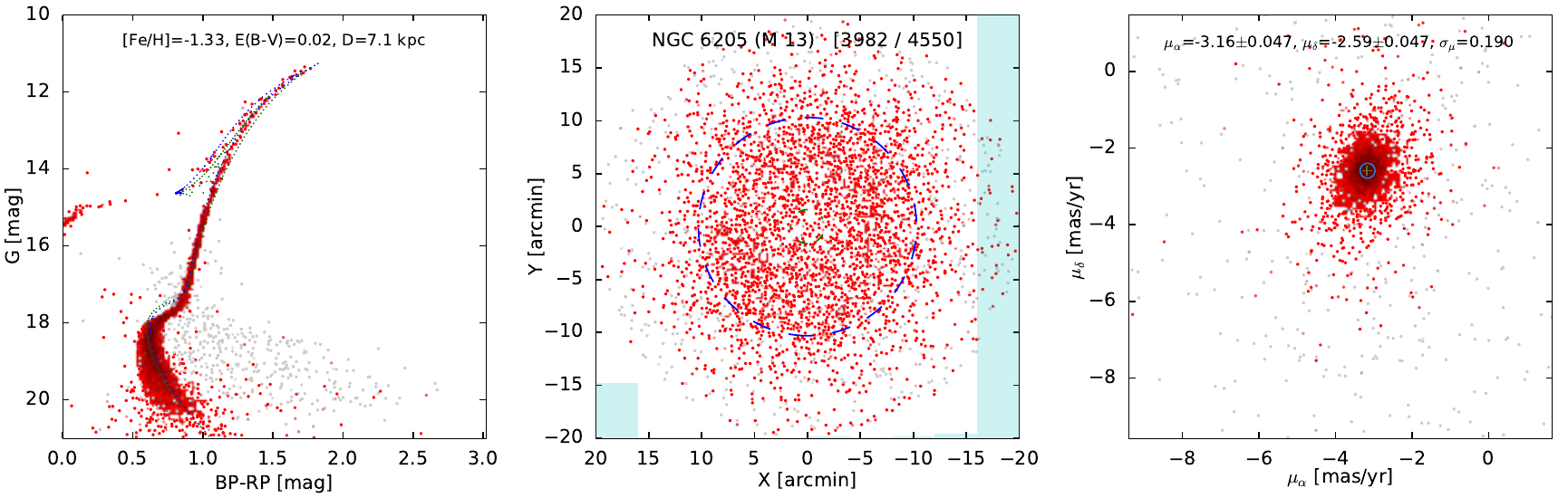}
\includegraphics{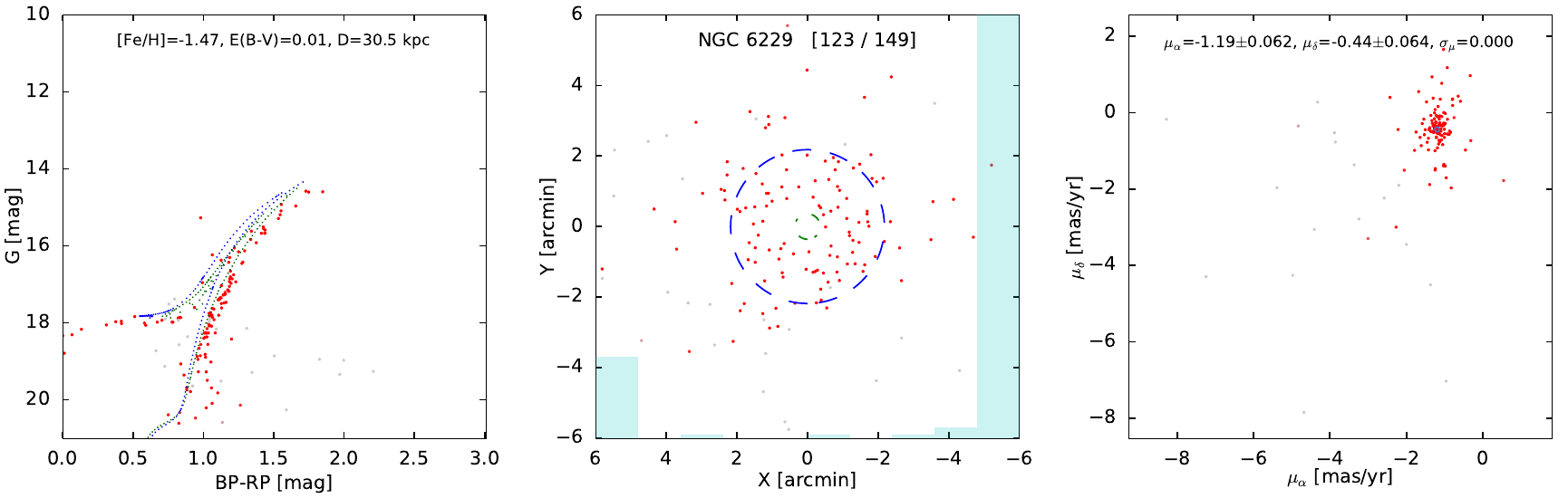}
\includegraphics{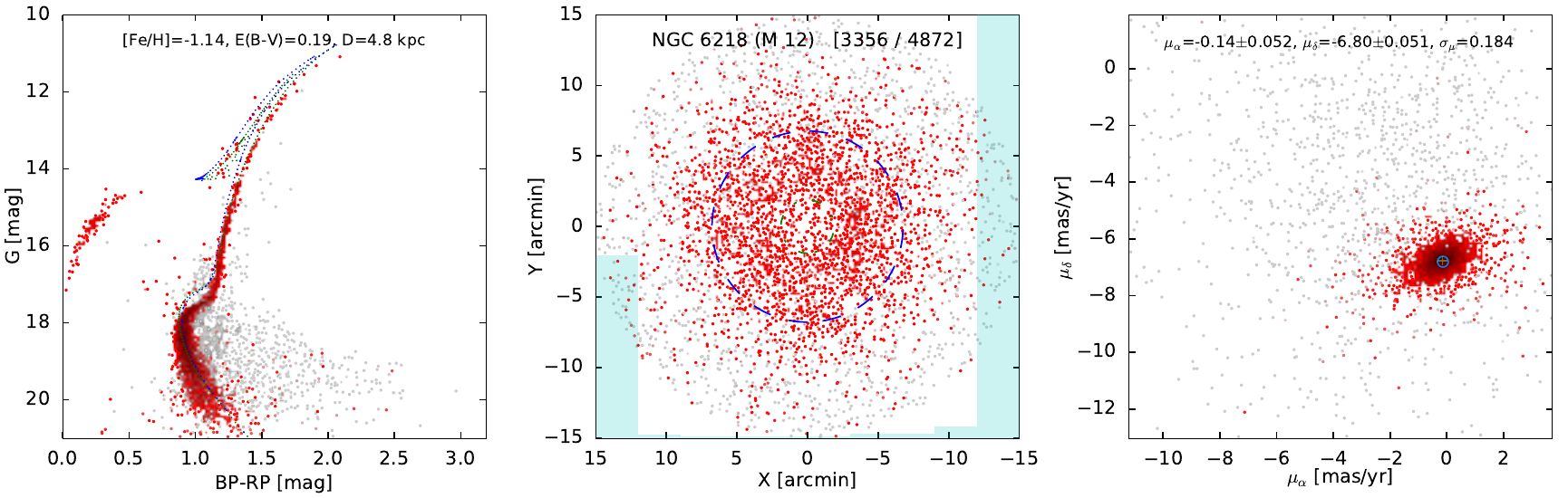}
\end{figure*}

\clearpage\begin{figure*}
\contcaption{}
\includegraphics{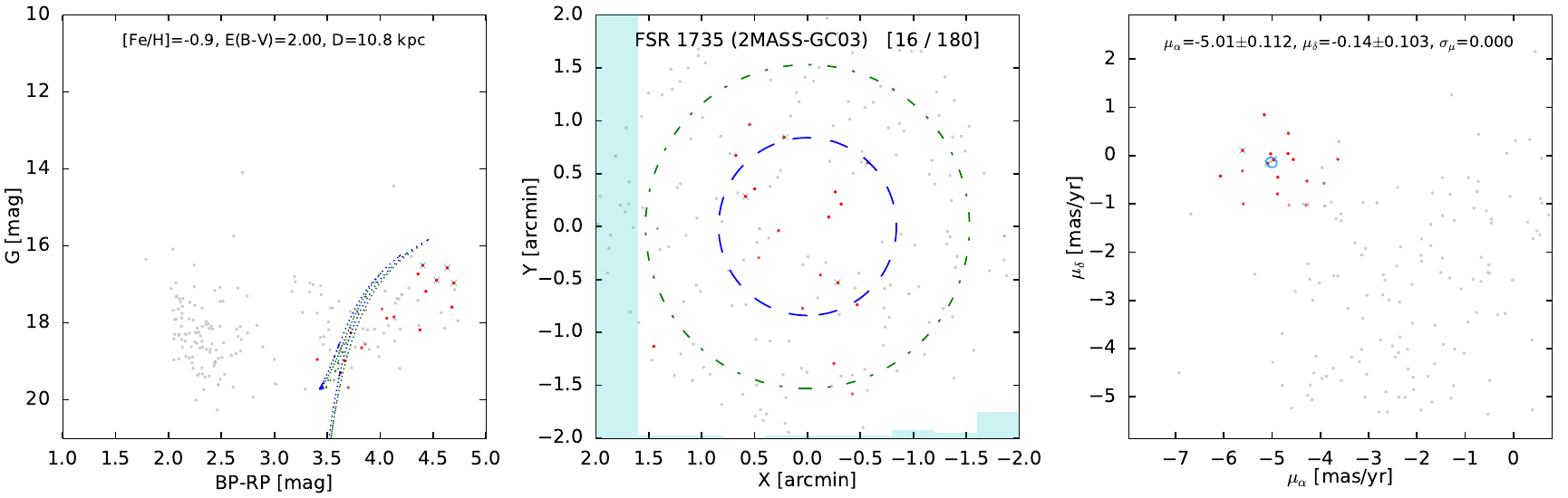}
\includegraphics{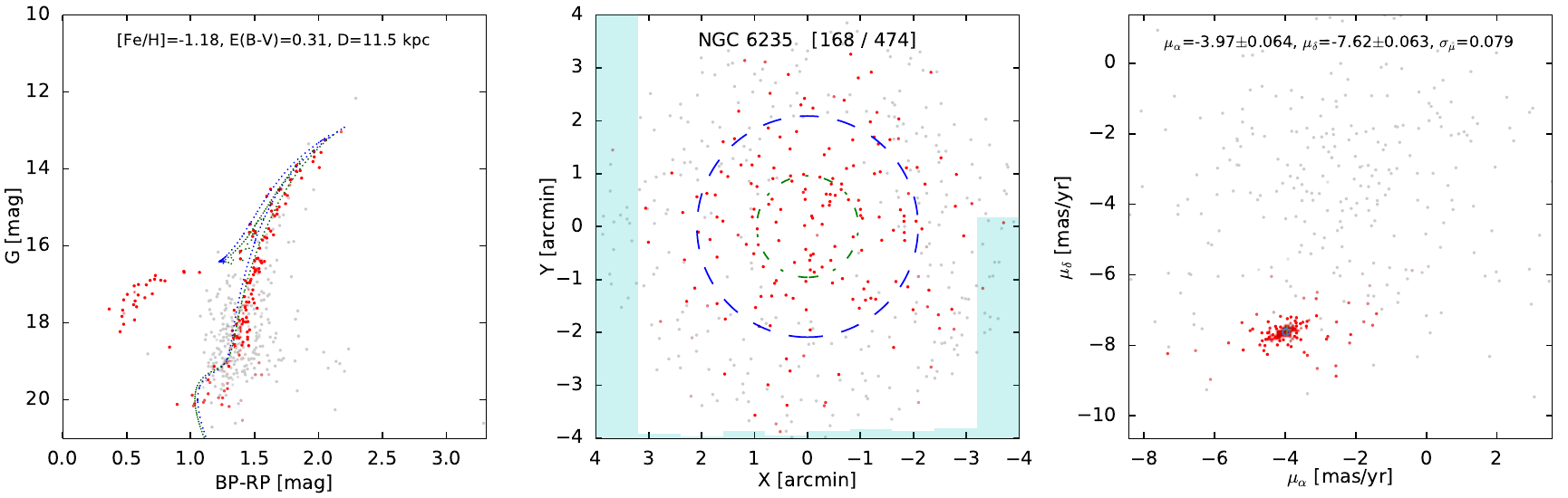}
\includegraphics{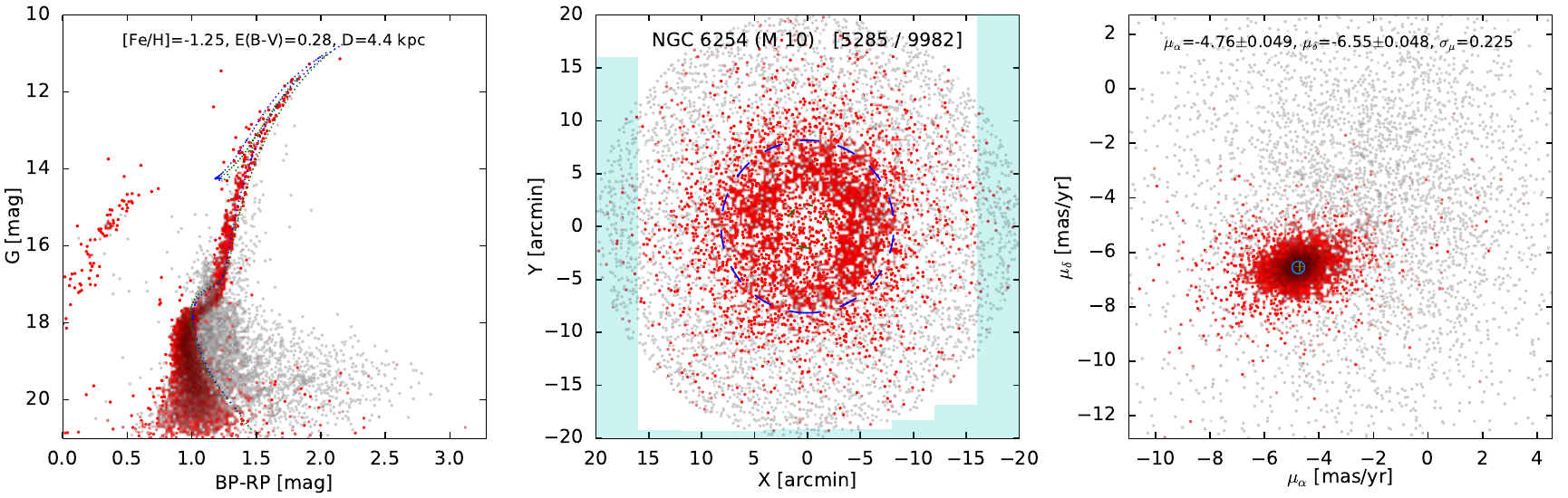}
\includegraphics{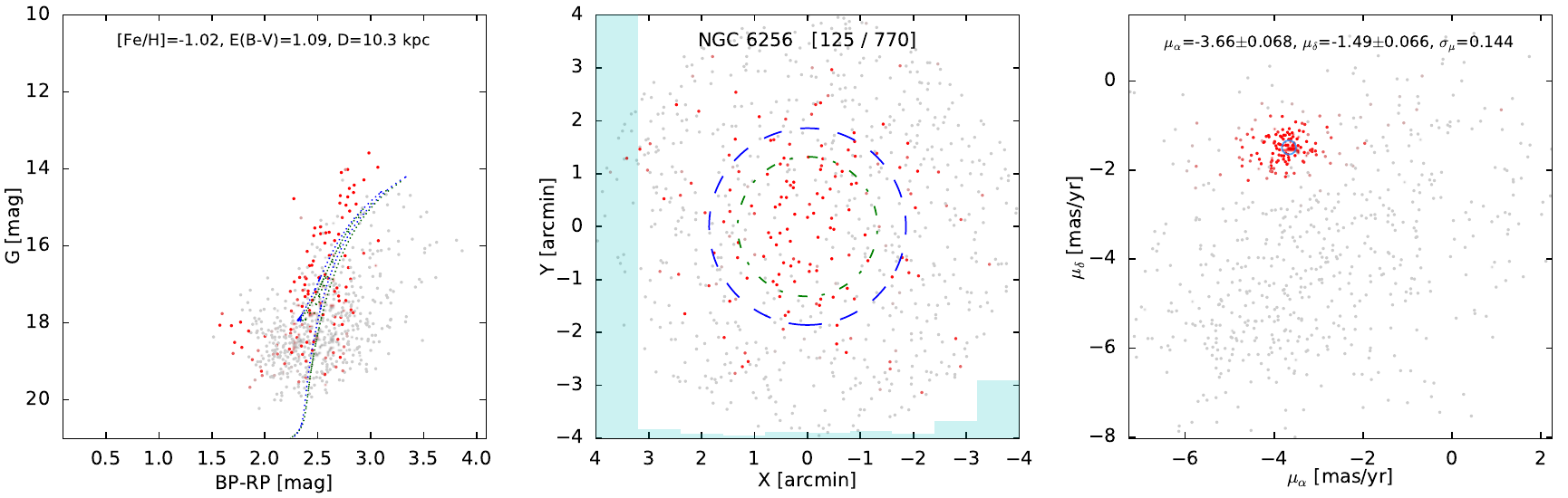}
\end{figure*}

\clearpage\begin{figure*}
\contcaption{}
\includegraphics{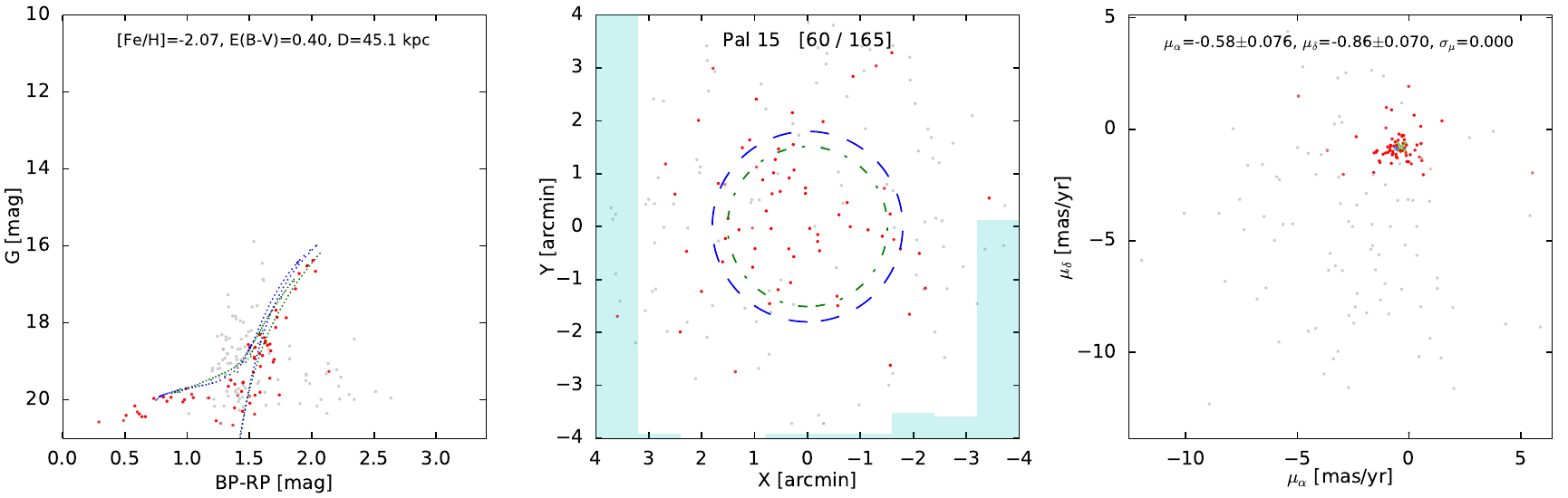}
\includegraphics{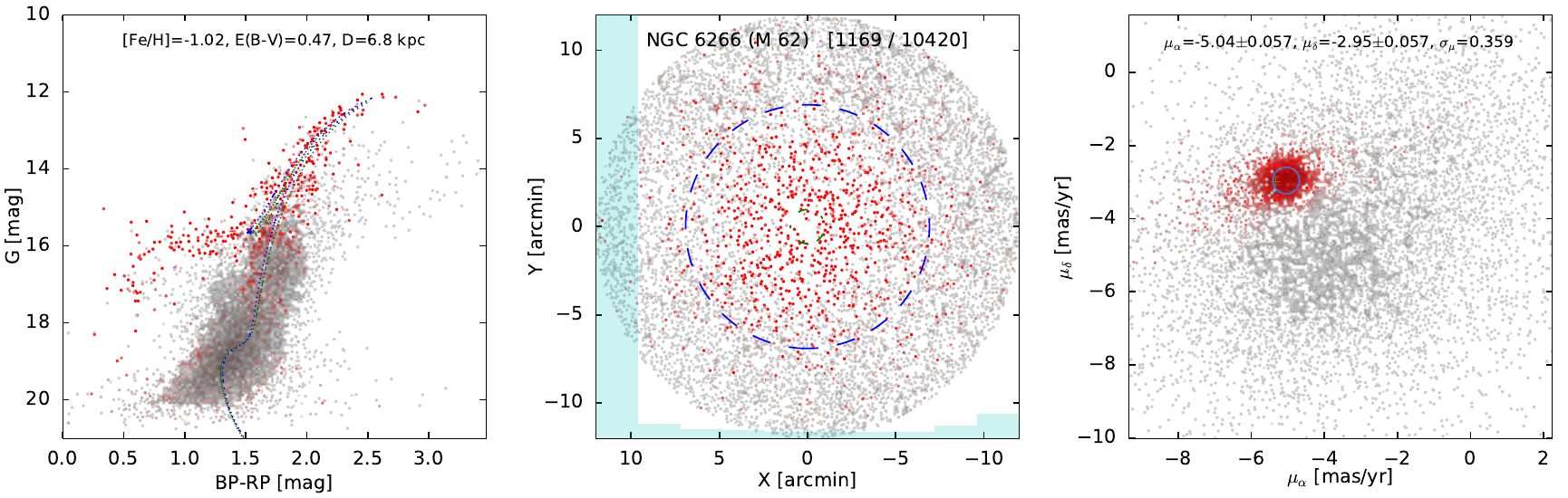}
\includegraphics{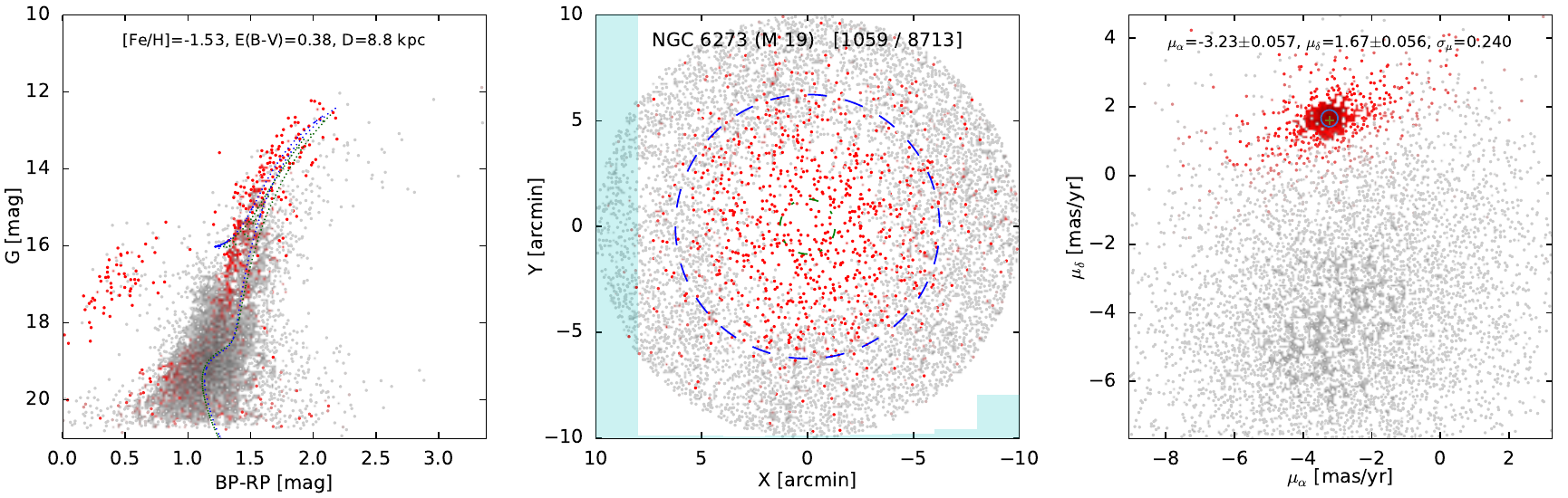}
\includegraphics{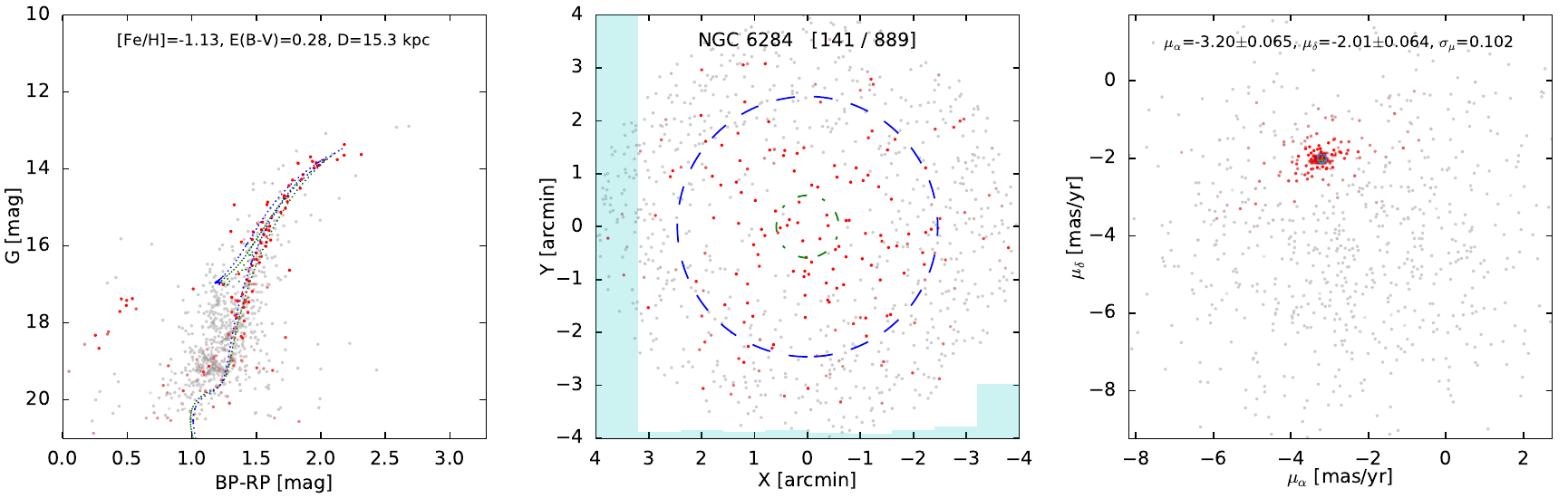}
\end{figure*}

\clearpage\begin{figure*}
\contcaption{}
\includegraphics{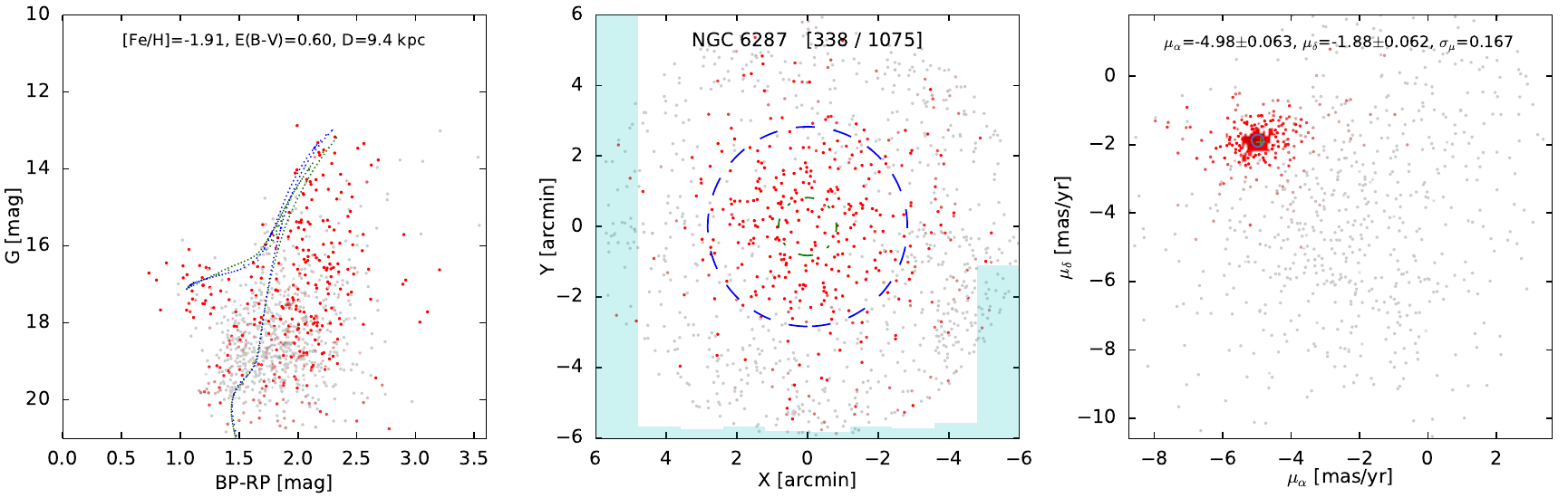}
\includegraphics{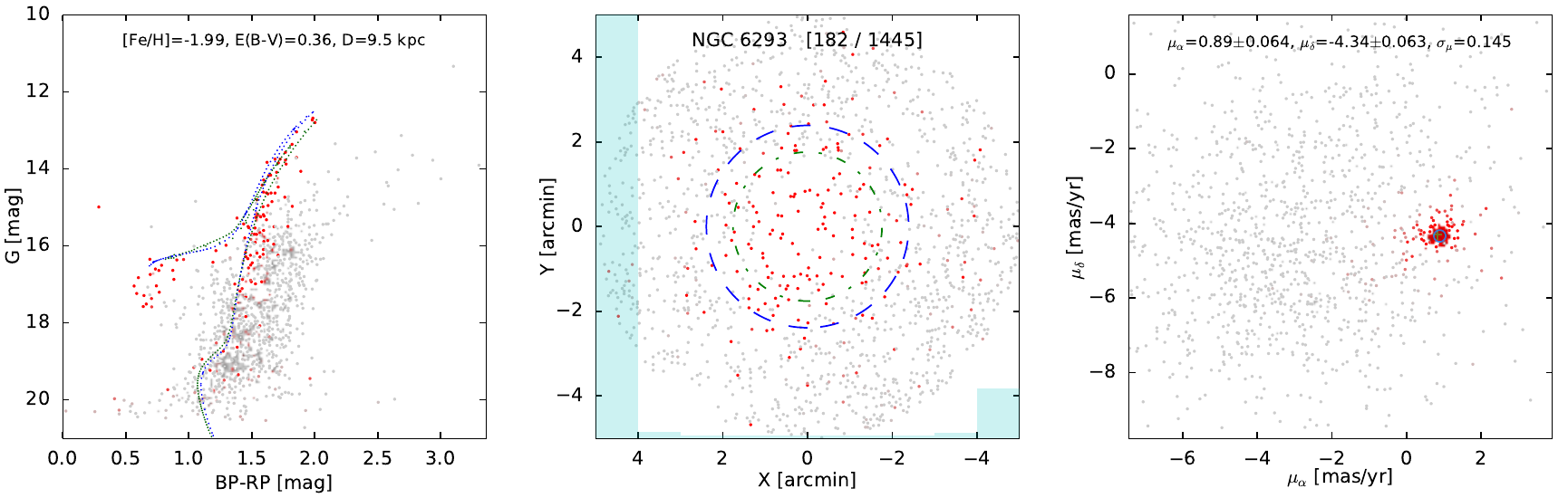}
\includegraphics{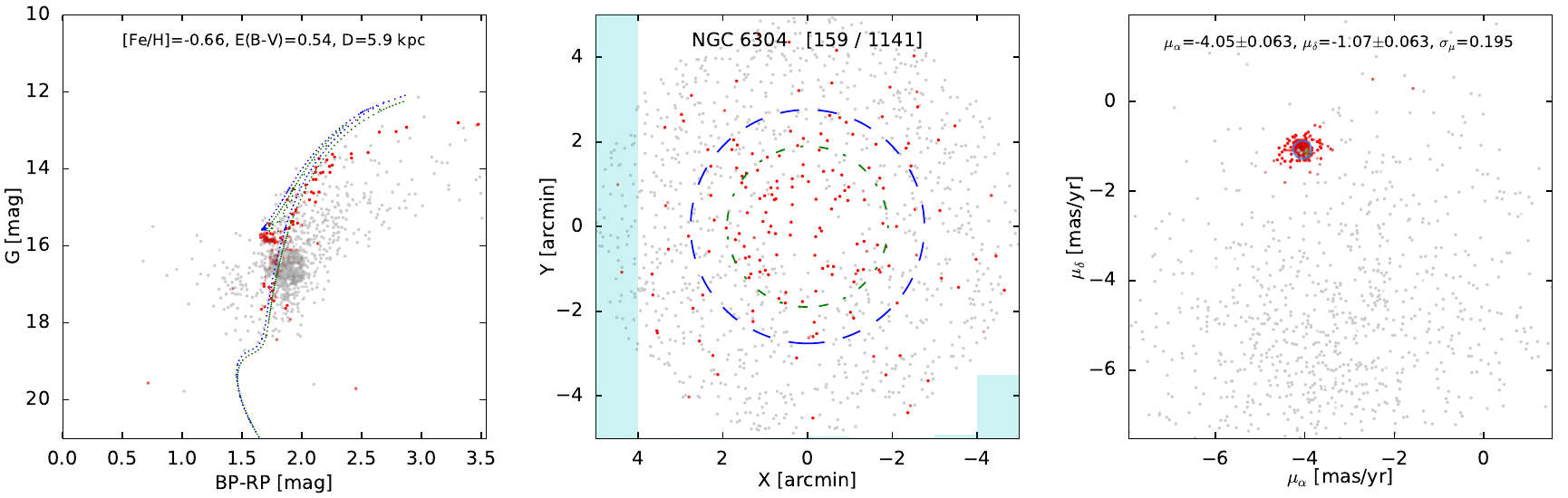}
\includegraphics{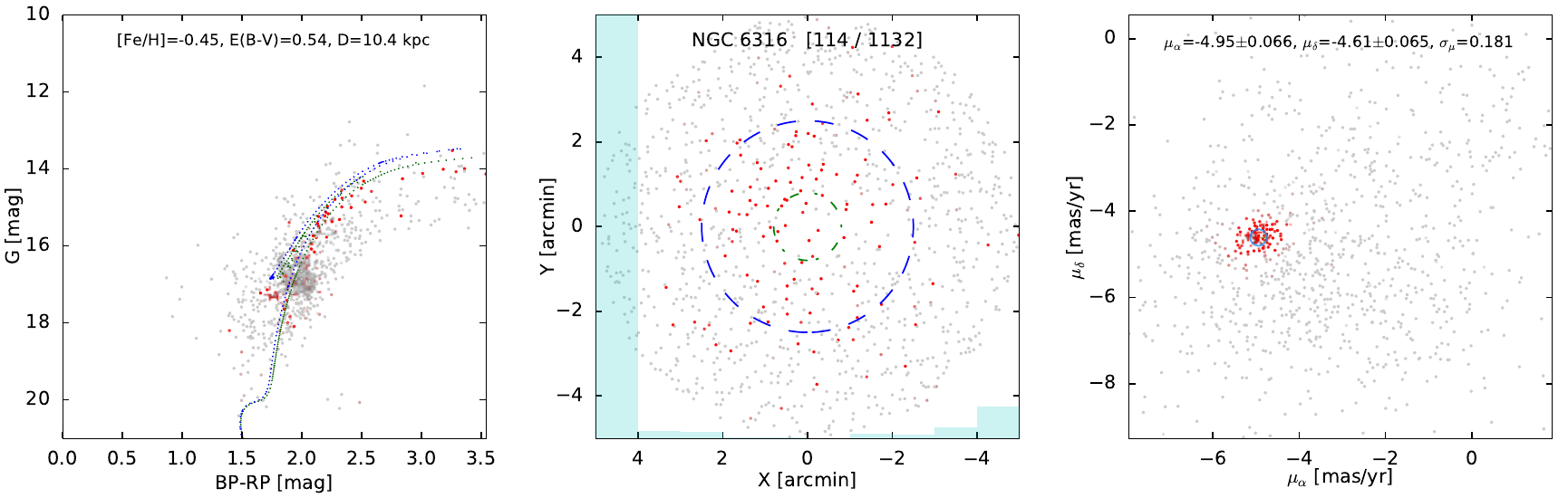}
\end{figure*}

\clearpage\begin{figure*}
\contcaption{}
\includegraphics{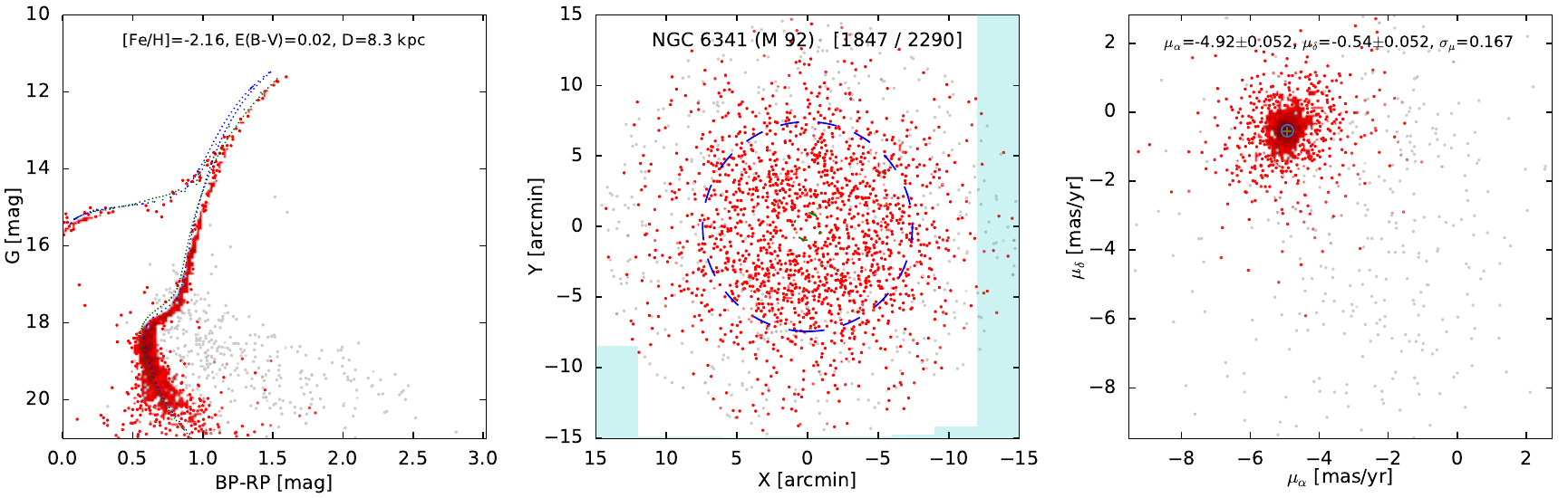}
\includegraphics{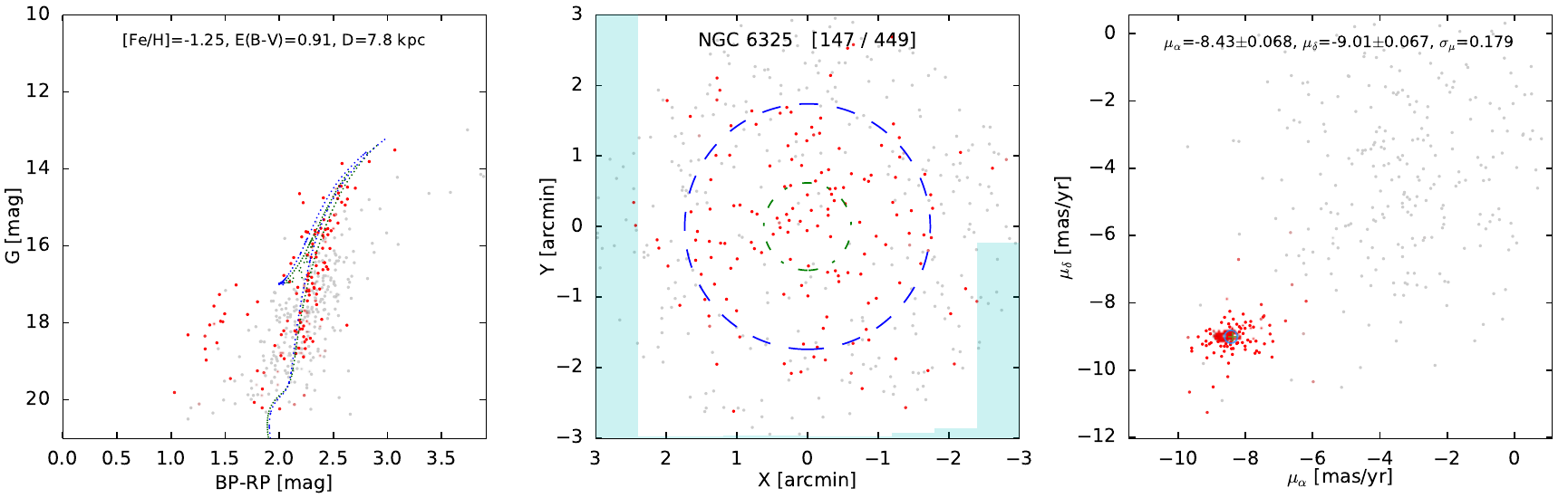}
\includegraphics{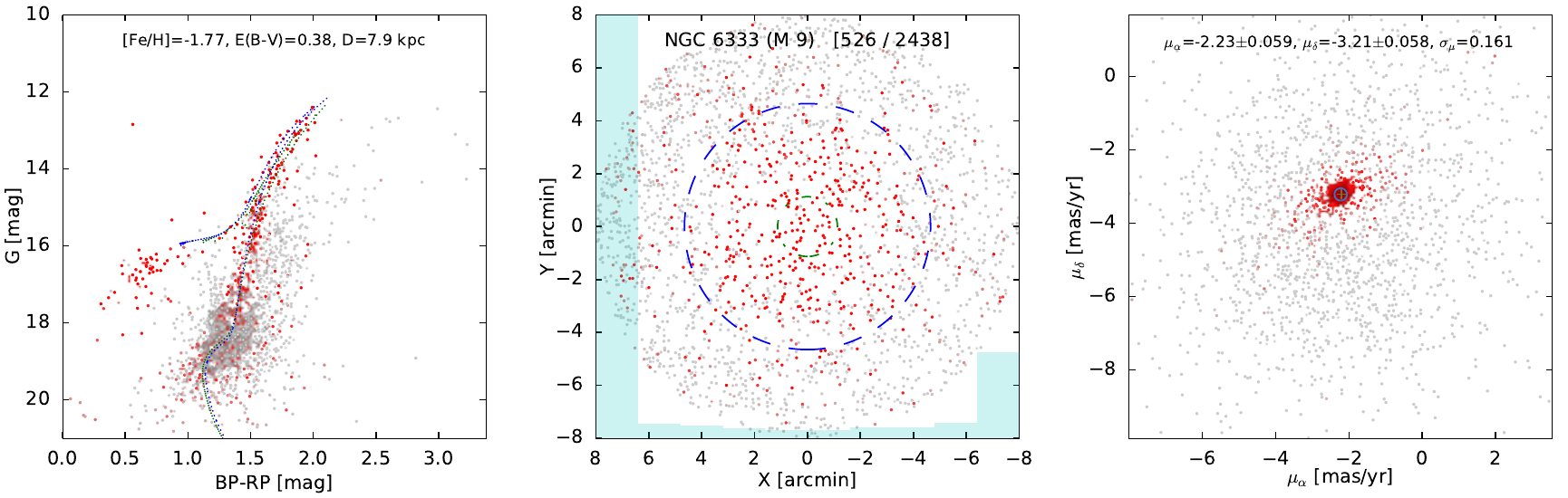}
\includegraphics{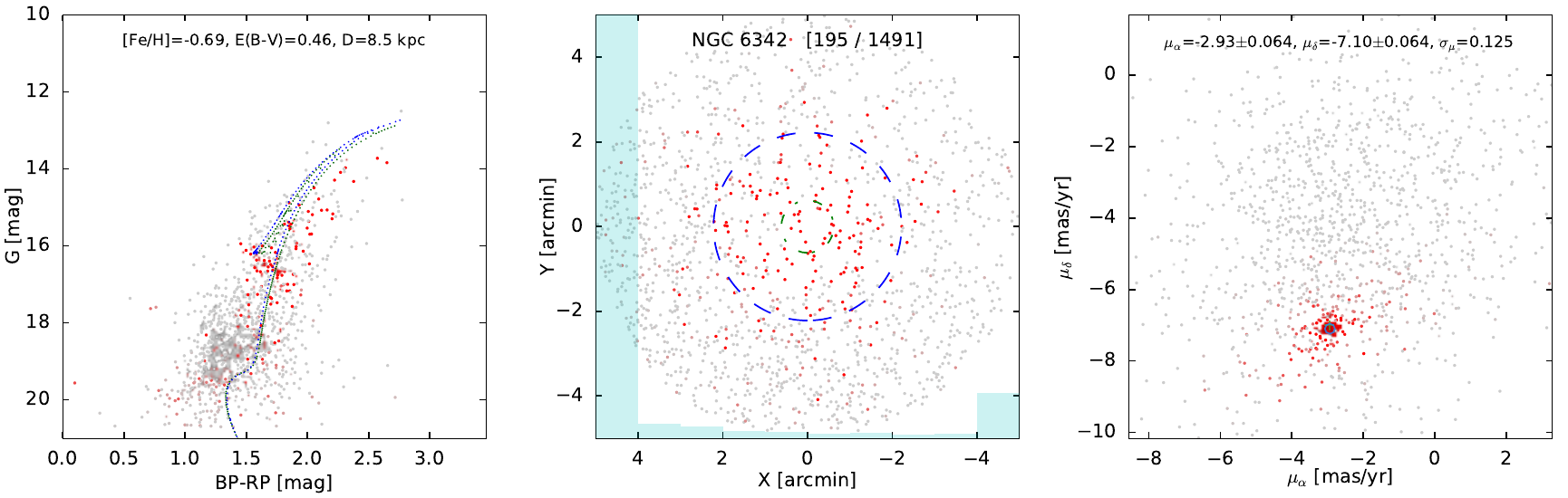}
\end{figure*}

\clearpage\begin{figure*}
\contcaption{}
\includegraphics{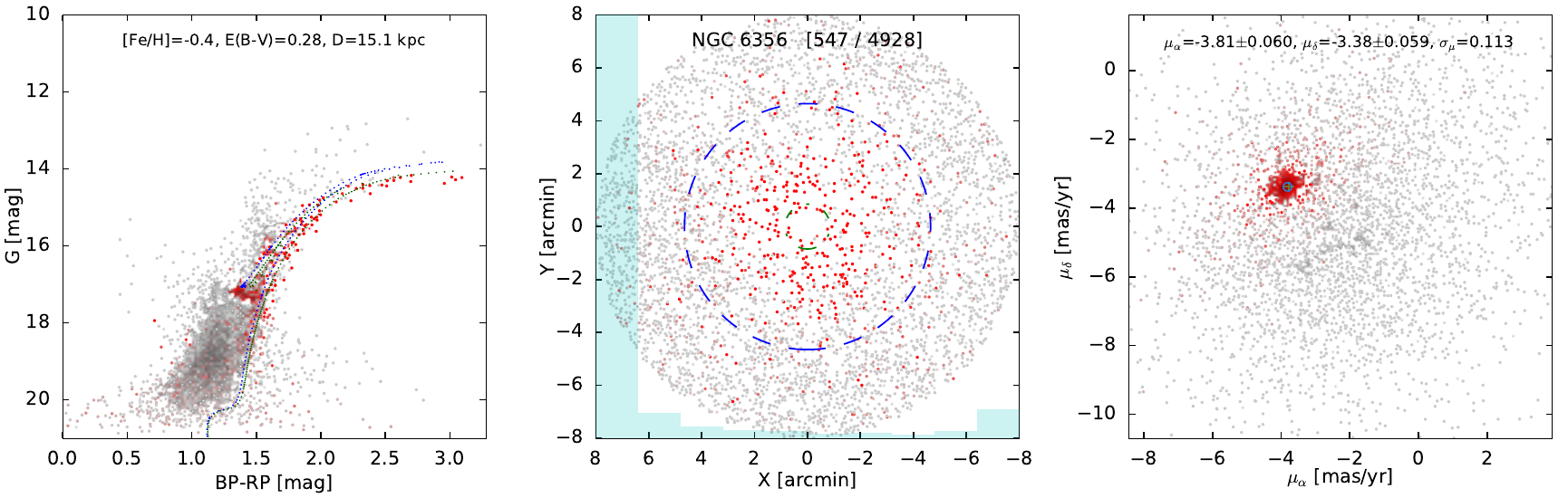}
\includegraphics{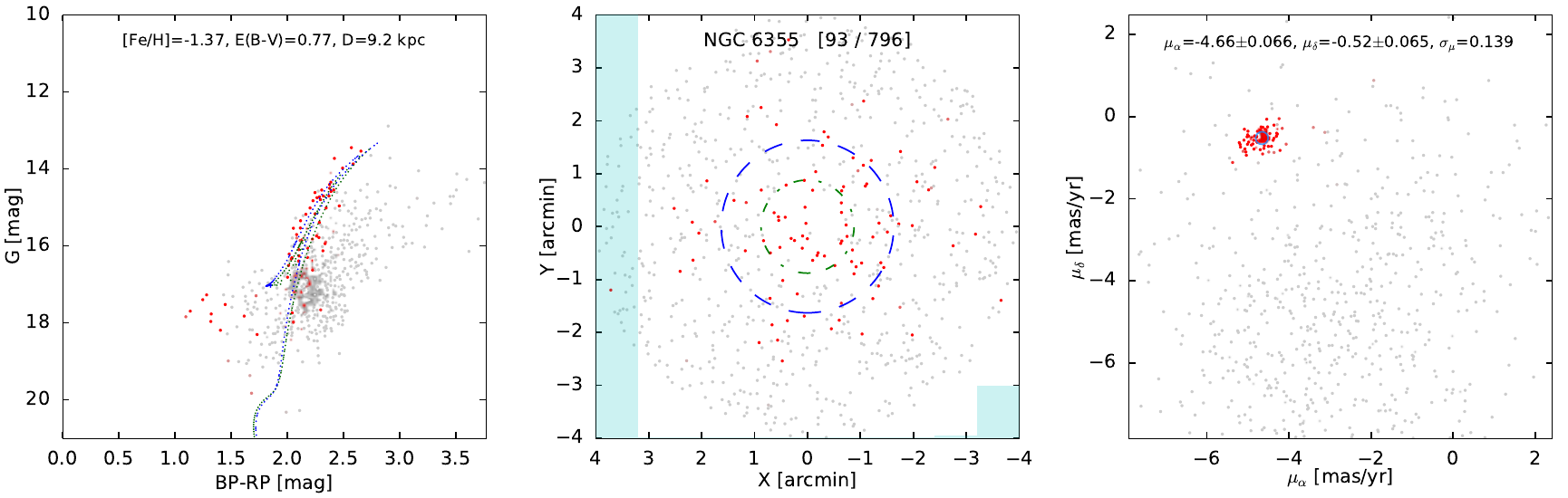}
\includegraphics{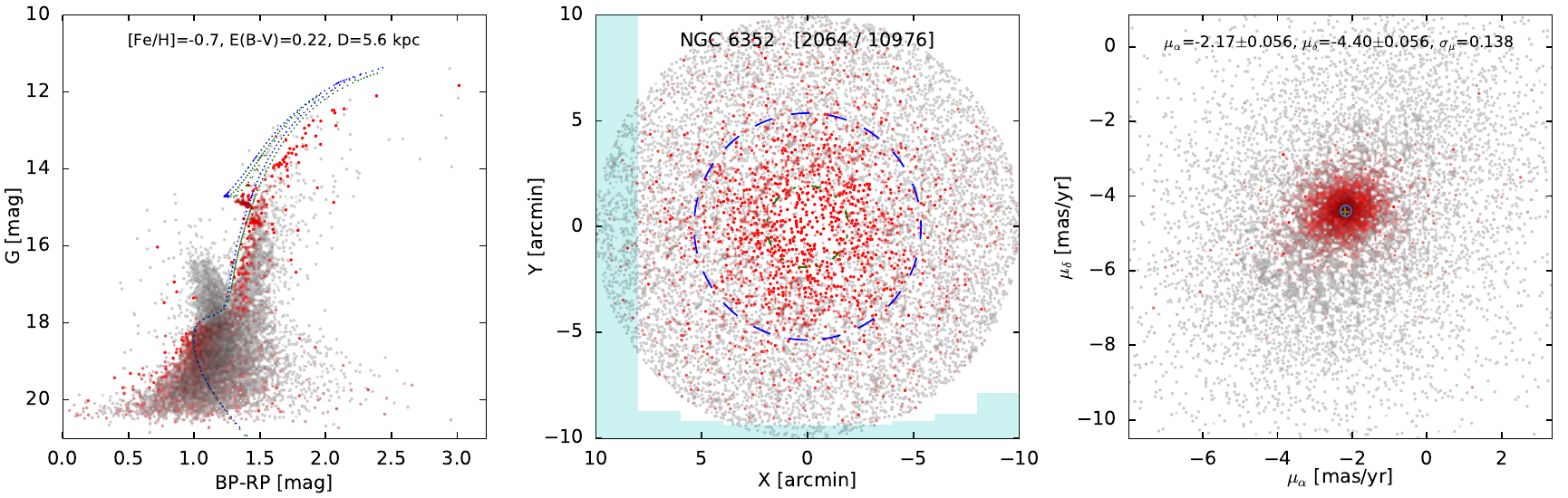}
\includegraphics{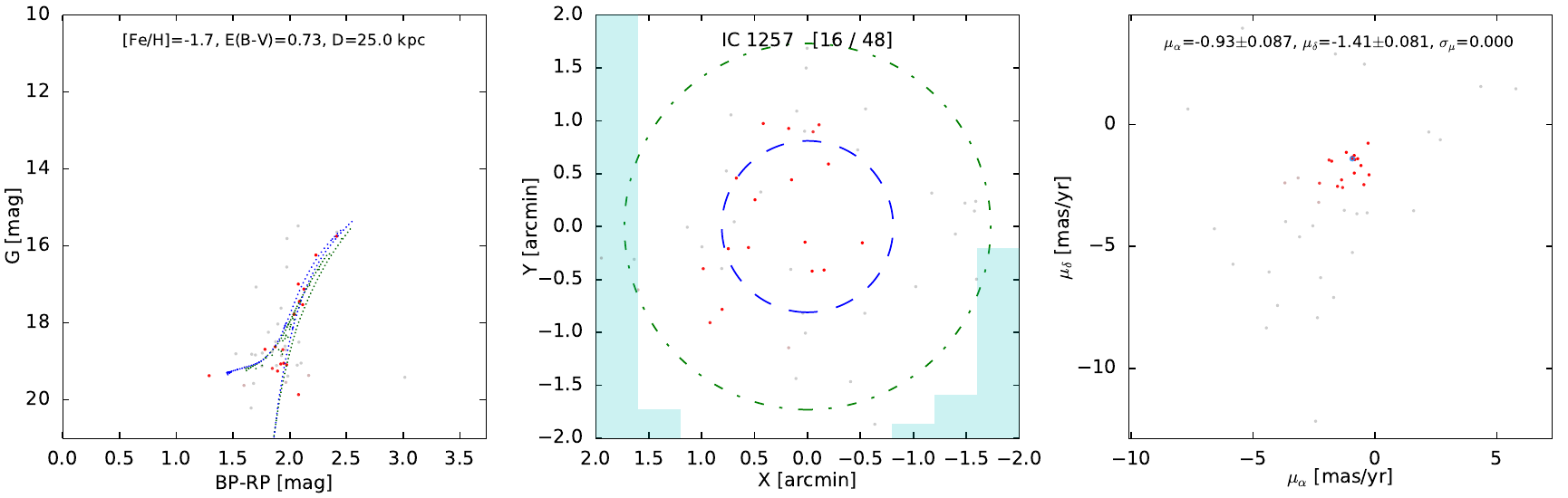}
\end{figure*}

\clearpage\begin{figure*}
\contcaption{}
\includegraphics{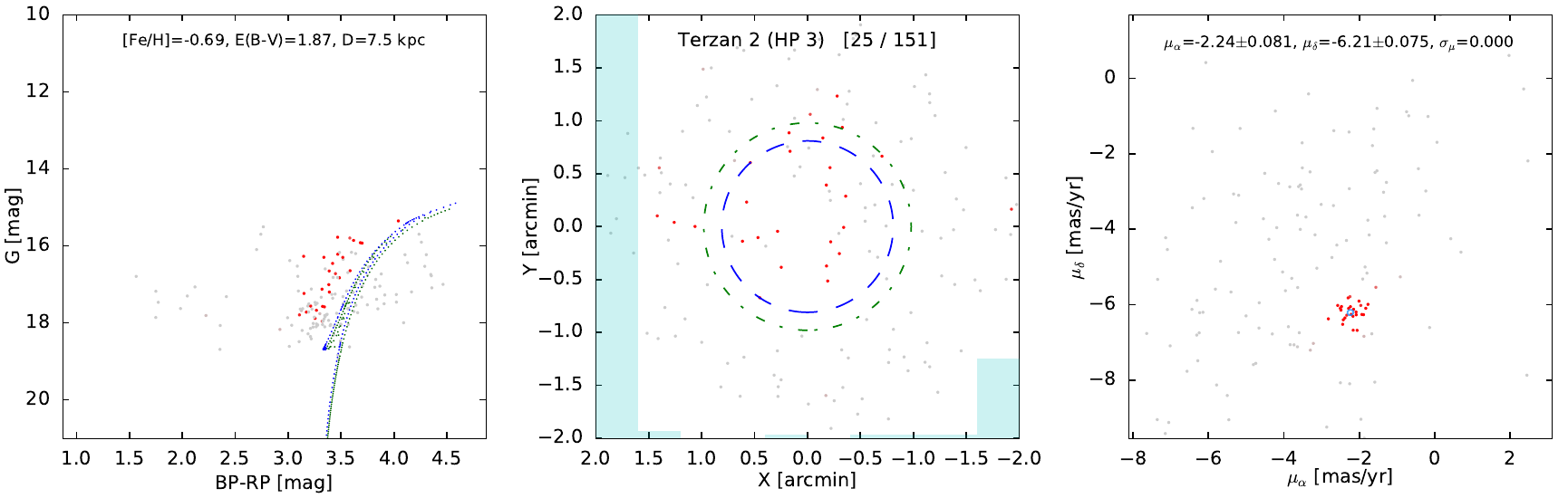}
\includegraphics{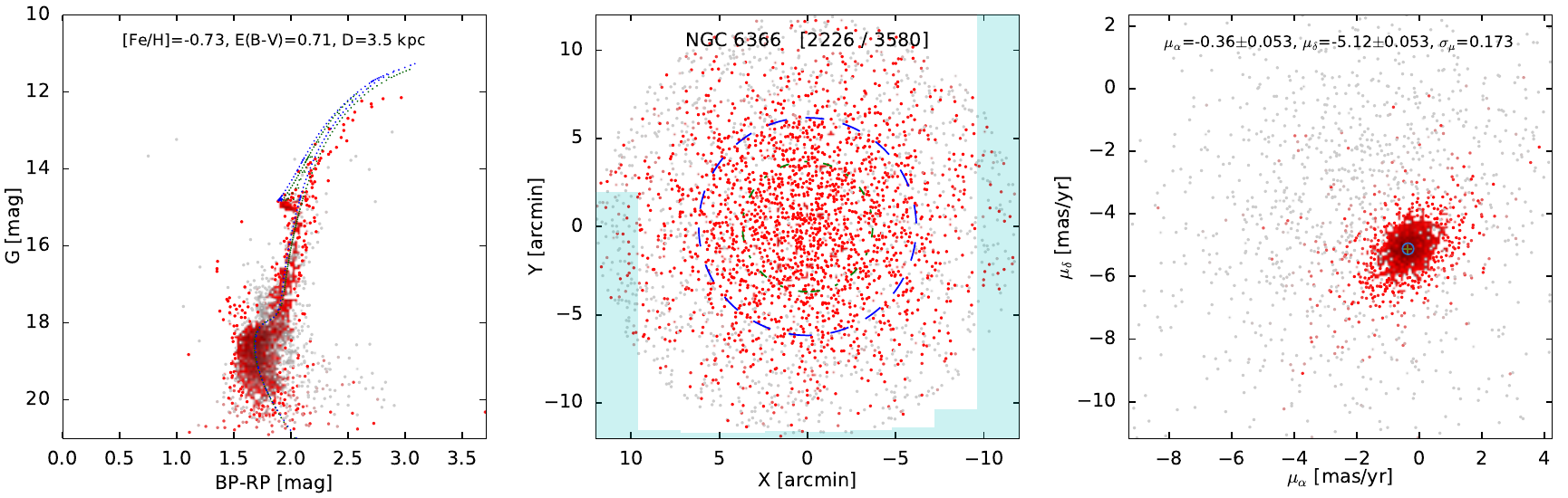}
\includegraphics{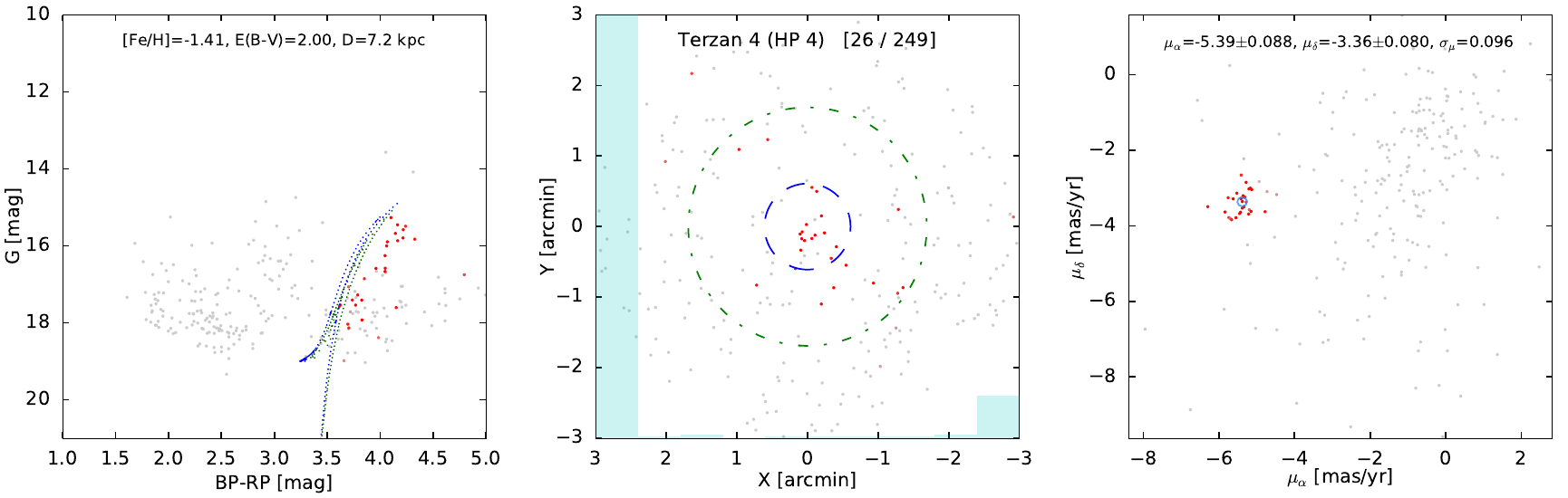}
\includegraphics{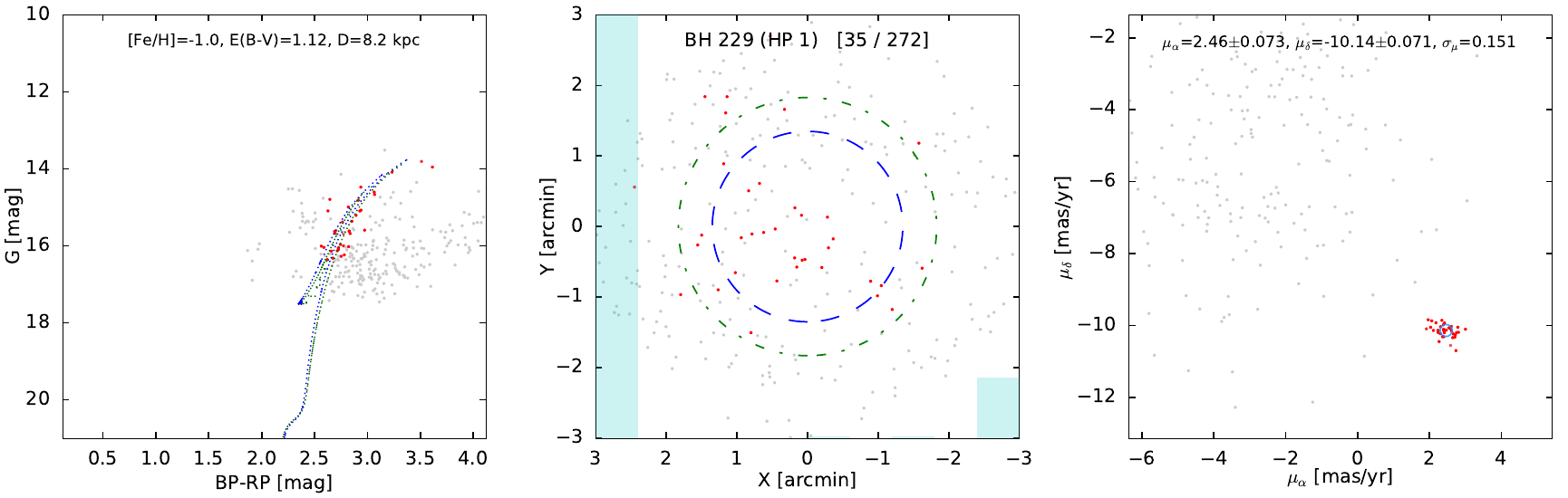}
\end{figure*}

\clearpage\begin{figure*}
\contcaption{}
\includegraphics{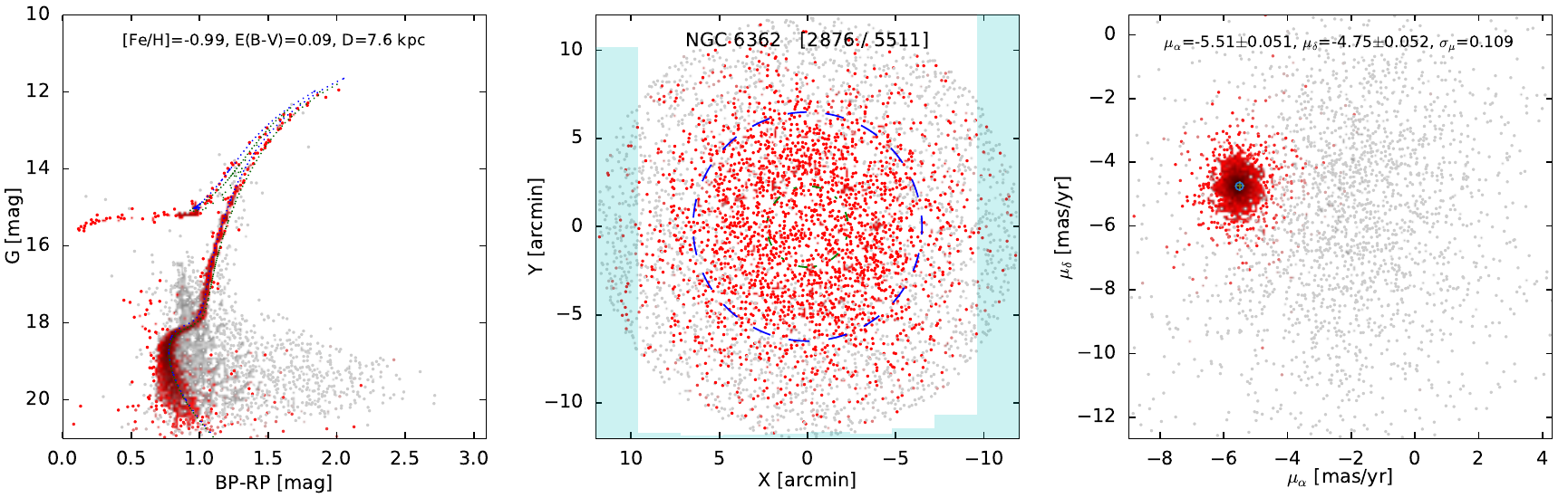}
\includegraphics{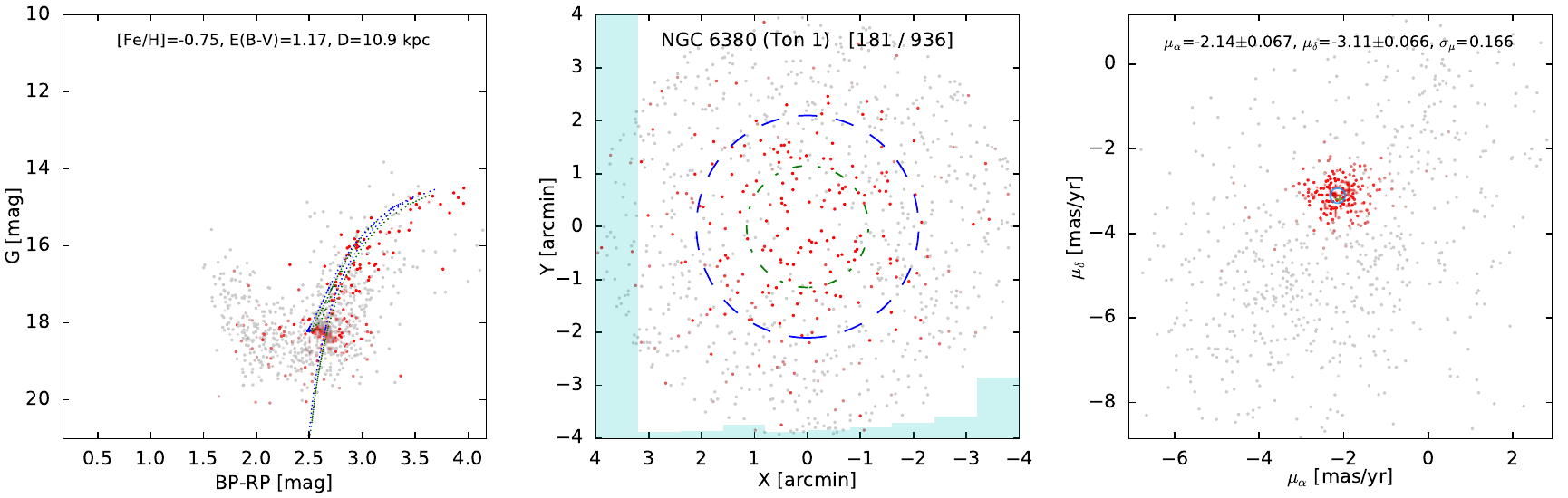}
\includegraphics{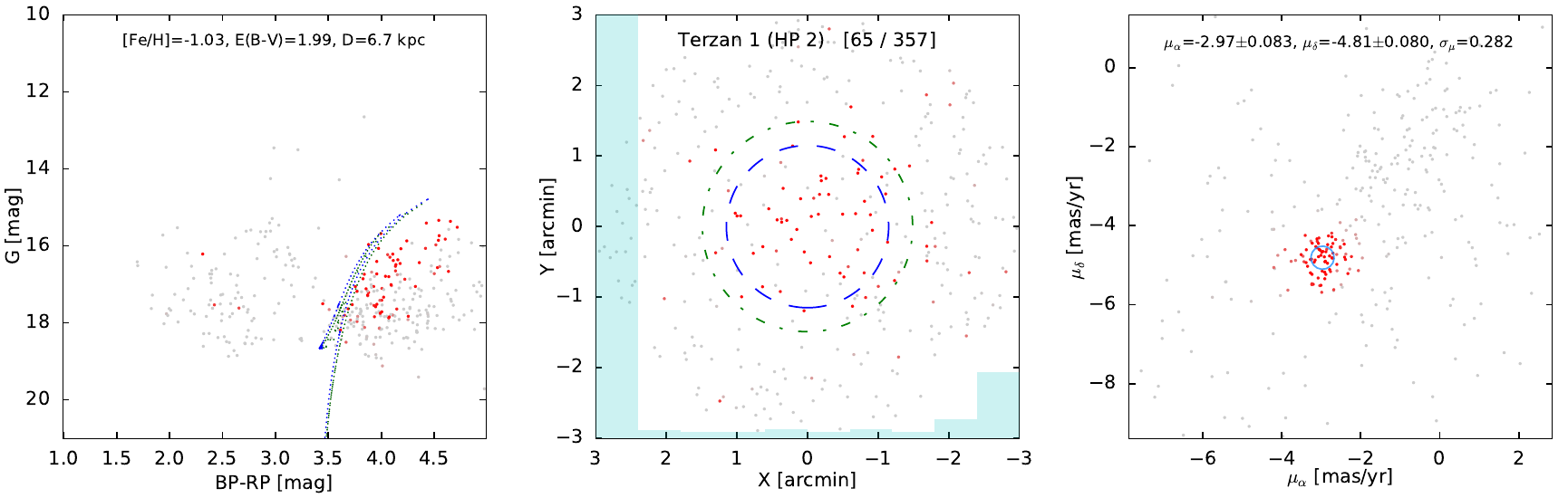}
\includegraphics{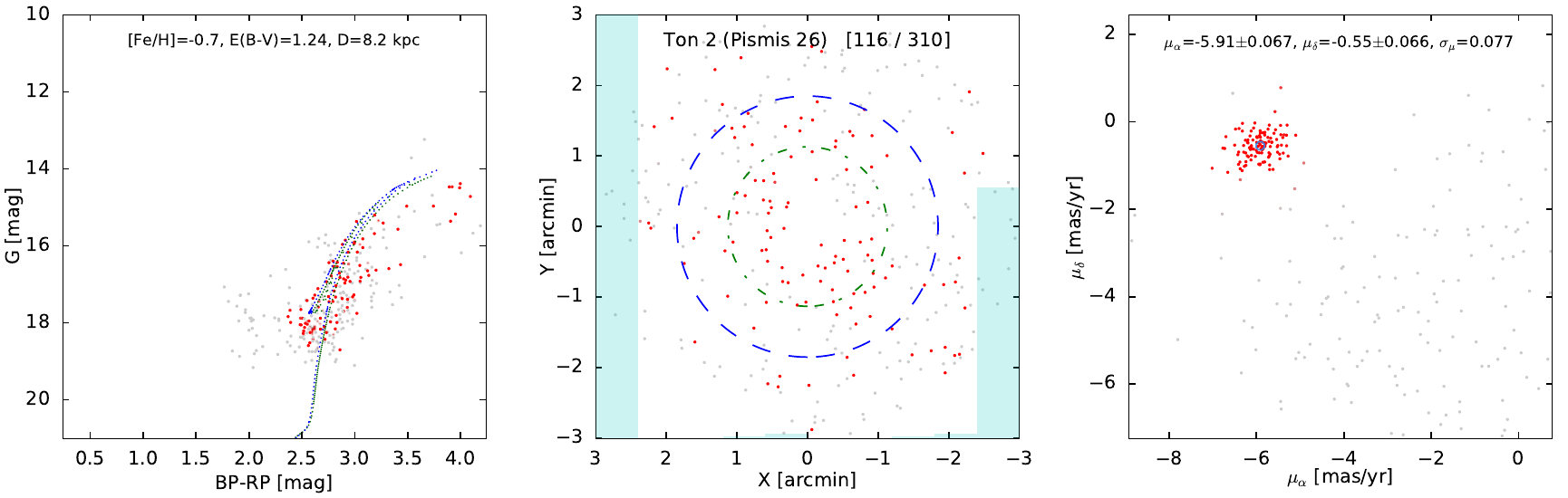}
\end{figure*}

\clearpage\begin{figure*}
\contcaption{}
\includegraphics{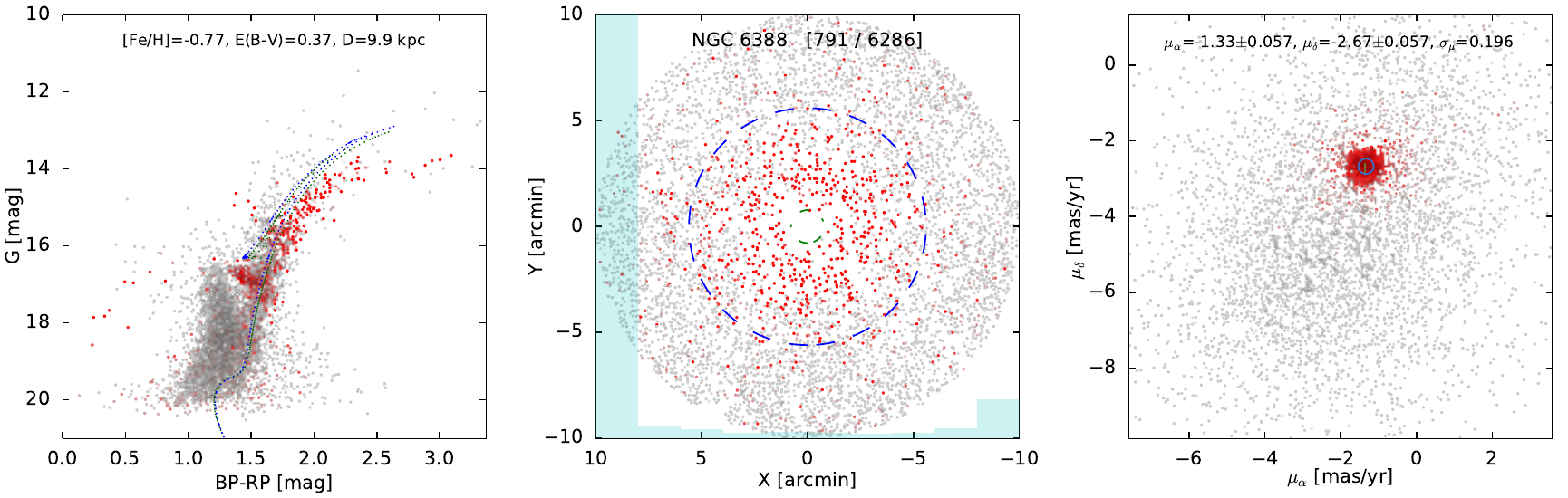}
\includegraphics{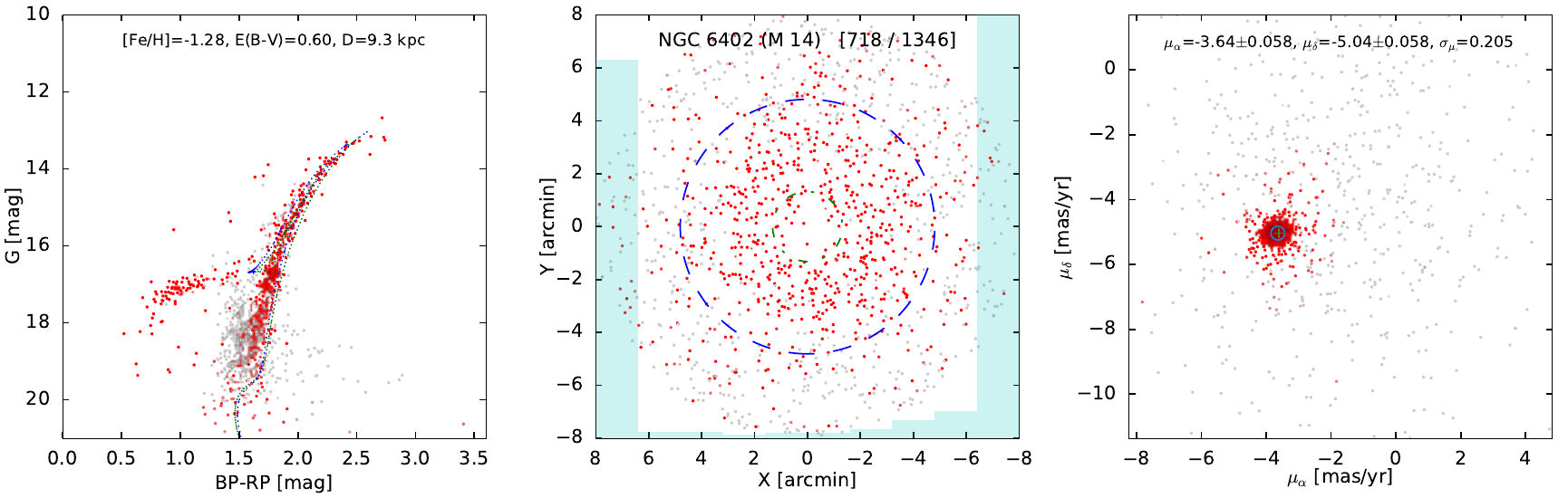}
\includegraphics{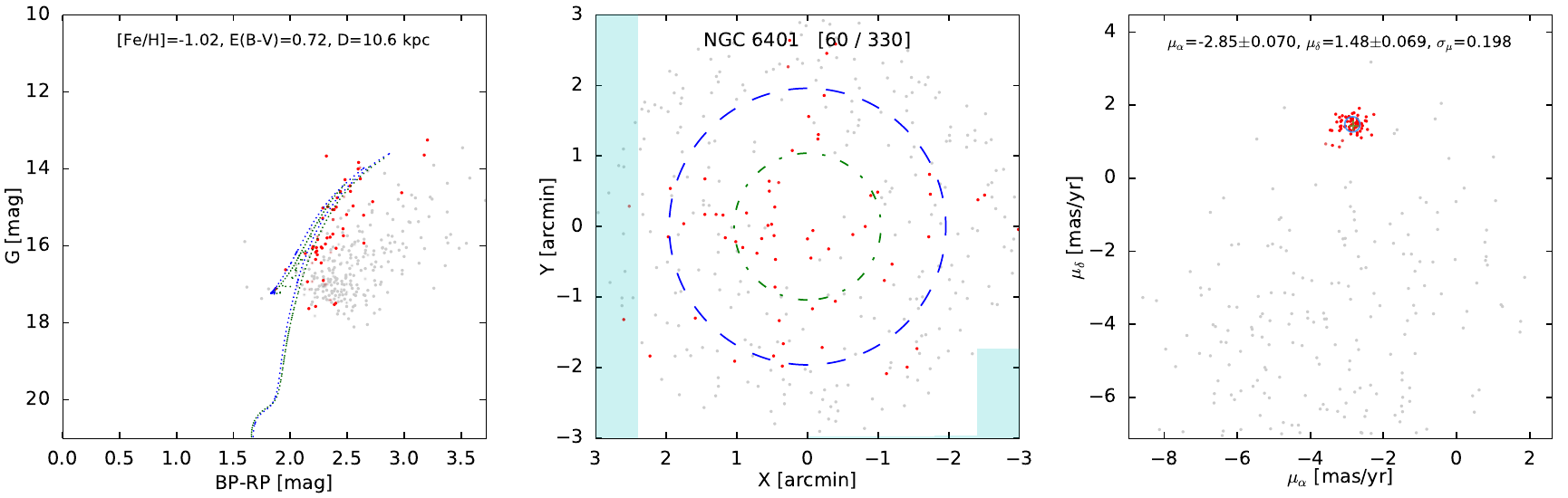}
\includegraphics{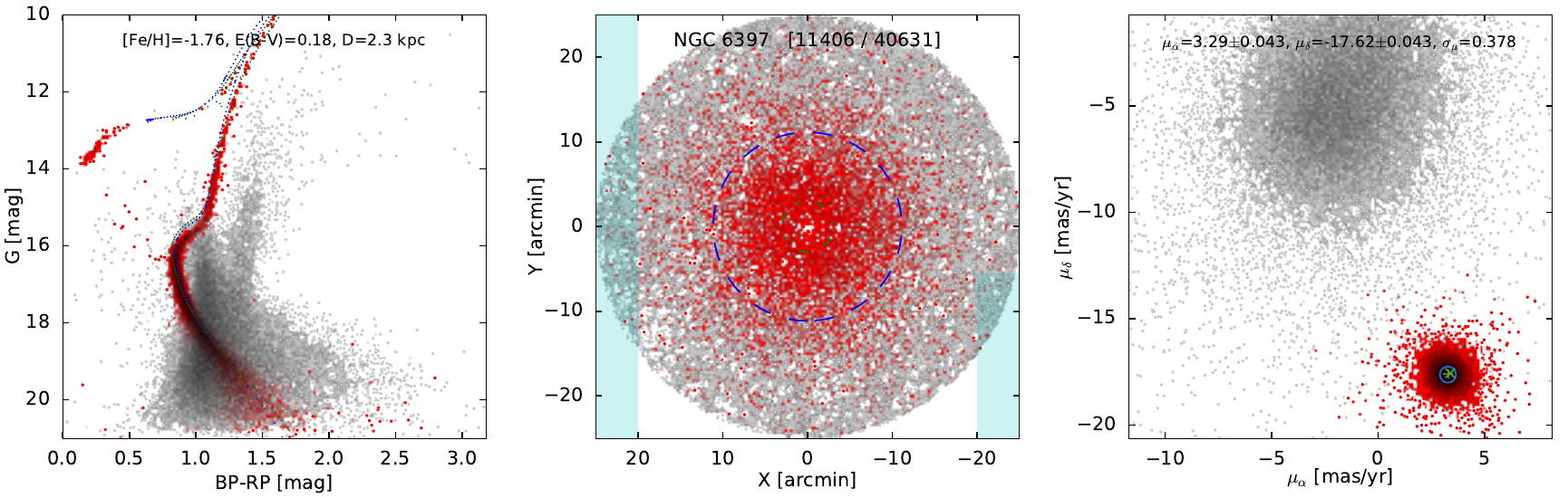}
\end{figure*}

\clearpage\begin{figure*}
\contcaption{}
\includegraphics{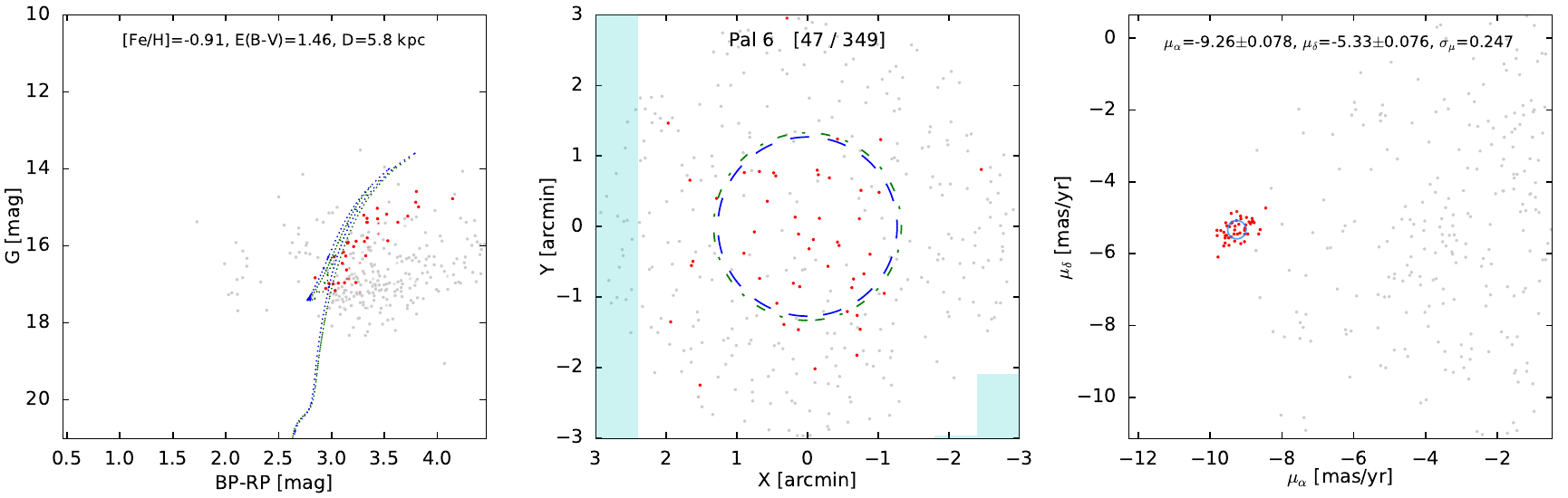}
\includegraphics{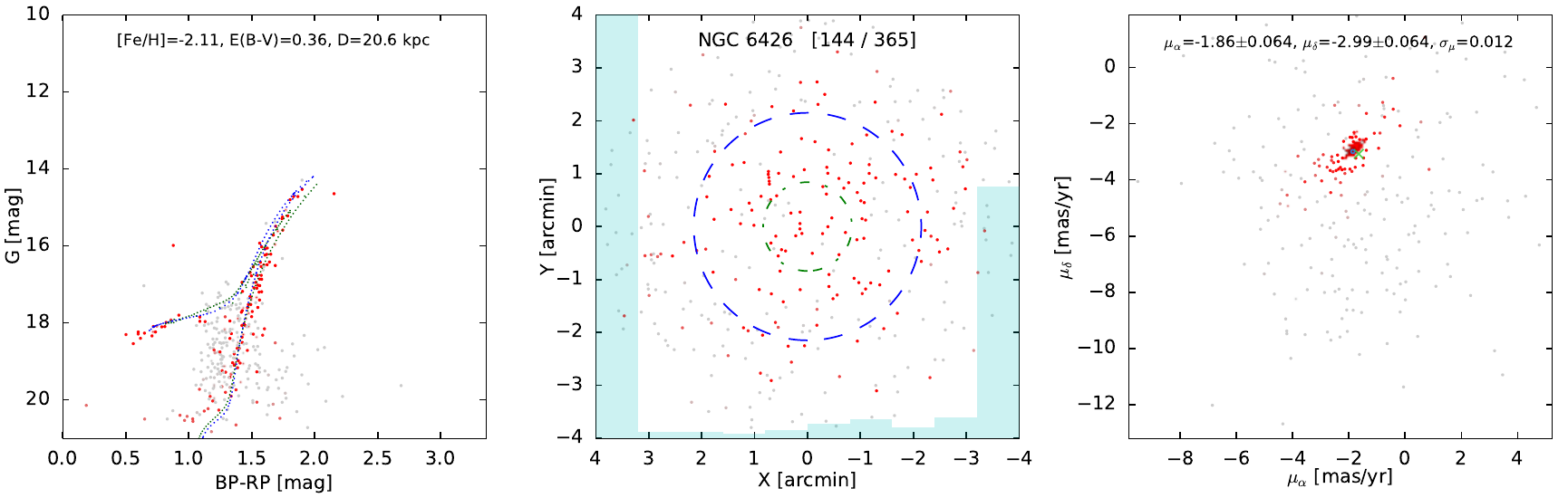}
\includegraphics{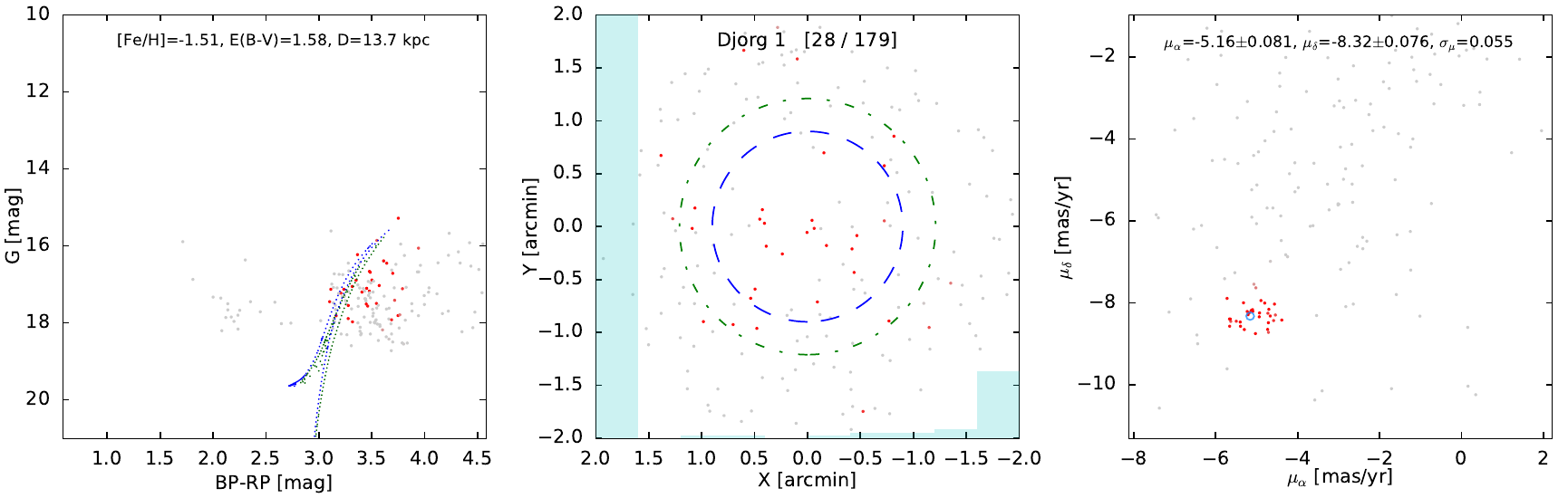}
\includegraphics{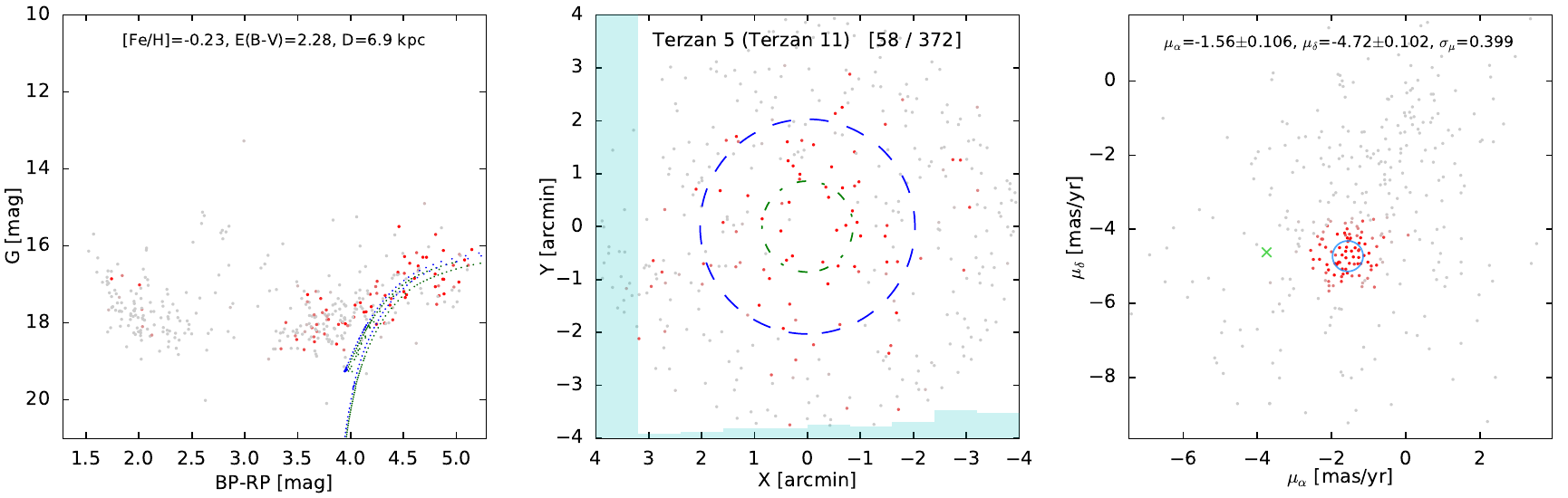}
\end{figure*}

\clearpage\begin{figure*}
\contcaption{}
\includegraphics{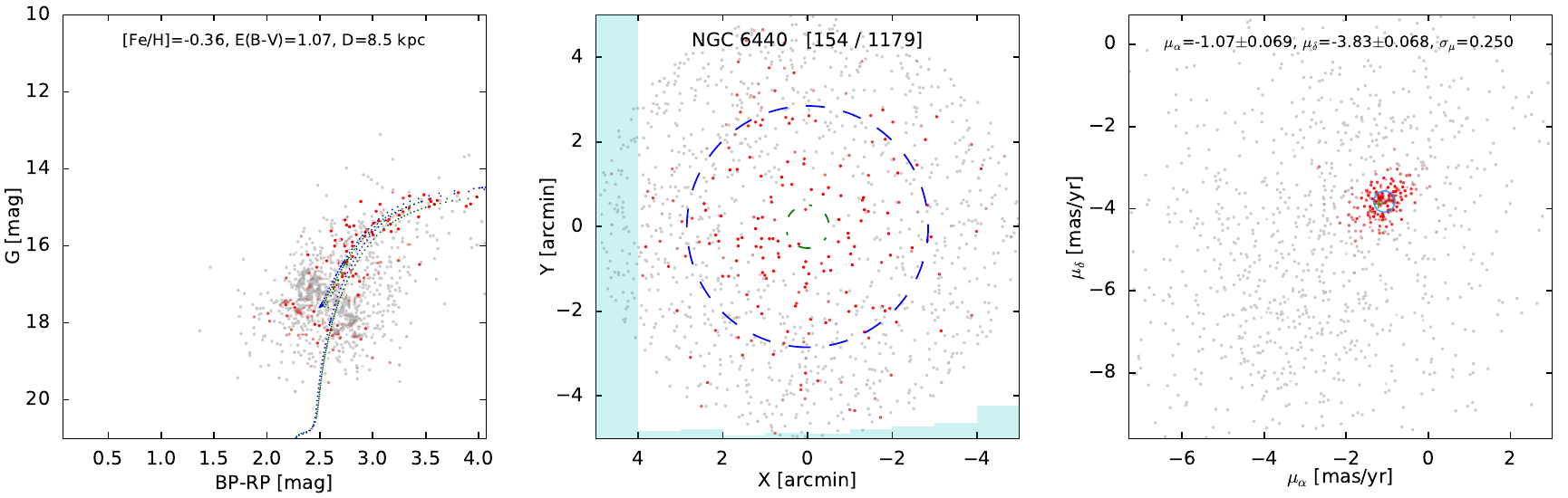}
\includegraphics{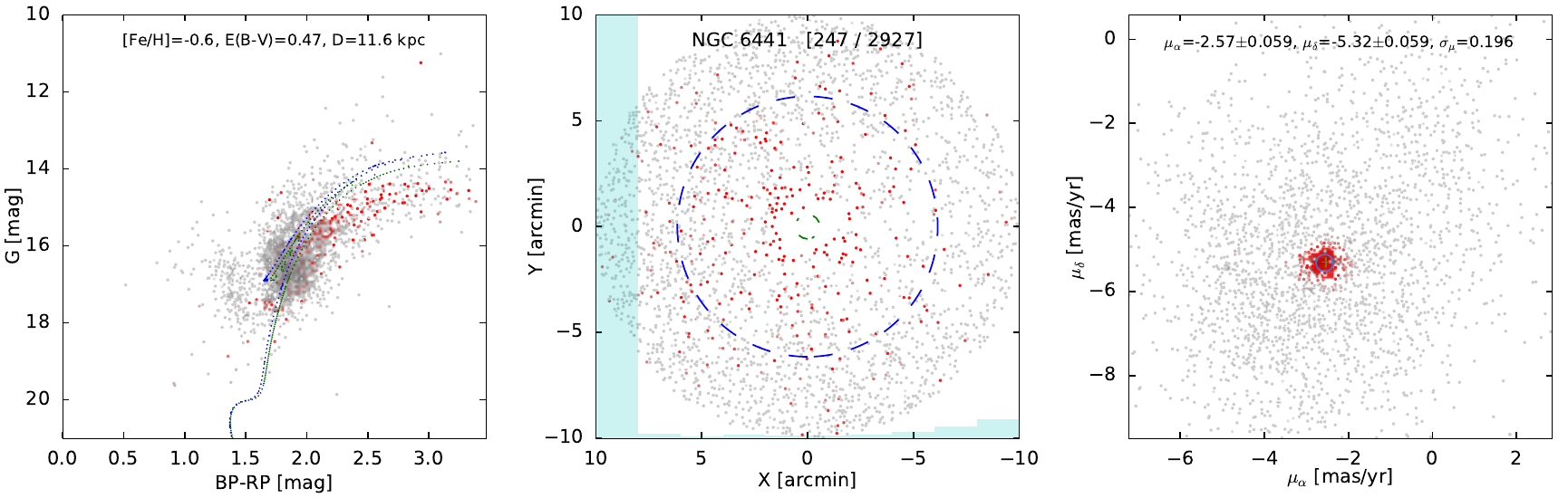}
\includegraphics{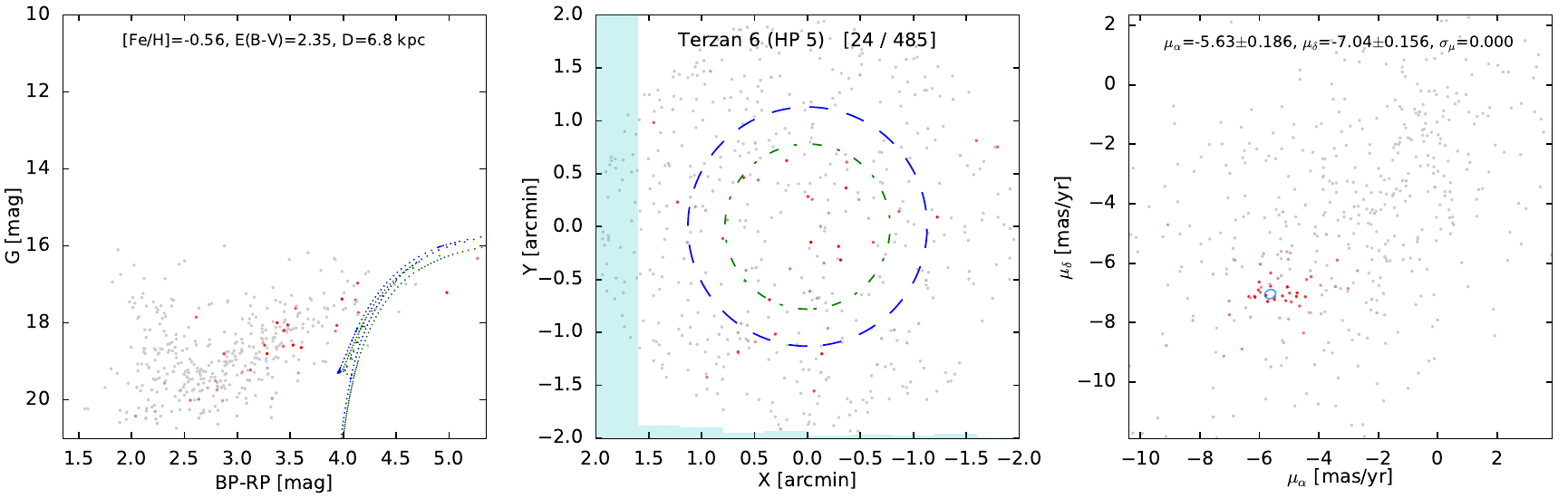}
\includegraphics{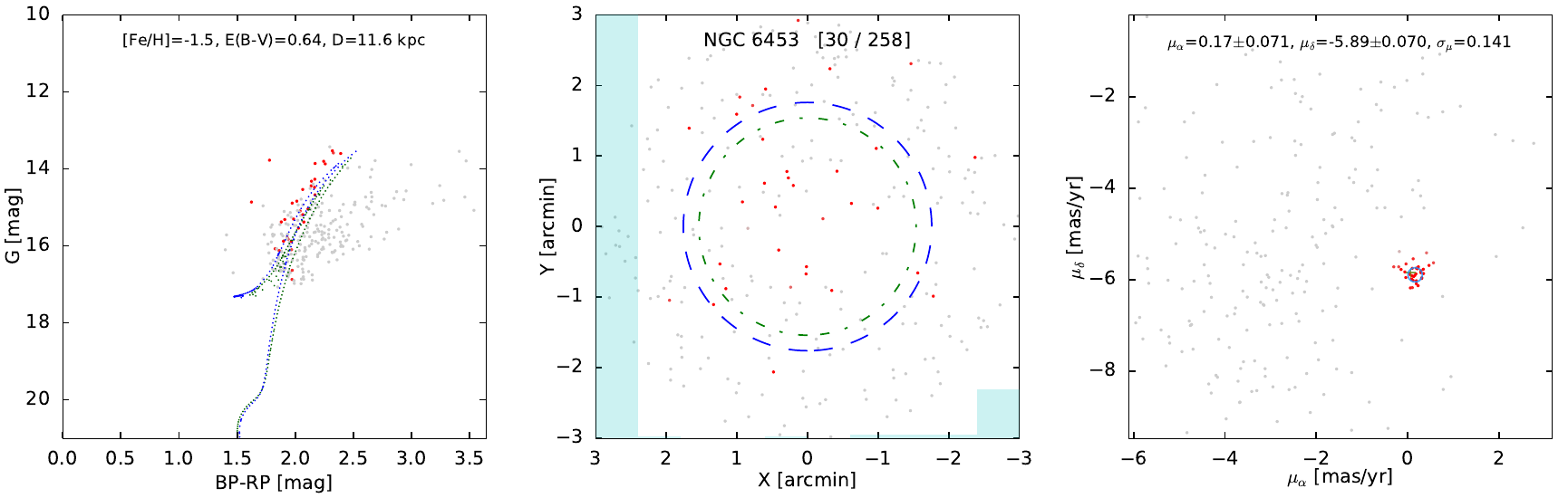}
\end{figure*}

\clearpage\begin{figure*}
\contcaption{}
\includegraphics{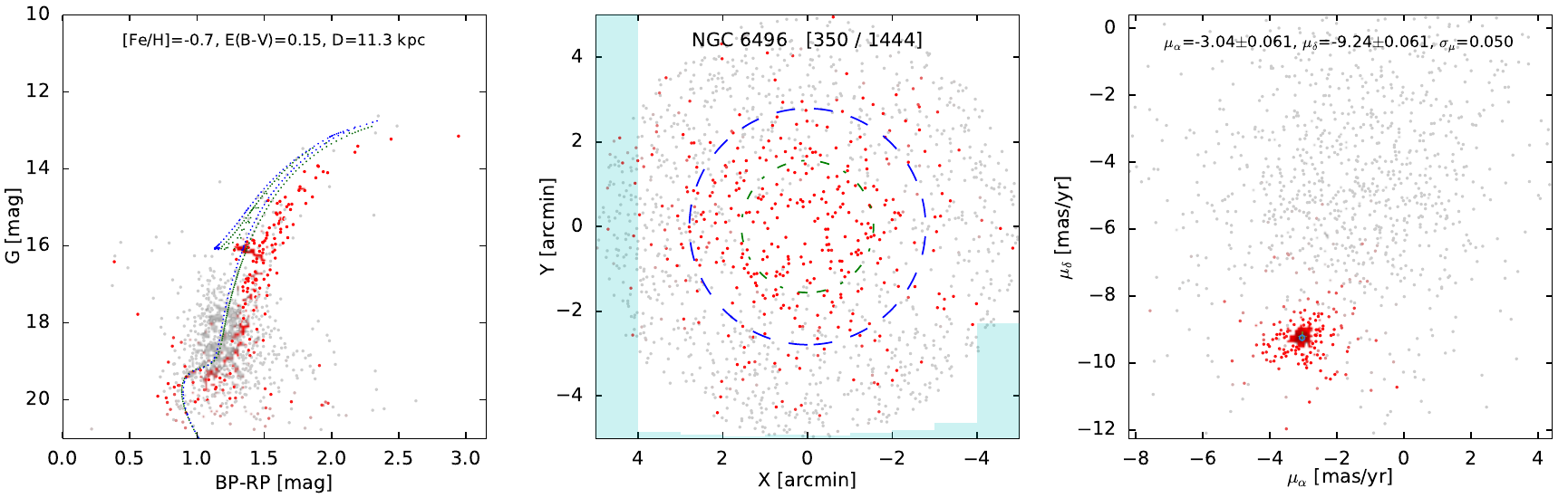}
\includegraphics{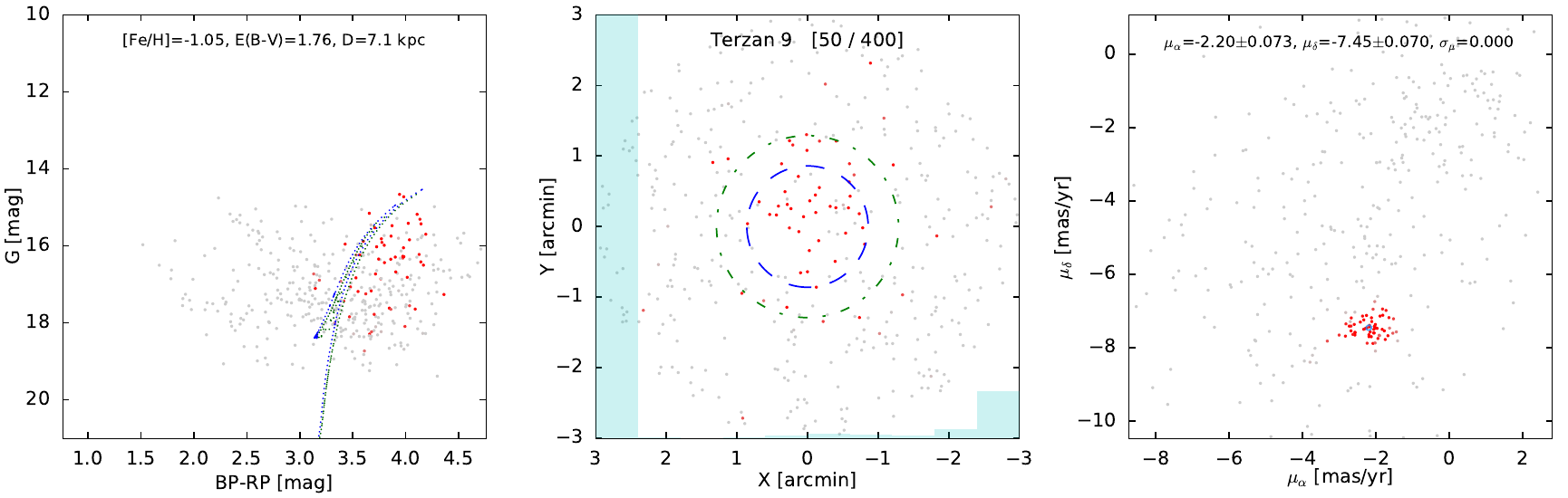}
\includegraphics{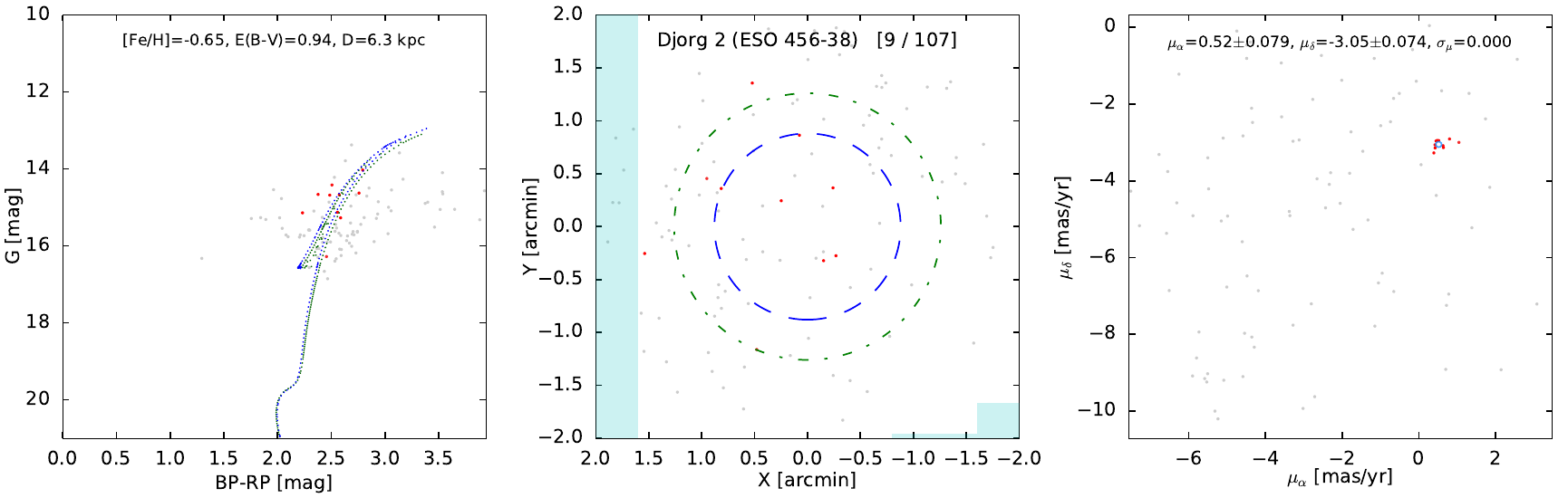}
\includegraphics{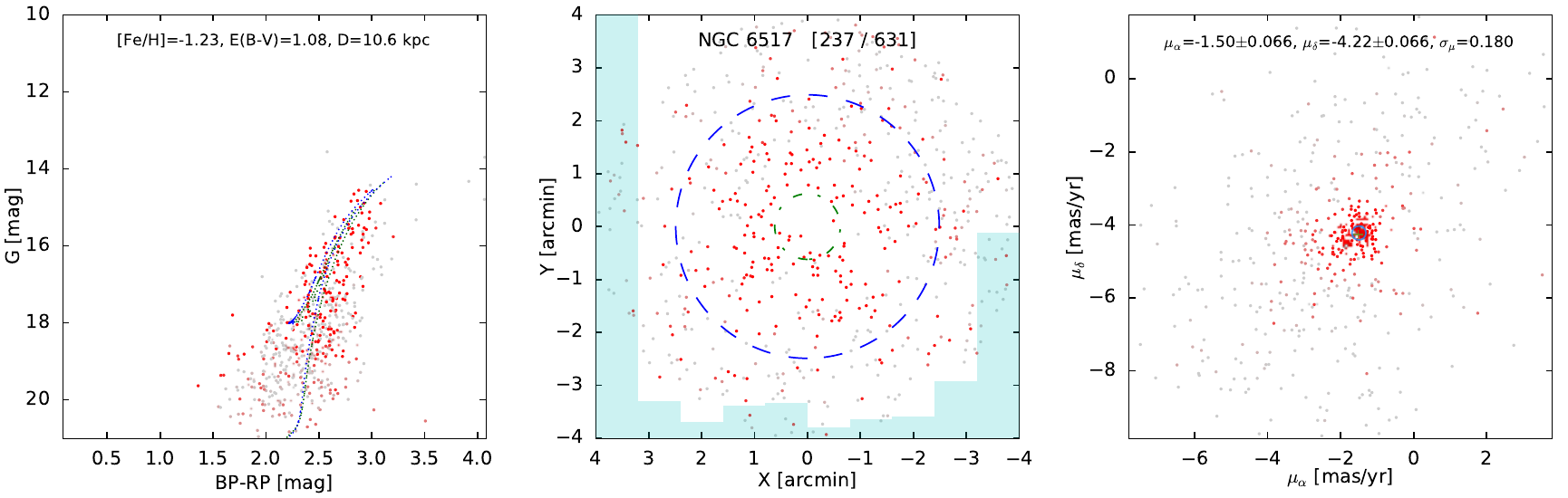}
\end{figure*}

\clearpage\begin{figure*}
\contcaption{}
\includegraphics{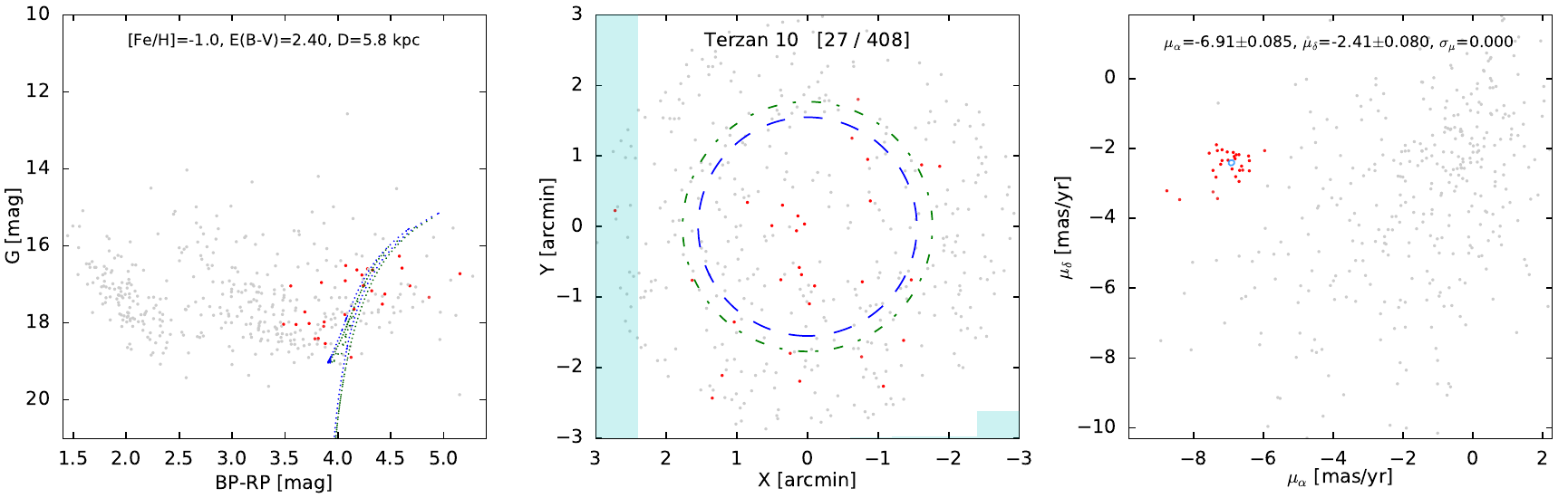}
\includegraphics{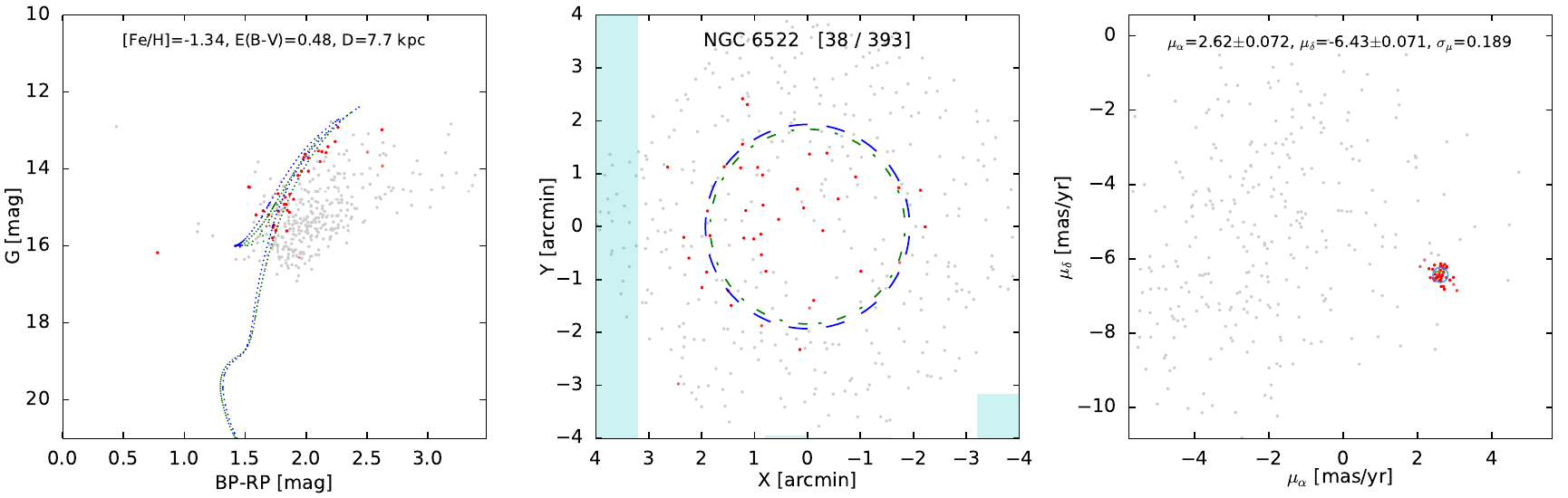}
\includegraphics{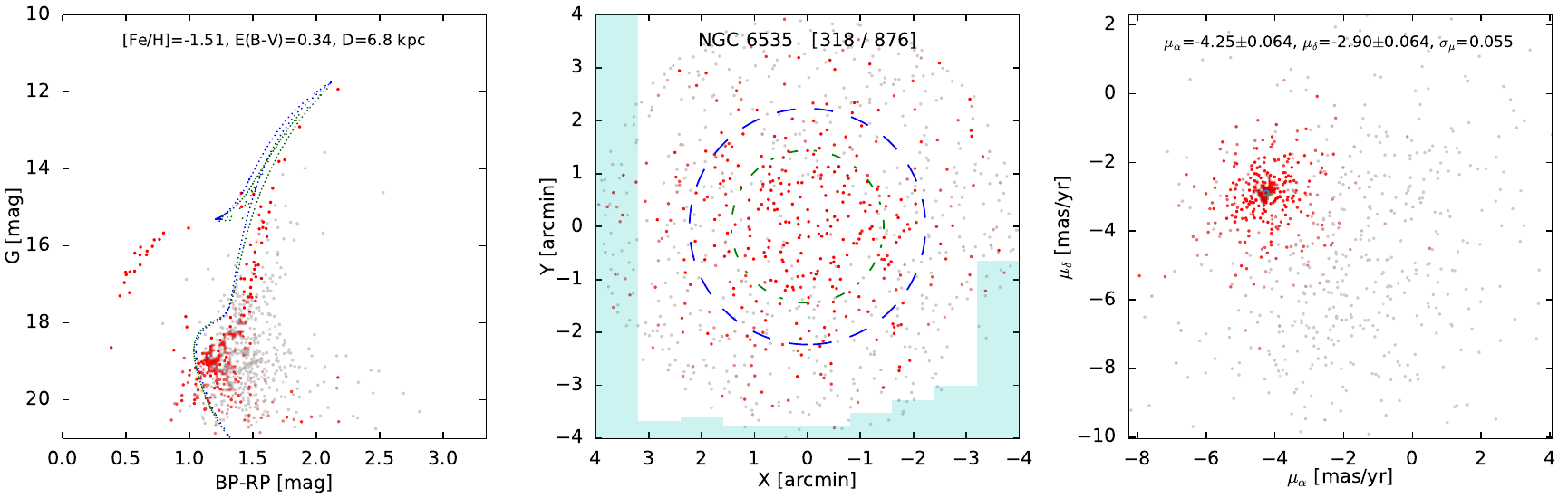}
\includegraphics{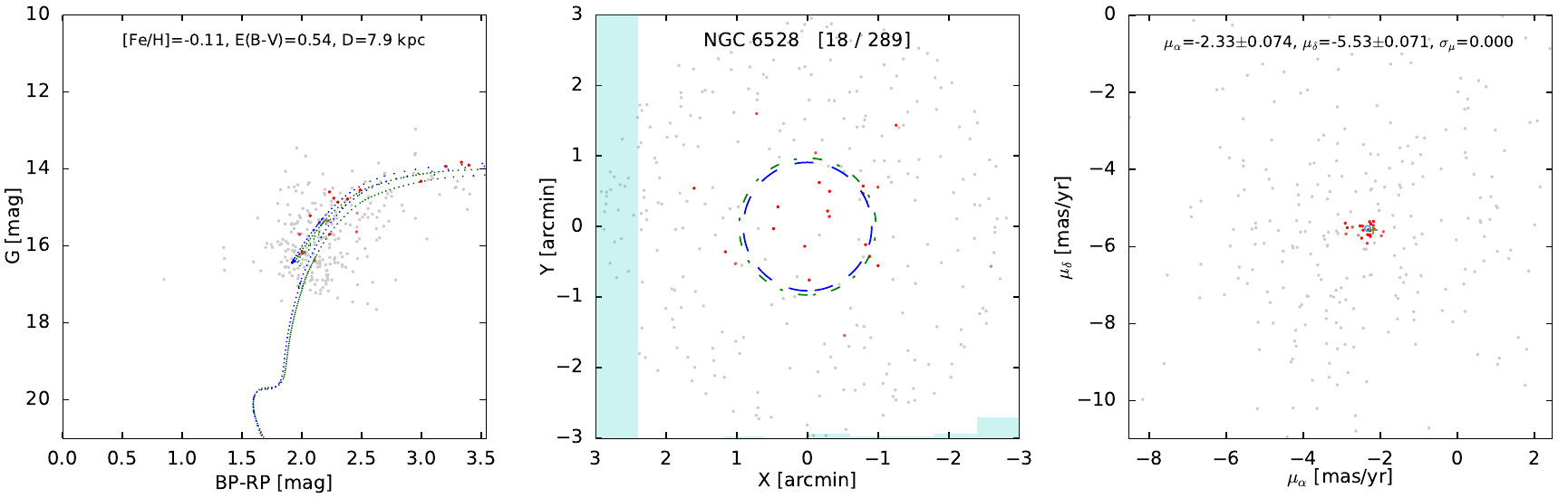}
\end{figure*}

\clearpage\begin{figure*}
\contcaption{}
\includegraphics{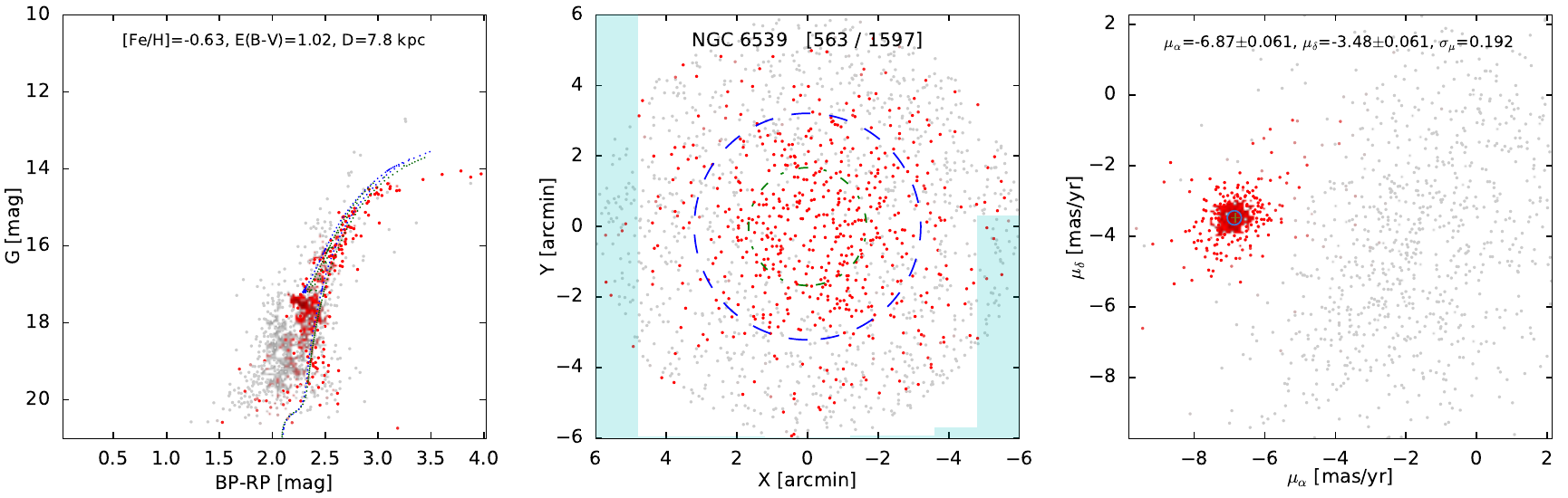}
\includegraphics{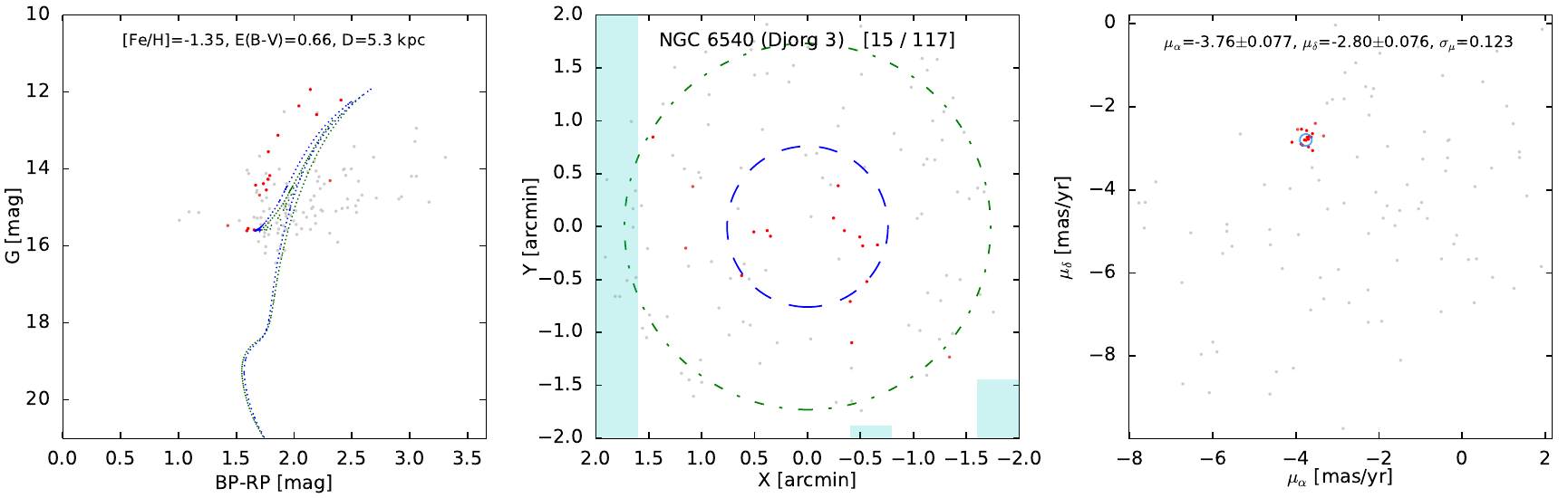}
\includegraphics{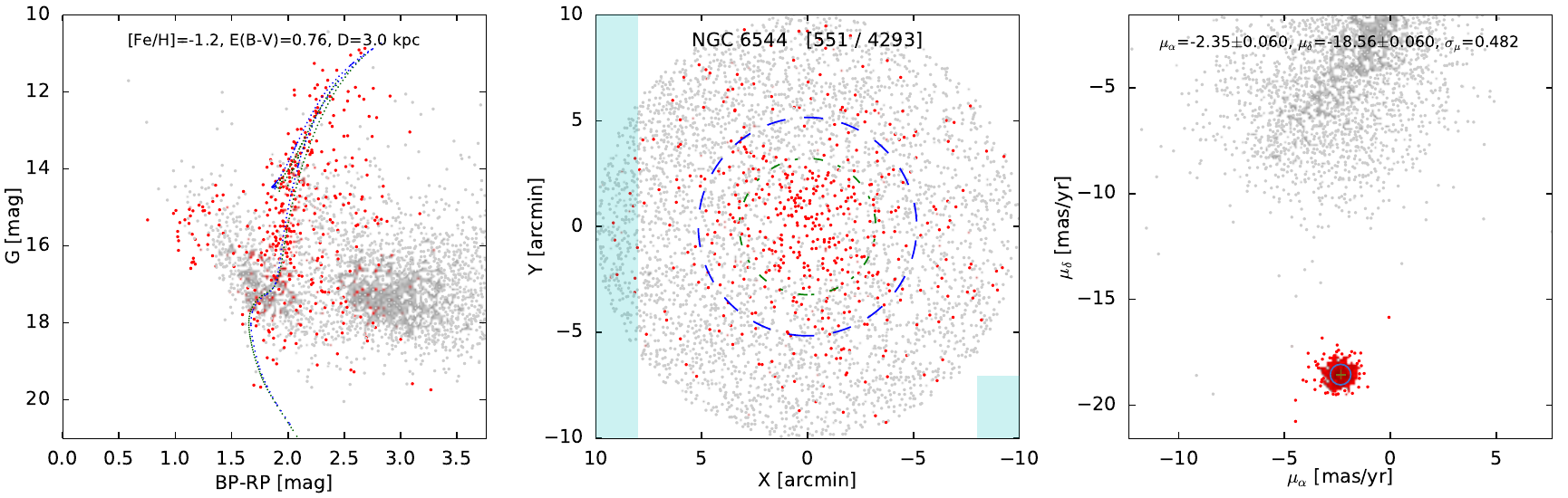}
\includegraphics{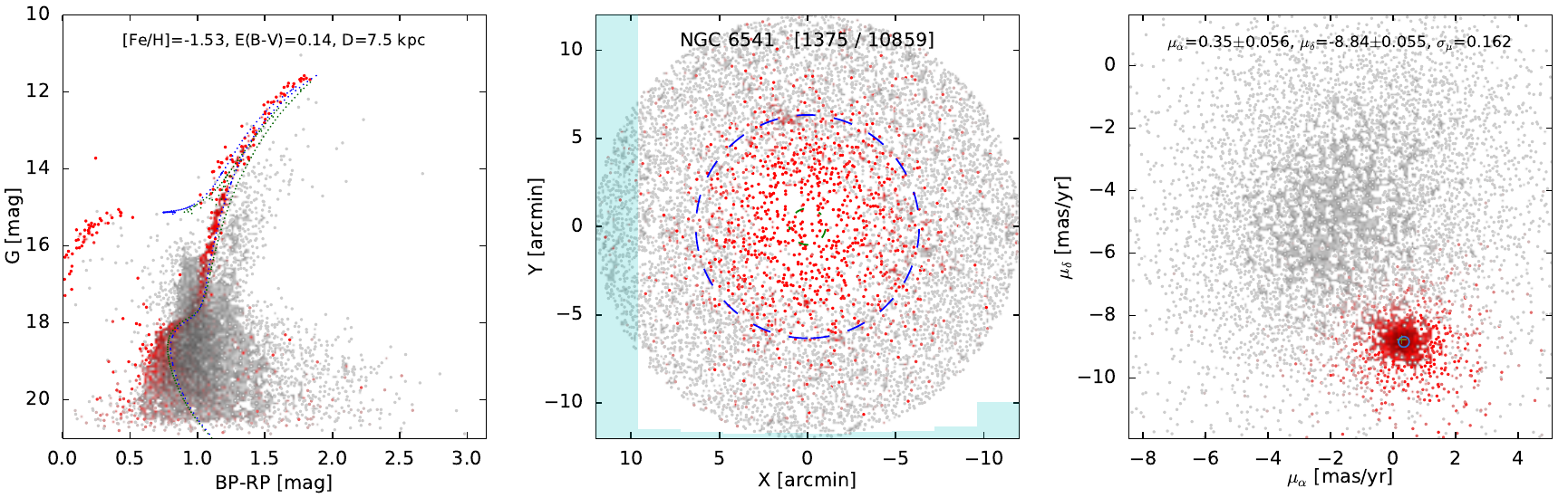}
\end{figure*}

\clearpage\begin{figure*}
\contcaption{}
\includegraphics{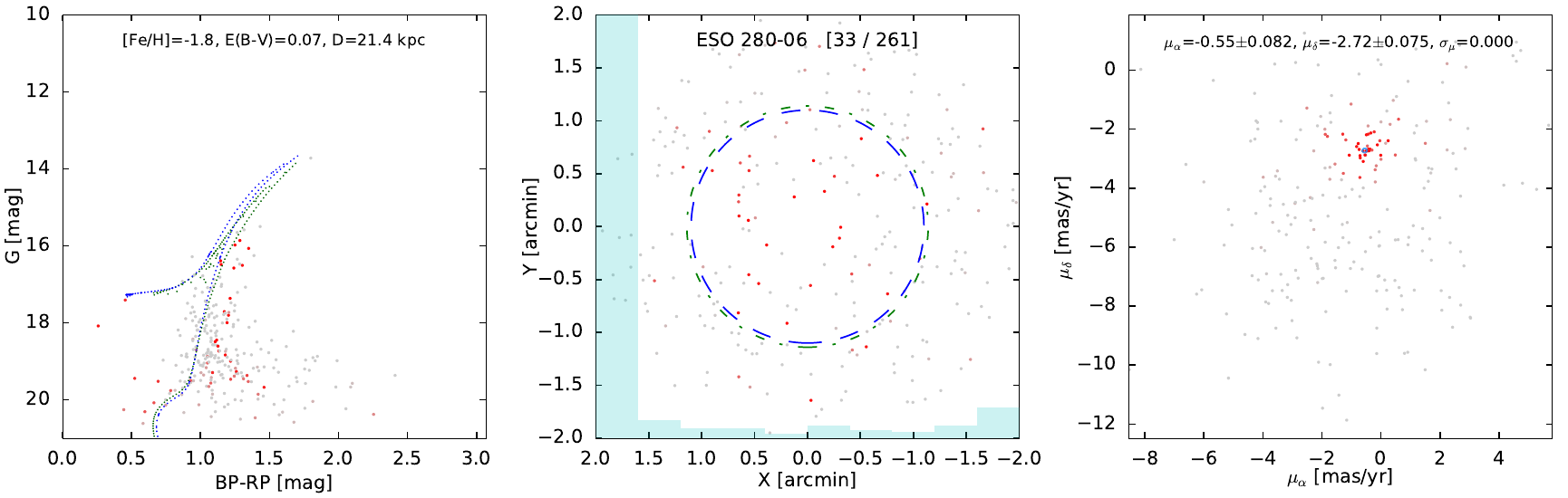}
\includegraphics{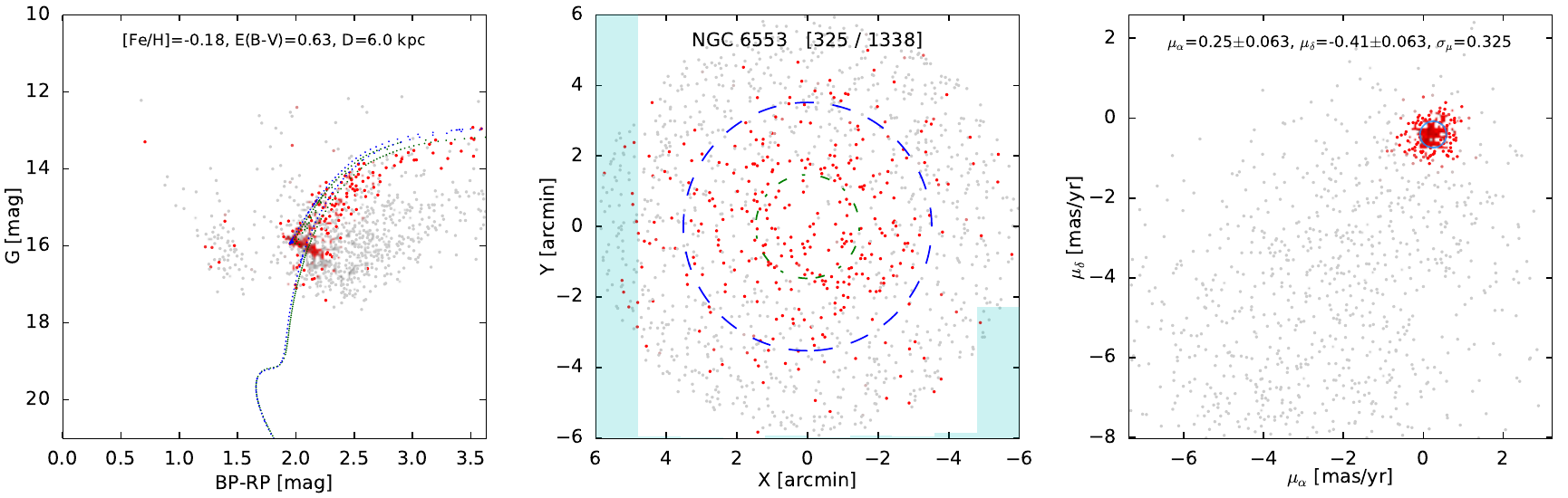}
\includegraphics{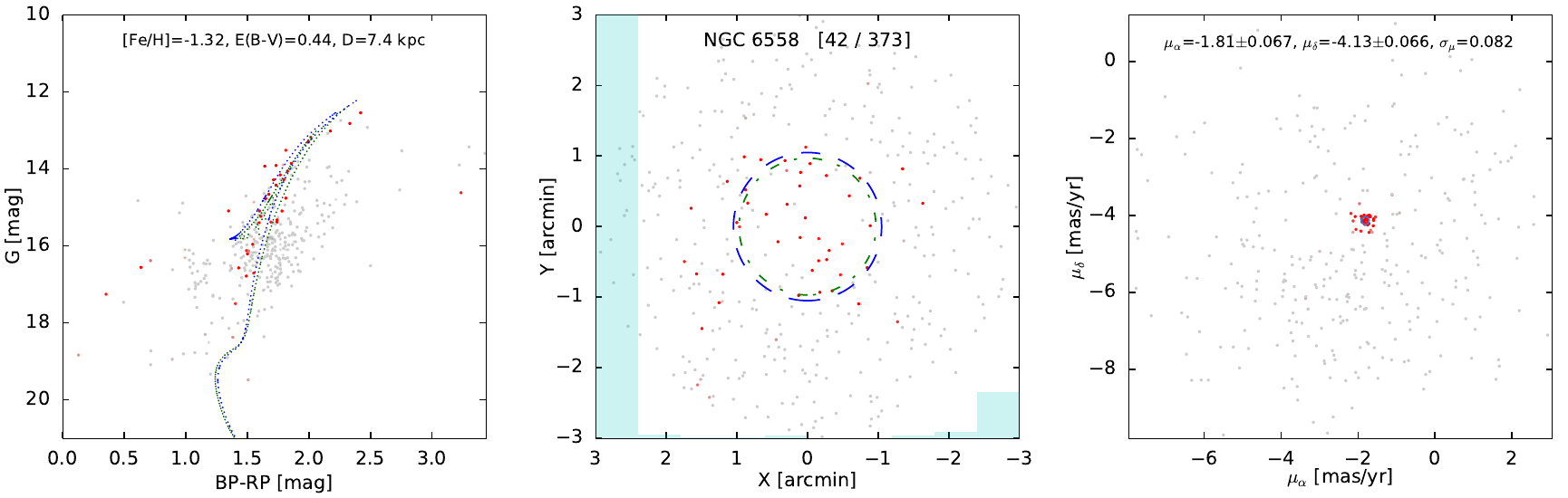}
\includegraphics{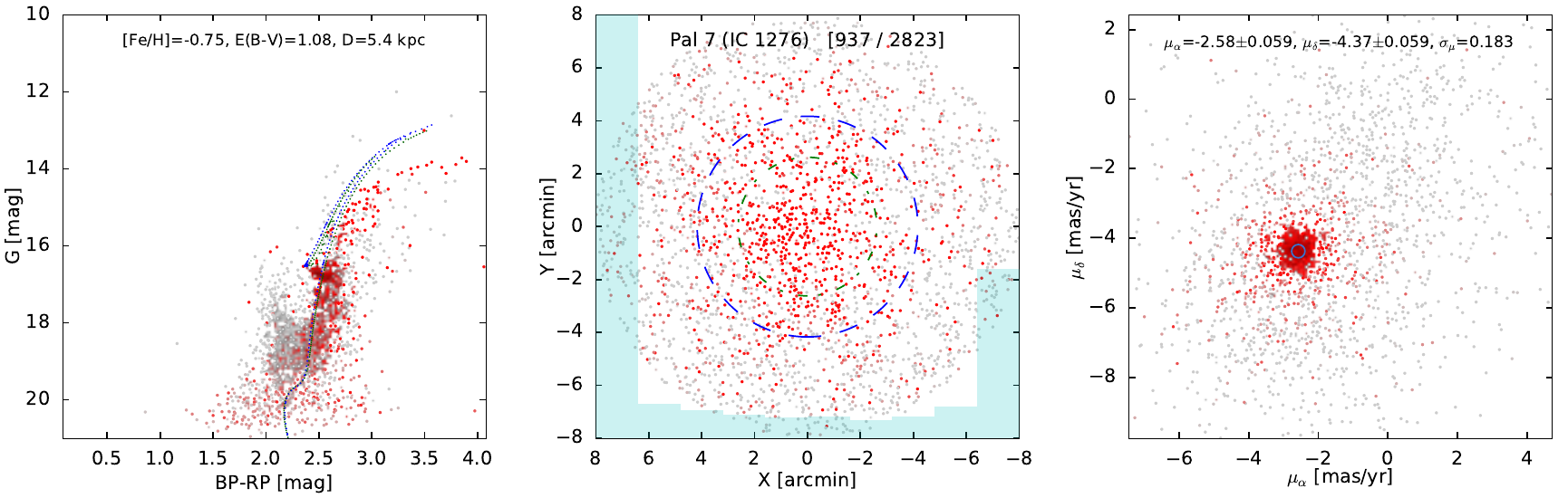}
\end{figure*}

\clearpage\begin{figure*}
\contcaption{}
\includegraphics{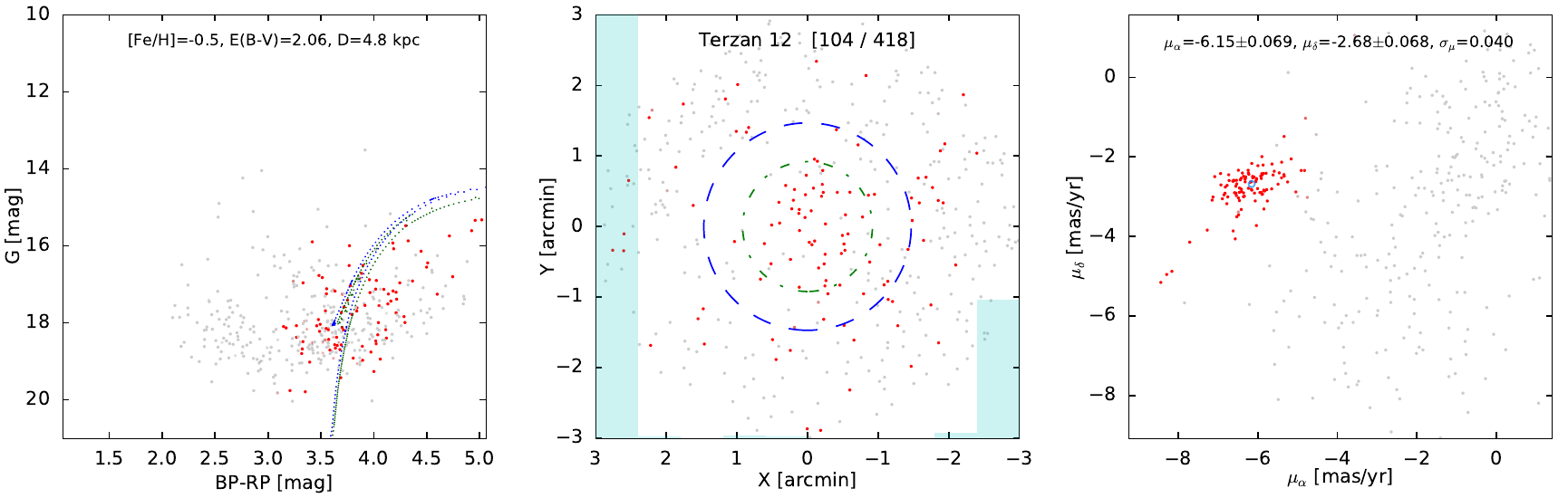}
\includegraphics{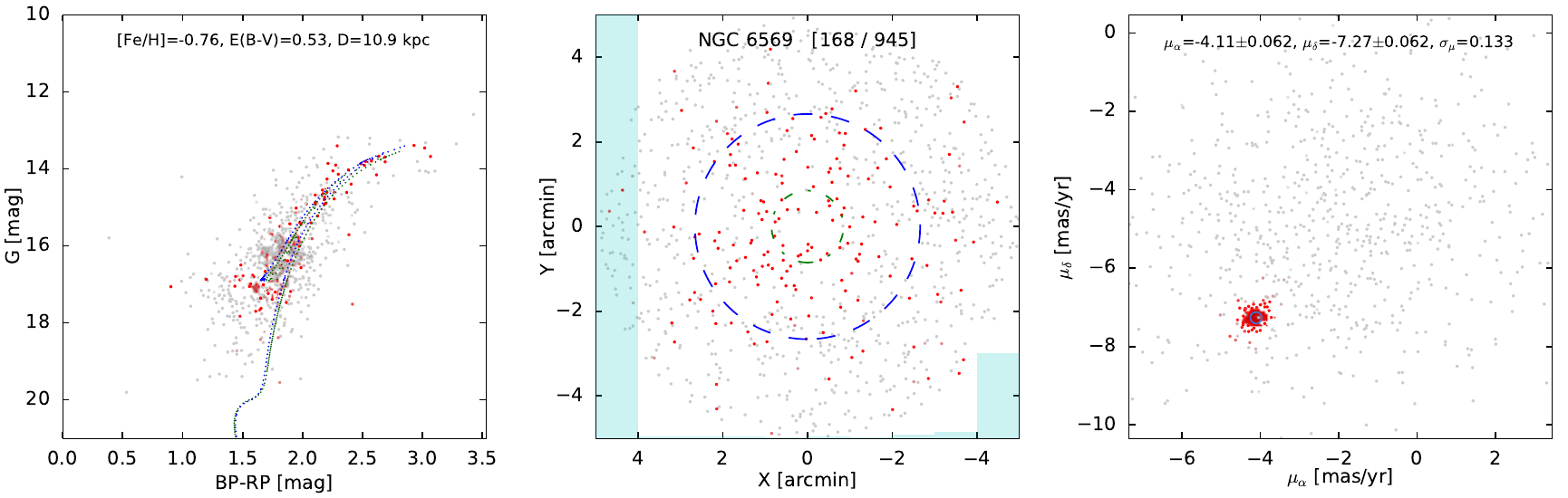}
\includegraphics{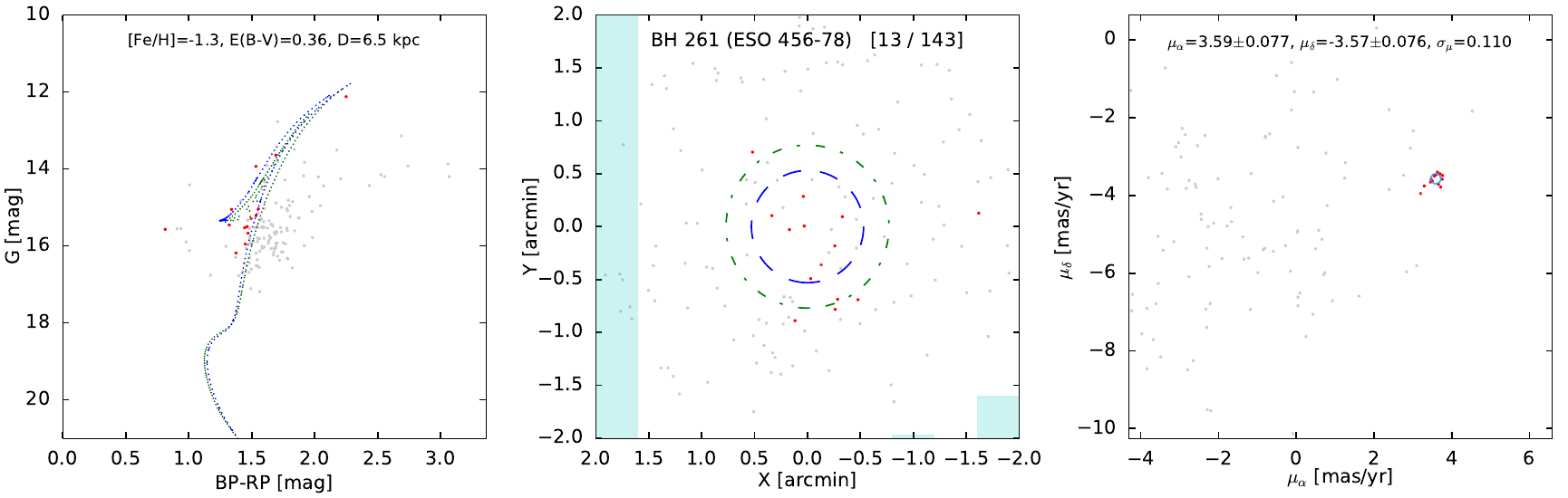}
\includegraphics{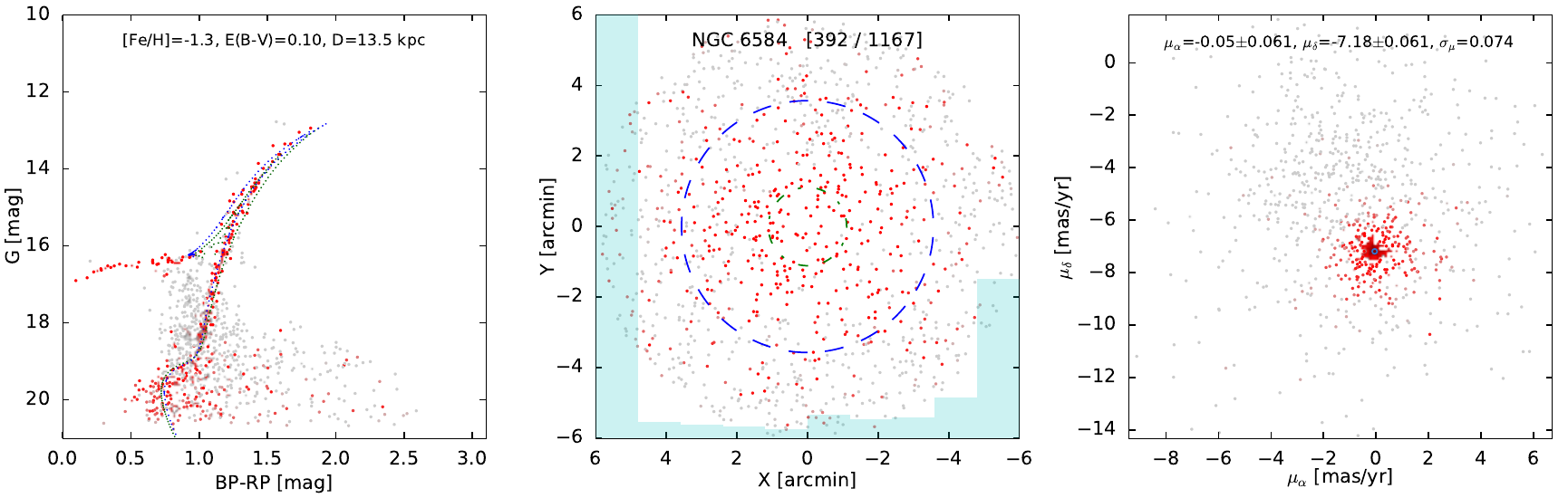}
\end{figure*}

\clearpage\begin{figure*}
\contcaption{}
\includegraphics{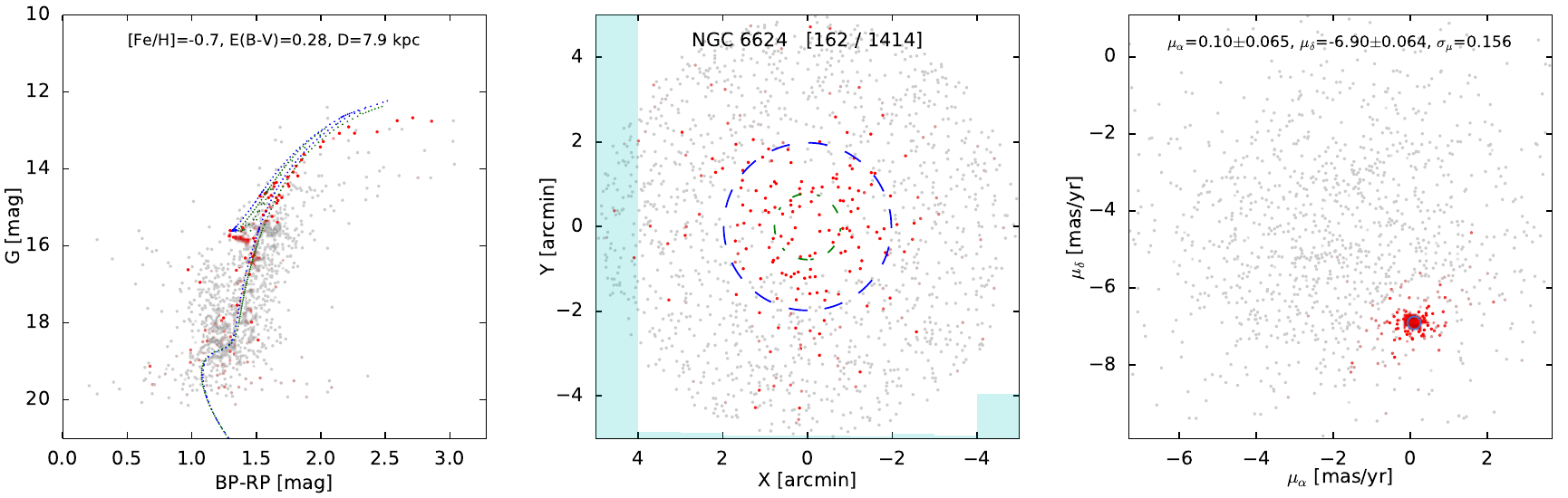}
\includegraphics{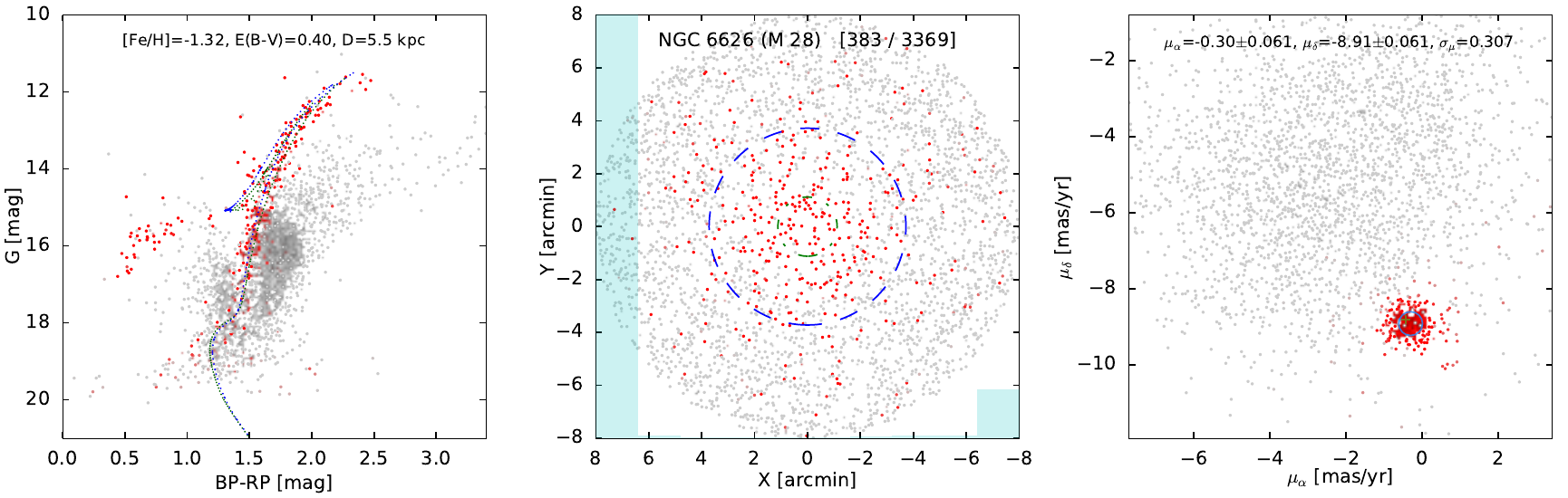}
\includegraphics{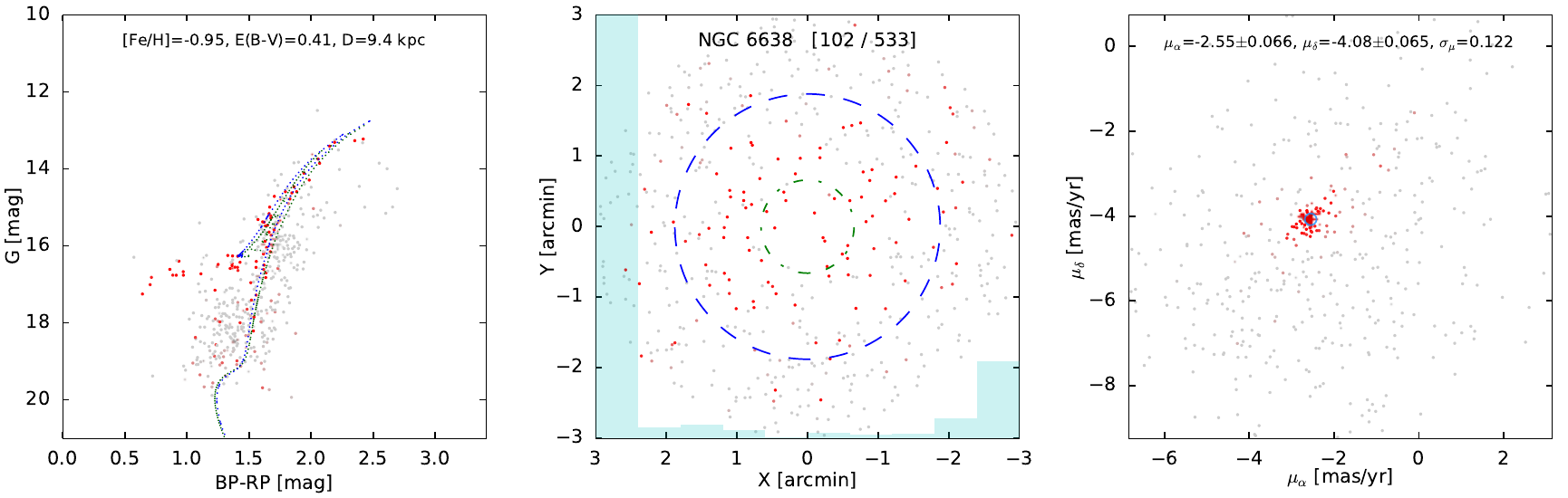}
\includegraphics{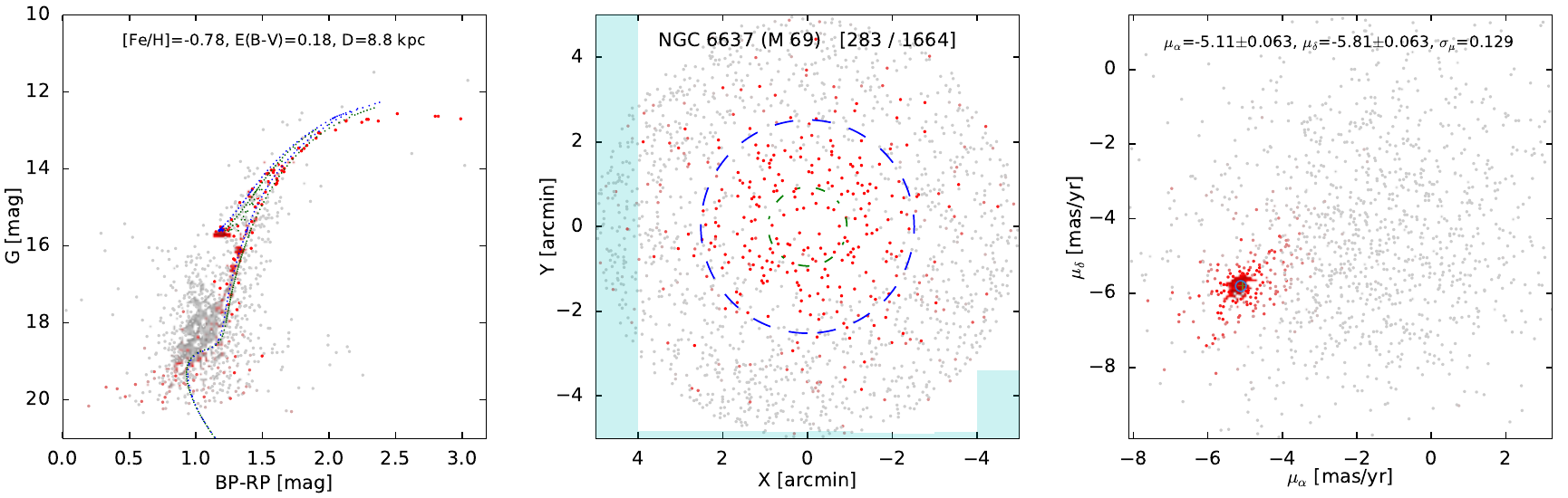}
\end{figure*}

\clearpage\begin{figure*}
\contcaption{}
\includegraphics{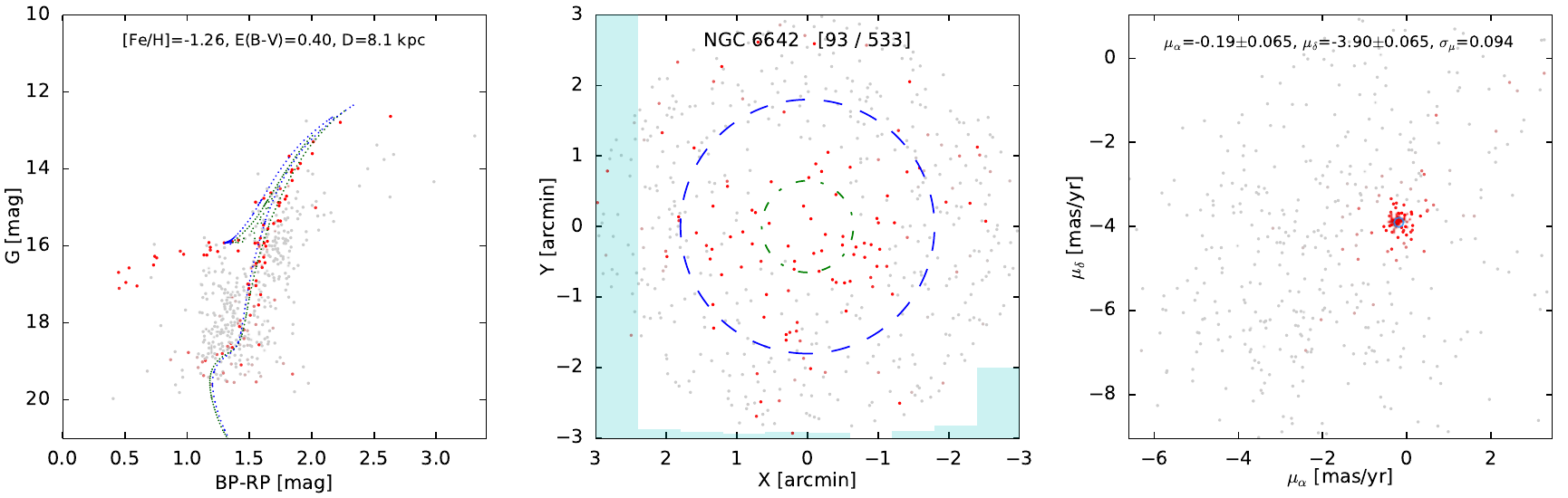}
\includegraphics{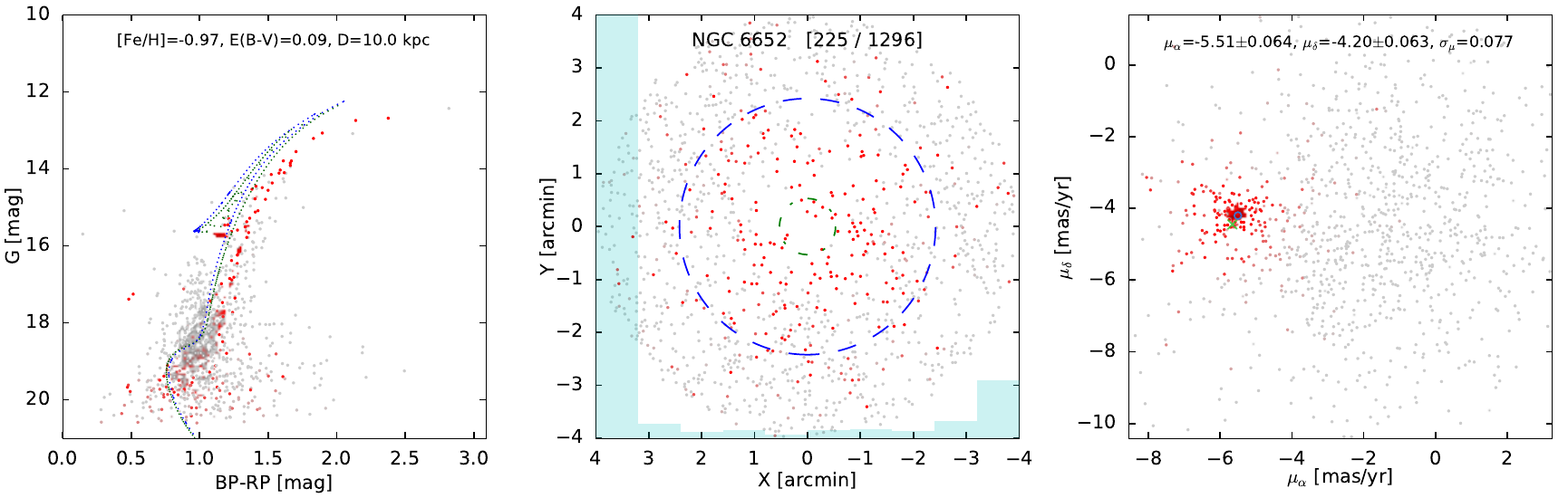}
\includegraphics{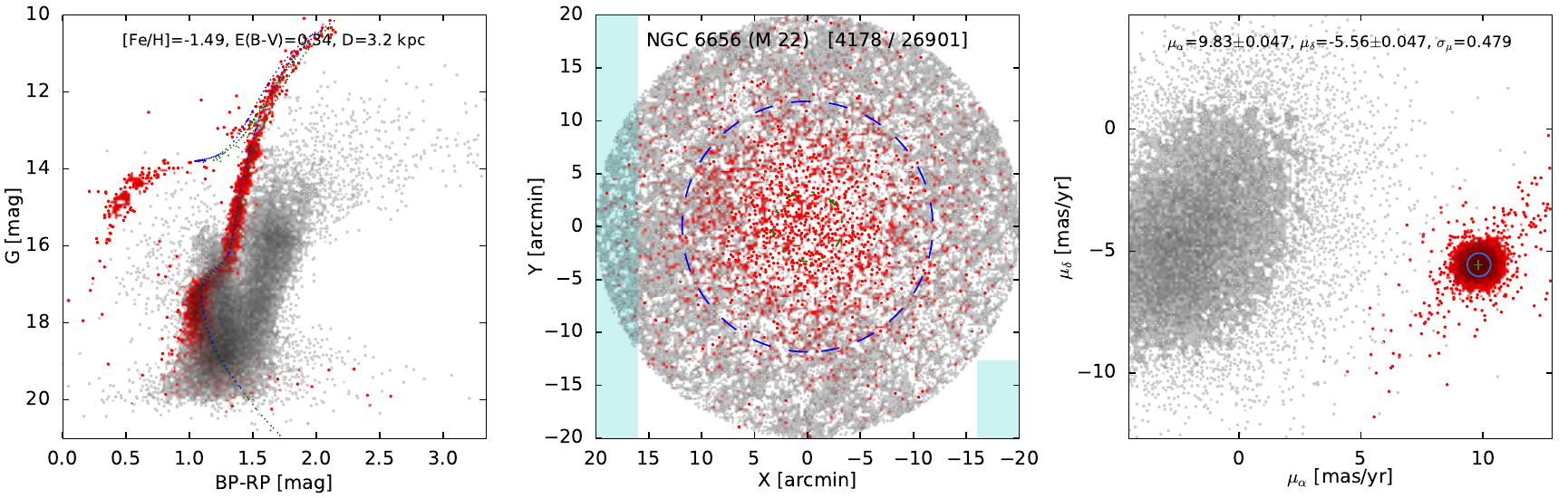}
\includegraphics{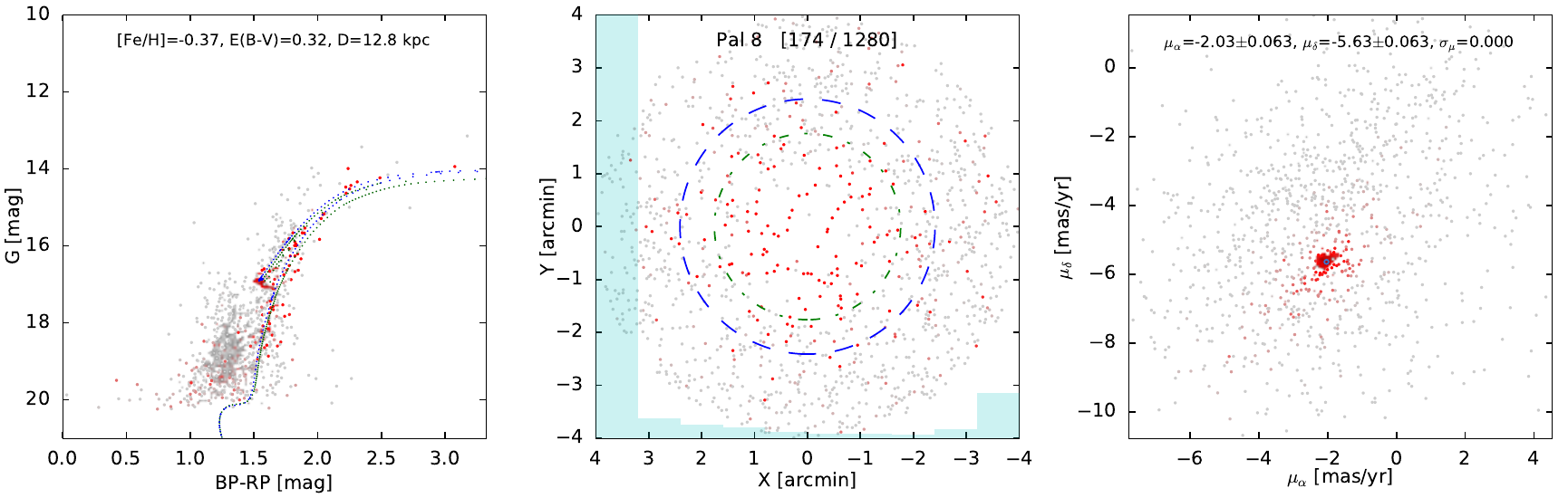}
\end{figure*}

\clearpage\begin{figure*}
\contcaption{}
\includegraphics{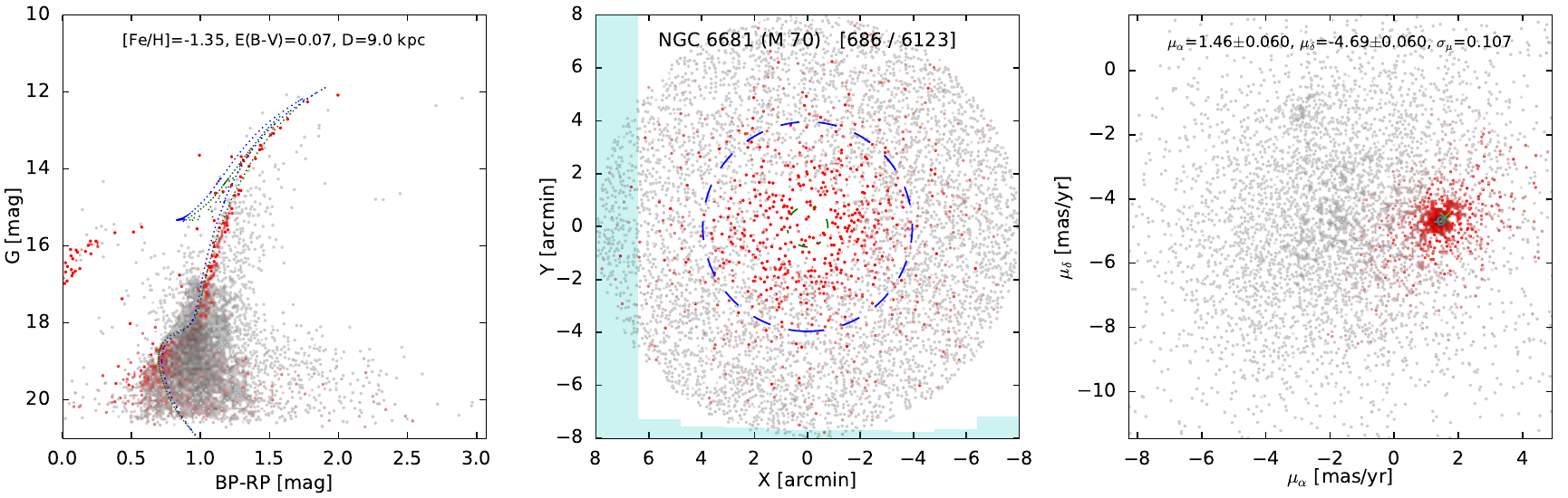}
\includegraphics{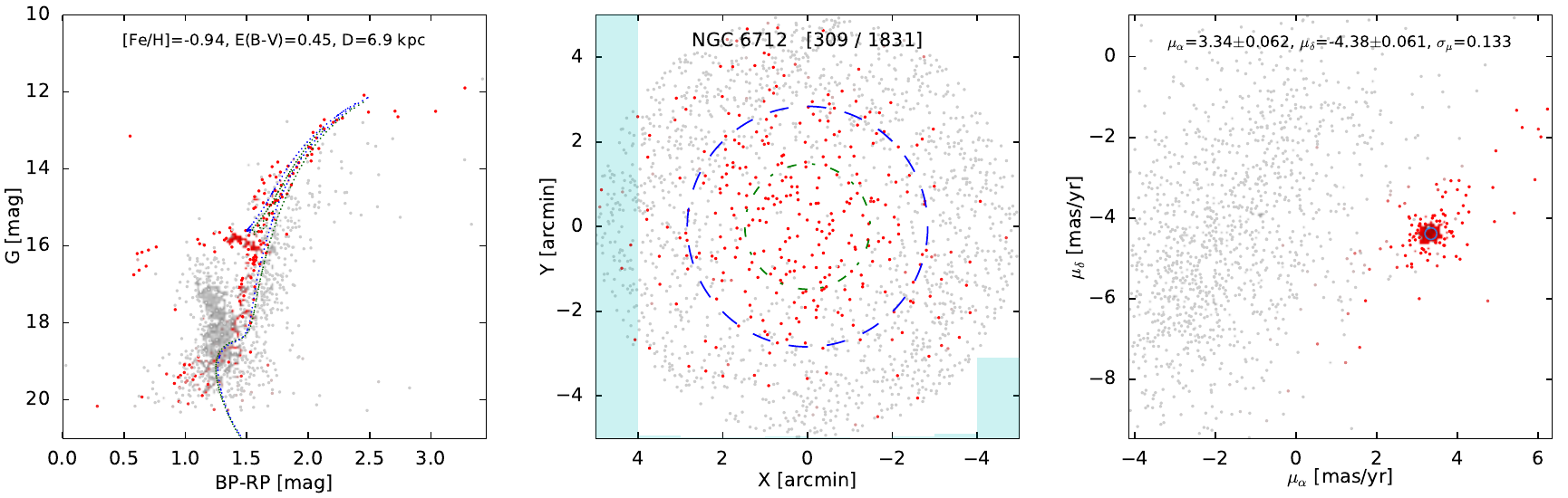}
\includegraphics{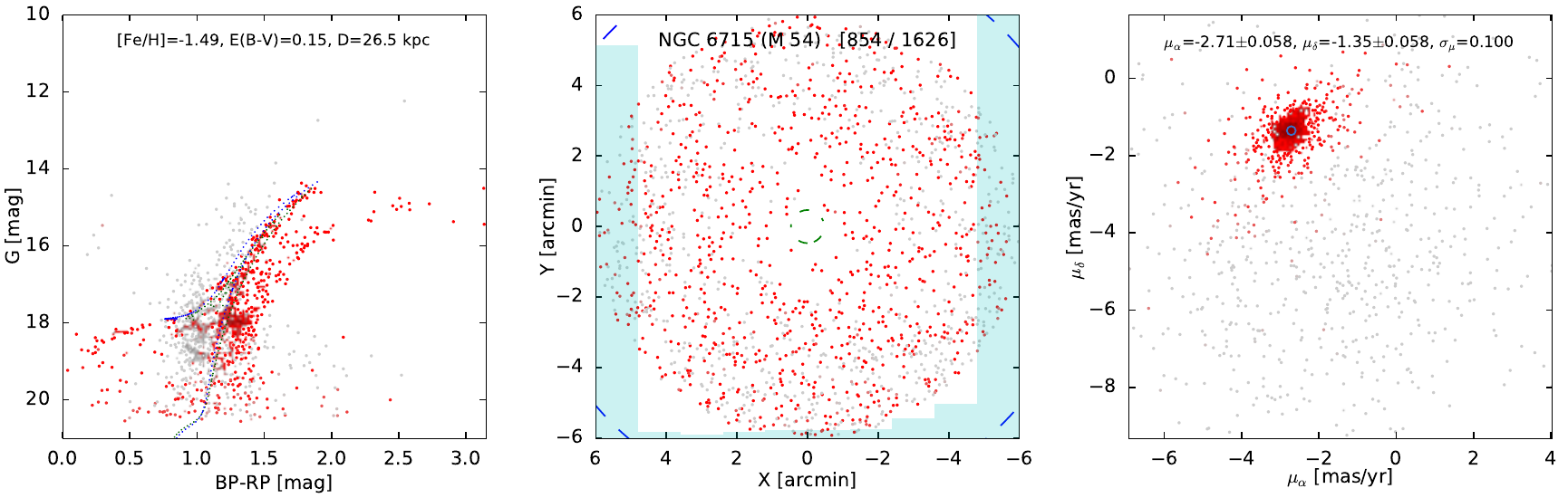}
\includegraphics{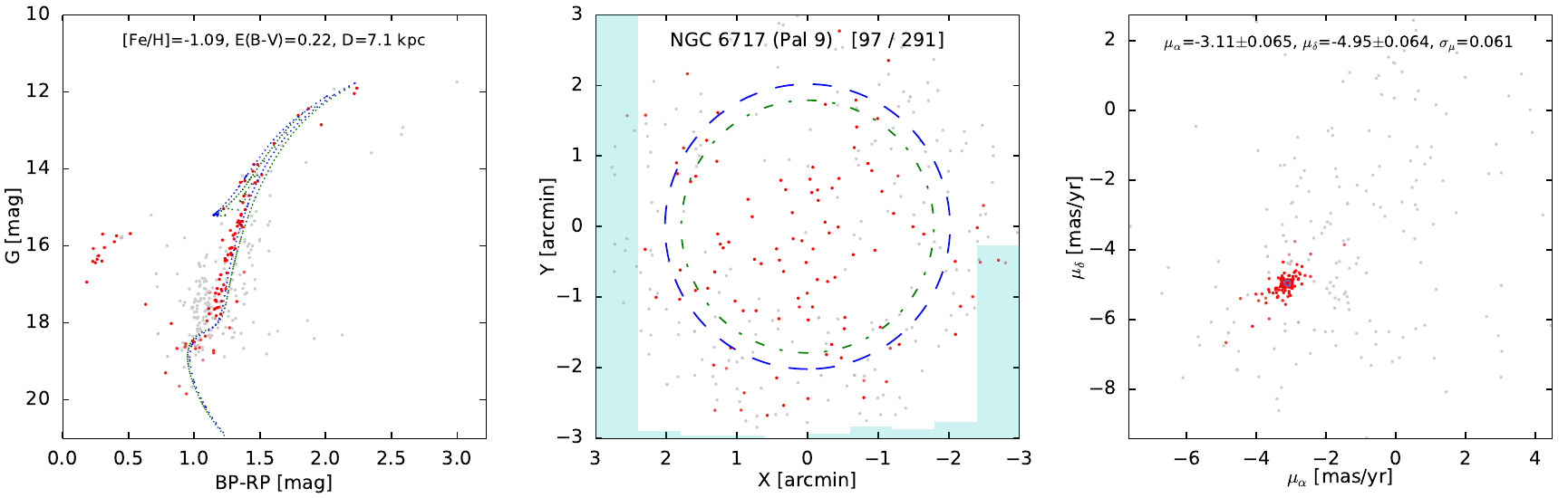}
\end{figure*}

\clearpage\begin{figure*}
\contcaption{}
\includegraphics{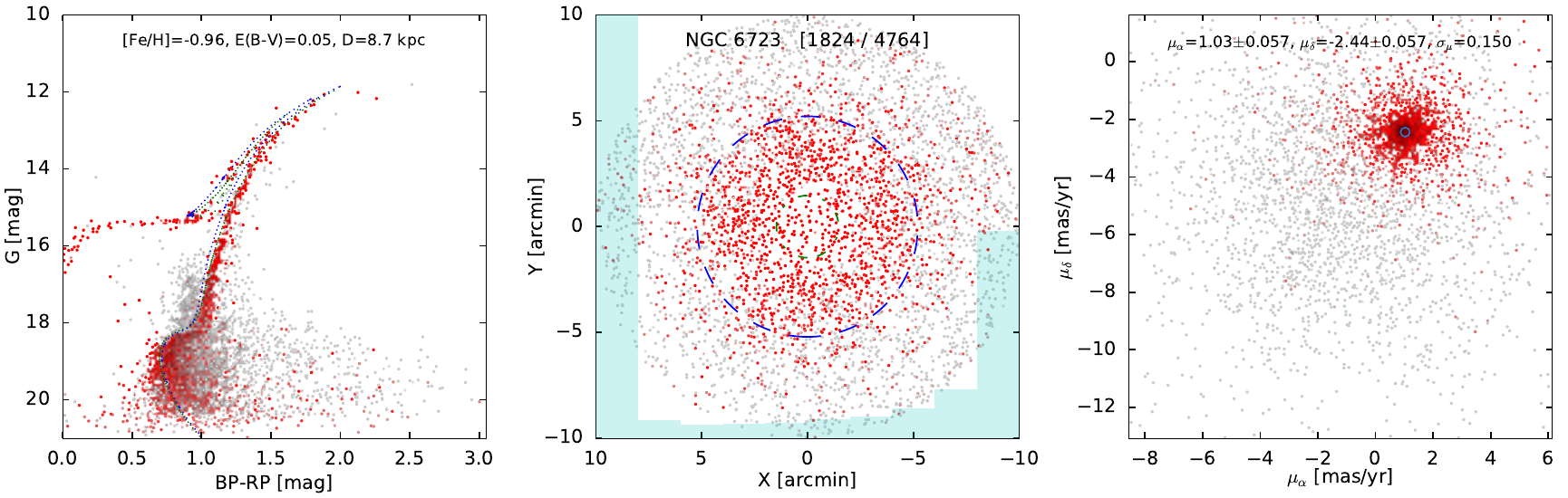}
\includegraphics{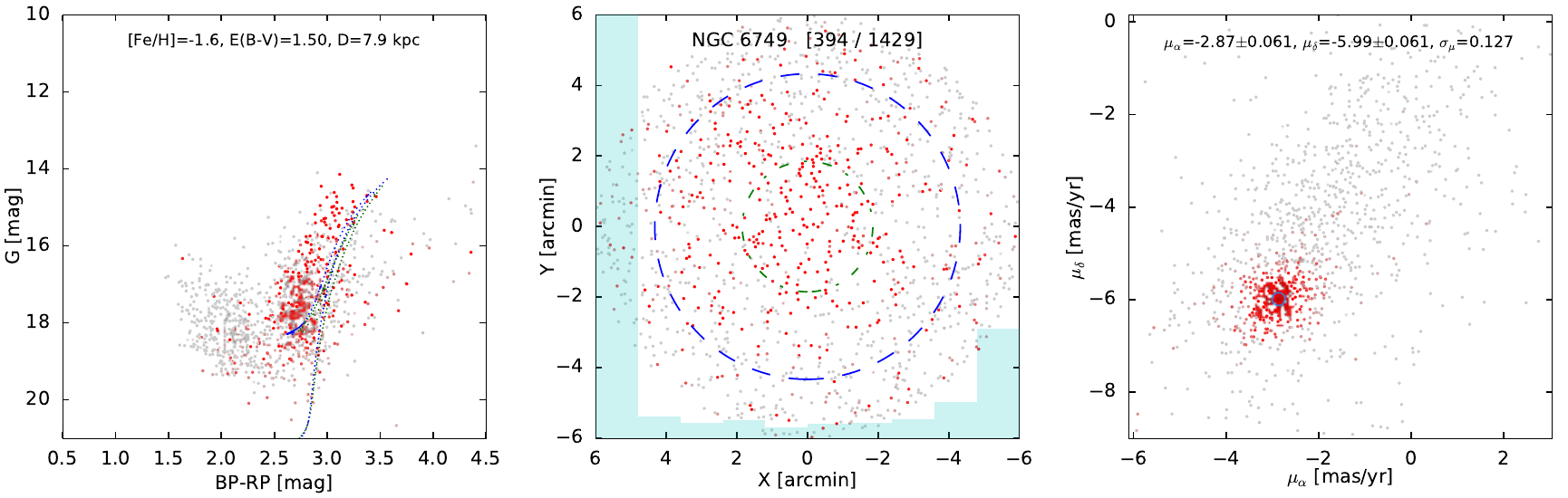}
\includegraphics{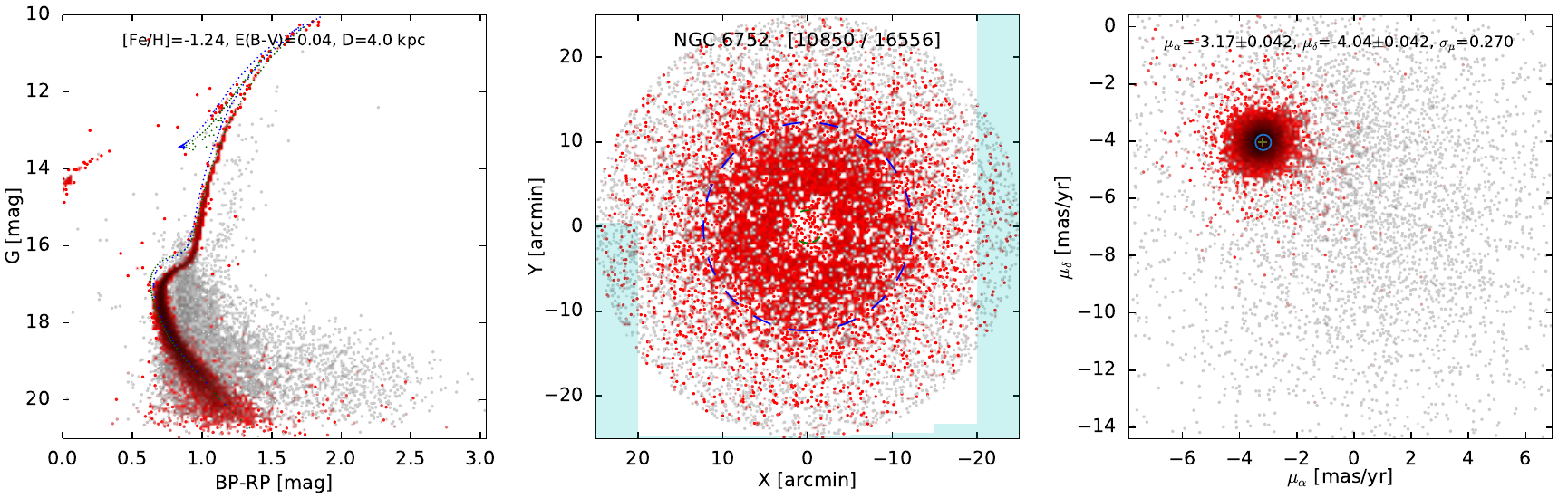}
\includegraphics{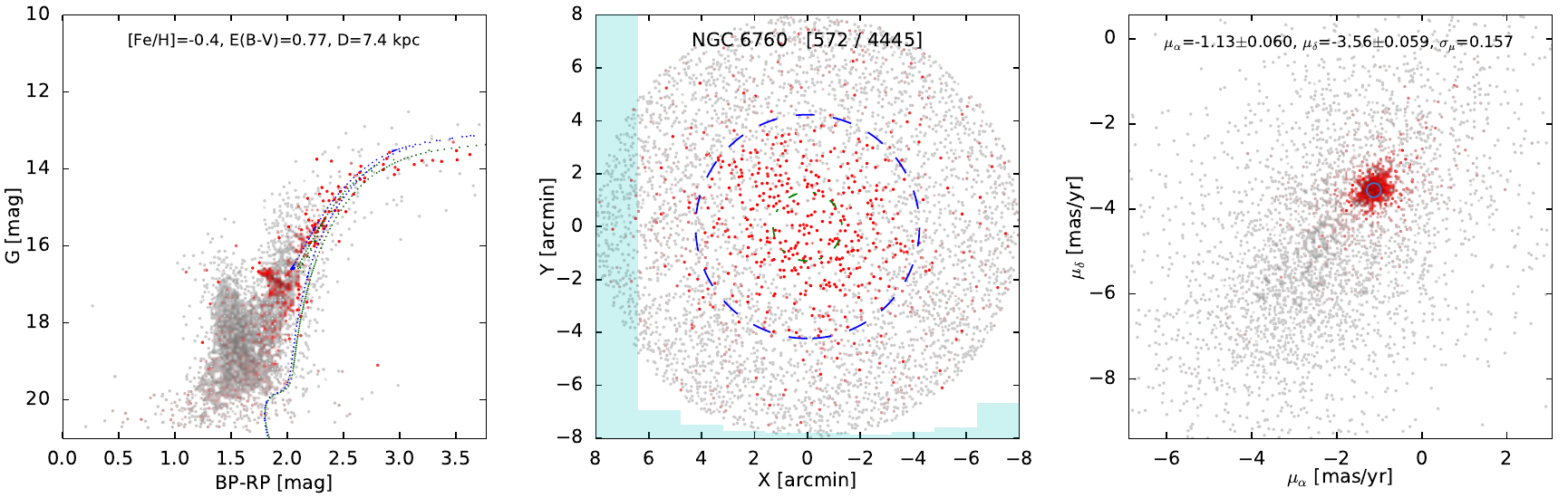}
\end{figure*}

\clearpage\begin{figure*}
\contcaption{}
\includegraphics{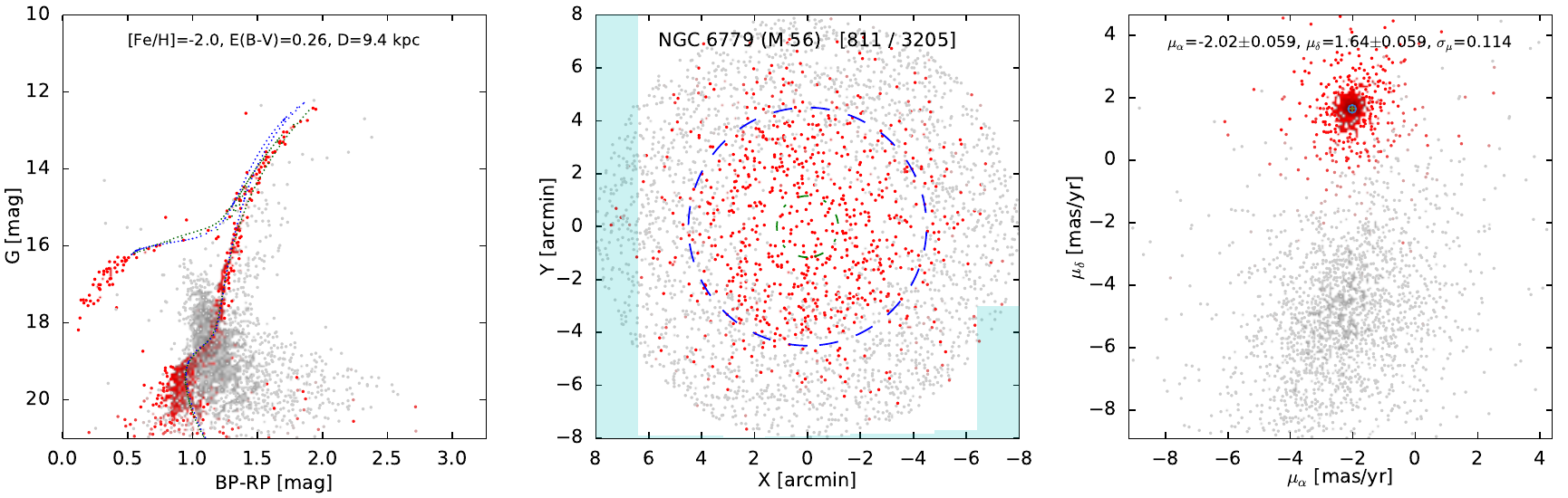}
\includegraphics{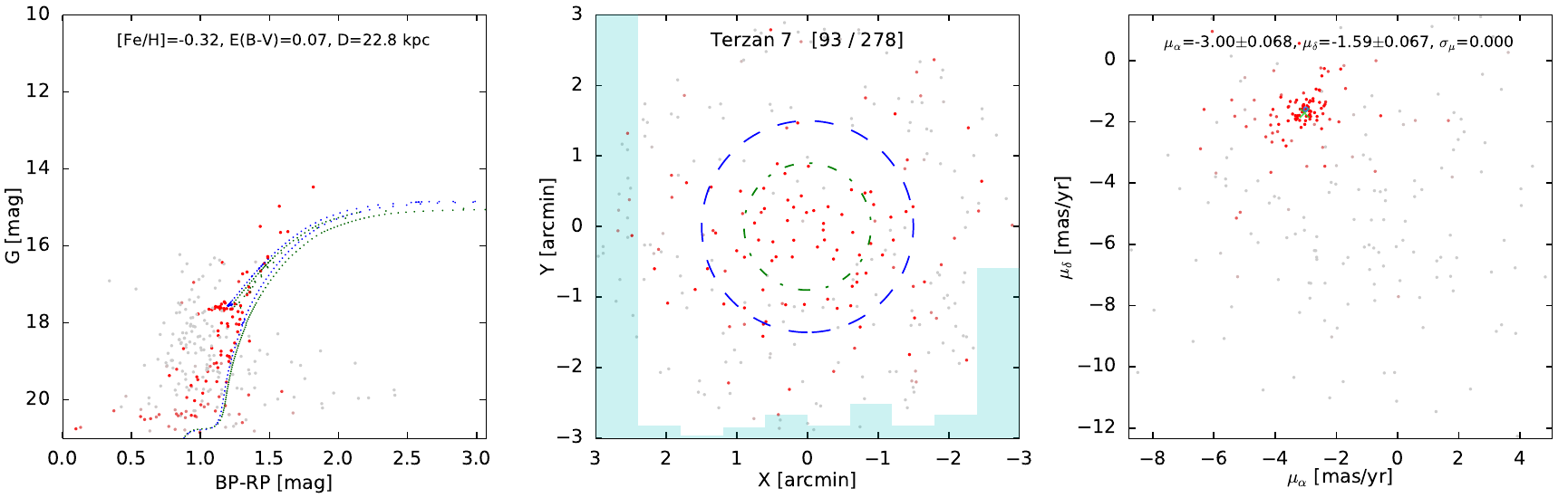}
\includegraphics{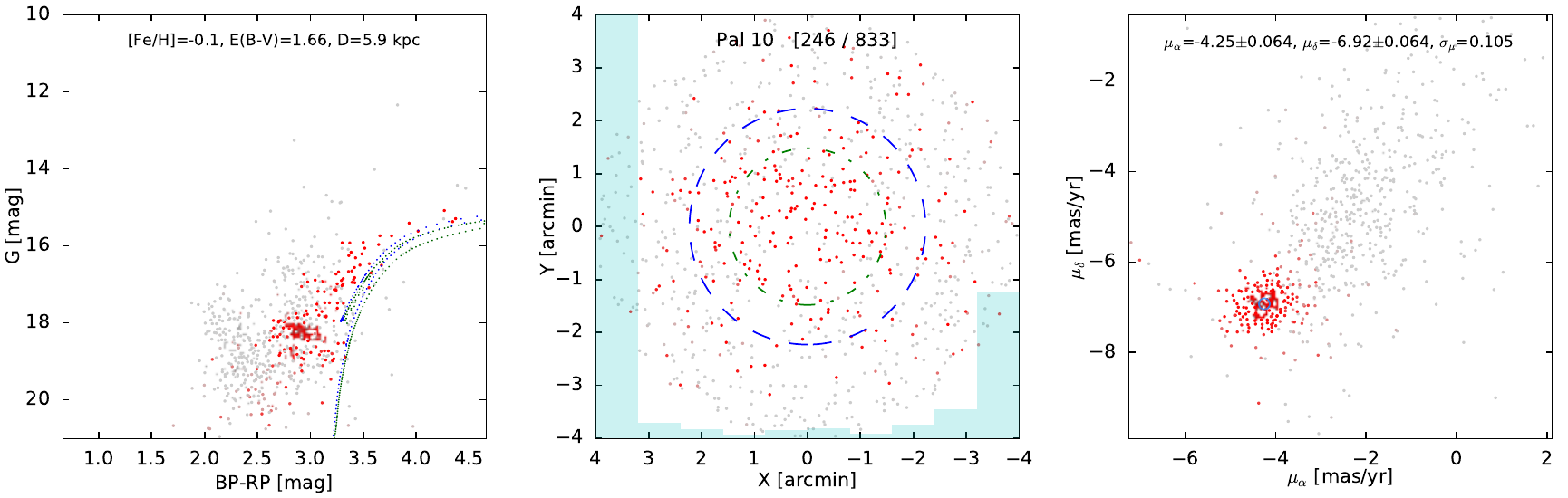}
\includegraphics{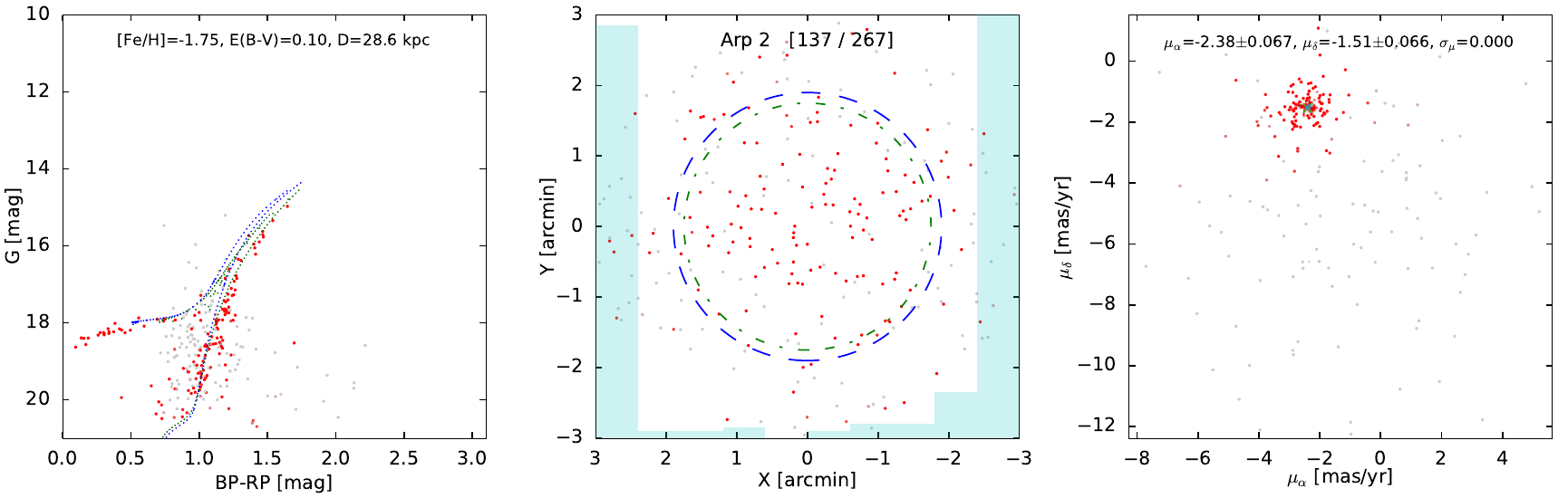}
\end{figure*}

\clearpage\begin{figure*}
\contcaption{}
\includegraphics{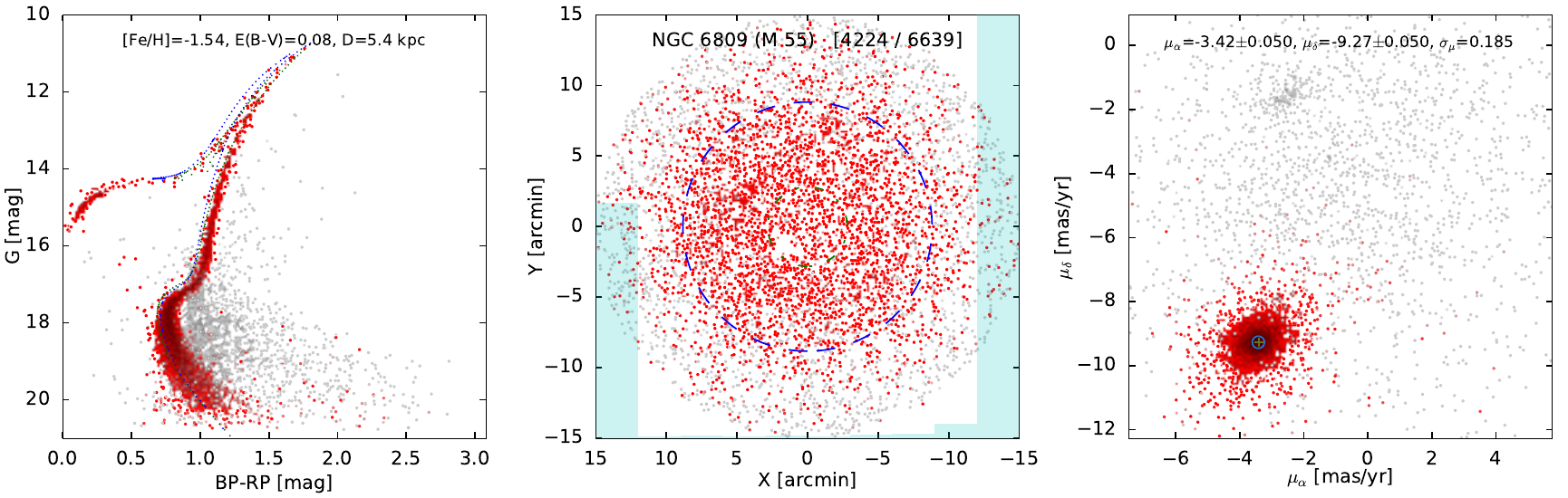}
\includegraphics{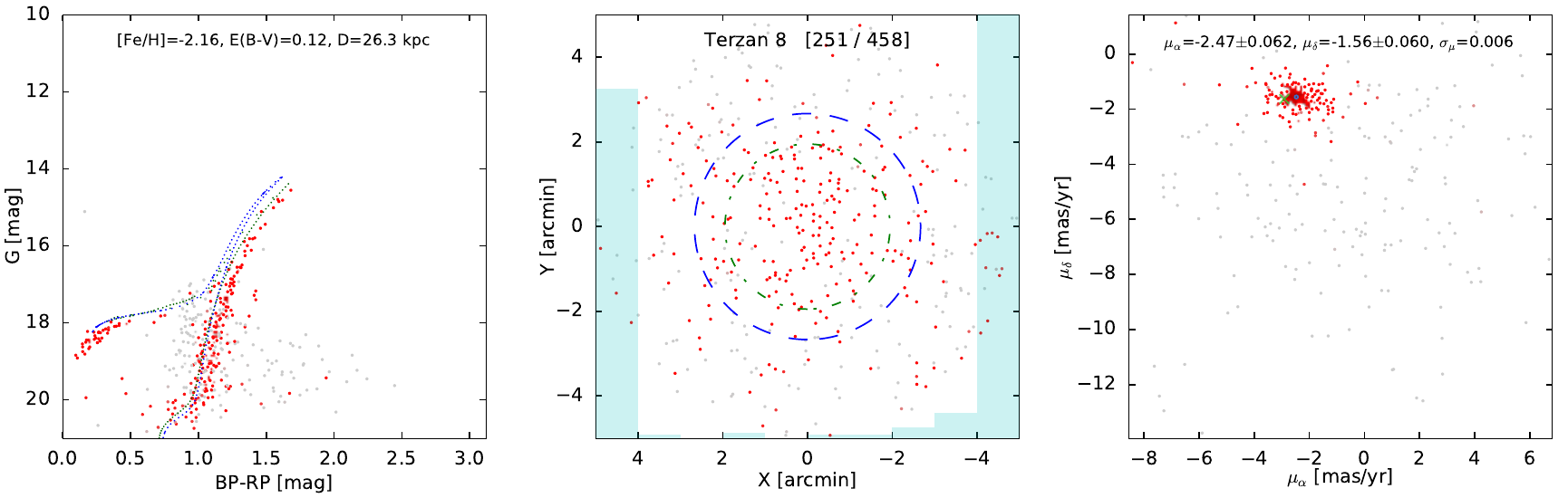}
\includegraphics{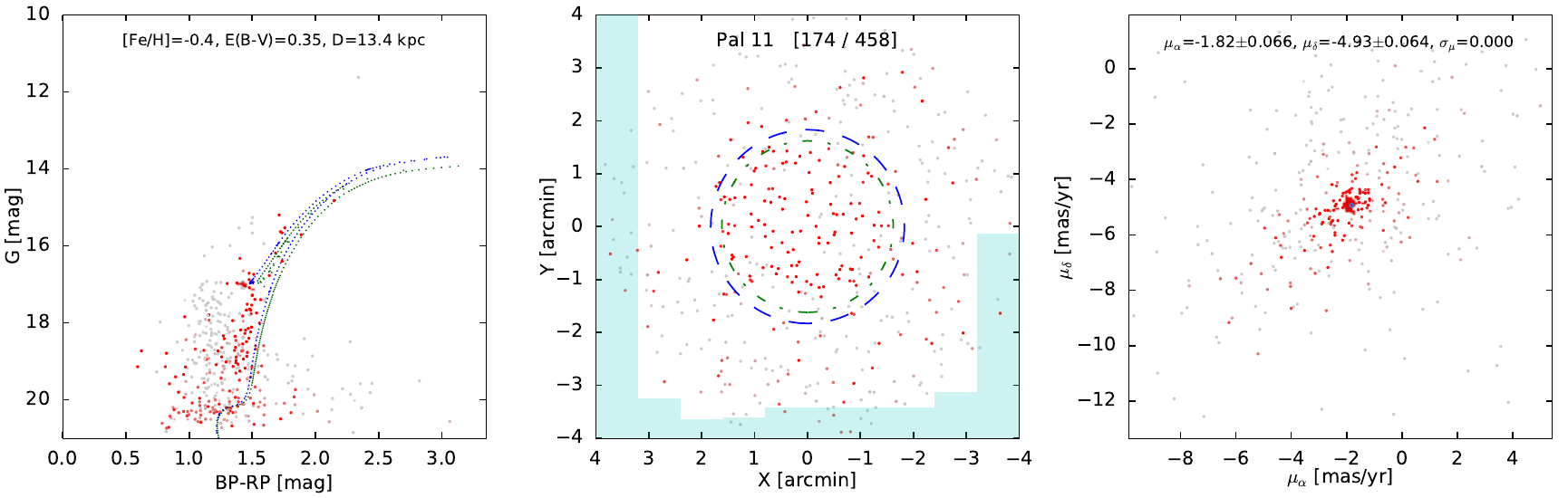}
\includegraphics{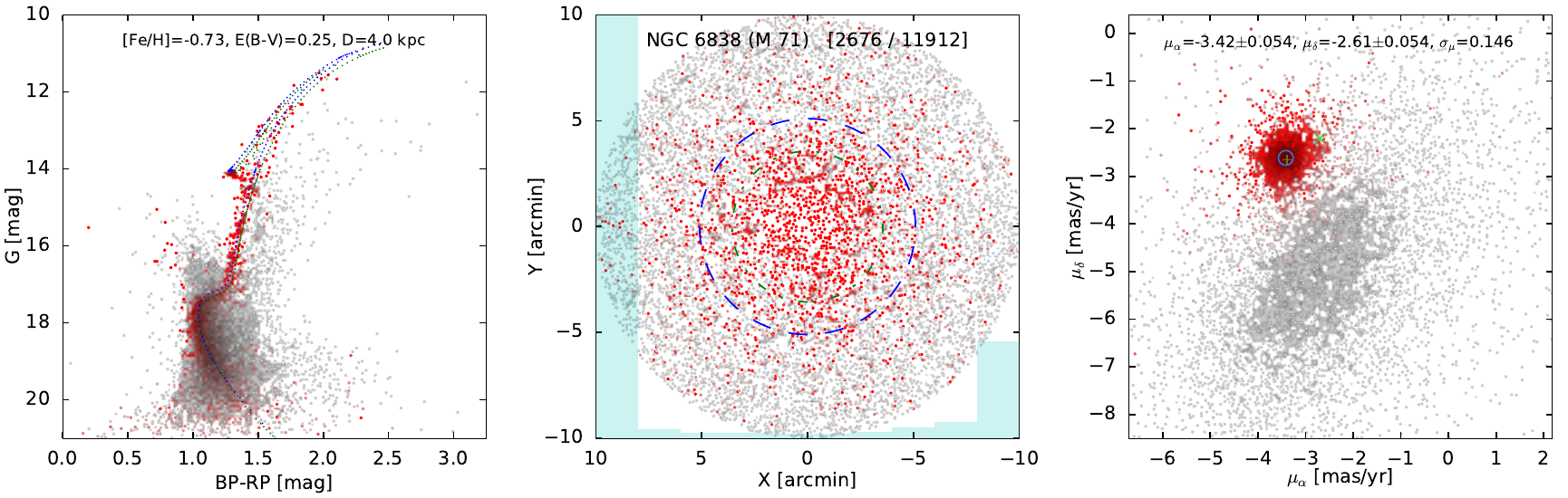}
\end{figure*}

\clearpage\begin{figure*}
\contcaption{}
\includegraphics{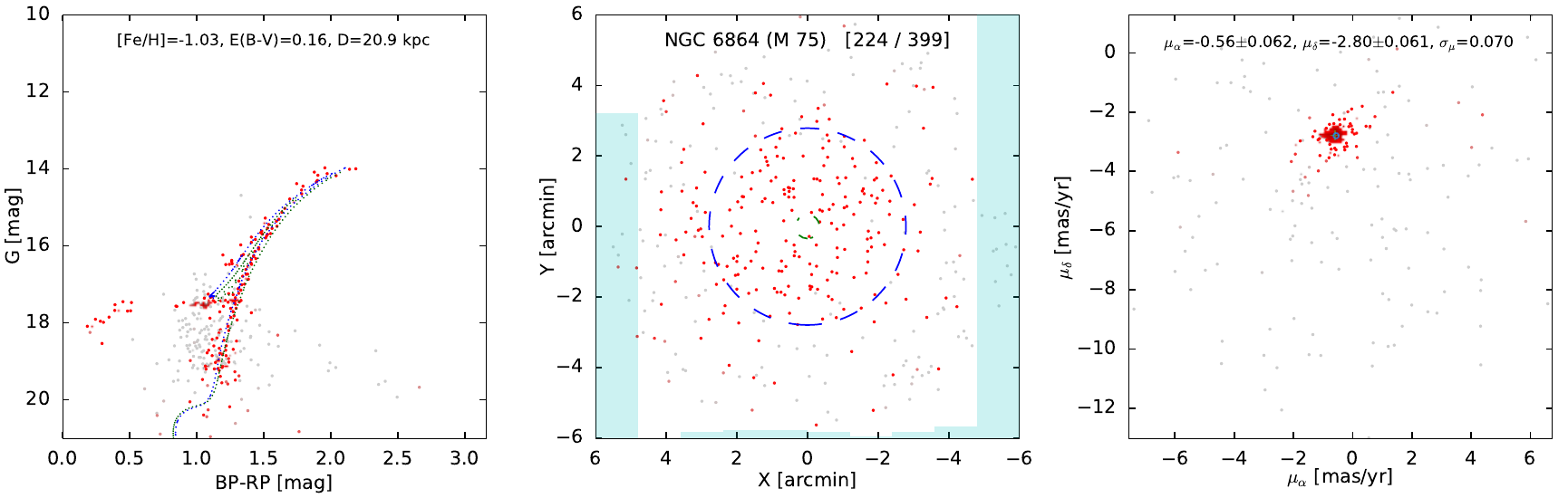}
\includegraphics{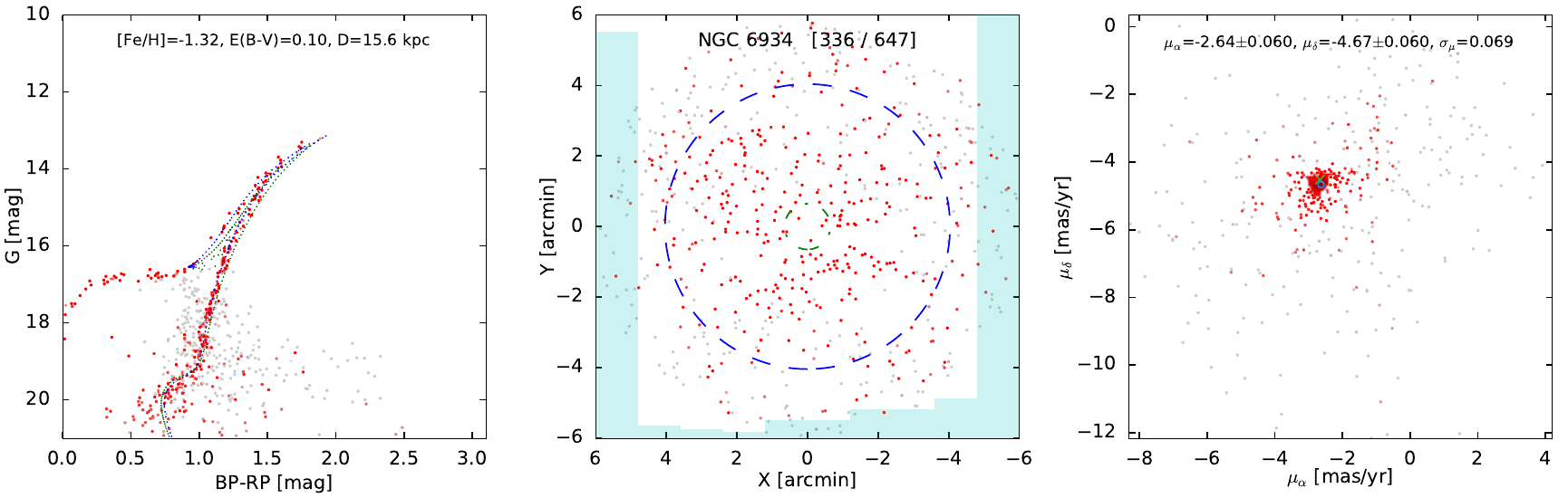}
\includegraphics{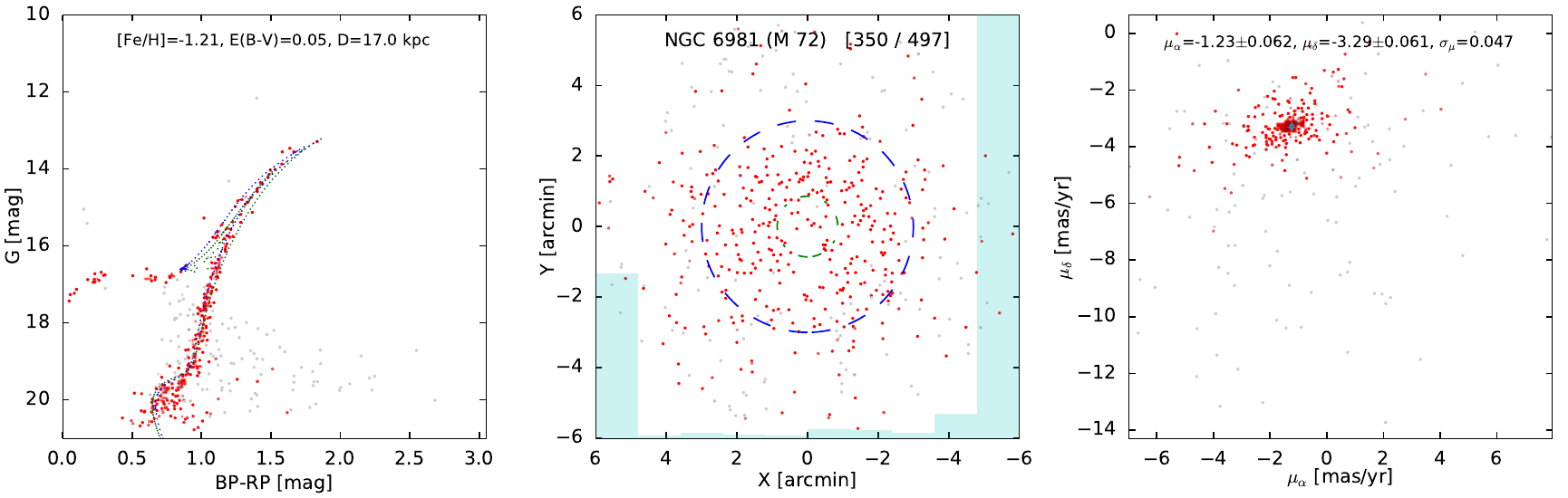}
\includegraphics{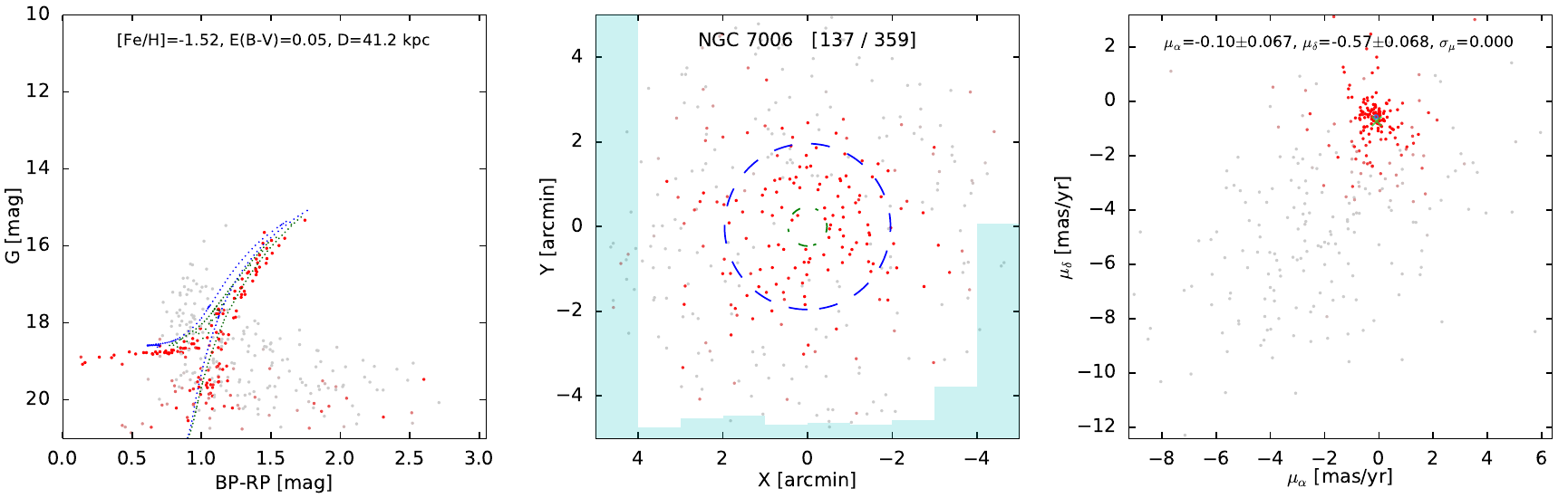}
\end{figure*}

\clearpage\begin{figure*}
\contcaption{}
\includegraphics{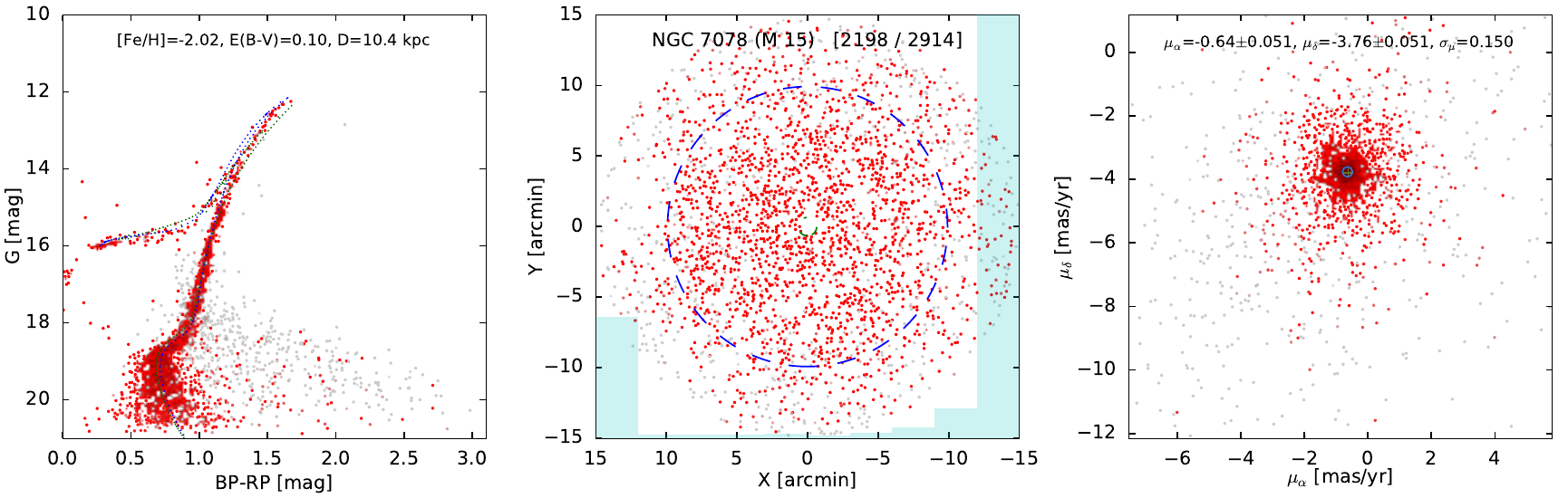}
\includegraphics{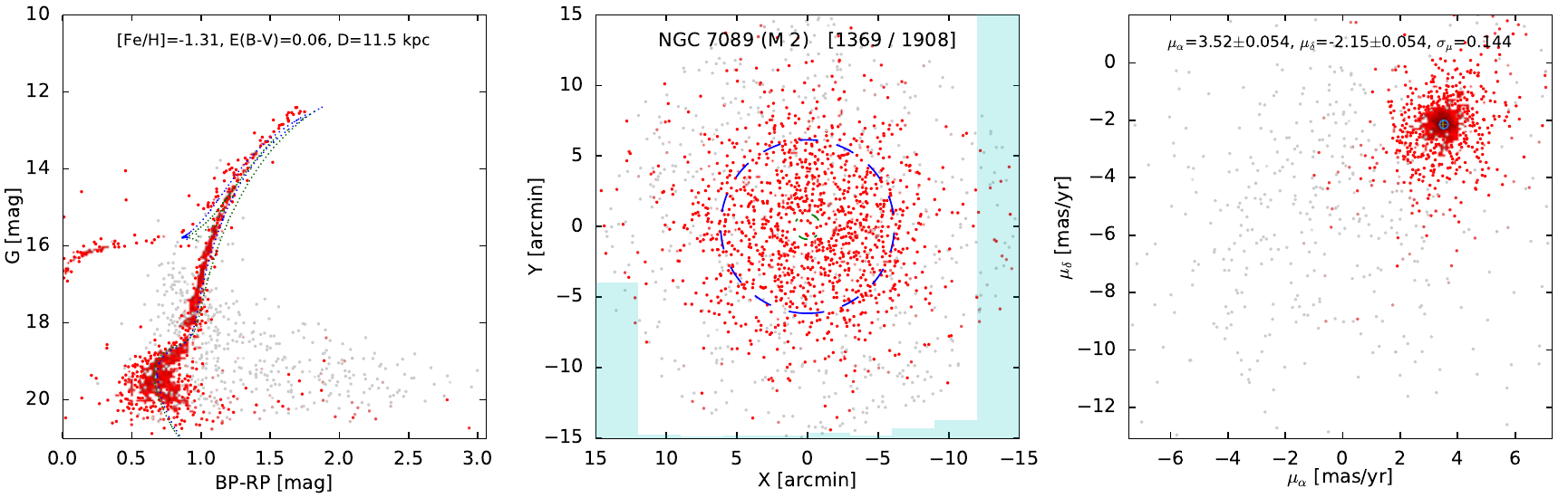}
\includegraphics{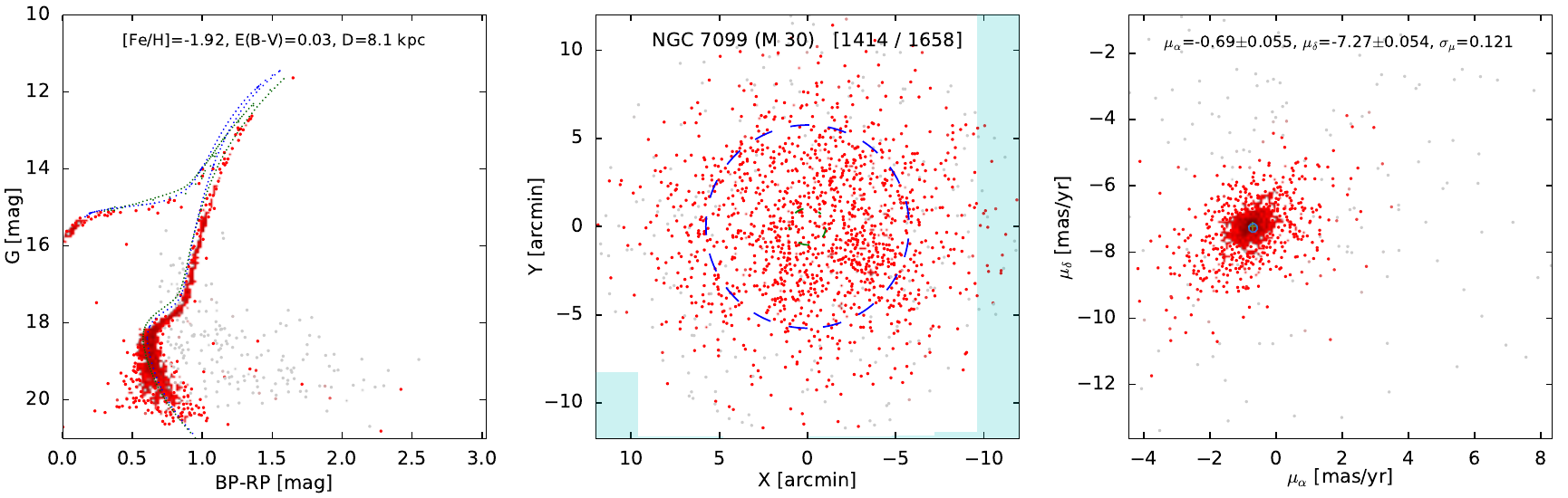}
\includegraphics{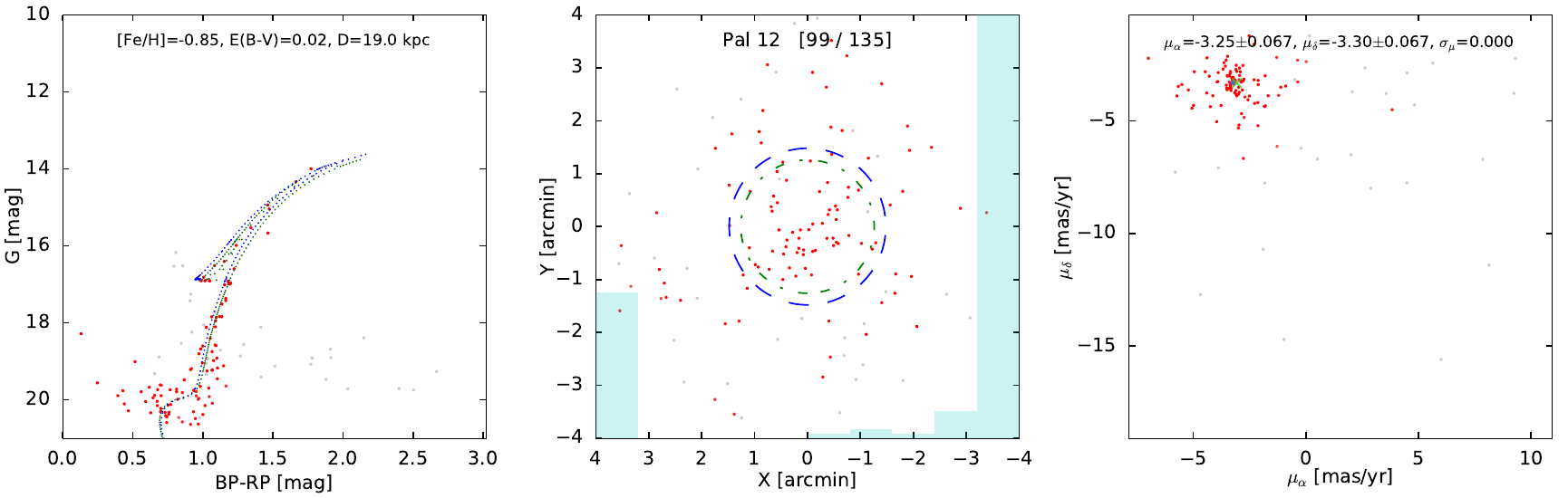}
\end{figure*}

\clearpage\begin{figure*}
\contcaption{}
\includegraphics{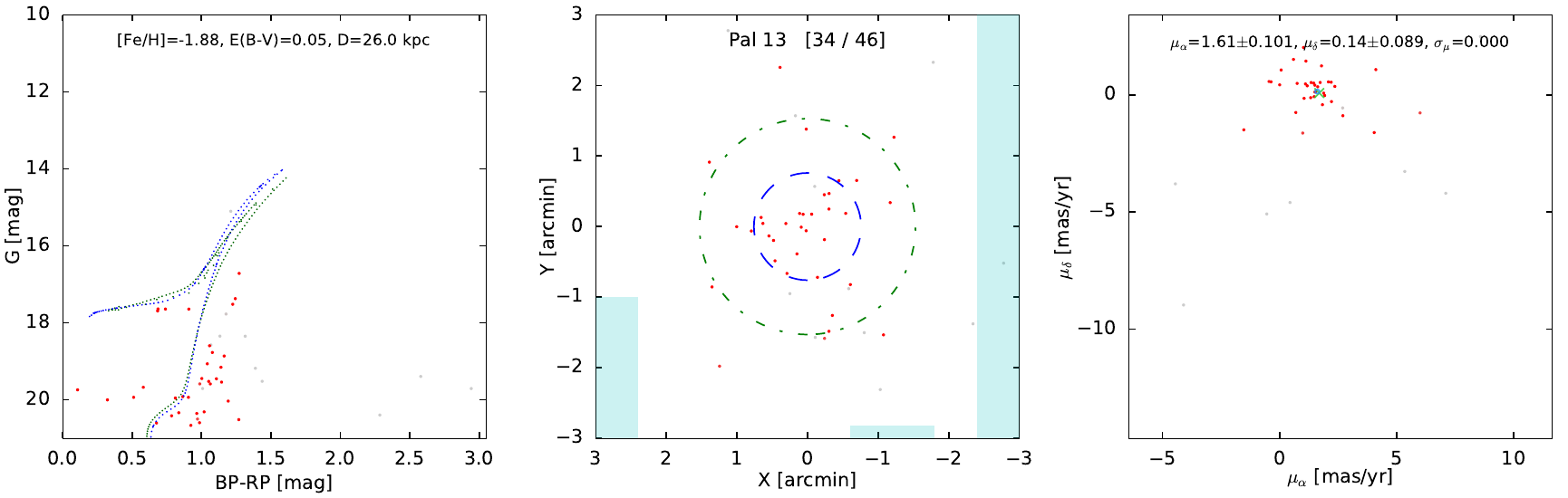}
\includegraphics{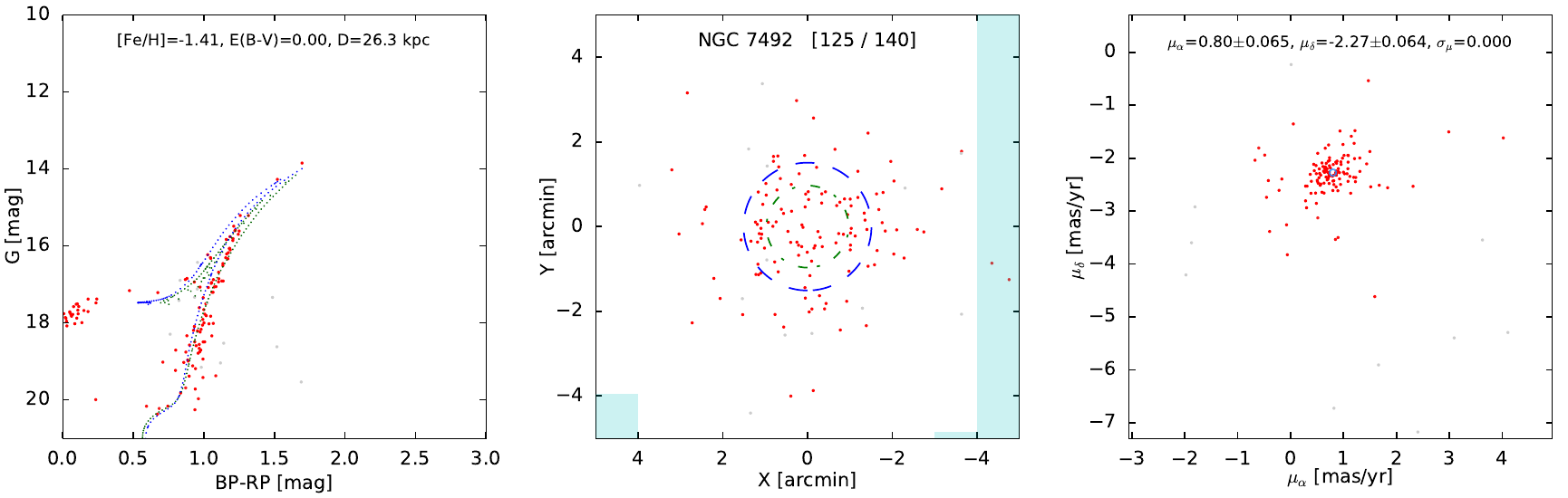}
\end{figure*}

\label{lastpage}
\end{document}